\documentclass[letter,11pt]{article}

\usepackage{jheppub}
\usepackage{hyperref}
\usepackage{graphicx}
\usepackage{amsmath}
\usepackage{amssymb}
\usepackage{xspace}
\usepackage{slashed}
\usepackage{subcaption}
\usepackage[normalem]{ulem}
\usepackage[dvipsnames]{xcolor}
\usepackage[utf8]{inputenc}
\usepackage[T1]{fontenc}
\usepackage{mathtools}
\usepackage{stmaryrd}
\usepackage{enumerate}
\usepackage{mathrsfs}

\newcommand{\eq}[1]{Eq.~\eqref{eq:#1}}
\newcommand{\eqs}[2]{Eqs.~\eqref{eq:#1} and \eqref{eq:#2}}
\renewcommand{\sec}[1]{Sec.~\ref{sec:#1}}
\newcommand{\tab}[1]{Tab.~\ref{tab:#1}}

\newcommand{\app}[1]{App.~\ref{app:#1}}
\newcommand{\fig}[1]{Fig.~\ref{fig:#1}}

\newcommand{\nocontentsline}[3]{}
\newcommand{\tocless}[2]{\bgroup\let\addcontentsline=\nocontentsline#1{#2}\egroup}

\newcommand{\df}{\mathrm{d}}

\def\veps{\varepsilon}
\newcommand{\vf}{\upsilon}

\newcommand{\ra}{\rightarrow}

\newcommand{\eps}{\epsilon}

\newcommand{\bn}{{\bar{n}}}

\newcommand{\nn}{\nonumber}

\newcommand{\Min}{\mathrm{min}}
\newcommand{\Max}{\mathrm{max}}

\newcommand{\cut}{\mathrm{cut}}
\renewcommand{\max}{\mathrm{max}}

\newcommand{\bea}{\begin{eqnarray}}
\newcommand{\eea}{\end{eqnarray}}

\newcommand{\C}{C\xspace}
\newcommand{\CS}{CS\xspace}

\newcommand{\CSm}{CS$_{\rm m}$\xspace}
\newcommand{\CSg}{CS$_{\rm g}$\xspace}

\newcommand{\SCETa}{\ensuremath{{\rm SCET}_{\rm I}}\xspace}

\newcommand{\zcut}{z_{\rm cut}}
\newcommand{\tzcut}{\tilde z_{\rm cut}}
\newcommand{\qcut}{Q_{\rm cut}}

\newcommand{\Pythia}{\textsc{Pythia}\xspace}
\newcommand{\Vincia}{\textsc{Vincia}\xspace}
\newcommand{\Herwig}{\textsc{Herwig}\xspace}
\newcommand{\Fastjet}{\textsc{FastJet}\xspace}
\newcommand{\cpp}{\textsc{c++}\xspace}
\newcommand{\SCETlib}{\textsc{SCETlib}\xspace}
\newcommand{\Mathematica}{\textsc{Mathematica}\xspace}
\newcommand{\PLHot}{\textsc{PLHot}\xspace}

\newcommand{\ee}{e^+e^-}

\def\ln{\textrm{ln}}
\def\df{\textrm{d}}
\def\nn{\nonumber}

\def\bndry{\varocircle}
\def\figeight{\circ\!\!\circ}

\def\Mb{M_{-1}^{\kappa\varocircle}}
\def\Mbq{M_{-1}^{q\varocircle}}

\allowdisplaybreaks[2]

\DeclareRobustCommand{\Ref}[1]{Ref.~\cite{#1}}
\DeclareRobustCommand{\Refs}[1]{Refs.~\cite{#1}}

\DeclareRobustCommand{\eq}[1]{Eq.~(\ref{eq:#1})}
\DeclareRobustCommand{\eqs}[2]{Eqs.~(\ref{eq:#1}) and (\ref{eq:#2})}

\usepackage[mathscr]{eucal}
\newcommand{\cS}{\mathscr{S}}

\preprint{\begin{flushright}
MIT--CTP 5235
\\ UWThPh 2020-31
\\ Nikhef 2020-031
\\ MAN/HEP/2020/011
\\ DESY 20-239
\end{flushright}}

\title{EFT for Soft Drop Double Differential Cross Section}

\author[a,b]{Aditya Pathak,}
\affiliation[a]{University of Vienna, Faculty of Physics, Boltzmanngasse 5, A-1090 Vienna, Austria}
\affiliation[b]{University of Manchester, School of Physics and Astronomy, Manchester, M13 9PL, United Kingdom}
\author[c]{Iain W.~Stewart,}
\author[c]{Varun Vaidya,}
\affiliation[c]{Center for Theoretical Physics, Massachusetts Institute of Technology, Cambridge, MA~02139, U.S.A.}
\author[d,e,f]{Lorenzo Zoppi}
\affiliation[d]{University of Amsterdam, Science Park 904, 1098 XH Amsterdam, The Netherlands}
\affiliation[e]{Nikhef, Theory Group, Science Park 105, 1098 XG Amsterdam, The Netherlands}
\affiliation[f]{Deutsches Elektronen-Synchrotron DESY, 22607 Hamburg, Germany}
\emailAdd{aditya.pathak@manchester.ac.uk}
\emailAdd{iains@mit.edu}
\emailAdd{vvaidya@mit.edu}
\emailAdd{lorenzo.zoppi@desy.de}

\abstract{
We develop a factorization framework to compute the double differential cross section in soft drop groomed jet mass and groomed jet radius. We describe the effective theories in the large, intermediate, and small groomed jet radius regions defined by the interplay of the jet mass and the groomed jet radius measurement. As an application we present the NLL$'$ results for the perturbative moments that are related to the coefficients $C_1$ and $C_2$ that specify the leading hadronization corrections up to three universal parameters. We compare our results with Monte Carlo simulations and a calculation using the coherent branching method.
}

\keywords{QCD, Factorization, Colliders}

\begin{document}
\maketitle

\section{Introduction}
\label{sec:Intro}

The physics of jets and their substructure has seen rapid development during the past decade. Originally designed for boosted object studies and pile-up mitigation at the LHC, the tools of jet grooming now play a key role in defining problems at the forefront of collider phenomenology~\cite{Altheimer:2013yza}. Recently, soft drop grooming~\cite{Larkoski:2014wba} (generalizing the modified mass drop algorithm~\cite{Butterworth:2008iy,Dasgupta:2013ihk}) has been widely studied, as the special nature of the algorithm enables a high precision description of measurement of infrared and collinear safe (IRC) observables in perturbative QCD. Furthermore, due to reduced sensitivity to nonperturbative effects of hadronization and contamination from the underlying event (UE), groomed jet observables are often more suited for LHC than their ungroomed counterparts. This has fueled a widespread interest in perturbative calculations of a variety of interesting jet based observables (see~\cite{Larkoski:2017jix} for a review) such as the groomed jet mass~\cite{Frye:2016aiz,Marzani:2017mva,Kang:2018jwa,Larkoski:2020wgx,Anderle:2020mxj}, groomed angularities~\cite{Kang:2018vgn}, soft drop thrust~\cite{Baron:2018nfz,Baron:2020xoi}, groomed multi-prong jet shapes~\cite{Larkoski:2017iuy}, groomed jet radius~\cite{Kang:2019prh}, energy drop~\cite{Cal:2020flh}, as well as calculations related to jets initiated by heavy quarks~\cite{Lee:2019lge,Hoang:2017kmk}.

Groomed jet observables have also found applications in the field of nonperturbative (NP) nuclear physics, such as in improving our understanding of hadron structure by accessing the NP initial state in the form of structure functions like the PDFs and the polarized/unpolarized TMDPDFs via scattering experiments such as DIS. This program entails that the observables chosen have small final-state nonperturbative effects, while at the same time be sensitive to the initial state hadronic physics. Groomed jets with an identified light/heavy hadron in the jet were proposed as probes of TMD evolution and distribution in \cite{Makris:2017arq,Makris:2018npl}. Other observables defined with jet grooming such as transverse momentum spectrum of groomed jets~\cite{Gutierrez-Reyes:2019msa} also meet this criteria. Soft drop jet observables have also recently been studied in the context of precision measurements at the LHC, such as the strong coupling constant $\alpha_s$~\cite{Marzani:2019evv} and the top mass~\cite{Hoang:2017kmk}. To obtain precision collider physics analyses using groomed observables, theoretical control and understanding of final state NP effects is as essential as an accurate description in the perturbative regime, since the NP effects can be as significant as higher order perturbative corrections.

Analyses of NP corrections for jet substructure have been carried out in Refs.~\cite{Dasgupta:2013ihk,Marzani:2017kqd,Frye:2016aiz,Hoang:2019ceu}.
In \Ref{Hoang:2019ceu} a field theory based formalism was developed to analyze nonperturbative (NP) corrections to soft drop jet mass due to hadronization. There the leading nonperturbative modes relevant for the largest hadronization correction to the groomed jet mass were identified, and were used to demarcate the regions modified by grooming in the jet mass spectrum into the soft drop nonperturbative region (SDNP), where the NP corrections are ${\cal O}(1)$, and the soft drop operator expansion region (SDOE), where perturbative contributions dominate, while NP corrections are small but still relevant for precision physics. It was shown that the leading nonperturbative corrections in the SDOE region are governed by the opening angle of the soft drop stopping pair, or the groomed jet radius $R_g$, at a given jet mass $m_J^2$, that determines the catchment area of the nonperturbative particles. This led to the observation that the dependence of the hadronization correction to the jet mass on kinematic parameters, namely the jet mass $m_J^2$ and jet energy $E_J$, and grooming parameters $\zcut$ and $\beta$, can be described in perturbation theory by specific moments of a multi-differential distribution involving the opening angle of the stopping pair. After this factorization of effects, one is then left with universal nonperturbative parameters that only depend on $\Lambda_{\rm QCD}$.

This appearance of the multi-differential groomed jet distributions motivates obtaining a more fundamental and precise description of these cross sections.
In recent years there has been significant progress on the analytic treatment of multi-differential\footnote{By multi-differential we refer to observables where the same set of particles in a given jet region is subjected to multiple measurements, and exclude cases that are essentially the overall kinematic information/property of the entire jet, such as jet rapidity or jet-$p_T$.} distributions. Some of the most significant recent advances~\cite{Larkoski:2014tva,Procura:2018zpn,Lustermans:2019plv} have been made using SCET~\cite{Bauer:2000ew,Bauer:2001ct,Bauer:2000yr,Bauer:2001yt,Bauer:2002nz}. In particular, double differential cross sections have allowed us to understand correlations between two different observables, such as simultaneous measurement of two angularities on a single jet~\cite{Larkoski:2014tva,Procura:2018zpn}. Doubly differential cross sections, such as 2 and 3-point energy correlation functions have been used to calculate the distribution of the groomed $D_2$ at NNLL accuracy~\cite{Larkoski:2015kga,Larkoski:2017cqq}, and to develop novel formalism for non-global logarithm resummation~\cite{Larkoski:2015zka}. In \Ref{Lustermans:2019plv} a joint resummation for 0-jettiness and total color singlet transverse momentum $q_T$ was performed at NNLL accuracy, where the novelty of this work lies in the fact that the two observables have different sensitivity to the recoil effects, leading to different logarithmic structures that are simultaneously resummed.

In \Ref{Hoang:2019ceu} two kinds of hadronization effects were identified that are specific to the groomed jet mass in the SDOE region: ``shift'' and ``boundary'' corrections. The shift correction results from contribution of the NP particles that survive the jet grooming, whereas the boundary correction captures the effect of modification of the soft drop test due to hadronization. Both effects were shown to be directly proportional to the parton level moments of the relative angle $\theta_{g}$ of the stopping pair, which is equivalent to the groomed jet radius (often called $R_g$ in the literature). We reserve the notation $\theta_{g}$ to refer to this opening angle and will use $R_g$ for the cumulative version of this variable. This SDOE region is dominated by resummation, and the modes with collinear-soft scaling responsible for setting $\theta_g$ play a crucial role in summing the logarithms, as well as in determining the hadronization corrections.

The SDOE region~\cite{Hoang:2019ceu}, where the parton level geometry determines catchment area of the NP particles, corresponds to jet mass values satisfying
\begin{align}\label{eq:SDOE}
\frac{Q \Lambda_{\rm QCD}}{m_J^2}
\Bigl( \frac{m_J^2}{Q Q_{\rm cut}}\Bigr)^{\frac{1}{2+\beta}}
\ll 1
\,,
\end{align}
where $\qcut = 2^\beta Q \zcut$ for hemisphere jets in $e^+e^-$ collisions and is defined in \eq{qcutDef} below for generic jet radius $R$ and for $pp$ collisions. $Q$ stands for the large momentum such that $Q = 2E_J$ for the $e^+e^-$ case and $Q = 2 p_T \cosh(\eta_J)$ for $pp$ collisions. $E_J$ is the energy of the groomed jet, $m_J$ is the measured jet mass, and $z_{\rm cut}, \beta$ are the Soft-Drop grooming parameters.
QCD perturbation theory implies we can write the cross section in terms of jets initiated by a light quark or gluon
\begin{align}\label{eq:dSigHad}
\frac{d \sigma^{\rm had}}{d m_J^2 d\Phi_J}
&= \sum_{\kappa=q,g} N_\kappa (\Phi_J, R, \zcut, \beta)
\frac{d \sigma^{\rm had}_\kappa}{d m_J^2d\Phi_J}
\,,
\end{align}
where $N_\kappa$ includes virtual corrections important for the normalization, as well as contributions from the rest of the event (such as parton distributions in the $pp$ case), and determines the relative contribution from quark and gluon jets.
It depends on the jet radius, $R$, the phase space variables of the jet denoted by $\Phi_J$, as well as the soft drop parameters, since it accounts for radiation that has been groomed away.
Equation~(\ref{eq:dSigHad}) applies for $\ee$ collisions in the dijet limit, where the phase space variable is simply the energy, $\Phi_J=\{E_J\}$, as well as for an inclusive jet with small $R$ in a $pp$ collision, where the phase space variables are transverse momentum and pseudorapidity, $\Phi_J=\{p_T,\eta_J\}$.

The leading hadronization corrections in $d\sigma^{\rm had}_\kappa/dm_J^2$ take the following form:
\begin{align}
\label{eq:sigfullk}
\frac{d \sigma^{\rm had}_\kappa}{d m_J^2d\Phi_J}
&= \frac{d \hat \sigma^\kappa}{d m_J^2d\Phi_J} - Q\, \Omega_{1\kappa}^{\figeight} \, \frac{d}{d m_J^2} \bigg(C^\kappa_1(m_J^2, Q, \zcut, \beta, R) \, \frac{d \hat \sigma^\kappa}{d m_J^2d\Phi_J}
\bigg) \\
&\qquad + \frac{Q\big(\Upsilon_{1,0}^\kappa + \beta \,\Upsilon_{1,1}^\kappa\big)}{m_J^2} \, C^\kappa_2(m_J^2, Q, \, \zcut, \beta, R) \, \frac{d \hat \sigma^\kappa}{d m_J^2d\Phi_J}
+ \ldots
\,, \nn
\end{align}
where $\hat \sigma_k$ is the partonic cross section. The three universal hadronic parameters $\Omega_{1\kappa}^{\figeight}$, $\Upsilon_{1,0}^\kappa$, and $\Upsilon_{1,1}^\kappa$ for $\kappa = q,g$ have dimensions of energy, are ${\cal O}(\Lambda_{\rm QCD})$, and encode the
nonperturbative information in the power corrections. In contrast, the dimensionless coefficients $C_{1,2}(m_J^2, Q, \zcut, \beta, R)$ determine the perturbative prefactors in the SDOE region, while also describing the dependence on the kinematic and grooming parameters.
A key point to note is that the factorization for power correction terms in \eq{sigfullk} was derived so far only at leading logarithmic (LL) accuracy~\cite{Hoang:2019ceu}.
The LL nature of the analysis allowed for the use of strong angular ordering to prove factorization of the nonperturbative matrix elements. Thus, while the partonic cross section $d\hat \sigma^\kappa$ can be improved independently order-by-order in perturbation theory, the factorization of the leading power corrections in \eq{sigfullk} beyond LL are likely to involve additional NP parameters with new perturbative coefficients,
which we indicate by the `$\ldots$' in \eq{sigfullk}.
By definition we include in $C_1$ and $C_2$ all terms beyond LL that are still proportional to the same combination of the 3 hadronic parameters shown in \eq{sigfullk}.

At LL accuracy, the Wilson coefficients $C_{1,2}^\kappa(m_J^2)$ are related to moments of the $\theta_{g}$ distribution for fixed $m_J$:
\begin{align}
\label{eq:C1C2def}
C_1^\kappa(m_J^2) & =
\frac{1}{\langle 1 \rangle (m_J^2)}\left \langle \frac{\theta_{g}}{2}\right \rangle \, ,
\qquad
C_2^\kappa(m_J^2) =
\frac{m_J^2/Q^2}{\langle 1 \rangle (m_J^2)}
\left\langle\frac{2}{\theta_{g}}\, \delta \big ( z_{g} - \zcut \theta_{g}^\beta \big) \right\rangle
\, ,
\end{align}
where the angle brackets represent averages that take into account the resummation of the large logarithms in the SDOE region. We have, for simplicity, suppressed the dependence on arguments other than $m_J^2$. In the second term the delta function ensures that the correction is only relevant for kinematic configurations that lie on the boundary of soft drop failing or passing test.
A calculation at LL accuracy of \eq{sigfullk} and the functions $C^\kappa_1(m_J^2)$ and $C^\kappa_2(m_J^2)$ in the coherent branching formalism was presented in \Ref{Hoang:2019ceu}. The LL results for these functions were shown to be in a reasonable agreement with the parton shower Monte Carlo (MC) simulations. From the point of view of phenomenological relevance, the terms appearing in the LL treatment are likely to suffice to capture the bulk of the NP corrections in the SDOE region, with higher order NP effects being below the level of perturbative uncertainty in parton-level predictions for groomed cross sections. However, since the coefficients $C_1^\kappa$ and $C_2^\kappa$ are defined such that they do not contain leading double logarithmic terms, $\sim (\alpha_s \ln^2)^k$, they are so far not known with even a leading treatment of resummation effects. This requires an analysis beyond LL order, which is a main goal of our work here. At the same time we intend to provide a more realistic assessment of perturbative uncertainties from higher order effects.

In \Ref{Kang:2019prh} results were presented for the cumulative soft drop $R_g$ distribution at NLL accuracy in SCET, including resummation of non-global logarithms due to CA clustering effects.
The factorization for groomed jet mass was developed and studied at NNLL accuracy in \Refs{Frye:2016aiz,Marzani:2017mva,Kang:2018jwa} and N$^3$LL accuracy for $\beta = 0$ in \Ref{Kardos:2020gty}.
In this paper we build upon these results and describe the joint resummation of the soft drop cross section differential in the jet mass and cumulative in $R_g$ at NLL $+$ NLO accuracy in the SDOE region.

Our analysis will involve so-called non-global large logarithms~\cite{Dasgupta:2001sh,Banfi:2002hw} that occur due to correlations between emissions across boundaries in phase space, and start at ${\cal O}(\alpha_s^2\ln^2 X)$, with $X$ involving a ratio of scales associated to the measurements or specifying the phase space boundaries. In our case a boundary is introduced by measurement of $\theta_g$ (as well as the jet boundary $R$).
Another type of logarithm can be introduced by the jet clustering algorithm~\cite{Delenda:2006nf} such as the C/A used in the definition of the soft drop reclustering. These are referred to as abelian clustering logarithms, and are induced by uncorrelated soft emissions close to the jet boundary.
Various techniques have been introduced to compute the non-global effects~\cite{Dasgupta:2001sh,Banfi:2002hw,Banfi:2010pa,Becher:2015hka,Larkoski:2015zka} and abelian clustering effects~\cite{KhelifaKerfa:2011zu,Delenda:2012mm,Dasgupta:2012hg,Kelley:2012kj,Kelley:2012zs}. For our analysis we will follow Ref.~\cite{Kang:2019prh}, where both effects were studied in detail and accounted for in the cross section single differential in groomed jet radius. We will find a very similar pattern on non-global effects and clustering effects for the double differential distribution considered here. We include them in the calculation of $C_1^\kappa$ and $C_2^\kappa$ below, though their contribution turns out to be relatively small.

As a first application of our double differential cross section framework, we compute the following partonic moments that are related to $C_{1,2}^\kappa(m_J^2)$:
\begin{align} \label{eq:C1C2multiDiffDef}
M_1^\kappa(m_J^2, \Phi_J, \zcut, \beta, R) &\equiv \Big(\frac{d\hat \sigma^\kappa}{dm_J^2d\Phi_J}\Big)^{-1}\!\!\int \!d \theta_{g} \frac{\theta_{g}}{2} \frac{d^2 \hat \sigma^\kappa}{d m_J^2 d \theta_{g}d\Phi_J}\,,
\\
\Mb(m_J^2, \Phi_J, \zcut, \beta, R) &\equiv
\Big(N_\kappa (\Phi_J, R, \zcut, \beta) \frac{d\hat \sigma^\kappa}{dm_J^2d\Phi_J}\Big)^{-1}
\nn
\\
&\qquad\times \int d\theta_{g}
\frac{m_J^2}{Q^2} \frac{2}{\theta_{g}}
\frac{d}{d \veps}
\Bigg[ N_\kappa (\Phi_J, R, \zcut, \beta, \veps) \frac{d^2 \hat \sigma^\kappa(\veps)}{d m_J^2 d \theta_{g}d\Phi_J} \Big|_{\theta_g \sim \theta_g^\star }\Bigg]
\bigg|_{\veps\to 0} \, .
\nn
\end{align}
The normalization factor $N_\kappa$ introduced in \eq{dSigHad} remains unchanged for the doubly differential distribution and hence drops out in the ratio defining the $M_1^\kappa(m_J^2)$ moment for the channel $\kappa$.
In the distribution $d\hat \sigma^\kappa(\veps)$ the soft drop test has been shifted by a small $\veps$ to implement the $\delta$ function in \eq{C1C2def}, which then also impacts the normalization $N_\kappa$.
For a single emission the modification for the Soft-Drop condition reads
\begin{align}\label{eq:ShiftSD}
\overline \Theta_{\rm sd} = \Theta (z - \zcut \theta_g^\beta) \rightarrow\overline \Theta_{\rm sd}(\veps) = \Theta( z - \zcut \theta_g^\beta + \veps) \,,
\end{align}
with the obvious generalization to the case of more emissions.
This modification preserves the IRC properties of the groomer for $\veps\to 0$.
The superscript `$\bndry$' in $\Mb(m_J^2)$ signifies that the double differential distribution is evaluated at the ``boundary'' of the soft drop constraint, which is implemented via this $\veps$-derivative.

From \eq{C1C2def} we have $C_1=M_1^\kappa$ and $C_2=\Mb$ at LL order.
We have used the notation for the moments displayed in \eq{C1C2multiDiffDef} to stress that these functions can be defined independently of their interpretation at LL accuracy as the Wilson coefficients $C_{1,2}$.
Furthermore, higher order resummed results for the corresponding double-differential cross sections immediately translate into perturbative predictions for $M_1^\kappa$ and $\Mb$ at the same accuracy.
At NLL order and beyond we anticipate that a significant portion of the higher order results for $C_{1,2}$ will still be captured by the moments $M_1^\kappa$ and $\Mb$, such that $C_1\simeq M_1^\kappa$ and $C_2\simeq \Mb$.
In the subscript $\theta_g \sim \theta_g^\star$ in the second line of \eq{C1C2multiDiffDef}, the angle $\theta_g^\star$, defined below in \eq{angularBoundsExplicit}, refers to the largest kinematically allowed groomed jet radius for a given measured jet mass value. The subscript signifies that the cross section is calculated only in the large $\theta_g$ region where the corrections related to $\theta_g$ measurement can be treated in the fixed-order perturbation theory, which is the most relevant region for $C_2$. There are other regions of the groomed jet radius spectrum that we will discuss, where this is not the case. We will show that carrying out calculations of these moments via the double differential cross section offers us unique insights as to which contributions to the cross section lead to deviations from the LL definitions of the Wilson coefficients in \eq{C1C2def}.
By identifying the relevant effective theories, we will see that these contributions come precisely from the regions of phase space that break the strong-ordering assumption of the partonic radiation.
Restricting ourselves to the large $\theta_g$ region allows us to determine the corrections that are consistent with the geometry which is a key part of the definition of the parameters $\Omega_{1\kappa}^{\figeight}$, $\Upsilon_{1,0}^\kappa$, and $\Upsilon_{1,1}^\kappa$ and thus the terms needed to determine higher order corrections to the coefficients $C_1$ and $C_2$.

Following the strategy formulated in \Ref{Kang:2019prh}, we preferentially work with the cumulant of the $R_g$ distribution:
\begin{align} \label{eq:dbleDiff}
\frac{d\Sigma^\kappa (R_g)}{dm_J^2d\Phi_J } = \int_0^{R_g} d\theta_{g} \: \frac{d^2\hat \sigma^\kappa}{d\theta_{g}dm_J^2d\Phi_J}
\, .
\end{align}
We will carry out the analysis for the cumulative cross-section $d\Sigma^\kappa (R_g)$ at next-to-leading-log-prime (NLL$'$) accuracy, where in accordance with the standard convention, the NLL resummation is supplemented by including all ${\cal O}(\alpha_s)$ terms in the factorization formula.
Note, however, that for the observable we are considering the leading non-trivial dependence on $R_g$ enters at ${\cal O}(\alpha_s)$.
This is because at least one emission off the energetic jet initiating parton is required to set a non-zero value of the groomed jet radius.
We will come across scenarios where this $R_g$ dependent contribution
appears in fixed-order perturbative terms.
In order to consistently implement resummation in these cases we will include additional ${\cal O}(\alpha_s^2)$ cross terms resulting from combining the aforementioned ${\cal O}(\alpha_s)$ fixed-order corrections with terms needed for a reasonable definition of NLL resummation.
In other cases, where the kinematic hierarchies allow for the $R_g$ dependence to be factorized (thus facilitating resummation of logarithms of $R_g$) the standard prescription for NLL$'$ resummation applies.

The outline for the rest of the paper is as follows: In \sec{kinematics} we discuss the kinematic constraints on the joint $R_g$ and $m_J$ distribution and the relevant effective theory modes that enter the analysis.
The EFT analysis identifies three different regions of resummation.
We present the factorization theorems for each of these regions in \sec{factAll} and discuss the calculation of the moments $M_1^\kappa$ and $\Mb$.
In \sec{resum} we describe the resummation in each of the three regimes, as well as the resummation of boundary corrections to the cumulative cross section upon shifting the soft drop condition as in \eq{ShiftSD}. This section also compiles the key formulae for the factorized cross sections in the
large, intermediate, and small $R_g$ regimes,
needed for the $C_1^\kappa$ and $C_2^\kappa$ calculations.
The various factorized cross sections are then combined in \sec{match} to arrive at the final result for the cumulative cross section that smoothly interpolates across these three regimes.
We will base the discussion on results for $\ee$ collisions and state appropriate generalizations for the $pp$ case at the end of each section.
In \sec{Num} we employ the results for matched $e^+e^-$ cross sections
to calculate and perform a numerical analysis of the moments
$C_1^q \simeq M_1^q $ and $C_2^q \simeq M_{-1}^{q\bndry} $, and compare the results with parton shower Monte Carlos and earlier coherent branching results.
Numerical results for $pp$ collisions are left to future work.
We then conclude in \sec{conclusion}.
Various more technical results and derivations are discussed in appendices.
\section{Kinematic constraints on the $R_g$ distribution and mode analysis}
\label{sec:kinematics}

In this section we discuss how the kinematic constraints imposed by the groomed jet mass measurement and the grooming condition result in the modes relevant for our effective field theory description of the double differential cross section.
We will state formulae for both $\ee$ and $pp$ collisions which have a similar structure. For $pp$ collisions we will limit ourselves to the case of inclusive jet measurement, where a single jet with a small jet radius $R/2 \ll 1$ is groomed and measured. In the case of $\ee$ collisions, we will, however, allow ourselves to consider large jet radius $R \sim 1$ with $R \leq \pi/2$.

In the following, we first briefly review the soft drop algorithm and set up the notation. We then proceed to discuss the various kinematic regimes for $R_g$ within constraints induced by jet mass measurement.
First off, we note that the exact meaning of the jet radius $R$ depends on whether we are considering $\ee$ or $pp$ collisions. In this paper we consider the $k_T$-type class of clustering algorithms. In an $\ee$ collider
the relevant distance measure $d_{ij}$ is proportional to $[1 - \cos(\theta_{ij})]/(1-\cos R)$, where $\theta_{ij}$ is the polar angle between the two particles $i$ and $j$.
On the other hand, in a hadron collider, the relevant distance measure $d_{ij}$ between two particles $i$ and $j$ is proportional to $\Delta R_{ij}^2 = 2(\cosh(\eta_i - \eta_j) - \cos(\phi_i - \phi_j)) \simeq (\eta_i - \eta_j)^2 + (\phi_i - \phi_j)^2$. For small jet radius $R \ll 1$ in $pp$ collisions, we can approximate $\Delta R_{ij} \simeq \theta_{ij} \cosh \eta_J$ and the pseudorapidities will be close, $\eta_i \sim \eta_j \sim \eta_J$. This then leads to a difference of a factor of $\cosh \eta_J$ in the two distance measures, and the polar angles for $pp$ are $\theta_{ij}\simeq \Delta R_{ij}/\cosh \eta_J$.

The soft drop procedure works as follows. First one reclusters the jet using the Cambridge/Aachen (C/A) algorithm~\cite{Dokshitzer:1997in,Wobisch:1998wt}, and navigates the clustering history backwards, testing at each node the condition\footnote{Here we do not explore the possibility to use soft drop as a tagger, and we will restrict our analysis to $\beta \geq 0$.}
\begin{align}\label{eq:SD}
&\frac{\min(E_i,E_j)}{E_i+E_j} > z_\cut \bigg( \sqrt{2}\frac{\sin(\theta_{ij}/2)}{\sin(R_0^{ee}/2)}\bigg)^\beta\,, &
&
(\textrm{$e^+e^-$ case})\, ,&
\\
&\frac{\min(p_{T_i},p_{T_j})}{p_{T_i}+p_{T_j}} > z_\cut \Big(\frac{\Delta R_{ij}}{R_0}\Big)^\beta\,, &
&
(\textrm{$pp$ case})\, ,&
\nn
\end{align}
where the parameters $R_0^{ee}$ and $R_0$ normalize the angular weight for the two scenarios.
So long as the tests keep failing, branches with lower energies or $p_T$'s are dropped, and the algorithm proceeds to the next node. As soon as one node passes the test, the algorithm stops and the remaining branches are kept in their entirety.
The pair of branches $i$ and $j$ that stop the groomer define the groomed jet radius $R_g$, which is $R_g=\theta_{ij}$ for $\ee$ and $R_g=\Delta R_{ij}=\theta_{ij} \cosh \eta_J $ for $pp$.
Following Refs.~\cite{Chien:2019osu,Hoang:2019ceu}, it is convenient to identify a common energy cut variable $\tzcut$ that can be employed for both $\ee$ and $pp$ collisions:
\begin{align}\label{eq:tzcut}
&\tzcut \equiv \zcut\bigg[\sqrt{2}\sin\Big(\frac{R^{ee}_0}{2}\Big)\bigg]^{-\beta}\,,&
&
(\textrm{$e^+e^-$ case})\,, &
\\
&\tzcut \equiv
\zcut\bigg(\frac{\cosh \eta_J}{R_0}\bigg)^\beta = \zcut' (\cosh \eta_J)^\beta \,, &
&
(\textrm{$pp$ case})\,, &
\nn
\end{align}
such that, in either case, the soft drop condition for small angles becomes: $\min(z_i, z_j) > \tzcut \theta_{ij}^\beta$. We have also introduced an auxiliary variable $\zcut' \equiv \zcut/R_0^\beta$ for later convenience~\cite{Hoang:2019ceu,Chien:2019osu}.
Additionally, we also identify the energy scale associated with the soft drop groomer:
\begin{align} \label{eq:qcutDef}
\qcut \equiv 2^\beta Q \tzcut \, .
\end{align}
With this definition, we note that the condition for the jet mass to be in the SDOE region in \eq{SDOE} now applies for both $\ee$ and $pp$ cases.

The groomed jet radius $R_g$ can fall anywhere between $0$ and $R$. The measurement of the jet mass on the groomed jet, however, kinematically restricts the possible range of $R_g$, and thus leads us to consider different regimes of the analysis of the double differential distribution. For the case of $\ee$ collisions, the bounds on range of $R_g$ resulting from jet mass measurement are given by:
\begin{align} \label{eq:angularBounds}
\theta_c(m_J^2) \leq \; R_{g} \;\leq\;
\min\big\{ \theta_{g}^{\star} (m_J^2, \qcut ,\beta) , R \}\,.
\end{align}
These bounds result from the jet mass measurement $m_J$ while demanding the soft drop passing condition, which constrains the values of
$R_g$ to the range shown. Up to the NLL accuracy these angles are given by
\begin{align} \label{eq:angularBoundsExplicit}
\theta_c(m_J^2) = \frac{2m_J}{Q}\,,
\qquad
\theta_{g}^\star (m_J^2, \qcut ,\beta) = 2\Big( \frac{m_J^2}{Q \qcut}\Big)^{\frac{1}{2+\beta}} \, .
\end{align}
If we had $R_{g} < \theta_c(m_J^2)$ then the jet mass would be constrained to be smaller than the desired value $m_J$, and hence kinematically forbidden. While, on the other hand, having $R_{g}>\theta^\star_{g}$ implies that the large-angle radiation passing the jet grooming constraint would necessarily yield a value of the jet mass larger than the one measured.
\begin{figure}[t!]
\centering
\includegraphics[width=0.495\textwidth]{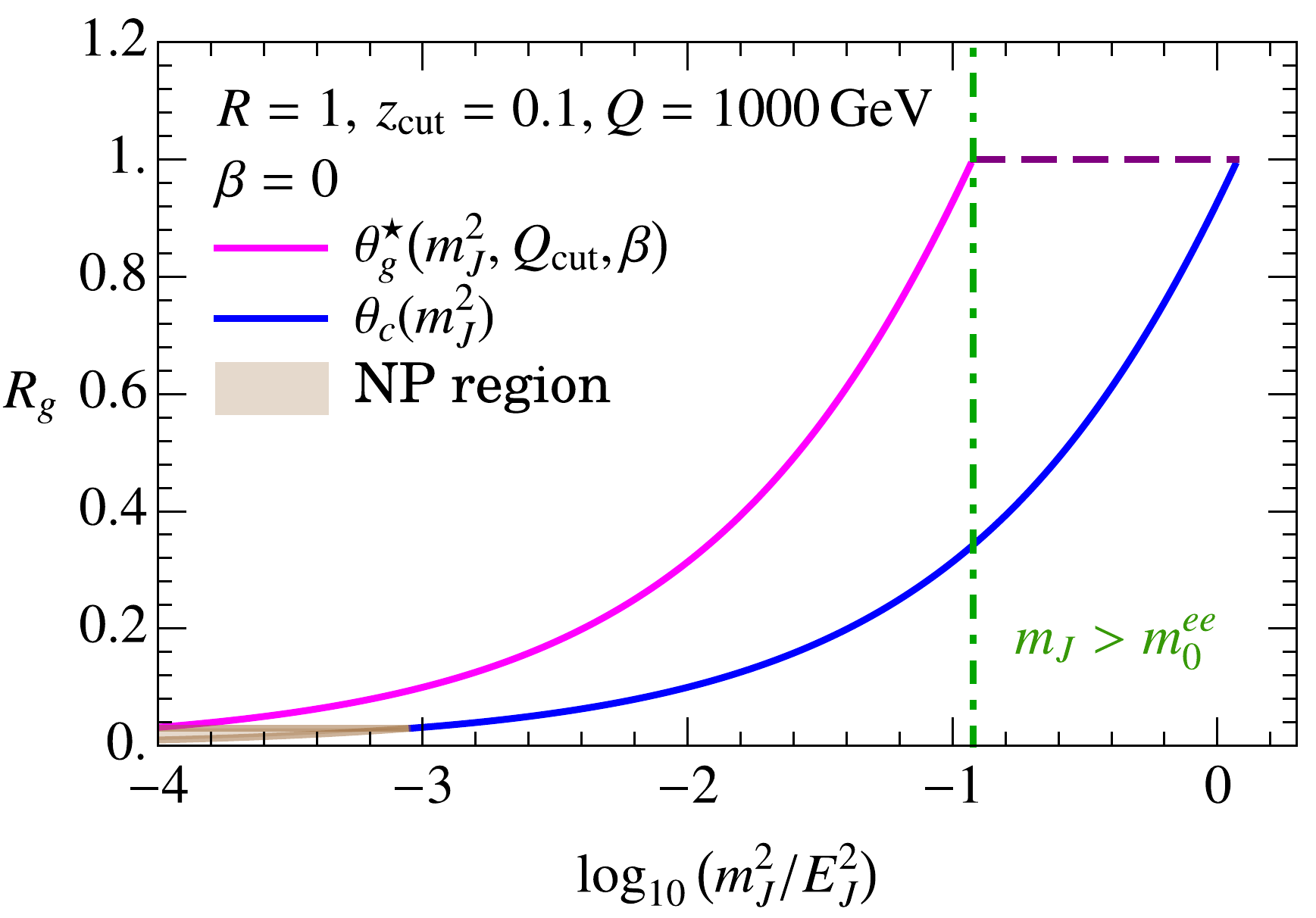}
\includegraphics[width=0.495\textwidth]{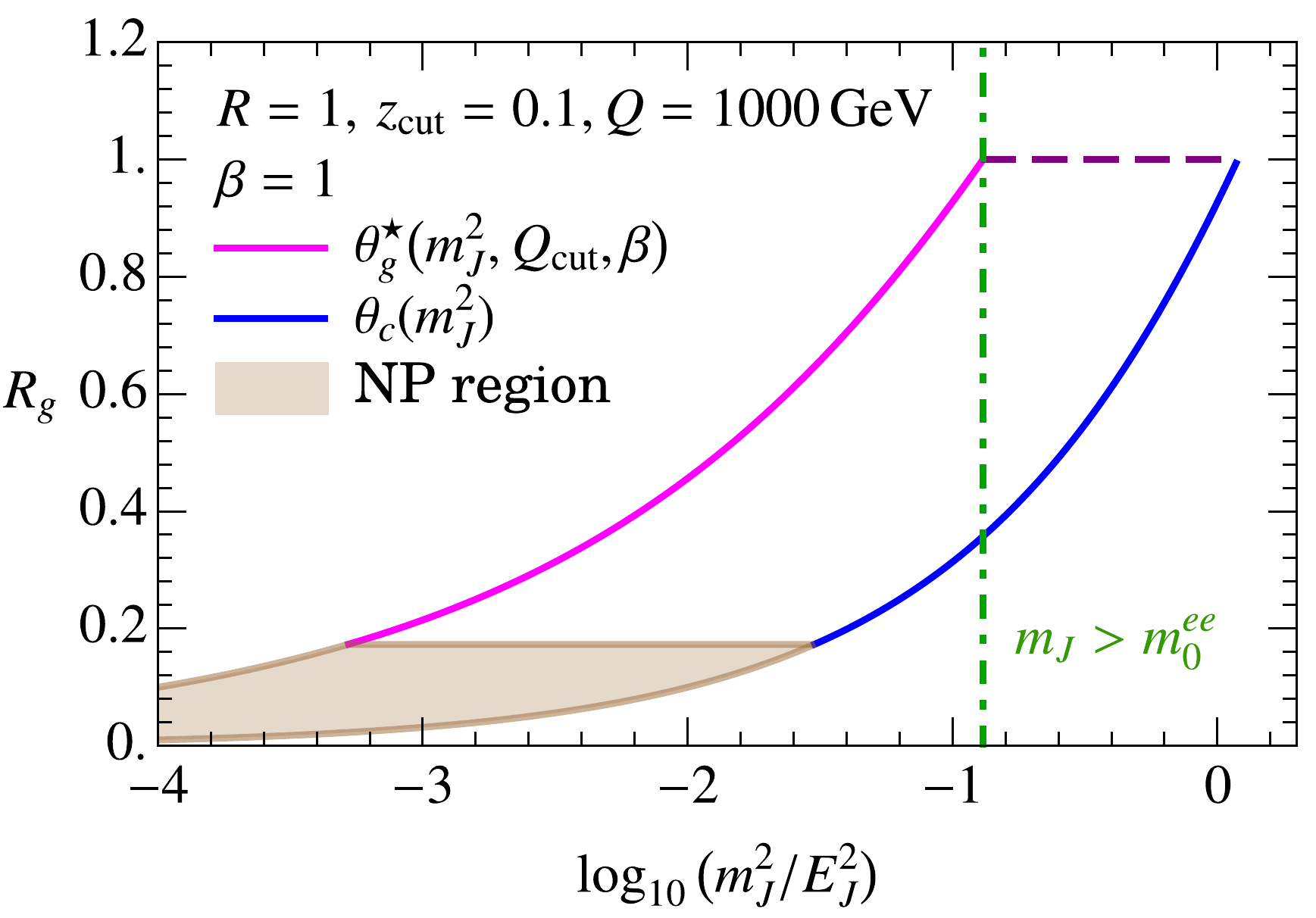}	\\
\includegraphics[width=0.495\textwidth]{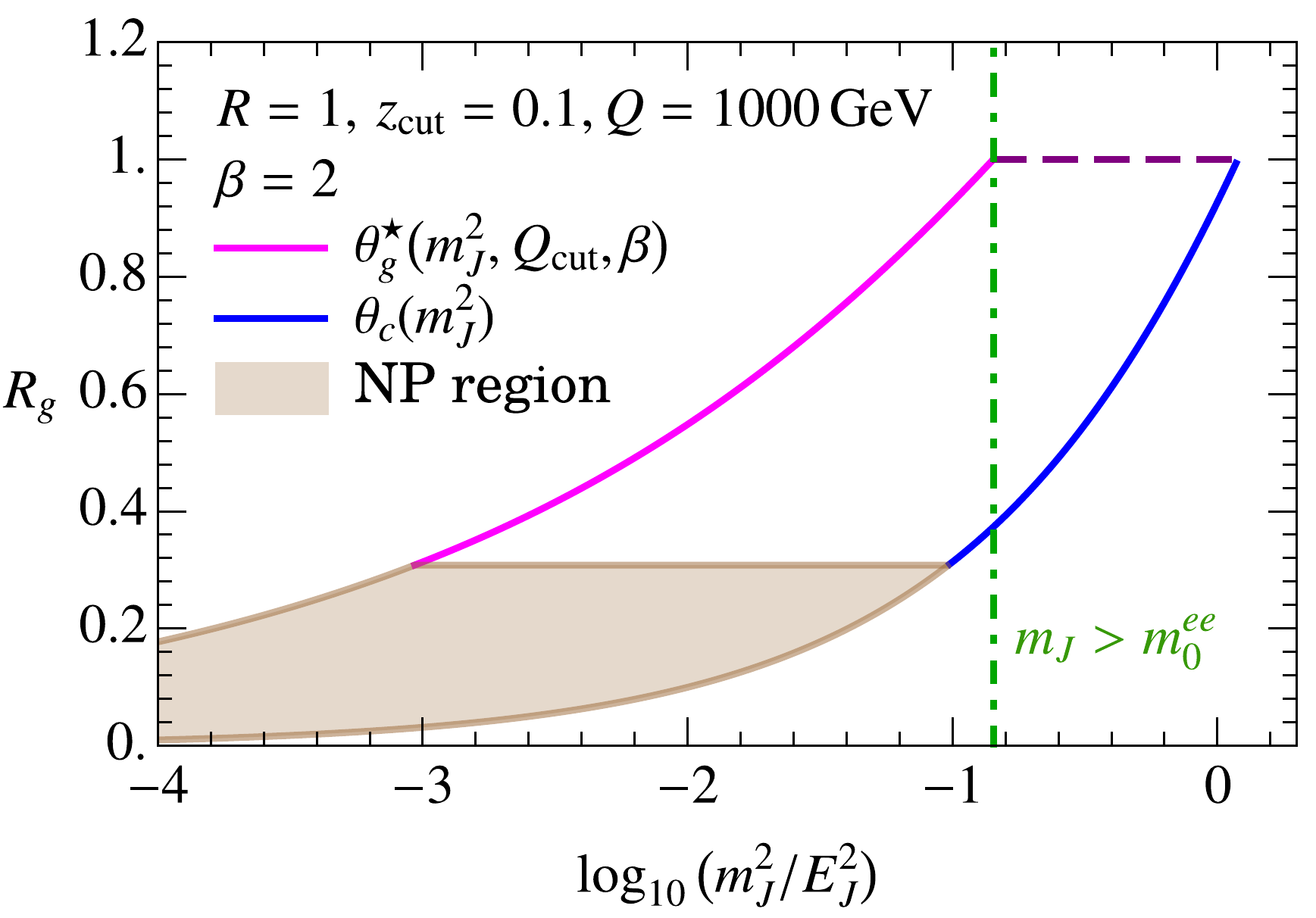}
\includegraphics[width=0.495\textwidth]{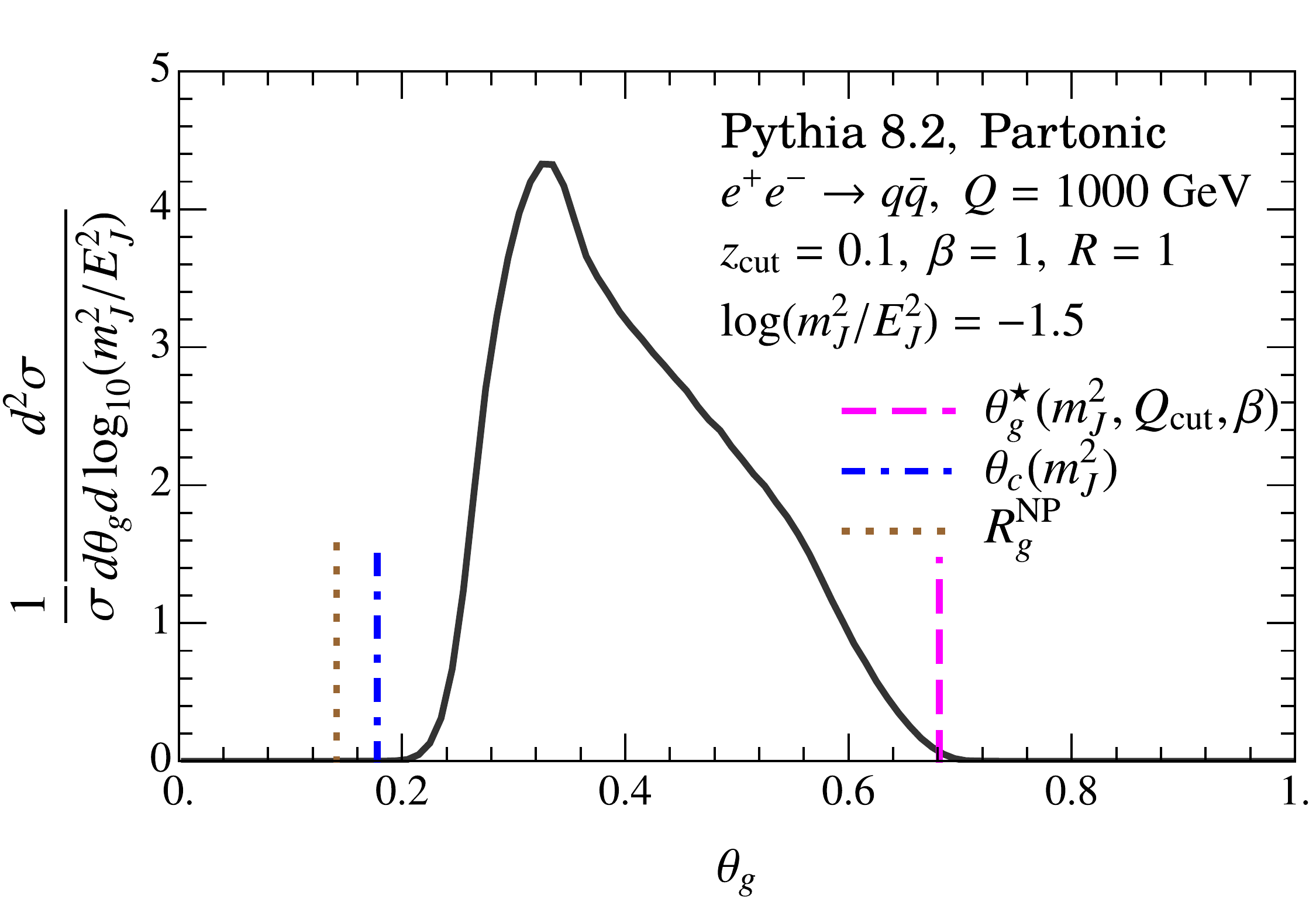}
\caption{Kinematic constraints on the range of $R_g$ for $\beta = 0,1,2$ with $Q=1000\,{\rm GeV}$ and $z_{\rm cut}=0.1$. The $R_g$ is constrained between $\theta_c(m_J^2)$ and $\theta_g^\star(m_J^2,\qcut,\beta)$. The dot-dashed vertical green line corresponds to the groomed to ungroomed transition. The shaded brown region demarcates the phase space where hadronization corrections are critical. In the bottom-right panel we show the distribution of the double differential distribution from \Pythia for representative values of $\zcut = 0.1$, $\beta = 1$ and $\log_{10}m_J^2/E_J^2 = -1.5$. We also display locations of the NLL kinematic boundaries and a rough estimate of the angle $R_g^{\rm NP}$ below which hadronization corrections become significant.}
\label{fig:phaseSpace}
\end{figure}
In the case of $pp$, the corresponding range for small $R$ jets is given by
\begin{align}
\theta_c^{(pp)} (m_J^2) \leq R_g \leq \min \bigl\{\theta_{g}^{(pp)\star} (m_J^2, p_T ,\beta) , R \bigr\}\, ,
\end{align}
with
\begin{align}\label{eq:angularBoundspp}
\theta_c^{(pp)} (m_J^2) \equiv \frac{m_J}{p_T} \, , \qquad
\theta_{g}^{(pp)\star} (m_J^2, p_T\zcut' ,\beta) = \Big(\frac{m_J^2}{p_T^2 \zcut'}\Big)^{\frac{1}{2+\beta}}
\, ,
\end{align}
where $\zcut'$ was given in \eq{tzcut}. These angular bounds are consistent with \eq{angularBounds} when we include the factors of $\cosh\eta_J$ in $R_g$ and $R$.

These kinematic bounds on $R_g$ in the $\ee$ case are displayed for $\beta = 0,1, 2$, $Q =1000$ GeV, $\zcut = 0.1$ and $R = 1$ in \fig{phaseSpace}. The allowed phase space is bounded by the solid magenta, solid blue, and dashed purple lines, and differs somewhat for the various choices of $\beta$.
(The brown shaded regions and green dashed line will be discussed below.) In the final panel of \fig{phaseSpace} we also show the parton level $\theta_g$ distribution of the
cross section with fixed $m_J$
from a \Pythia8.2 simulation. We see that the kinematic end points $\theta_c(m_J^2)$ and $\theta_g^\star(m_J^2,\qcut,\beta)$ in \eq{angularBoundsExplicit} are consistent with the results from \Pythia8.2 parton shower.

For larger jet masses where $\theta_g^\star (m_J^2, \qcut, \beta)> R$, the jet radius constraint limits $R_g\le R$. This happens for jet masses given by
\begin{align}\label{eq:m0}
&m_J^2 \ge	(m_0^{(ee)})^2 \equiv Q \qcut \sin^{2+\beta}\frac{R}{2} \,,&
&(\text{$\ee$ case})\, ,& \\
&m_J^2 \ge	(m_0^{(pp)})^2 \equiv p_T^2 \zcut' R^{2+\beta} \,,&
&(\text{$pp$ case})\, .&
\nn
\end{align}
The green dashed lin in \fig{phaseSpace} shows the result for $m_0^{(ee)}$.
This boundary was discussed in~\Ref{Chien:2019osu} for the $pp$ case, where as before, the $R/2\ll 1$ limit is assumed.
This region is referred to as the \textit{ungroomed} region in the context of single differential jet mass cross section, and here power corrections to the jet mass spectrum of ${\cal O}(m_J^2/(Q\qcut))$ cannot be neglected. Technically, in this region the grooming is still active but here the effects of grooming on the single differential soft drop $m_J^2$ cross section are described via a fixed order treatment. Having an additional measurement of groomed jet radius, however, further necessitates resummation for logarithms of $R_g$, which are also important in the ungroomed region.

In the following we will focus on the SDOE region defined by \eq{SDOE} where the following small angle approximation can be assumed:
\begin{align} \label{eq:smallMassAssumption}
\theta_{g}^\star (m_J^2, \qcut ,\beta) \ll R
\,,
\qquad
\theta_g^{(pp)\star} (m_J^2, p_T\zcut', \beta) \ll R
\, .
\end{align}
For the purpose of setting up the effective theory description, we will interpret the inequality \eq{smallMassAssumption} in a hierarchical sense.
With the inclusion of grooming, we therefore have 4 expansion parameters in our EFT namely $\theta_c, \theta_g^*, R_g$, and $z_{\text{cut}}$.
In the SDOE region we also have $\theta_{c} \ll \theta^\star_{g}$, which then leads us to consider three possible hierarchies between our expansion parameters.
\begin{enumerate}
\item{\makebox[6cm]{Large groomed jet radius:\hfill}} $\theta_c \ll R_g \lesssim \theta^\star_{g} \ll R $ ,
\item{\makebox[6cm]{Intermediate groomed jet radius:\hfill}} $\theta_c\ll R_{g} \ll \theta^\star_{g} \ll R$ ,
\item{\makebox[6cm]{Small groomed jet radius:\hfill}} $\theta_c\lesssim R_{g} \ll \theta^\star_{g} \ll R$ .
\end{enumerate}
The main analysis in this work will focus on these scenarios, and we will extrapolate the results to the cases where
$\theta_g^\star \lesssim R$ or $m_J\ge m_0$ by encoding a basic description of these regions, but without trying to enforce the same resummation precision that we achieve for regions 1, 2, and 3.
Since the perturbative moments in \eq{C1C2multiDiffDef} are naturally computed through an integral over the groomed jet radius, their determination for a specific value $m_J$ require either combining or considering transitions between contributions from all three regimes.

We now return to discuss the brown shaded region in \fig{phaseSpace}. With the addition of the $R_g$ measurement, the $m_J$ dependent condition in \eq{SDOE} for the single-differential spectrum should be further qualified to provide a restriction in the two-dimensional $m_J$--$R_g$ plane.
For low jet masses, nonperturbative effects become ${\cal O}(1)$ and the SDOE approximation in \eq{SDOE} no longer holds. This happens for jet masses in the soft drop nonperturbative region (SDNP) defined by
\begin{align}\label{eq:mJSDNP}
m_J^2 \lesssim (m_J^2)_{\rm SDNP}
\equiv
Q \Lambda_{\rm QCD}\Big(\frac{\Lambda_{\rm QCD}}{\qcut}\Big)^{\frac{1}{1+\beta}} \, .
\end{align}
For such low jet masses, any emission that stops soft drop is nonperturbative with a virtuality $p^2\sim \Lambda_{\rm QCD}^2$.
When we include an additional measurement of $R_g$ within the bounds in \eq{angularBounds}, the emission that stops soft drop can become non-perturbative for small $R_g$, irrespective of the value of $m_J$. The corresponding angle is obtained by solving the constraints $p^2\sim \Lambda_{\rm QCD}^2$ and $z=\tzcut \theta^\beta$ which implies that the emission will be nonperturbative when
\begin{align}\label{eq:RgNP}
& R_g \lesssim
(R_g)_{\rm NP} \sim 2 \Big(\frac{\Lambda_{\rm QCD}}{\qcut}\Big)^{\frac{1}{1+\beta}} \,,&
&(\text{$\ee$ case})\, ,& \\
&R_g \lesssim
(R_g)_{\rm NP} \sim \Big(\frac{\Lambda_{\rm QCD}}{p_T \zcut'}\Big)^{\frac{1}{1+\beta}} \,,&
&(\text{$pp$ case})\, .& \nn
\end{align}
The angle $(R_g)_{\rm NP}$ is the same as the angle $\theta_{\Lambda \rm CS}$ defined in \Ref{Hoang:2019ceu} in the context of the SDNP region.
In \fig{phaseSpace} we shade the regions $ R_g \lesssim R_g^{\rm NP}$ brown for the various values of $\beta$ in order to highlight that in these regions hadronization corrections are ${\cal O}(1)$ and cannot be ignored.
The impact of the nonperturbative region grows with increasing $\beta$, so while the perturbative constraints defined in \eq{angularBounds} are applicable for the entire range for $\beta = 0$, the NP region already covers a large part of the allowed phase space for $\beta=2$.
Thus for the combined $R_g$ and $m_J$ measurements, the SDNP region is defined by \eq{RgNP} which automatically implies \eq{mJSDNP}, and the SDOE region is defined by
\begin{align}
\biggl( \frac{R_g}{2}\biggr)^{1+\beta}
\gg \biggl( \frac{(R_g)_{\rm NP}}{2}\biggr)^{1+\beta} \,,
\end{align}
which automatically implies \eq{SDOE}.
Since the partonic moments $M_1^\kappa$ and $\Mb$ defined in \eq{C1C2multiDiffDef} involve integration over the entire allowed range of $R_g$ for a given jet mass, terms in the perturbative expression can become sensitive to small renormalization scales, which must be frozen at a scale $\mu\gtrsim 1\,{\rm GeV}$ to ensure they remain valid. If we studied hadronic versions of the moments $M_1^\kappa$ and $\Mb$, then in some cases (like $\beta=2$) they would differ substantially from their partonic counterparts (which are related to $C^{\kappa}_{1,2}$), due to significant contributions from the brown region of \fig{phaseSpace}.
In the next section, we take detailed look at each of the regimes 1,2,3 and determine the corresponding factorization formulae utilizing a mode analysis in SCET.

\section{Factorization}
\label{sec:factAll}
\begin{figure}[t!]
\centering
\includegraphics[width=\textwidth]{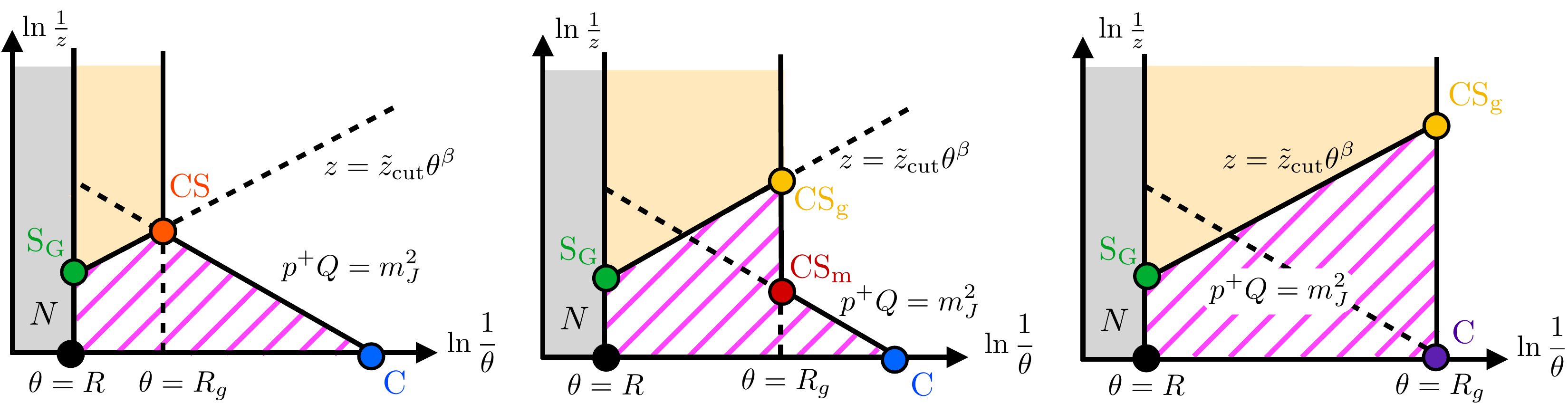}
\caption{The mode picture for three regimes of large (left), intermediate (center) and small (right) groomed jet radius, labeling emissions with their energy fraction $z$ and angle $\theta$ relative to the emitting particle. The area shaded in yellow corresponds to soft dropped radiation, while the area hatched in magenta is vetoed by the measurement (black lines), where dashed segments do not enforce vetoes. Emissions in the area shaded in grey lie out of the initial (ungroomed) jet.}
\label{fig:modePicture}
\end{figure}
In this section we discuss the modes that arise in the three regimes discussed above.
These modes are best depicted in the Lund plane shown in \fig{modePicture}, where we have chosen the variables $\theta$, the angle of a soft/collinear emission relative to the jet axis, and $z$, the energy fraction relative to the total jet energy $E_J$ (or $p_T$ in case of $pp$ collisions).
Every point on the phase space corresponds to an emission off the jet at a given angle and with given energy fraction.
The choice of $\ln \theta^{-1}$-$\ln z^{-1}$ helps us to focus on the soft-collinear emissions that are far away from the origin~\cite{Frye:2016aiz}. In these coordinates distinct points correspond to emissions that are exponentially far apart.
Here the jet mass measurement corresponds to a line with slope $-2$, the equation of the soft drop constraint is a line with slope $\beta$, and the equation $\theta = R_g$ is a vertical line.
The three regimes are shown in the three panels of \fig{modePicture}, and differ due to the location of the vertical $R_g$-line.
At LL accuracy, in the soft-collinear limit, the probability of an emission in a certain region of the Lund plane is uniform in the logarithmic coordinates.
Emissions that are vetoed at LL accuracy correspond to the area hatched in magenta, while the area shaded in yellow identifies soft dropped emissions.
The various modes that enter the factorization analysis are located on the boundary of the vetoed region, at the intersections of the various constraint lines, and are indicated by colored circles.
Since these modes are far from the origin they are either soft, collinear, or both soft and collinear.

It is useful to determine the momentum scaling of the modes represented by colored dots in \fig{modePicture}.
We start with the modes that are common to all the scenarios, and in subsequent subsections describe the additional collinear-soft modes that are specific for each of the three regimes.
For $\ee$ collisions, in the dijet limit, the differential jet mass cross section cumulative in $R_g$ can be expressed as
\begin{align}\label{eq:quarkCrossSection}
\frac{d \Sigma(R_g)}{d m_J^2 d\Phi_J}
&= N_q^{e^+e^-}(Q, R, \zcut, \beta,\mu) \frac{d \Sigma^q(R_g,\mu)}{d m_J^2 d\Phi_J}
\, .
\end{align}
This notation holds for both hadron and parton level cross sections.
The normalization factor $N_q^{e+e^-}$ obtains contribution both from hard modes with $p^\mu\sim Q$, as well as from the hard-collinear mode with
$\theta = R$ and $z \sim 1$ which scales as
\begin{align}
p^\mu\sim Q \tan \frac{R}{2} \bigg(\tan \frac{R}{2} ,
\frac{1}{\tan \frac{R}{2}}, 1\bigg) \,,
\qquad
(\text{$\ee$ case})
\, .
\end{align}
Here and below we display the momenta scaling in the light-cone coordinates $p^\mu = (p^+, p^-, p_\perp)$ defined with respect to a light-like reference vector $n_J=(1,\hat n_J)$ in the direction of the jet axis $\hat n_J$, such that
\begin{align}
p^\mu = \frac{\bn_J^\mu}{2} p^+ + \frac{n_J^\mu}{2} p^- + p_\perp^\mu \, ,
\end{align}
where $\bn_J^\mu$ is an auxiliary light-like four-vector satisfying $\bn_J^2 = 0$ and $\bn_J \cdot n_J = 2$.
The hard and hard-collinear modes are the same mode for $R/2\sim 1$.
We assume that the jet has been identified via an IRC safe jet algorithm and the details of auxiliary measurements outside the jet will be irrelevant to our analysis here.

In hadron collisions, we will have jets initiated by both quarks and gluons in the final state, and thus the cross section will involve sum over $\kappa = q,g$. The equivalent of the normalization factor in \eq{quarkCrossSection} will now also account for parton distribution functions and the hard scattering cross section.
In particular, the following analog of \eq{quarkCrossSection} will now hold true for inclusive jet measurement in the $pp$ collisions:
\begin{align}\label{eq:ppCrossSection}
\frac{d \Sigma(R_g)}{d m_J^2d\Phi_J}
&= \sum_{\kappa = q,g} N^{pp}_\kappa( \Phi_J, R, \zcut, \beta,\mu ) \frac{d \Sigma^\kappa(R_g,\mu)}{d m_J^2d\Phi_J}
\, ,
\end{align}
where we sum over the contributions to the cross section for the jet initiating parton $\kappa = q,g$.
As before, the normalization $N_\kappa^{(pp)}$ includes contributions from hard modes with $p\sim p_T$ and
the hard-collinear mode with the following scaling:
\begin{align}
p^\mu\sim p_T R
\bigg(\frac{R}{2\cosh \eta_J} , \, \frac{2\cosh \eta_J}{R} , \, 1 \bigg) \,,
\qquad
(\text{$pp$ case})	\, .
\end{align}
Here, the large momentum component of the jet now must be related to the jet $p_T$ as $Q = 2 p_T \cosh\eta_J$.

The global-soft $({\rm S_G})$ mode is located at the intersection of the soft drop line and the jet radius constraint at $\theta = R$, and also contributes to $N_\kappa^{(pp,\ee)}$. These ${\rm S_G}$ modes account for the radiation that was initially clustered in the jet but failed to pass soft drop. These modes have momentum that scales as
\begin{align}
\label{eq:pGS}
&p_{\rm S_G}^\mu \sim Q_{\rm cut} \tan^{1+\beta} \frac{R}{2}
\bigg(\tan\frac{R}{2},\frac{1}{\tan\frac{R}{2}} , 1 \bigg) &
&(\text{$\ee$ case}) \,, & \\
&p_{\rm S_G}^\mu \sim p_T \zcut' R^{1+\beta}
\bigg(\frac{R}{2\cosh \eta_J},\, \frac{2\cosh \eta_J}{R} , \, 1 \bigg)
&
&(\text{$pp$ case})\, .&
\nn
\end{align}
We find it helpful to isolate the hard and global-soft contributions as follows:
\begin{align}
\label{eq:Gngl}
N^{(pp,\ee)}_\kappa(\Phi_J, R, \zcut, \beta,\mu)
= \sum_{j}\widehat N^j_\kappa(\Phi_J, R, \mu)\,
\otimes_{\Omega} \widehat S^{\kappa,j}_G(\qcut,R, \beta, \mu)
\,,
\end{align}
where the effect of grooming is entirely accounted by the global-soft function $\widehat S_G^{\kappa,j}$, and only $\kappa = q$ is relevant for $\ee$ collisions in the dijet limit.
Here $\otimes_{\Omega}$ indicates that the angular integrals in the two functions involved cannot be done independently since both the modes have the same angular scaling, i.e. they cannot be distinguished by their angular separation.
The sum $j$ in the convolution in \eq{Gngl} sums over the individual Hard partons.
This means that there are new logarithms, known as non-global logarithms (NGLs) in the ratio of the virtualities of the two modes that are involved in the angular convolution, and which can only be resummed by doing a more sophisticated resummation or RG evolution.
Analytical resummation using the complete factorization formula is difficult and Monte Carlo techniques are usually used for resumming these NGLs.

Following \cite{Kang:2019prh}, at NLL accuracy, the NGLs can be accounted for by writing
\begin{align}\label{eq:Fngl}
& \sum_j \widehat N^j_\kappa(\Phi_J, R, \mu)\,
\otimes_{\Omega} \widehat S^{\kappa,j}_G(\qcut,R, \beta, \mu) \\
&\qquad =N_\kappa(Q, R, \mu)\,
S_G^\kappa\Big(\qcut\tan^{1+\beta}\frac{R}{2}, \beta, \mu ,R \Big) \,
\Xi_{G}^\kappa(t_G) \, , &&\quad (\text{$\ee$ case}) \, ,\nn
\\
&\qquad = N_\kappa(\Phi_J, R, \mu)\,
S_G^\kappa\Big(p_T\zcut' R^{1+\beta}, \beta, \mu\Big) \,
\Xi_{G}^\kappa(t_G) \, , &&\quad (\text{$pp$ case})
\nn
\, ,
\end{align}
where the new functions $N_\kappa$ and $S_G^\kappa$ are obtained by doing independent solid angle integration and the leading effect of the NGLs is described by the function $\Xi_G^\kappa$.
The global-soft function $S_G^\kappa$ is a matrix element of soft Wilson lines with the modes of global-soft scalings, and has been extracted at two loops for $\kappa = q$~\cite{Frye:2016aiz,Bell:2018vaa}. In the $\ee$ formula in \eq{Fngl} we have shown explicitly how the scale $\qcut$ appears in combination with the jet radius. The additional argument $R$ on $S_G^\kappa$ denotes fixed order corrections, relevant for $R\sim1$, which do not contribute to the anomalous dimension; these corrections can be dropped for the $pp$ case we consider, as we indicate by omitting the last argument in $S_G^\kappa$ in the corresponding formula.

The hard function $N_\kappa$ appears universally for dijet production in $e^+e^-$ collisions where it is known up to three loops for the hemisphere case where it is determined by the quark form factor~\cite{vanNeerven:1985xr,Matsuura:1988sm,Gehrmann:2005pd,Moch:2005id,Baikov:2009bg,Lee:2010cga}, while for calculations with a jet of radius $R$ see Refs.~\cite{Ellis:2010rwa,Dasgupta:2014yra,Kaufmann:2015hma,Kang:2016mcy,Dai:2016hzf}.
The $N_\kappa$ is also universal for small $R$ inclusive jets in the $pp$ case, see \Refs{Chien:2015ctp,Dai:2016hzf}.

Finally, following \cite{Kang:2019prh}, in \eq{Fngl} we introduced the variable
\begin{align}\label{eq:tVariableNGL}
t_G &= t [\mu_{gs}, \mu_N] \, ,
\end{align}
which appears as the first argument of $\Xi_G^\kappa$. Here
\begin{align}\label{eq:tNGL}
t[ \mu_0, \mu_1] \equiv \frac{1}{2\pi} \int_{\mu_{0}}^{\mu_1} \frac{d\mu'}{\mu'}\alpha_s(\mu') \,.
\end{align}
In \eq{Fngl} the integral runs between the global-soft scale $\mu_{gs}$ and the hard scale $\mu_N$, whose explicit expressions are given in \eqs{naturalScales}{naturalScalespp} below.

We note that the NGLs associated with the function $\widehat S^\kappa_G$ do not affect the measurement of our observables directly but only modify the overall normalization. While this is important in determining the relative fractions of quark and gluon jets in $pp$ collisions, in our numerical analyses we only consider the $e^+e^-$ case and hence only deal with quark jets.
At the same time, the observables we are interested in are the {\it normalized} moments of the double differential cross section which actually do not depend on $\widehat S_G^q$ and $\widehat N^q$ functions.
We will therefore ignore the $\Xi_G^q$ function in our numerical analyses and discuss only the other angular averaged functions.
For consistency we will leave $\Xi_G^q$ explicit in the formulae in this section.

Next we describe the collinear mode (C) represented by a blue/purple dot in \fig{modePicture}. This is the largest energy mode, $z\sim 1$ that contributes to the jet mass. Requiring $m_J^2 \ll E_J^2$ results in the mode being at the smallest angle. Thus, the energetic emissions represented by this mode always pass grooming, and have momenta scaling as
\begin{align} \label{eq:pC}
p_{\rm C}^\mu \sim \Bigl( \frac{m_J^2}{Q}, Q, m_J \Bigr) \,.
\end{align}
Note that the corresponding angle for these modes is $\theta_c = 2m_J/Q$. The scaling holds for both $\ee$ and $pp$ cases.

As shown in \fig{modePicture}, additional soft-collinear modes arise from the interplay between the jet mass measurement, the grooming condition, and the $R_g$ constraint, and depend on whether we are in the regime with large, intermediate, or small groomed jet radius. We will discuss these modes and formulate the factorization formulae for the cross section in \eq{dbleDiff} for each of these regimes in turn. We will base our discussion of the factorization formulae on the $\ee$ case and compile the analogous results for $pp$ case at the end of each section.
In general the ungroomed fat jet of radius $R$ is typically initially isolated with the anti-$k_T$~\cite{Cacciari:2008gp} algorithm and is subsequently reclustered in the soft drop procedure using the C/A algorithm~\cite{Dokshitzer:1997in,Wobisch:1998wt}. Thus, in addition to NGLs, sequential clustering algorithms give rise to ``clustering logarithms'' associated with independent emissions that appear at the same order as the NGLs. They can also be accounted for using a multiplicative function in the manner of \eq{Fngl}. We describe them in detail in \app{Abel} and discuss them in each region of factorization in the following subsections.

\subsection{Large groomed jet radius}
\label{sec:LargeRgFact}
We start for the case represented on the left of \fig{modePicture}, where the groomed jet radius is comparable with the maximum angle $R_{g} \lesssim \theta_{g}^\star(m_J^2, \qcut, \beta)$. Here soft drop is stopped by a wide-angle emission with relatively large virtuality, described by a collinear-soft mode (orange), with $(+,-,\perp)$ momentum scaling
\begin{align}
&p^\mu_{\rm CS} \sim \frac{m_J^2}{Q(\theta_g^\star/2)}
\Big(\frac{\theta_g^\star}{2}, \frac{2}{\theta_g^\star}, 1\Big)
\,&
&(\text{$\ee$ case})\,,&
\\
&p^\mu_{\rm CS} \sim \frac{m_J^2}{p_T \theta_g^{(pp)\star}}
\bigg(\frac{\theta_g^{(pp)\star}}{2\cosh\eta_J}, \frac{2\cosh \eta_J}{\theta_g^{(pp)\star}}, 1\bigg)
\,&
&(\text{$pp$ case})\, .&
\nn
\end{align}
If other such emissions pass soft drop, their virtuality is large enough to contribute to the groomed jet mass. We state here the results for $\ee$ collisions that can be straightforwardly generalized to the $pp$ case as we show below.
In this regime there is a single collinear-soft mode in addition to the modes discussed above, and the factorization formula is given by a generalization of the jet mass factorization formula~\cite{Frye:2016aiz} to incorporate a more differential collinear soft function:
\begin{align}
\label{eq:FactLargeThetaCS}
\frac{d \Sigma(R_g)}{d m_J^2 } &=
\sum_{j}\widehat N^j_q(Q, R, \mu)\,
\otimes_{\Omega} \widehat S^{q,j}_G(\qcut,R, \beta, \mu)
\\
&\qquad \times
\,\qcut^{\frac{1}{1+\beta}}
\! \int \!d \ell^+ J_q (m_J^2 - Q\ell^+, \mu) \,
S_c^q
\Big[
\ell^+ \qcut^{\frac{1}{1+\beta}},\frac{R_g}{2}\qcut^{\frac{1}{1+\beta}}, \beta, \mu \Big]
\,, \nn
\\
&=
N_q(Q, R, \mu)\,
S_G^q\Big(\qcut\tan^{1+\beta}\frac{R}{2}, \beta, \mu ,R\Big) \,
\Xi_{G}^{q}(t_G)
\nn
\\
&\qquad \times
\,\qcut^{\frac{1}{1+\beta}}
\! \int \!d \ell^+ J_q (m_J^2 - Q\ell^+, \mu) \,
S_c^q
\Big[
\ell^+ \qcut^{\frac{1}{1+\beta}},\frac{R_g}{2}\qcut^{\frac{1}{1+\beta}}, \beta, \mu \Big]
\,. \nn
\end{align}
In the second equality we have made use of \eq{Fngl} to perform the angular integrations.
In \eq{FactLargeThetaCS} the contributions from collinear regions are encoded in the jet function $J_q (m_J^2, \mu)$, which is universal across various ungroomed and groomed event shapes and known up to three loops~\cite{Bauer:2003pi,Bosch:2004th,Becher:2006qw,Bruser:2018rad}, and by the collinear-soft function $S_c^q\bigl[\ell^+ \qcut^{\frac{1}{1+\beta}}, \frac{R_g}{2}\qcut^{\frac{1}{1+\beta}}, \beta, \mu \bigr]$ that we introduce here, together with renormalization group evolution for the scales between these functions. Since these modes contribute additively to the groomed jet mass measurement, they enter with a convolution in \eq{FactLargeThetaCS}. For this integration we use $\ell^+ = m_J^2/Q$, the small momentum component of collinear(-soft) radiation. Extracting the overall factor of $\qcut^{\frac{1}{1+\beta}}$ allows for rewriting $S_c^q$ as a combination of only the arguments in brackets.

The collinear-soft function $S_c^\kappa\bigl[\ell^+ \qcut^{\frac{1}{1+\beta}}, \frac{R_g}{2}\qcut^{\frac{1}{1+\beta}}, \beta, \mu \bigr]$ in \eq{FactLargeThetaCS} is closely related to the collinear-soft function that appears for the single differential jet mass distribution $S_c^\kappa \bigl[ \ell^+ \qcut^{\frac{1}{1+\beta}},\beta, \mu \bigr ]$, but with an additional constraint from the groomed jet radius.
Including the $R_g$ measurement at one loop for $e^+e^-$ collisions yields the following result:
\begin{align}\label{eq:ScLargeRgFull}
&
S_c^\kappa
\Bigl[
\ell^+ \qcut^{\frac{1}{1+\beta}}, \beta, \mu \Bigr]
-\frac{2}{2+\beta}\frac{\alpha_s C_\kappa}{\pi}
\Theta\bigg(
\Big(\frac{\ell^+}{\qcut}\Big)^{\frac{1}{2+\beta}}
-\tan\frac{R_g}{2}
\bigg)
\frac{1}{\ell^+ \qcut^{\frac{1}{1+\beta}} }
\log\bigg(
\frac{\ell^+}{\qcut \tan^{2+\beta} \frac{R_g}{2}}
\bigg)
\, ,
\end{align}
where $C_q = C_F=4/3$, $C_g = C_A=3$ are the standard SU(3) fundamental and adjoint quadratic Casimirs. In general, we note that results for $e^+e^-$ collisions can be expressed in terms of $\tan\frac{R_g}{2}$, which allows for a smooth transition to $R_g \lesssim R$ region for large jet masses.
For the kinematic range we explore, $\tan\frac{R_g}{2} \sim \frac{R_g}{2}$ and we recover the functions of combinations $\ell^+ \qcut^{\frac{1}{1+\beta}}$ and $\frac{R_g}{2}\qcut^{\frac{1}{1+\beta}}$, which applies for both $e^+e^-$ and $pp$ collisions,
\begin{align}\label{eq:ScLargeRg}
& S_c^\kappa
\Big [
\ell^+ \qcut^{\frac{1}{1+\beta}}, \frac{R_g}{2}\qcut^{\frac{1}{1+\beta}}, \beta, \mu
\Big]
=
S_c^\kappa
\Big[
\ell^+ \qcut^{\frac{1}{1+\beta}}, \beta, \mu \Big]
\\
&\qquad +\frac{\alpha_s C_\kappa}{\pi}
\Bigg[
-\frac{2}{2+\beta}
\Theta\bigg(
\ell^+\qcut^{\frac{1}{1+\beta}}
-\Big(\frac{R_g}{2}\qcut^{\frac{1}{1+\beta}}\Big)^{2+\beta}
\bigg)
\frac{1}{\ell^+ \qcut^{\frac{1}{1+\beta}} }
\log\bigg(
\frac{\ell^+\qcut^{\frac{1}{1+\beta}}}{ \big(\qcut^{\frac{1}{1+\beta}}\frac{R_g}{2}\big)^{2+\beta}}
\bigg)
\Bigg]
\,. \nn
\end{align}
We elaborate on this function further in \sec{LargeRg} when we discuss the resummation of large logarithms in this regime.

For the single differential jet mass cross section in the SDOE region, \Ref{Frye:2016aiz} showed that the $Q$ and $\zcut$ dependence enters the collinear-soft function $S_c^\kappa$ through the combination $\ell^+ \qcut^{\frac{1}{1+\beta}}$, as we have displayed in \eqs{FactLargeThetaCS}{ScLargeRg}. For the double differential cross section we now show that the additional measurement $R_g$ appears in the combination $2\ell^+/R_g$, or equivalently $R_g\, \qcut^{\frac{1}{1+\beta}}$. For a given set of momenta $\{p_i^\mu\}$ that enter the calculation of the collinear-soft function we perform the rescaling
\begin{align}
\label{eq:qcutrescaling}
p_i^+ = (\ell^+)\ k_i^a \,
\qquad
p_{i\perp} = \Bigl(\qcut^{\frac{1}{2+\beta}}\big(\ell^+\big)^{\frac{1+\beta}{2+\beta}} \Bigr) \ k_{i\perp} \,,
\qquad
p_i^- = \Bigl( \qcut^{\frac{2}{2+\beta}}\big(\ell^+\big)^{\frac{\beta}{2+\beta}} \Bigr) \ k_i^b \, .
\end{align}
Hence, the angle of these subjets or particles in the new (dimensionless) coordinates $\{k_i^a,k_i^b,k^\perp_i\}$ is given by
\begin{align}
\frac{\theta_i}{2} = \frac{p_{i\perp}}{p_i^-} = \bigg(\frac{\ell^+}{\qcut}\bigg)^{\frac{1}{2+\beta}} \, \frac{k_{i\perp}}{k_i^b} \, .
\end{align}
Similarly the relative angle between any two particles or subjets, $\theta_{ij}$, after this rescaling becomes
$\theta_{ij} = (\ell^+/\qcut)^{\frac{1}{2+\beta}} \, \hat\theta_{ij}$, where the rescaled relative angle is only a function of the new coordinates, $\hat\theta_{ij}=\hat\theta_{ij}(k_i,k_j)$.
In addition to the soft drop test, and the jet mass measurement, that already appear for the single differential distribution, we now have an additional comparison of angles $\theta_{ij}$ with $R_g$, such that the measurement function now additionally involves
\begin{align} \label{eq:RgThetas}
&\prod_{i,j\in J} \Theta(R_g - \theta_{ij})
= \prod_{i,j\in J} \Theta \bigg(\frac{R_g}{2} - \bigg(\frac{\ell^+}{\qcut}\bigg)^{\frac{1}{2+\beta}}
\hat \theta_{ij}(k)
\bigg) \\
&\qquad = \prod_{i,j\in J}
\Theta \bigg(\frac{R_g}{2\ell^+} - \big(\ell^+\qcut^{\frac{1}{1+\beta}}\big)^{-\frac{1+\beta}{2+\beta}}
\hat \theta_{ij}(k)
\bigg) \ = \ \prod_{i,j\in J}
\Theta \bigg(\frac{R_g\,\qcut^{\frac{1}{1+\beta}}}{2} - \big(\ell^+\qcut^{\frac{1}{1+\beta}}\big)^{\frac{1}{2+\beta}}
\hat \theta_{ij}(k)
\bigg) \, . \nn
\end{align}
Here the $i,j\in J$ product is over all subjets $i$ and $j$ which are present when the jet grooming has terminated.
\eq{RgThetas} together with the arguments for the single differential case from~\cite{Frye:2016aiz} demonstrate that only the variable combinations $\frac{R_g}{2}\qcut^{\frac{1}{1+\beta}}$ and $\ell^+ \qcut^{\frac{1}{1+\beta}}$ appear.

Finally, we note that the NGLs for the modes with $\theta \sim R$ described by the function $\Xi_{G}^{q}(t_G)$ only affect the overall normalization, they do not affect the normalized moments that we are interested in for $C_1$ and $C_2$, and hence we ignore $\Xi_G^q$ for the rest of the paper. NGLs usually appear due to correlation between emissions at the jet boundary. In our case, we also have a boundary associated with the groomed jet and it is natural to ask whether there are NGLs that appear here. A calculation in \app{NGLcluster} shows that for the large $R_g$ regime, the associated logarithms are not large logarithms, and hence can be treated in fixed order perturbation theory. The same will also apply to the abelian clustering logarithms described in \app{Abel}.

We now briefly discuss the generalization for the $pp$ case. We write the factorization formula for $pp$ collisions in a notation such that all the results from $\ee$ case can be directly applied using the following substitutions:
\begin{align}\label{eq:dictionary}
Q\ell^+ \ra p_T r^+
\, ,
\qquad
\ell^+ \qcut^{\frac{1}{1+\beta}} \ra r^+ (p_T \zcut')^{\frac{1}{1+\beta}}
\, ,
\qquad
\frac{R_g}{2} \qcut^{\frac{1}{1+\beta}} \ra R_g (p_T \zcut')^{\frac{1}{1+\beta}}
\, ,
\end{align}
where $\zcut'$ was defined in \eq{tzcut}.
The $r^+$ variable in \eq{dictionary} differs from $\ell^+$ in the $\ee$ case by a factor of $2\cosh\eta_J$, and hence cancels against the ones from $R_g/\cosh \eta_J$ and those in $\qcut$. With this change of variables the same result for $S_c^\kappa$ as in \eq{ScLargeRg} applies for the $pp$ case.
The result for the factorization formula in the large $R_g$ regime for $pp$ collisions is then given by
\begin{align}
\label{eq:FactLargeThetaCSpp}
\frac{d \Sigma(R_g)}{d m_J^2d\Phi_J} &=
\sum_{\kappa = q,g}
N_\kappa(\Phi_J, R, \mu)\,
S_G^\kappa\bigl(p_T \zcut' R^{1+\beta}, \beta, \mu\bigr) \, \Xi_{G}^{\kappa}(t_G)
\\
&\qquad \times
\,(p_T\zcut')^{\frac{1}{1+\beta}}
\! \int \!d r^+ \: J_\kappa (m_J^2 - p_T r^+, \mu) \,
S_c^\kappa
\Bigl[
r^+ (p_T\zcut')^{\frac{1}{1+\beta}},R_g(p_T\zcut')^{\frac{1}{1+\beta}}, \beta, \mu \Bigr]
\,. \nn
\end{align}
Note that here we use the global-soft function result expanded in the small angle limit in \eq{SG1loopExpand}.
As a result for $pp$ there's no further dependence on $R$ apart from the one in combination with $p_T\zcut'$ as shown in the first argument of $S_G^\kappa$ in \eq{FactLargeThetaCSpp}.

\subsection{Intermediate groomed jet radius}
\label{sec:Intermediate}

We next turn to the case of groomed jet radius measurement in the intermediate regime, whose modes are depicted in the central panel of \fig{modePicture}.
Because here $R_{g} \ll \theta^\star_{g}$, the soft drop is now stopped by an emission with smaller angle (marked \CSg in yellow). The virtuality of these emissions is too low to contribute to the jet mass measurement; however, the groomed jet mass is affected by emissions with larger virtuality that see the groomed jet boundary (marked \CSm, shown in red).
We can think of these two modes as a refactorization of the collinear-soft, \CS, mode from the large groomed jet radius regime into \CSm and \CSg modes at the same angle $R_g$ with the following $(+,-,\perp)$ momentum scaling:
\begin{align}
&p_{\rm CS_m} \sim \frac{m_J^2}{Q (R_g/2)} \Big(\frac{R_g}{2}, \frac{2}{R_g}, 1\Big) \, ,&
&p_{\rm CS_g} \sim \qcut\Big( \frac{R_g}{2}\Big)^{1+\beta}\Big(\frac{R_g}{2}, \frac{2}{R_g}, 1\Big)\, &
&\hspace{-15pt}(\text{$\ee$ case})\, ,& \nn \\
&p_{\rm CS_m} \sim \frac{m_J^2}{p_T R_g} \Big(\frac{R_g}{2\cosh\eta_J}, \frac{2\cosh\eta_J}{R_g}, 1\Big) \, ,&
&p_{\rm CS_g} \sim p_T \zcut' R_g^{1+\beta}\Big(\frac{R_g}{2\cosh\eta_J}, \frac{2\cosh\eta_J}{R_g}, 1\Big)\, &
&(\text{$pp$ case}) \, .&
\end{align}
Given the scaling in these formulae, we see that both the resulting collinear-soft functions will give emissions at the same angle while being hierarchically separated in their energy. This is a similar situation to the one described in \cite{Becher:2015hka} with the result that the factorization for this process is complicated by the presence of NGLs.
Specifically, it leads to a multi-Wilson line structure for the matrix element of the \CSg modes, with a Wilson line for $each$ final state \CSm emission. We have already encountered for a similar hierarchy of energies between the global soft and the hard function in \eq{Gngl}, which accounted for wide angle emissions ($\theta \sim 1$) leading to a multi-Wilson line structure for the function $\widehat S_G^\kappa$ in \eq{Gngl}. Hence, following \eq{Fngl}, the factorization formula for this regime for $\ee$ collisions is given by
\begin{align} \label{eq:FactIntNGL}
\Sigma(m_J^2, R_g) &=
N_q(Q, R, \mu)\, S_G^q\Big(\qcut\tan^{1+\beta}\frac{R}{2}, \beta, \mu ,R \Big)
\, \Xi_{G}^{q}(t_G)
\\
&\qquad \times
Q \int_0^{m_J^2/Q} d \ell^c
J_q \big(m_J^2 - Q \ell^c, \mu \big)
\,\sum_k \widehat{\cS}_{c_m}^{q,k}\Big(\frac{\ell^c}{R_g/2}, \mu\Big)\otimes_{\Omega} \widehat S_{c_g}^{q,k}\!\Big(\frac{R_g}{2}\qcut^{\frac{1}{1+\beta}},\beta, \mu\Big)
\nn \,.
\end{align}
The hard, global-soft, and jet functions are the same as in \eq{FactLargeThetaCS}.
Consistent with the mode picture, we now find two collinear-soft functions, where the subscripts $c_g$ and $c_m$ specify the modes that are responsible for stopping the groomer and contributing to the mass measurement, respectively.
Note that we have written the factorization formula for the cumulant of the jet mass cross section. It was shown in Ref.~\cite{Dasgupta:2001sh} that NGLs in the presence of a jet mass measurement can be described by a multiplicative function in cumulant space.
Here, as in \eq{Gngl}, $\otimes_{\Omega}$ indicates that the angular integrals in two functions involved cannot be done independently and the sum $k$ in the convolution sums over contributions from various \CSm partons. However, unlike the NGLs in the global soft function, the NGLs associated with the $\widehat{\cS}^q_{c_m},\widehat S^q_{c_g}$ functions do affect the measured quantities $R_g$ and $m_J$ and hence must be more carefully accounted for. To do this we write
\begin{align}
\label{eq:NGLInt}
& \sum_k \widehat{\cS}_{c_m}^{q,k}\Big(\frac{\ell^c}{R_g/2}, \mu\Big)\otimes_{\Omega} \widehat S_{c_g}^{q,k}\!\Big(\frac{R_g}{2}\qcut^{\frac{1}{1+\beta}},\beta, \mu\Big)
\\
&\quad
= {\cS}_{c_m}^{q}\Big(\frac{\ell^c}{R_g/2}, \mu\Big)\:
S_{c_g}^{q}\!\Big(\frac{R_g}{2}\qcut^{\frac{1}{1+\beta}},\beta, \mu\Big)
\:
\Xi_S^q\Bigg(t \Big[ \frac{R_g}{2}\qcut^{\frac{1}{1+\beta}}, \frac{\ell^c}{R_g/2}\Big ]
\Bigg)
\,,\nn
\end{align}
where the functions ${\cS}_{c_m}^{q},S_{c_g}^q$ are obtained by doing independent solid angle integration over each function and the NGLs are included in the function $\Xi_S^q$, and the function $t$ was defined in \eq{tNGL}.
The double differential cross section is given by
\begin{align} \label{eq:FactIntThetaCS}
\Sigma(m_J^2, R_g) &=
N_q(Q, R, \mu)\,
S_G^q\Big(\qcut\tan^{1+\beta}\frac{R}{2}, \beta, \mu ,R\Big) \, \Xi_{G}^{q}(t_G)\,
S_{c_g}^q\!\Big(\frac{R_g}{2}\qcut^{\frac{1}{1+\beta}},\beta, \mu\Big)
\\
&\ \times
\Xi_S^q (t_S)\:
Q \int_0^{m_J^2/Q} d \ell^c
J_q \big(m_J^2 - Q \ell^c, \mu \big)
\, {\cS}_{c_m}^{q}\Big(\frac{\ell^c}{R_g/2}, \mu\Big)\,
\nn .
\end{align}
Here the resummation of the non-global logarithms is carried out independently of the global-log resummation, and hence the NGLs are included multiplicatively as shown. The variable $t_{S}$ is defined by
\begin{align}\label{eq:tSdef}
t_S = t[\mu_{cs_g},\mu_{cs_m}] \, ,
\end{align}
where the expressions for canonical values of $\mu_{cs_g}$ and $\mu_{cs_m}$ scales are given in \eqs{naturalScales}{moreNaturalScales}.
A detailed calculation of the lowest order result for $\Xi_S^q$ is presented in \app{NGLcluster}, yielding
\begin{align} \label{eq:NGLFull}
\Xi_S^q(t_S)	\,= \, 1- \frac{4}{9}C_FC_A \left(\frac{\alpha_s}{2\pi}\right)^2\frac{\pi^2}{3}\ln^2\bigg( \Big(\frac{\theta^\star_g}{R_g}\Big)^{2+\beta} \bigg) + \ldots
\,,
\end{align}
while the calculation of the leading abelian clustering logarithms is presented in \app{Abel}.
We discuss in \sec{resum} our approach to including the NGLs which includes terms beyond the one shown explicitly in \eq{NGLFull}.

The extension to the $pp$ case is straightforward, with the factorization formula now given by
\begin{align} \label{eq:ppFactIntThetaCS}
\frac{d \Sigma(R_g)}{d m_J^2d\Phi_J} &=
\sum_{\kappa = q,g}
N_\kappa(\Phi_J, R, \mu)\,
S_G^\kappa(p_T \zcut' R^{1+\beta}, \beta, \mu) \, \Xi_{G}^{\kappa}(t_G)
\,S_{c_g}^\kappa\!\big(R_g (p_T \zcut')^{\frac{1}{1+\beta}} ,\beta, \mu\big) \nn
\\
& \times
\Xi_S^\kappa(t_S)
p_T \int_0^{m_J^2/p_T} d r^c
J_\kappa \big(m_J^2 - p_T r^c, \mu \big)
\, {\cS}_{c_m}^{\kappa}\Big(\frac{r^c}{R_g}, \mu\Big)\,
\, .
\end{align}
Here we have made use of the change of variables described in \eq{dictionary} to recycle all the results for $\ee$ case. Additionally, the argument of ${\cS}_{c_m}^\kappa$, in \eq{FactIntThetaCS} is changed as $2\ell^c/R_g \ra r^c/R_g$. The factors of $\cosh \eta_J$ implicit in the definition of $r^c$ above cancel against those from $R_g/\cosh \eta_J$ and in $Q$. The expressions for natural scales for $\mu_{cs_g}$ and $\mu_{cs_m}$ for $pp$ case
are given in \eqs{naturalScalespp}{muCSmPPNatural}.

Next, the relation between the differential collinear-soft function $S_{c_m}^\kappa$ and the cumulative function ${\cS}_{c_m}^\kappa$ is given by
\begin{align}
{\cS}_{c_m}^\kappa (\ell^c , \mu ) = \int_0^{\ell^c} d \ell^+ \: S_{c_m}^\kappa (\ell^+, \mu ) \, ,
\end{align}
such that, at one loop the results for the two collinear-soft functions in \eq{FactIntThetaCS} are given by
\begin{align}
\label{eq:CSintOneLoop}
S_{c_g}^{\kappa}\Big(\frac{R_g}{2}\qcut^{\frac{1}{1+\beta}},\beta, \mu\Big) &=
1 - \frac{\alpha_s C_\kappa}{2 \pi}\frac{1}{1+\beta} \bigg[\frac{1}{2} \log^2 \bigg(\frac{\mu^2}{\qcut^2 (R_g/2)^{2(1+\beta)}}\bigg) - \frac{\pi^2}{12}\bigg] \, , \\
S_{c_m}^\kappa \Big(\frac{2\ell^+}{R_g}, \mu\Big) &=
\delta\Big(\frac{2\ell^+}{R_g}\Big) + \frac{\alpha_s C_\kappa}{\pi} \bigg[\frac{-2}{\mu} {\cal L}_1\bigg(\frac{1}{\mu}\frac{2\ell^+}{R_g}\bigg) + \frac{\pi^2}{24} \delta \Big(\frac{2\ell^+}{R_g}\Big)\bigg] \,.
\nn
\end{align}
Here we define the plus functions
\begin{align}
{\cal L}_ n (x ) \equiv \Big[\frac{\Theta(x)\ln^n x }{x} \Big]_+ \,
\end{align}
to integrate to zero over the interval $x \in [0, 1]$.
The result in \eq{CSintOneLoop} for $S_{c_g}^{\kappa}$ was presented in \cite{Kang:2019prh}, while the calculation for $S^\kappa_{c_m}$ is analogous to that of jet mass (or jet angularities) with a specified jet radius $R$, here replaced by $R_g$~\cite{Ellis:2010rwa}.
Expanding away terms that are needed for a large $R_g$ but are power corrections for this intermediate $R_g$ region, we have the following relation between functions in the large and intermediate $R_g$ regions
\begin{align}
\label{eq:CSrefactorization}
S_c^\kappa
\Bigl (
\ell^+ \qcut^{\frac{1}{1+\beta}}, \frac{R_g}{2}\qcut^{\frac{1}{1+\beta}}, \beta, \mu
\Bigr )=
S_{c_g}^\kappa\!\Big(\frac{R_g}{2}\qcut^{\frac{1}{1+\beta}},\beta, \mu\Big)\,
S_{c_m}^\kappa\Big(\frac{2\ell^+}{R_g}, \mu\Big)\,
\bigg[1 +\mathcal{O}\Big(\frac{R_g^{2+\beta}\qcut}{\ell^+}\Big) \bigg]\,.
\end{align}
This refactorization can be checked from \eq{CSintOneLoop} by making a comparison with with \eq{ScLargeRg}.

\subsection{Small groomed jet radius}
\label{sec:Small}

Last, we consider the regime $\theta_c \lesssim R_g \ll \theta_g^\star $, where the groomed jet radius is set by the smallest opening angle compatible with the jet mass measurement. Here the collinear-soft radiation responsible for stopping grooming \CSg has an angle comparable to the more energetic collinear emissions but does not contribute to the jet mass measurement.
The modes \CSm and \C thus collapse into a single mode that has the following scaling set by the groomed jet radius:
\begin{align}
&p_{\rm C} \sim \frac{QR_g}{2} \Big(\frac{R_g}{2}, \frac{2}{R_g}, 1 \Big)
\sim \Bigl( \frac{m_J^2}{Q}, Q, m_J \Bigr)
\, &
&(\text{$\ee$ case})\,,&
\\
&p_{\rm C} \sim p_T R_g \bigg(\frac{R_g}{2\cosh \eta_J}, \frac{2\cosh \eta_J}{R_g}, 1\bigg)
\sim \Bigl( \frac{m_J^2}{Q}, Q, m_J \Bigr)
\, &
&(\text{$pp$ case})\, .&
\nn
\end{align}
The rightmost expressions show that in the small $R_g$ region, the scaling of the mode is equivalent to the momentum scaling of the collinear mode in \eq{pC}.
For this case the factorization formula for the cross section for $\ee$ collisions reads
\begin{align}
\label{eq:FactSmallThetaCS}
\frac{d \Sigma(R_g)}{d m_J^2} &=
\frac{1}{\big(Q\frac{R_g}{2}\big)^2}
N_q(Q, R, \mu)\,
S_G^q\Big(\qcut\tan^{1+\beta}\frac{R}{2}, \beta, \mu ,R\Big)
\, \Xi_{G}^{q}(t_G)\,\\
&\qquad \times \sum_{k} \widehat S_{c_g}^{q,k} \Big(\frac{R_g}{2}\qcut^{\frac{1}{1+\beta}}, \beta, \mu\Big)\,
\otimes_{\Omega} \widehat {\cal C}_k^q\bigg [\frac{4m_J^2}{Q^2R_g^2}, \frac{QR_g}{2}, \mu\bigg ]
\nn \\
&=
\frac{1}{\big(Q\frac{R_g}{2}\big)^2}
N_q(Q, R, \mu)\, \nn
S_G^q\Big(\qcut\tan^{1+\beta}\frac{R}{2} ,R, \beta, \mu \Big)
\, \Xi_{G}^{q}(t_G)\,\\
&\qquad \times S_{c_g}^{q} \Big(\frac{R_g}{2}\qcut^{\frac{1}{1+\beta}}, \beta, \mu\Big)\,
{\cal C}^q\bigg [\frac{4m_J^2}{Q^2R_g^2}, \frac{QR_g}{2}, \mu\bigg ]
\:
\Xi_C^q(t_C)
\,. \nn
\end{align}
Following the same argument that lead to \eq{FactIntNGL} for the case of intermediate groomed jet radius, the factorization formula given here also contains NGLs associated with the function pair $\widehat S_{c_g},\widehat {\cal C}^q$ and the function pair $\widehat S_G, \widehat N_q$ which each encode emissions at two different sets of angles, $R$ and $R_g$ respectively, but are hierarchically separated in energy.
As for the intermediate case, we can write the factorization in terms of the angular averaged functions and a function $\Xi_C^q(t_C)$. The calculation for $\Xi_C^q$ is very analogous to that of $\Xi_S^q$ in \eq{NGLInt} and is discussed in \app{NGLcluster}. Here the variable $t_C$ is given by
\begin{align}\label{eq:tCdef}
t_C = t[\mu_{cs_g}, \mu_{c}] \, .
\end{align}
From expressions of the canonical scales $\mu_{cs_g}$ and $\mu_c$ in \eq{naturalScales} we see that $t_C$ depends only on $R_g$. In this regime the ratio $4m_J^2/(Q^2R_g^2) \sim 1$ and hence the NGLs involving jet mass can be treated as subleading.
The jet mass dependent terms are included in a fixed order expansion in the ${\cal C}^q$ function.

In this EFT regime, power corrections in the ratio $\theta_c/R_g = 2m_J/(QR_g)$ cannot be neglected, and are captured by the new dimensionless collinear function ${\cal C}^q [4m_J^2/(Q^2R_g^2), QR_g/2, \mu]$ in \eq{FactSmallThetaCS}, which describes energetic emissions that set the groomed jet mass. The collinear-soft radiation with similar opening angle, but much lower energy, responsible for stopping grooming, is still described by the same $S_{c_g}^q$ function as in the regime of intermediate groomed jet radius. Consistency of factorization with the case of intermediate groomed radius requires
\begin{align}\label{eq:CFact}
\frac{1}{\big(Q\frac{R_g}{2}\big)^2} {\cal C}^q\bigg [\frac{4m_J^2}{Q^2R_g^2}, \frac{QR_g}{2}, \mu\bigg ] =
\int \frac{d \ell^+}{R_g/2}
J_q \big(m_J^2 - Q \ell^+, \mu \big)
\, S_{c_m}^q\Big(\frac{\ell^+}{R_g/2}, \mu\Big)\,
\bigg[1 +\mathcal{O}\Big(\frac{4m^2_J}{Q^2R^2_g}\Big) \bigg]\,.
\end{align}
We have computed this new jet function at ${\cal O}(\alpha_s)$, finding for quarks
\begin{align} \label{eq:Cq}
\mathcal{C}^q\Big[\frac{4m_J^2}{Q^2R_g^2},\frac{QR_g}{2},\mu\Big] & =
\delta\Big(\frac{4m_J^2}{R_g^2Q^2}\Big)
+ \frac{\alpha_s C_F}{2\pi} \Bigg\{
\delta\Big(\frac{4m_J^2}{Q^2R_g^2}\Big)
\bigg( \frac{1}{2} \ln^2 \frac{4\mu^2}{Q^2 R_g^2} + \frac{3}{2}\ln\frac{4\mu^2}{Q^2R_g^2}+\frac{7}{2}-\frac{5\pi^2}{12}\bigg)
\nonumber \\
& +\theta(QR_g-4m_J)\mathcal{L}_0\Big(\frac{4m_J^2}{Q^2R_g^2}\Big)
\bigg[4\,\ln\bigg(\frac{1}{2}+\frac{1}{2}\sqrt{1-\frac{16m_J^2}{Q^2R_g^2}}\bigg)-\frac{3}{2}\sqrt{1-\frac{16m_J^2}{Q^2R_g^2}}\,\,\bigg]
\nonumber\\
&-2\theta(QR_g-4m_J)\mathcal{L}_1\Big(\frac{4m_J^2}{Q^2R_g^2}\Big)\Bigg\}\,.
\end{align}
This result provides an explicit check on the refactorization in \eq{CFact} when expanding away power corrections in the ratio $m_J/(QR_g)$. Note that $\mathcal{C}^q$ depends on the renormalization scale $\mu$ only through the Dirac delta function term in the first line of \eq{Cq}. Therefore, as expected by RG consistency
with the remaining $m_J$ independent functions in \eq{FactSmallThetaCS},
the anomalous dimension for $\mathcal{C}^q$ is independent of the groomed jet mass.
Since in this small $R_g$ regime we have $4 m_J^2/(Q^2 R_g^2) = \theta_c^2/R_g^2\sim 1$, the terms in $\mathcal{C}^q$ with more non-trivial $m_J$ dependence do not involve potentially large logarithms.
We also note that these non $\delta$-function terms in \eq{Cq} vanish for $R_g < 2\theta_c$ (rather than $\theta_c$, as expected from \eq{angularBounds}). This is a one-loop accident due to the single-splitting geometry of the event at this order, so the phase-space region $\theta_c<R_g<2\theta_c$ will be filled by higher-order corrections and RG evolution.

The extension of the factorization formula in \eq{FactSmallThetaCS} to $pp$ collisions is given by
\begin{align}
\label{eq:ppFactSmallThetaCS}
\frac{d \Sigma(R_g)}{d m_J^2d\Phi_J} &=
\frac{1}{(p_T R_g)^2}
\sum_{\kappa = q,g}
N_\kappa(\Phi_J, R, \mu)\,
S_G^\kappa(p_T \zcut' R^{1+\beta}, \beta, \mu) \, \Xi_{G}^{\kappa}(t_G)\,\\
&\qquad \times
S_{c_g}^\kappa\!\big(R_g (p_T \zcut')^{\frac{1}{1+\beta}} ,\beta, \mu\big)\,
{\cal C}^\kappa\bigg [\frac{m_J^2}{p_T^2R_g^2}, p_T R_g, \mu\bigg ]\,
\Xi_C^\kappa(t_C)
\,. \nn
\end{align}
Here as before we utilize \eq{dictionary} and substitute $4m_J^2/(QR_g)^2 \ra m_J^2/(p_TR_g)^2,\,QR_g/2 \ra p_TR_g$ in the arguments of the collinear function ${\cal C}^\kappa$. The $t_C$ argument in $\Xi_C$ involves the canonical values of the $\mu_{cs_g}$ and $\mu_{C}$ scales provided in \eq{naturalScalespp}.

\subsection{Implementation of moments $M_1^q$ and $\Mbq$}
\label{sec:C1calc}

Having discussed the factorization for the double differential cross section $d\Sigma(R_g)/(dm_J^2d\Phi_J)$ in the three regimes we now turn to how we use this cross section to determine the moments $M_1^q$ and $\Mbq$. Here we focus on the quark case relevant for $\ee$ collisions in the dijet limit.

According to the definition in \eq{C1C2multiDiffDef}, the moment $M_1^\kappa$ (related to the coefficient $C_1^\kappa$ of the shift power correction), simply requires taking the normalized first angular moment of the double differential distribution.
Explicitly,
\begin{align} \label{eq:C1intermediateExplicit}
M_1^q(m_J^2) =\bigg[\int_{\theta_\Min}^{\theta_\Max}
\!\!\!\!\!\! d\theta_{g}\, \frac{\theta_{g}}{2}\,
\bigg(\frac{d}{dR_g}\frac{d \Sigma(R_g)}{d m_J^2 }\bigg)_{R_g=\theta_g}\bigg]
\bigg/
\int_{\theta_\Min}^{\theta_\Max}
\!\!\!\!\!\! d\theta_{g}\,
\bigg(\frac{d}{dR_g}\frac{d \Sigma(R_g)}{d m_J^2 }\bigg)_{R_g=\theta_g}\,.
\end{align}
We can remove the need to explicitly take the $R_g$ derivative in \eq{C1intermediateExplicit} by integrating by parts, such that
\begin{align} \label{eq:IBPC1}
M_1^q(m_J^2) =\bigg[\frac{\theta_\Max}{2}\,
\frac{d \Sigma^q(\theta_\Max)}{d m_J^2 }-\frac{1}{2}\int_{\theta_\Min}^{\theta_\Max}
\!\!\!\!\!\! d\theta_{g}
\frac{d \Sigma^q(\theta_g)}{d m_J^2 }\bigg]
\bigg/
\frac{d \sigma^q}{d m_J^2}\,.
\end{align}
We observe that the prefactor $N^q_{\rm evol}$ in \eq{quarkCrossSection} cancels out in the ratio shown in \eq{IBPC1}, since it is independent of the groomed jet radius.
Note that despite $C_1^q$ being determined by contributions from the large $R_g$ region, we are free to use the cross section which interpolates smoothly between all regions of $R_g$ here. This is because when we integrate this moment starting from the lower limit $\theta_{\rm min}$ contributions from the small $R_g$ or intermediate $R_g$ regions are power suppressed.
We will describe the resummation of the large logarithms of jet mass and groomed jet radius present in the double differential distribution in three regimes in the next section, and combine the results from each regime in \sec{match} to obtain a consistent description across the entire $m_J$-$R_g$ phase space of interest. This result will then be used to calculate the moment $M_1^q(m_J^2)$ via \eq{IBPC1} and will be interpreted as a result for $C_1^q(m_J^2)$ at leading power.

Next we turn to the moment $\Mbq$ that is related to the coefficient $C_2^\kappa(m_J^2)$ which is the Wilson coefficient for the power corrections at the boundary of the groomed region and hence is evaluated when the soft drop condition is \textit{just} satisfied. This was defined above in \eq{C1C2multiDiffDef} using the soft drop condition with a small shift in the threshold energy cut by $\veps$ in \eq{ShiftSD}.
The derivative with respect to $\veps$ evaluated at $\veps =0$ allows us to to implement the $\delta$ function in \eq{C1C2def} as can be seen by expanding the shifted measurement in \eq{ShiftSD} to first order in $\veps$.
This is the effective soft drop condition that we will implement on our final state. For the purpose of considering the renormalization group consistency, it is simplest to first do the resummation, and only take the derivative with respect to $\veps$ after obtaining the RG evolved functions.

In contrast to $M_1^\kappa(m_J^2)$, since the computation of $\Mb(m_J^2)$ involves an an average value of the inverse groomed jet radius it appears to be more sensitive to the small angle regime of the $\theta_g$ spectrum.
However, as part of the definition in \eq{C1C2multiDiffDef} we also include the restriction to the large $R_g$ region ($\theta_g\sim\theta_g^*$), since as discussed earlier, the non-perturbative parameters and Wilson coefficient $C_{2}^\kappa$ encode the geometry associated to this region as part of their intrinsic definitions.
Therefore to compute the partonic coefficient $C_2^\kappa$, we must restrict ourselves solely to the region where large $R_g$ regime contributes.
We will describe in \sec{trans} how this is accomplished.
Note that this is also consistent with the fact that the nonperturbative brown shaded region in \fig{phaseSpace} should not contribute to the perturbative Wilson coefficient $C_2^\kappa$. From the mode analysis in \fig{modePicture} we saw that the large $R_g$ and the small $R_g$ regimes constitute the upper and lower boundaries of the $R_g$ values with intermediate $R_g$ filling the bulk, so the large $R_g$ region is separated from the NP region shaded in brown. Finally, we will show in \sec{LLCB} that taking LL limit of our calculation of the soft drop boundary cross section in the large $R_g$ regime derived in \sec{LargeRgShift} reproduces the LL result for $C_2^\kappa$ from \Ref{Hoang:2019ceu}.

\section{Resummation}
\label{sec:resum}

Having setup the EFT factorization in the three regions, with the formulae in Eqs.~(\ref{eq:FactLargeThetaCS}, \ref{eq:FactIntThetaCS}, \ref{eq:FactSmallThetaCS}), we next implement resummation of large logarithms by exploiting these equations.
The resummation in SCET is carried out via a UV renormalization in the EFT and a subsequent renormalization group evolution (RGE) of these matrix elements. The RGE evolves these matrix elements from their natural scales $\mu_i$ where logarithms are minimized to a common final scale $\mu$ and results in resummations of large logarithms of ratios of the $\mu_i$ scales.
The resummed cross sections will then allow us to extract the moments $M_1^\kappa(m_J^2)$ and $\Mb (m_J^2)$ introduced in \eq{C1C2multiDiffDef}.

As we mentioned above in \sec{Intro} we will carry out the resummation at NLL$'$ accuracy, where in addition to the NLL resummation, we will include NLO fixed order corrections to the perturbative functions in the factorization formulae in Eqs.~(\ref{eq:FactLargeThetaCS}), (\ref{eq:FactIntThetaCS}) and (\ref{eq:FactSmallThetaCS}).
In \tab{RGE} we summarize the orders to which the cusp anomalous dimension $\Gamma^{\rm cusp}[\alpha_s]$, various non-cusp anomalous dimensions $\gamma_i[\alpha_s]$, and $\beta$ function are needed to achieve LL, NLL and NLL$'$ accuracy. We collect expressions for the evolution kernels, anomalous dimensions, and details of evolution in \app{formulae}.
In the intermediate $R_g$ regime, the $R_g$ dependence appears directly through the evolution scales, and there are no further complications.
However, in the large $R_g$ regime we see from \eq{ScLargeRg} that the $R_g$-dependence is only seen at one loop in the collinear-soft function through a fixed-order term.
Thus in the large $R_g$ regime we must include additional $R_g$ dependent fixed order terms as indicated in the next to last column of \tab{RGE}. Similarly, in the small $R_g$ regime, the measurement of the jet mass appears as a fixed order correction on top of the $R_g$ measurement. As we can see from \eq{Cq}, the leading order (LO) distribution for $m_J >0$ starts at one-loop. Thus, in our NLL$'$ treatment we will include additional fixed order terms relevant for these regimes, as indicated by entries in the last two columns of \tab{RGE}.

Here a generic function ${\cal F}$ in the factorization theorem is assumed to have a fixed order expansion of the following form:
\begin{align}
{\cal F}(x,\mu) = \sum_{m = 0} \sum_{n = 1}^{2m} a_{mn}^{\cal F}\Big(\frac{\alpha_s}{4\pi}\Big)^m L_{\cal F}^n \, ,
\end{align}
where $L_{\cal F}$ are logarithms (transformed to Laplace space, $m_J^2\to x$, see \app{formulae} for details) and the $a_{mn}^{\cal F}$ are coefficients which may contain non-logarithmic dependence on the variables.
For the large $R_g$ factorization formula, the collinear-soft function $S_c	\bigl[ \ell^+ \qcut^{\frac{1}{1+\beta}}, \frac{R_g}{2}\qcut^{\frac{1}{1+\beta}}, \beta, \mu \bigr]$ receives a one-loop $R_g$ dependent correction that does not involve logarithms of $\mu$, which we indicate by $\alpha_s a_{10}^{S_c}\big(R_g/\theta_g^\star\big)$. This one-loop term constitutes the LO distribution of $R_g$ in this regime and depends on $R_g$ through the dimensionless ratio $R_g/\theta_g^\star$, where $\theta_g^\star$ was defined above in \eq{angularBoundsExplicit}. At NLL$'$ order one would additionally
add the $R_g$ dependent ${\cal O}(\alpha_s^2)$ terms in the large $R_g$ regime, in order to include the dependence on $R_g$ to one higher order. In our analysis we only include the ${\cal O}(\alpha_s^2)$ terms indicated in \tab{RGE} which are determined by lower order anomalous dimensions and the one-loop constant terms, which omits the currently unknown ${\cal O}(\alpha_s^2)$ $R_g$-dependent term in $S_c^\kappa$ (we will estimate the impact of this term in our numerical uncertainty analysis). Finally, as $R_g \ra 0$, the ratio $R_g/\theta_g^\star$ becomes a power correction leading to factorization in the intermediate regime, where the standard treatment of NLL$'$ counting applies.

A similar reasoning applies to the regime of small $R_g$, where the collinear function ${\cal C}^q$ defined in \eq{Cq} receives one-loop, non-logarithmic corrections depending on $m_J$. As mentioned above, for $m_J > 0$, this is the LO distribution in the small-$R_g$ regime, which we denote by $\alpha_s a_{10}^{\cal C}(\theta_c^2/R_g^2)$, with $\theta_c$ defined above in \eq{angularBoundsExplicit}. We include the additional ${\cal O}(\alpha_s^2)$ fixed-order terms in this regime at NLL$'$ in the same multiplicative way, as summarized in the table. Thus we note that the treatment of terms is slightly different in the three regimes. We will further justify our specific approach through the smooth matching between regimes as well as parametrize the uncertainty due to the unknown non-logarithmic 2-loop pieces in \sec{match}. Furthermore, we note that, up to NLL$'$ accuracy, we do not need to consider ${\cal O}(\alpha_s^2)$ corrections to the $\mu_i$-dependent SCET functions parametrized by ${\cal F}$ in \tab{RGE} (such as the jet, collinear-soft and global-soft functions) as their inclusion with ${\cal O}(\alpha_s)$ LO distributions in the small or large $R_g$ regimes will only enter at ${\cal O}(\alpha_s^3)$.

\begin{table}[t!]
\begin{center}
\begin{tabular}{ c | c c c c c c}
& $\!\!\Gamma^{\rm cusp}
\!\!$
& $\!\!\gamma \!\!$
& $\!\!\beta (\alpha_s)\!\!$
& $\alpha_s^m L_{\cal F}^{n\geq 0}$
& Large $R_g$ additions
& Small $R_g$ additions
\\[0.5ex]
\hline \\[-12pt]
LL 	& $\alpha_s$ &		- 	& $\alpha_s$
& - & 	$+\alpha_s \, a_{10}^{S_c}\big(\frac{R_g}{\theta_g^\star}\big) $
&	$+\alpha_s \, a_{10}^{\cal C}\big(\frac{\theta_c^2}{R_g^2}\big) $
\\[5pt]
NLL 	 & $\alpha_s^2$ & 	$\alpha_s$ 	 & $\alpha_s^2$
& - &
$+\alpha_s \, a_{10}^{S_c}\big(\frac{R_g}{\theta_g^\star}\big)$
&
$+\alpha_s \, a_{10}^{\cal C}\big(\frac{\theta_c^2}{R_g^2}\big)$
\\[5pt]
NLL$'$	 & $\alpha_s^2$ & 	$\alpha_s$ 	 & $\alpha_s^2$
&\;$\alpha_s\,${\small $\displaystyle{ \sum_{j=0}^2}$ } $\!\!\! a_{1j}^{\cal F} L_{\cal F}^j$
&\; $+\alpha_s^2\, a_{10}^{S_c}\big(\frac{R_g}{\theta_g^\star}\big) \displaystyle{\sum_{j=0}^2} a_{1j}^{\cal F} L_{\cal F}^j$
&\;$+\alpha_s^2\, a_{10}^{\cal C}\big(\frac{\theta_c^2}{R_g^2}\big) \displaystyle{\sum_{j=0}^2}a_{1j}^{\cal F} L_{\cal F}^j$
\\
\end{tabular}
\end{center}
\caption{Highest order of the anomalous dimensions and fixed order terms that are needed for the given accuracy. The last two columns indicate additional fixed order terms that are added in the large $R_g$ and small $R_g$ regime respectively. }
\label{tab:RGE}
\end{table}

Lastly, we note that the double differential cross section is affected by non-global logarithms,
which require resummation for a complete treatment of the double differential cross section in all regions.
In \sec{factAll} we have shown how the leading NGLs affect the factorization formulae for our cross section in each of the regimes, giving rise to the functions $\Xi_G^q$, $\Xi_S^q$, and $\Xi_C^q$.
First of all, since we are interested in the moments $C_1^q$ and $C_2^q$, the NGLs related to $\Xi_G$ in the normalization cancel out as described in \sec{C1calc}.
Next, recall that we will focus on the case where the approximation $R_g/2 \ll1 $ can be assumed.
For the choice of kinematic parameters that we explore in \sec{Num} we will find that the variables $t_S$ and $t_C$ in \eqs{tSdef}{tCdef} are at most 0.2 across the allowed range of $R_g$ for $m_J^2$ in the SDOE region.
As was shown in \Ref{Kang:2019prh}, the result for LL large-$N_c$ resummed NGLs for $R_g/2\ll 1$ can be well approximated by the two-loop fixed order term when written in terms of the variable $t$, for $t \lesssim 0.3$ (see Fig.~4 in \Ref{Kang:2019prh}).
Hence, the two loop term will suffice for our purposes, and we do not need to carry out a Monte Carlo LL resummation of the NGLs in our analysis.

Below in sections~\ref{sec:smallResum}, \ref{sec:resummation}, \ref{sec:Small} we therefore explain how to carry out the resummation of global logarithms associated to the scales appearing in the remaining functions in the factorization formulae.
In the following, we first discuss the simpler cases of resummation in the small and intermediate $R_g$ regimes and set up the relevant notation. Then we turn to the large $R_g$ region, where the interplay between the LO term depending on $R_g/\theta^\star_g$ and RG evolution requires more complicated calculations with Laplace transforms. Finally, we describe the resummation for the boundary corrections relevant to the calculation of $C_2^q(m_J^2)$, where we must account for the shifted soft drop condition in \eq{ShiftSD}.

\subsection{Small $R_g$}
\label{sec:smallResum}

Resummation takes on the simplest form in the regime of small $R_g$, where the factorization structure is purely multiplicative. As written in \eq{FactSmallThetaCS}, each function depends on a common and arbitrary renormalization scale $\mu$. Any one specific choice of $\mu$ induces in the fixed-order ingredients large logarithms of the ratio of widely separated scales (indicated by the function arguments), which can be minimized by choosing instead using a different scale $\mu_i$ for each function. These scales are then connected to a common scale $\mu$ via renormalization group (RG) running, which is how the resummation of large logarithms is carried out in SCET.

Including the appropriate resummation factors, as discussed in \app{transforms}, the RG-evolved cross section in the small $R_g$ regime for $\ee$ collisions reads
\begin{align}
\label{eq:xsecRGsmallRg}
&\frac{d \Sigma(R_g)}{d(\ln\, m_J^2) }
= \frac{m_J^2}{\big(Q\frac{R_g}{2}\big)^2}\,
N_q^{\rm evol}(Q, \qcut,\beta, R, \mu ; \mu_N, \mu_{gs})\:
e^{\big [K_{cs_g} (\mu, \mu_{cs_g} ) + K_{c} (\mu, \mu_c )\big] }
\\
& \times
\Big(\frac{\mu_{cs_g}}{\qcut (\frac{R_g}{2})^{1+\beta} }\Big)^{\omega_{cs_g} (\mu, \mu_{cs_g})}
\Big(\frac{\mu_c}{Q\frac{R_g}{2}}\Big)^{\omega_c(\mu, \mu_c)}
S_{c_g}^q\!\Big [\frac{R_g}{2} \qcut^{\frac{1}{1+\beta}},\mu_{cs_g}\Big]
{\cal C}^q\bigg [\frac{m_J^2}{Q^2R_g^2}, QR_g, \mu\bigg ]
\Xi_C^q(t_C)
\,,\nn
\end{align}
where the dimensionless prefactor $N_q^{\rm evol}$ is given by
\begin{align}
\label{eq:HSGevol}
N_q^{\rm evol}(Q, \qcut, & \beta,R, \mu ; \mu_N, \mu_{gs})
\equiv N^q(Q, R, \mu_Q) \,
S_G^q\Big(\qcut\tan^{1+\beta}\frac{R}{2},R, \beta, \mu_{gs}\Big)
\,\Xi_G^q(t_G)
\\
&\qquad \times
e^{K_{N^q}(\mu, \mu_N) +K_{gs}(\mu, \mu_{gs}) }
\Big(\frac{\mu_N}{Q\tan\frac{R}{2}}\Big)^{\omega_{N^q}(\mu, \mu_N) } \Big(\frac{\mu_{gs}}{\qcut\tan^{1+\beta}\frac{R}{2}}\Big)^{\omega_{gs}(\mu, \mu_{gs})} \, .
\nn
\end{align}
The resummation kernels are defined in \eq{KOmega}. For compactness, in the subscripts of scales $\mu_X$ and kernels $K_X$ and $\omega_X$, we use the shorthand notation $gs$, $cs_g$ and $c$ for the symbols $S_G^q$, $S_{c_g}^q$ and $\mathcal{C}^q$ respectively. The functions are RG evolved from their respective scales $\mu_i$ to a common scale $\mu$, where natural scales for $\ee$ case are
\begin{align}
\label{eq:naturalScales}
\mu_N^{\rm can.} = Q \tan\frac{R}{2}
\,,\:\:\,\;\;
\mu_{gs}^{\rm can.} = \qcut \tan^{1+\beta}\frac{R}{2}
\,,\:\:\,\;\;
\mu_{cs_{g}}^{\rm can.} = \qcut \Big(\frac{R_g}{2}\Big)^{1+\beta}
\,,\:\:\,\;\;
\mu_{c}^{\rm can.} =Q \frac{R_g}{2}\,.
\end{align}
Here the superscript `can.' denotes the canonical choice of the scale.

The function $\mathcal{C}^\kappa$ that appears in \eq{xsecRGsmallRg} is a generalization of \eq{Cq} which includes two-loop terms, given by
\begin{align} \label{eq:CqFiniteMJ}
\frac{m_J^2}{\big(Q\frac{R_g}{2}\big)^2}\, {\cal C}^q\bigg [\frac{4m_J^2}{Q^2R_g^2}, \frac{QR_g}{2}, \mu\bigg ] &\equiv
\frac{\alpha_sC_F}{\pi} a_{10}^{\cal C}\Big (\frac{\theta_c^2}{R_g^2}\Big)
\Bigg[
1 + \frac{\alpha_sC_F}{\pi} \bigg(\frac{1}{4} \ln^2 \frac{4\mu^2}{Q^2 R_g^2} + \frac{3}{4}\ln\frac{4\mu^2}{Q^2R_g^2}
\bigg)
\\
&\qquad+ \frac{\alpha_s \beta_0}{4\pi} \ln \frac{4\mu^2}{Q^2R_g^2}\Bigg] + \Big(\frac{\alpha_s}{\pi} a_{20}^{\cal C} \Big) \frac{\alpha_sC_F}{\pi}a_{10}^{\cal C}\Big (\frac{64}{25}\frac{(\theta_c/2)^2}{R_g^2}\Big) \,
+{\cal O}(\alpha_s^2)
\nn
\, .
\end{align}
Following the notation in \tab{RGE} we have identified the LO small $R_g$ distribution for non-zero jet masses as
\begin{align} \label{eq:a10C}
a_{10}^{\cal C}\Big (\frac{\theta_c^2}{R_g^2}\Big)\equiv
\Theta\Big(\frac{1}{2}-\frac{\theta_c}{R_g}\Big)\Bigg[
2\,\ln\bigg(\frac{1}{2}+\frac{1}{2}\sqrt{1-\frac{4\theta_c^2}{R_g^2}}\bigg)
-\frac{3}{4}\sqrt{1-\frac{4\theta_c^2}{R_g^2}}
-\ln\Big(\frac{\theta_c^2}{R_g^2}\Big)\Bigg]\,.
\end{align}
In the square brackets of \eq{CqFiniteMJ} we have included the logarithms of $\mu/(QR_g/2)$ that are generated at two-loops due to the anomalous dimension of the ${\cal C}^q$ function and the running coupling. Furthermore, we have switched to the distribution differential in $\ln\, m^2_J$, where the Jacobian factor $m_J^2$ is absorbed (in the case of quarks) in the combination
turning the distributions in \eq{Cq} into ordinary functions, and leading to the expression in \eq{a10C} for the LO distribution.
The parameter $a_{20}^{\cal C}$ is included in order to parameterize the uncertainty resulting from our lack of knowledge of the two-loop non-logarithmic corrections in the small $R_g$ regime. For this uncertainty analysis we are making an approximation that the non-logarithmic corrections have the same functional form as the leading order distribution. At the same time, the modified argument of the function $a_{10}^{\cal C}$ in the second line reflects the difference in the end points of the LO and NLO distributions. In the LO distribution in \eq{Cq} the end point of the $R_g$ distribution is given by $R_g^{\rm min}|_{\rm LO} = 2\theta_c$. In contrast, the calculation of minimization of $R_g$ for a given jet mass for three particle configuration at NLO yields an end point at $R_g^{\rm min}|_{\rm NLO} = (8/5) \theta_c$ in the small angle limit, which was computed in appendix B of \Ref{Marzani:2017mva}. In \sec{Num} we will vary the parameter $a_{20}^{\cal C}$ in the range $[-2\pi, 2\pi]$ to obtain an estimate of the uncertainty from missing ${\cal O}(\alpha_s^2)$ terms.

We now state the result for $pp$ collisions:
\begin{align}
\label{eq:ppxsecRGsmallRg}
&\frac{d \Sigma(R_g)}{d(\ln\, m_J^2) d\Phi_J}
= \frac{m_J^2}{\big(p_T R_g\big)^2}
\sum_{\kappa=q,g}
N^\kappa_{\rm evol}(\Phi_J, \qcut,\beta, R, \mu ; \mu_N, \mu_{gs})\:
e^{\left[K_{cs_g} (\mu, \mu_{cs_g} ) + K_{c} (\mu, \mu_c )\right] }
\\
& \times
\Big(\frac{\mu_{cs_g}}{R_g^{1+\beta}p_T\zcut' }\Big)^{\omega_{cs_g} (\mu, \mu_{cs_g})}
\Big(\frac{\mu_c}{p_T R_g}\Big)^{\omega_c(\mu, \mu_c)}
S_{c_g}^\kappa\!\Big [R_g (p_T\zcut')^{\frac{1}{1+\beta}},\mu_{cs_g}\Big]
{\cal C}^\kappa\bigg [\frac{m_J^2}{p_T^2R_g^2}, p_TR_g, \mu\bigg ]
\Xi_C^\kappa(t_C)
\,,\nn
\end{align}
where we follow the same substitutions of the arguments as described below \eq{ppFactSmallThetaCS}.
Additionally, the first argument $\theta_c/R_g$ in ${\cal C}_q$ in \eq{CqFiniteMJ} can be simply replaced by $\theta_c^{(pp)}/R_g$ with $\theta_c^{(pp)}$ defined in \eq{angularBoundspp}.
The normalization factor is now given by
\begin{align}\label{eq:Nevolpp}
N_\kappa^{\rm evol}(\Phi_J, \qcut, & \beta,R, \mu ; \mu_N, \mu_{gs})
\equiv N^\kappa(\Phi_J, R, \mu_N) \,
S_G^\kappa\Big(p_T \zcut' R^{1+\beta} , \beta, \mu_{gs}\Big)
\,\Xi_G^\kappa(t_G)	\nn
\\
&\qquad \times
e^{K_{N^\kappa}(\mu, \mu_N) +K_{gs}(\mu, \mu_{gs}) }
\Big(\frac{\mu_N}{p_T R}\Big)^{\omega_{N^\kappa}(\mu, \mu_N) } \Big(\frac{\mu_{gs}}{R^{1+\beta} p_T \zcut'}\Big)^{\omega_{gs}(\mu, \mu_{gs})} \, .
\end{align}
Accordingly, the natural scales for the $pp$ case are
\begin{align}
\label{eq:naturalScalespp}
\mu_N^{\rm can.} = p_T R
\,,\:\:\,\;\;
\mu_{gs}^{\rm can.} =p_T \zcut' R^{1+\beta}
\,,\:\:\,\;\;
\mu_{cs_{g}}^{\rm can.} =p_T \zcut' R_g^{1+\beta}
\,,\:\:\,\;\;
\mu_{c}^{\rm can.} =p_TR_g\,.
\end{align}
\subsection{Intermediate $R_g$}
\label{sec:resummation}

Resummation in this regime is conveniently performed in Laplace space, where convolutions over the groomed jet mass reduce to ordinary products. For the intermediate $R_g$ regime we have
\begin{align} \label{eq:LaplaceDbleDiff}
\Sigma(m_J^2, R_g) &=
N_q(Q, R, \mu)\,
S_G^q\Big(\qcut\tan^{1+\beta}\frac{R}{2}, \beta, \mu ,R\Big) \, \Xi_{G}^{q}(t_G)\,
S_{c_g}^q\!\Big(\frac{R_g}{2}\qcut^{\frac{1}{1+\beta}},\beta, \mu\Big)
\nn
\\
&\quad \times
\Xi_S^q(t_S)
\int_0^{m_J^2} d s \int_0^{s/Q} \frac{d\ell^+}{R_g/2} \: J_q \big(s - Q \ell^+, \mu \big) S_{c_m}^q \Big(\frac{\ell^+}{R_g/2}, \mu\Big)\,
\nn
\\
&=
N_q(Q, R, \mu)\,
S_G^q\Big(\qcut\tan^{1+\beta}\frac{R}{2}, \beta, \mu ,R\Big) \, \Xi_{G}^{q}(t_G)\,
S_{c_g}^q\!\Big(\frac{R_g}{2}\qcut^{\frac{1}{1+\beta}},\beta, \mu\Big)
\nn
\\
&\quad \times
\Xi_S^q(t_S)
\int_0^{m_J^2} d s \int \frac{dx}{2\pi i}\, e^{xs }
\:\tilde{J}_q (x, \mu)
\, \tilde S_{c_m}^q \Big(\frac{x Q R_g}{2}, \mu\Big)
\, ,
\end{align}
where $\tilde{J}_q$ and $\tilde{S}^q_{c_m}$ are the Laplace transforms of $J_q$ and $S^q_{c_m}$ respectively. The one-loop results for $\tilde{J}_\kappa$ and $\tilde{S}^\kappa_{c_m}$ are provided in \eqs{JLapDef}{ScmLap} respectively.
Details of the Laplace transform between momentum and position space are presented in \app{transforms}.

At NLL$'$ accuracy, the simple one-loop expressions of the functions in \eq{LaplaceDbleDiff} make it straightforward to take the inverse Laplace transform.
To simplify the final formulae after taking inverse Laplace transform, we introduce an alternative notation for the Laplace space expressions,
\begin{align}\label{eq:altNotation}
\tilde J_\kappa \big [ \log (e^{\gamma_E} x \mu_J^2), \alpha_s (\mu_J)\big] &\equiv \tilde J_\kappa (x, \mu_J) \, ,
\nn \\
\tilde S_{c_m}^\kappa \big[\log (e^{\gamma_E} u \mu_{cs_m}), \alpha_s(\mu_{cs_m})\big]
&\equiv
\tilde S_{c_m}^\kappa (u, \mu_{cs_m}) \, ,
\end{align}
where we notice that the dependence of the functions on the r.h.s. on their first argument is purely logarithmic.
In \eq{altNotation} we have made explicit the Laplace space logarithms and the factors of running coupling.
In terms of this notation, the cross section (for $m_J^2 > 0$) with RG evolution made explicit reads
\begin{align}
\label{eq:xsecRG0}
\frac{d \Sigma(R_g)}{d m_J^2}
&= N_q^{\rm evol}(Q, \qcut,\beta, R, \mu ; \mu_N, \mu_{gs})
\Bigg(
\Xi_S^q(t_S)
\frac{d \Sigma_{\rm int}^q(R_g, \partial_\Omega)}{d m_J^2}
\frac{e^{\gamma_E \Omega}}{\Gamma(- \Omega)}
\\
&\qquad + \bigg(m_J^2 \frac{d}{dm_J^2} \Xi_S^q(t_S)	\bigg)
\frac{d \Sigma_{\rm int}^q(R_g, \partial_\Omega)}{d m_J^2}
\frac{e^{\gamma_E \Omega}}{\Gamma(1- \Omega)}
\Bigg) \bigg|_{\Omega = \tilde \omega(\mu_{cs_m}, \mu_J)}
\, ,
\nn
\end{align}
where
\begin{align}
\label{eq:xsecRG}
&\frac{d \Sigma_{\rm int}^q(R_g, \partial_\Omega)}{d m_J^2}
\equiv \frac{1}{m_J^2}
\:
e^{\big [K_{cs_g} (\mu, \mu_{cs_g} ) + K_{cs_m} (\mu, \mu_{cs_m} ) + K_{J} (\mu, \mu_{J} )\big] }
\\
&\quad\times
\Big(\frac{\mu_{cs_g}}{\qcut (R_g/2)^{1+\beta} }\Big)^{\omega_{cs_g} (\mu, \mu_{cs_g})}
\Big(\frac{\mu_J^2}{m_{J}^2}\Big)^{ \omega_J(\mu, \mu_J)}
\Big(\frac{QR_g\mu_{cs_m}}{ 2m_J^2}\Big)^{\omega_{cs_m} (\mu, \mu_{cs_m})}
\nn
\\
&\quad\times
\,
S_{c_g}^q\!\Big [\frac{R_g}{2}\qcut^{\frac{1}{1+\beta}},\mu_{cs_g}\Big]
\tilde J_q\Big [ \partial_{\Omega} + \log\Big(\frac{\mu_J^2}{m_{J}^2}\Big),\, \alpha_s(\mu_J) \Big] \,
\tilde S_{c_m}^q\Big [\partial_{\Omega} + \log\Big(\frac{QR_g\mu_{cs_m} }{2m_J^2}\Big) , \alpha_s(\mu_{cs_m})\Big ]
\nn
\,,
\end{align}
with the same prefactor $N^\kappa_{\rm evol}$ defined in \eq{HSGevol}.
We have identified the RG invariant single logarithmic kernel $\tilde \omega$ as
\begin{align} \label{eq:combinationOfKernels}
\tilde \omega (\mu_{cs_m}, \mu_J) =
\omega_{cs_m} (\mu, \mu_{cs_m})
+ \omega_J (\mu, \mu_J)
\,.
\end{align}
The first argument involving the derivative $\partial_{\Omega}$ in the Laplace transforms of the jet and the collinear-soft functions in \eq{xsecRG} is understood to replace the logarithms that appear in the corresponding Laplace space expressions in the notation described in \eq{altNotation}. In the subscript $\mu_{cs_m}$ we introduced the shorthand label $cs_m$ for the symbol $S_{c_m}^q$. The collinear scale in \eq{naturalScales} is now replaced by the two mass-dependent canonical scales
\begin{align}
\label{eq:moreNaturalScales}
\mu_{cs_{m}}^{\rm can.} = \frac{2m_J^2}{QR_g}
\,,\qquad
\mu_J^{\rm can.} = m_J\,.
\end{align}
In the small $R_g$ limit when $R_g = \theta_c = 2m_J/Q$ both of these scales merge such that $\mu_{cs_m}^{\rm can}=\mu_J^{\rm can.} = \mu_c^{\rm can.}= m_J$.
Indeed, as required by RG consistency, the evolution factors in \eq{xsecRG} reduce to the ones in \eq{xsecRGsmallRg} in this limit.

Furthermore, we can simplify the derivative of $\Xi_S(t_S)$ in \eq{xsecRG0} using \eqs{tNGL}{tSdef} to arrive at the following simple expression:
\begin{align}
m_J^2 \frac{d}{m_J^2} \Xi_S \big(t[\mu_{cs_g}, \mu_{cs_m}]\big)
&= \Big(m_J^2\frac{d}{m_J^2} \mu_{cs_m} \Big)
\frac{\partial }{\partial \mu_{cs_m}} t[\mu_{cs_g}, \mu_{cs_m}]
\frac{\partial}{\partial t} \Xi_S (t)\bigg|_{t = t_S}
\\
&= \frac{\alpha_s(\mu_{cs_m})}{2\pi} 	\frac{\partial}{\partial t} \Xi_S (t) \bigg|_{t = t_S}
\nn \, ,
\end{align}
where in the second line we made use of the canonical scale choice for $\mu_{cs_m}$ in \eq{moreNaturalScales}. One obtains a yet simpler expression when approximating $\Xi_S (t)$ by the two-loop result in \eq{NGLFull2}.

Lastly, the generalization for $pp$ collisions is straightforward, with the factorization formula given by
\begin{align}
\label{eq:ppxsecRG}
\frac{d \Sigma(R_g)}{d m_J^2 d\Phi_J}
&=
\sum_{\kappa=q,g}
N^\kappa_{\rm evol}(\Phi_J, \qcut,\beta, R, \mu ; \mu_N, \mu_{gs})
\Bigg(
\Xi_S^\kappa(t_S)
\frac{d \Sigma_{\rm int}^{\kappa(pp)}(R_g, \partial_\Omega)}{d m_J^2d \Phi_J}
\frac{e^{\gamma_E \Omega}}{\Gamma(- \Omega)}
\\
&\qquad + \bigg(m_J^2 \frac{d}{dm_J^2} \Xi_S^\kappa(t_S)	\bigg)
\frac{d \Sigma_{\rm int}^{\kappa(pp)}(R_g, \partial_\Omega)}{d m_J^2d \Phi_J}
\frac{e^{\gamma_E \Omega}}{\Gamma(1- \Omega)}
\Bigg)
\bigg|_{\Omega = \tilde \omega(\mu_{cs_m}, \mu_J)}
\, ,
\nn
\end{align}
with
\begin{align}
&\frac{d \Sigma_{\rm int}^{\kappa(pp)}(R_g, \partial_\Omega)}{d m_J^2 d\Phi_J}
= \frac{1}{m_J^2}
\:
e^{\big [K_{cs_g} (\mu, \mu_{cs_g} ) + K_{cs_m} (\mu, \mu_{cs_m} ) + K_{J} (\mu, \mu_{J} )\big] }
\nn
\\
&\quad\times
\Big(\frac{\mu_{cs_g}}{p_T \zcut' R_g^{1+\beta} }\Big)^{\omega_{cs_g} (\mu, \mu_{cs_g})}
\Big(\frac{\mu_J^2}{m_{J}^2}\Big)^{ \omega_J(\mu, \mu_J)}
\Big(\frac{p_TR_g\mu_{cs_m}}{m_J^2}\Big)^{\omega_{cs_m} (\mu, \mu_{cs_m})}
S_{c_g}^\kappa\!\Big [R_g (p_T \zcut')^{\frac{1}{1+\beta}},\mu_{cs_g}\Big]
\nn
\\
&\quad\times
\,
\tilde J_\kappa\Big [ \partial_{\Omega} + \log\Big(\frac{\mu_J^2}{m_{J}^2}\Big),\, \alpha_s(\mu_J) \Big] \,
\tilde S_{c_m}^\kappa\Big [\partial_{\Omega} + \log\Big(\frac{p_TR_g\mu_{cs_m} }{m_J^2}\Big) , \alpha_s(\mu_{cs_m})\Big ]
\,,
\end{align}
where we use the normalization factor for the $pp$ case in \eq{Nevolpp}.
The generalization for canonical scale $\mu_{cs_m}$ for $pp$ case is given by
\begin{align}\label{eq:muCSmPPNatural}
\mu_{cs_m}^{\rm can.} = \frac{m_J^2}{p_T R_g}
\, \qquad (\text{$pp$ case}) \,,
\end{align}
and the same formula for the jet scale in \eq{moreNaturalScales} applies here. We note that given our convention for Laplace transforms in \eqs{JLapTrans}{ScmLapTrans}, precisely the same functions $\tilde J_\kappa$ and $\tilde S_{c_m}^\kappa$ as in \eqs{xsecRG0}{xsecRG} appear here (with a possibility for $\kappa = g$).
\subsection{Large $R_g$}
\label{sec:LargeRg}

We now derive a formula for the resummed cross section in the region $\theta_c^2 \ll R_g \lesssim \theta_g^\star(m_J^2)$. The factorization structure and the one-loop ingredients relevant to this regime were discussed in \sec{LargeRgFact}. NLL$'$ resummation requires introducing a number of new functions, whose notation we describe below, while leaving presentation of the full expressions to \app{LargeRg}.

The key ingredient of our analysis is the collinear-soft function displayed at one loop in \eq{ScLargeRg}. Since its anomalous dimension is unaffected by the $R_g$ measurement, it has the same RG evolution as the collinear-soft function that enters the single differential jet mass distribution. At the same time, the $R_g$ dependence only arises at one loop, which makes the $\mathcal{O}(\alpha_s)$ correction to the collinear-soft function the leading order contribution to the double differential cross section. This motivates rewriting the collinear-soft function in the following multiplicative manner:
\begin{align}\label{eq:ScConvLargeRg}
S_c^\kappa \Bigl[\ell^+ \qcut^{\frac{1}{1+\beta}}, \frac{R_g}{2}\qcut^{\frac{1}{1+\beta}}, \beta, \mu\Bigr]
&=\int d \ell^{\prime +}
\qcut^{\frac{1}{1+\beta}}
S_c^\kappa\Bigl[(\ell^+ - \ell^{\prime+})\qcut^{\frac{1}{1+\beta}}, \beta, \mu\Bigr]
\\
&\qquad \times
\Delta S_c^\kappa
\Bigl [\ell^{\prime+}\qcut^{\frac{1}{1+\beta}}, \frac{R_g}{2}\qcut^{\frac{1}{1+\beta}}, \beta \Bigr]\,,
\nn
\end{align}
where the collinear-soft function $S_c^\kappa[\ell^+\qcut^{\frac{1}{1+\beta}}, \beta, \mu]$ is singly differential in the jet mass and is responsible for RG evolution, whereas the $R_g$ dependent corrections are incorporated via $\Delta S_c^\kappa$, which is by construction independent of the renormalization scale.
\eq{ScConvLargeRg} is just a reorganization of the perturbative series. However, including one-loop corrections in both functions on the r.h.s. allows us to supplement our predictions with terms of the form $\{$leading RG logarithms$\} \times \{$leading $R_g$ corrections$\}$, which when taken together first appear at $\mathcal{O}(\alpha^2_s)$. In fact, we define NLL$'$ accuracy for the double differential distribution in the large $R_g$ region based on this reorganization. However, we stress that such a prescription does not capture all $\mathcal{O}(\alpha_s^2)$ terms in the collinear-soft function and in particular misses the genuine two-loop $R_g$ dependent corrections.
We obtain an estimate for the uncertainty from these missing two-loop terms via the procedure described below.
To make the $\Delta S_c^\kappa$ term in \eq{ScConvLargeRg} explicitly independent of $\mu$, we can further rewrite it as
\begin{align}
\label{eq:ScLargeRgFO}
& \Delta S_c^\kappa
\Bigl [\ell^+\qcut^{\frac{1}{1+\beta}}, \frac{R_g}{2}\qcut^{\frac{1}{1+\beta}}, \beta \Bigr]
=
\delta\Bigl( \ell^+ Q_{\rm cut}^{\frac{1}{1+\beta}}\Bigr)
\\
&\qquad +
\int \frac{d \ell^{\prime +} \qcut^{\frac{1}{1+\beta}}}{\big(\frac{R_g}{2}\qcut^{\frac{1}{1+\beta}}\big)^{2+\beta}}
\:
\Delta S_c^{[\beta_0]}
\Bigl[(\ell^+ - \ell^{\prime+})\qcut^{\frac{1}{1+\beta}}, \beta, \mu\Bigr]
\:
\Delta S_c^{[C_\kappa]}
\Bigg[
\frac{\ell^{\prime +} \qcut^{\frac{1}{1+\beta}}}{\big(\frac{R_g}{2}\qcut^{\frac{1}{1+\beta}}\big)^{2+\beta}},
\beta , \mu
\Bigg]
\, ,
\nn
\end{align}
where we have defined the one-loop contributions
\begin{align}\label{eq:ScLargeRgFO2}
\Delta S_c^{[\beta_0]}
\Bigl[\ell^+ \qcut^{\frac{1}{1+\beta}}, \beta, \mu\Bigr]
= \delta\Bigl( \ell^+ Q_{\rm cut}^{\frac{1}{1+\beta}}\Bigr)
&- \frac{\alpha_s(\mu)\beta_0}{2\pi}
\Big(	\frac{1+\beta}{2+\beta}	\Big)
\frac{1}{\mu^{\frac{2+\beta}{1+\beta}}}{\cal L}_0\!\biggl[ \frac{\ell^+ Q_{\rm cut}^{\frac{1}{1+\beta}}}{\mu^{\frac{2+\beta}{1+\beta}}} \biggr]
+ {\cal O}(\alpha_s^2)
\,,
\\
\label{eq:ScLargeRgFO3}
\Delta S_c^{[C_\kappa]}
\Bigg[
\tilde \ell =
\frac{\ell^{ +} \qcut^{\frac{1}{1+\beta}}}{\big(\frac{R_g}{2}\qcut^{\frac{1}{1+\beta}}\big)^{2+\beta}},
\beta , \mu
\Bigg]
=
&-
\frac{\alpha_s(\mu)C_\kappa}{\pi}
\frac{2}{2+\beta}
\Theta(\tilde \ell - 1 )
\frac{\log (\tilde \ell)}{\tilde \ell }
\bigg(1
+ \frac{\alpha_s(\mu)}{\pi} a_{20}^{S_c}
\bigg)
+ {\cal O}(\alpha_s^2)
\,.
\end{align}
The function $\Delta S_c^{[C_\kappa]}$ carries the $R_g$ dependent one-loop term in \eq{ScLargeRg}, while $\Delta S_c^{[\beta_0]}$ renders the convolution in \eq{ScLargeRgFO} $\mu$ independent at NLL$'$ accuracy by compensating for the NLL running of the coupling in $\Delta S_c^{[C_\kappa]}$. This separation is again just a reshuffling of the perturbative series, as the $\mathcal{O}{(\alpha_s^2)}$ cross terms are predicted from RG consistency of the double differential cross section. The construction to make the $R_g$ dependent correction $\Delta S_c^\kappa$ in \eq{ScLargeRgFO} explicitly independent of $\mu$ can similarly be generalized at higher orders.
Finally, we note that expanding \eq{ScConvLargeRg} to ${\cal O}(\alpha_s)$ we recover the one-loop expression of the collinear-soft function given above in \eq{ScLargeRg}. As a further check, the $R_g$ dependent term vanishes when $R_g = \theta_g^\star$ ($\tilde{\ell}=1$ in \eq{ScLargeRgFO3}), and the collinear-soft function cumulative in $R_g$ of \eq{ScConvLargeRg} reduces in this limit to the collinear-soft function single differential in the groomed jet mass.

Since our treatment lacks the 2-loop non-logarithmic $R_g$ dependent corrections, we have parametrized the resulting uncertainty in \eq{ScLargeRgFO3} via the parameter $a_{20}^{S_c}$ which we will vary in our uncertainty analysis. This estimate utilizes an approximation that the two-loop corrections have the same functional form as that of the one-loop term, but with an unknown normalization. In the numerical studies presented below we will consider variation of $a_{20}^{S_c}$ in the range $[-2\pi,2\pi]$. We will return to discussing this when we study the numerical results in \sec{Num}.

To carry out resummation in Laplace space the transform of the collinear-soft function is defined with respect to the variable $\ell^+\qcut^{\frac{1}{1+\beta}}$:
\begin{align}\label{eq:ScTransformDef}
\tilde S_c^\kappa\Bigl(s, \frac{R_g}{2}\qcut^{\frac{1}{1+\beta}},\beta , \mu \Bigr)
&=
\qcut^{\frac{1}{1+\beta}}\int d \ell^+
e^{-s\big( \ell^+\, Q_{\rm cut}^{\frac{1}{1+\beta}}\big)}
S_c^\kappa\Bigl[\ell^+\, Q_{\rm cut}^{\frac{1}{1+\beta}}, \frac{R_g}{2}\qcut^{\frac{1}{1+\beta}},\beta, \mu \Bigr] \,.
\end{align}
Equations \eqref{eq:ScConvLargeRg} and \eqref{eq:ScLargeRgFO} in Laplace space read
\begin{align}\label{eq:ScConvLargeRgLap}
\tilde S_c^\kappa\Bigl(s, \frac{R_g}{2}\qcut^{\frac{1}{1+\beta}},\beta,\mu \Bigr)
& =
\tilde S_c^\kappa \bigl(s,\beta,\mu \bigr)
\Delta \tilde S_c^\kappa\Bigl(s, \frac{R_g}{2}\qcut^{\frac{1}{1+\beta}},\beta \Bigr)\,,
\\
\label{eq:ScConvLargeRgLap2}
\Delta \tilde S_c^\kappa\Bigl(s, \frac{R_g}{2}\qcut^{\frac{1}{1+\beta}},\beta \Bigr)
& = 1 +
\Delta \tilde S_c^{[\beta_0]} \bigl(s,\beta,\mu \bigr)
\Delta \tilde S_c^{[C_\kappa]} \Bigl(s\, \qcut^{\frac{2+\beta}{1+\beta}} \Big(\frac{R_g}{2}\Big)^{2+\beta} , \beta, \mu\Bigr)
\,,
\end{align}
where the Laplace transforms of the individual terms are given in \eqs{LapSc}{ScCalLaplace}. Hence, the cross section for $\ee$ collisions in the Laplace space takes on the form
\begin{align}\label{eq:FactLargeRgLaplace}
\int_0^\infty \!\! dm_J^2\, \,e^{-x\,m_J^2}
& \frac{d \Sigma(R_g)}{d m_J^2}\bigg|_{R_g \lesssim \theta_g^\star}
=
N_q(Q, R, \mu)\,
S_G^q\Big(\qcut\tan^{1+\beta}\frac{R}{2}, \beta, \mu ,R\Big)
\, \Xi_G^q(t_G)\,
\nn
\\
& \times
\tilde{J}_q (x, \mu)
\, \tilde S_c^q\Bigl(x Q \qcut^{-\frac{1}{1+\beta}},\beta,\mu \Bigr)
\: \Delta \tilde S_c^q
\Bigl(x Q \qcut^{-\frac{1}{1+\beta}}, \frac{R_g}{2}\qcut^{\frac{1}{1+\beta}}, \beta\Bigr)
\,.
\end{align}

In \app{deriveQ} we show that carrying out RG evolution in Laplace space and transforming back to momentum space yields the following result for the resummed cross section at NLL$'$ accuracy for the product of terms on the second line of \eq{FactLargeRgLaplace},
\begin{align}
\label{eq:xsecRGLargeRg}
\frac{d \Sigma(R_g)}{d m_J^2}\bigg|_{R_g \lesssim \theta_g^\star}
&=
N^q_{\rm evol}\big (Q, \qcut,\beta, R, \mu ; \mu_N, \mu_{gs}\big)
\frac{d\sigma^q\big [\partial_\Omega \big]}{dm_J^2}
\Bigg(
\frac{e^{\gamma_E\Omega}}{\Gamma(- \Omega)}
+
{\cal Q}^q\Big [ \Omega, \frac{R_g}{\theta_g^\star}, \beta \Big]
\Bigg)
\Bigg|_{\Omega = \tilde \omega(\mu_{cs}, \mu_J)}
\,.
\end{align}
The function of the differential operator $\partial_\Omega$ in \eq{xsecRGLargeRg} (along with the first term in brackets in \eq{xsecRGLargeRg}) coincides with the single differential jet mass cross section in its fully factorized form:
\begin{align}\label{eq:mJOperator}
\frac{d \sigma^q\big [\partial_\Omega \big]}{d m_J^2}\:
&\equiv
\frac{1}{m_{J}^2}
e^{\big [K_{cs} (\mu, \mu_{cs} ) + K_{J} (\mu, \mu_{J} )\big] }
\Big(\frac{\mu_J^2}{m_{J}^2}\Big)^{ \omega_J(\mu, \mu_J)}
\bigg( \frac{Q \mu_{cs}}{m_J^2} \Big(\frac{\mu_{cs}}{\qcut}\Big)^{\frac{1}{1+\beta}}\bigg)^{\omega_{cs} (\mu, \mu_{cs})}
\nn
\\
& \times
\tilde J_q\Big [ \partial_{\Omega} + \log\Big(\frac{\mu_J^2}{m_{J}^2}\Big),\, \alpha_s(\mu_J) \Big] \,
\tilde S_c^q \bigg [\partial_{\Omega} + \log\bigg( \frac{Q \mu_{cs}}{m_J^2} \Big(\frac{\mu_{cs}}{\qcut}\Big)^{\frac{1}{1+\beta}}\bigg) ,\beta, \alpha_s(\mu_{cs})\bigg ] \, ,
\end{align}
where the Laplace space logarithms in the jet and collinear-soft functions are replaced by the argument in brackets.
Here, in analogy with \eq{altNotation}, we have defined the alternative notation for the Laplace transform of the collinear-soft function
\begin{align}\label{eq:ScNewOldLap}
\tilde S_c^\kappa \Big[\log \big( s e^{\gamma_E}\mu^{\frac{2+\beta}{1+\beta}} \big) , \beta, \alpha(\mu_{cs})\Big] \equiv
\tilde S_c^\kappa\bigl(s, \beta,\mu_{cs} \bigr)\,,
\end{align}
with the explicit one-loop expression given in \eq{ScLap2},
while in analogy with \eq{combinationOfKernels}, we have introduced in \eq{xsecRGLargeRg} the RG invariant combination
\begin{align}
\tilde \omega (\mu_{cs}, \mu_J) \equiv \omega_J(\mu, \mu_J) + \omega_{cs}(\mu, \mu_{cs}) = \omega_J(\mu_{cs}, \mu_J ) \, ,
\end{align}
with the definition of $\omega_{J, S_c}$ provided in \eq{KOmega}.
The RG evolved normalization factor $N^\kappa_{\rm evol}$ in \eq{mJOperator} is the same as \eq{HSGevol}. Along with the scales that are in common with the small and intermediate $R_g$ regimes in \eqs{naturalScales}{moreNaturalScales}, the logarithms in the $S_c^\kappa$ and the corresponding RG evolution factors are minimized by the following choice of the collinear-soft scale:
\begin{align}\label{eq:muCSNatural}
\mu_{cs}^{\rm can.} = \Big(\frac{m_J^2}{Q}\Big)^{\frac{1+\beta}{2+\beta}} \qcut^{\frac{1}{2+\beta}} \, .
\end{align}
Note that since all the scales and anomalous dimensions for this regime are independent of $R_g$, so is the function of the derivative operator in \eq{mJOperator}.

The effect of the cumulative $R_g$ measurement on \eq{xsecRGLargeRg} is accounted for entirely by the kernel ${\cal Q}^\kappa$ defined as
\begin{align}\label{eq:UJScOmega}
{\cal Q}^\kappa\Big [ \Omega, \Big(\frac{R_g}{\theta_g^\star}\Big)^{2+\beta}, \beta \Big]
&\equiv e^{\gamma_E\Omega}\Delta \tilde S_c^{[\beta_0]} \bigg [\partial_{\Omega} + \log\bigg( \frac{Q \mu_{cs}}{m_J^2} \Big(\frac{\mu_{cs}}{\qcut}\Big)^{\frac{1}{1+\beta}}\bigg) ,\beta, \alpha_s(\mu_{cs})\bigg ]
\nn \\
&\qquad \times
\int \frac{d\varrho }{2\pi i} \: e^{\varrho }
\varrho^{\Omega}
\Delta \tilde S_c^{[C_\kappa]}
\Bigg(\varrho \Big(\frac{R_g}{\theta_g^\star}\Big)^{2+\beta}, \beta, \mu_{cs}\Bigg)
\, ,
\end{align}
which is $\mu$-independent by construction and vanishes at $R_g = \theta_g^\star(m_J^2,\qcut, \beta)$.
Here $\Delta \tilde S_c^{[\beta_0]}$ is expressed in the notation equivalent to that in \eq{ScNewOldLap},
\begin{align}
\Delta\tilde S_c^{[\beta_0]} \Big[\log \big( s e^{\gamma_E}\mu^{\frac{2+\beta}{1+\beta}} \big) , \beta, \alpha(\mu_{cs})\Big] \equiv
\Delta\tilde S_c^{[\beta_0]}\bigl(s, \beta,\mu_{cs} \bigr)\,,
\end{align}
see \eq{ScBeta0Lap}.
The explicit expression for ${\cal Q}^\kappa$ is given in in \eq{UJScOmegaFull}. To make a connection with the notation introduced in \tab{RGE}, we can identify the LO distribution in the large $R_g$ regime that multiplies the NLL evolution factors in \eq{mJOperator} as
\begin{align}\label{eq:a10Sc}
\frac{\alpha_sC_\kappa}{\pi}a_{10}^{ S_c} \Big(\frac{R_g}{\theta_g^\star}, \beta \Big) = {\cal Q}^\kappa\Big [ \tilde \omega(\mu_{cs}, \mu_J), \frac{R_g}{\theta_g^\star(m_J^2)}, \beta \Big] + {\cal O}(\alpha_s^2)\,.
\end{align}

Next, the result for $pp$ case is given by
\begin{align}
\label{eq:ppxsecRGLargeRg}
\frac{d \Sigma(R_g)}{d m_J^2 d\Phi_J }\bigg|_{R_g \lesssim \theta_g^\star}
&=
\sum_{\kappa = q,g}
N^\kappa_{\rm evol}\big (\Phi_J, \qcut,\beta, R, \mu ; \mu_N, \mu_{gs}\big)
\\
&\qquad
\times \frac{d\sigma^\kappa\big [\partial_\Omega \big]}{dm_J^2 d\Phi_J}
\Bigg(
\frac{e^{\gamma_E\Omega}}{\Gamma(- \Omega)}
+{\cal Q}^\kappa\Big [ \Omega, \frac{R_g}{\theta_g^{(pp)\star}(m_J^2)}, \beta \Big]
\Bigg)
\Bigg|_{\Omega = \tilde \omega(\mu_{cs}, \mu_J)}
\,,
\nn
\end{align}
where
\begin{align}\label{eq:ppmJOperator}
\frac{d \sigma^\kappa\big [\partial_\Omega \big]}{d m_J^2d\Phi_J }\:
&\equiv
\frac{1}{m_{J}^2}
e^{\big [K_{cs} (\mu, \mu_{cs} ) + K_{J} (\mu, \mu_{J} )\big] }
\Big(\frac{\mu_J^2}{m_{J}^2}\Big)^{ \omega_J(\mu, \mu_J)}
\bigg( \frac{p_T \mu_{cs}}{m_J^2} \Big(\frac{\mu_{cs}}{p_T \zcut'}\Big)^{\frac{1}{1+\beta}}\bigg)^{\omega_{cs} (\mu, \mu_{cs})}
\\
& \times
\tilde J_\kappa\Big [ \partial_{\Omega} + \log\Big(\frac{\mu_J^2}{m_{J}^2}\Big),\, \alpha_s(\mu_J) \Big] \,
\tilde S_c^\kappa \bigg [\partial_{\Omega} + \log\bigg( \frac{p_T \mu_{cs}}{m_J^2} \Big(\frac{\mu_{cs}}{p_T \zcut'}\Big)^{\frac{1}{1+\beta}}\bigg) ,\beta, \alpha_s(\mu_{cs})\bigg ] \, .
\nn
\end{align}
Additionally, the same formula for ${\cal Q}^\kappa$ in \eq{UJScOmega} applies but the arguments of $\Delta \tilde S_c^{[\beta_0]}$ being the same as that of $\tilde S_c^\kappa$ in the last line of the equation above. In deriving this result we have used the following convention for Laplace transform of the collinear-soft function for $pp$ case
\begin{align}\label{eq:ScTransformDefpp}
&\tilde S_c^\kappa\Bigl(s, R_g(p_T\zcut')^{\frac{1}{1+\beta}},\beta , \mu \Bigr)
\\
&\qquad =
(p_T\zcut')^{\frac{1}{1+\beta}}
\int d r^+
e^{-s\big( r^+\, (p_T\zcut')^{\frac{1}{1+\beta}}\big)}
S_c^\kappa\Bigl[r^+ (p_T\zcut')^{\frac{1}{1+\beta}}, R_g (p_T\zcut')^{\frac{1}{1+\beta}},\beta, \mu \Bigr]
\nn
\,,
\end{align}
and the analogs of \eqs{ScConvLargeRgLap}{ScConvLargeRgLap2} can be derived by following the substitutions in \eq{dictionary}.
Finally, we see that the natural scale choice for $\mu_{cs}$ in $pp$ collisions is given by
\begin{align} \label{eq:xsecRGLargeRgPP}
\mu_{cs}^{\rm can.} =
\Big(\frac{m_J^2}{p_T}\Big)^{\frac{1+\beta}{2+\beta}} (p_T\zcut')^{\frac{1}{2+\beta}}
\, \qquad (\text{$pp$ case}) \, .
\end{align}

\subsection{Boundary corrections in the Large $R_g$ region}
\label{sec:LargeRgShift}

Next we discuss boundary corrections to the soft drop cross section in the regime where $R_g \lesssim \theta_g^\star(m_J^2,\qcut,\beta)$,
relevant for computing the $\varepsilon$ derivative in \eq{C1C2multiDiffDef}.
We first compute the bare collinear-soft and global-soft functions with the shifted soft drop condition in \eq{ShiftSD}. This will allow us to identify the relevant corrections to the double differential cross section at first order in the shift parameter $\veps$. We will find that both the function of the derivative operator $d \sigma(\partial_\Omega)$ and the evolution kernel ${\cal Q}^\kappa$ in \eqs{xsecRGLargeRg}{ppxsecRGLargeRg} receive ${\cal O}(\veps)$ modifications.
\tocless\subsubsection{Boundary corrections to bare soft matrix elements}
The ${\cal O}(\veps)$ corrections to the soft matrix elements lead to logarithmic and non-logarithmic terms proportional to $R_g^{-\beta} Q\veps/\qcut$.\footnote{Note that we use $\veps$ for the shift to soft drop condition and $\eps$ for the dimensional regularization parameter.}
This can be seen from the bare result for the collinear-soft function with the shifted soft drop condition defined in \eq{ShiftSD} given by
\begin{align}\label{eq:ScBareShifted0}
&\qcut^{\frac{1}{1+\beta}}S_{c}^{\kappa [1], \rm bare} \bigl [
\ell^+ , R_g, \qcut ,\overline \Theta_{\rm sd}(\veps), \beta, \mu \bigr ]
\\
&\qquad
\equiv
\frac{\alpha_s C_\kappa}{\pi}
\frac{(\mu^2 e^{\gamma_E})^{\eps}}{\Gamma(1-\eps)}
\int \frac{dp^+ dp^-}{(p^+p^-)^{1+\eps}}
\overline \Theta_{\rm sd}(\veps)
\big[
\overline \Theta_{R_g}
\delta\big (\ell^+ - p^+\big )
- \delta (\ell^+ )
\big]
\, ,
\nn
\end{align}
where the condition to pass or fail the $R_g$ constraint are written respectively as
\begin{align}\label{eq:ThetaRg}
\overline \Theta_{R_g} \equiv \Theta \bigg ( \frac{R_g}{2} - \Big(\frac{p^+}{p^-} \Big)^{\frac{1}{2}} \bigg) \, ,
\qquad
\Theta_{R_g} \equiv 1 - \overline \Theta_{R_g} \, .
\end{align}
Expanding
\begin{align} \label{eq:sdShiftExpand}
\overline \Theta_{\rm sd}(\veps ) = \overline \Theta_{\rm sd} + \veps \delta_{\rm sd} + {\cal O}(\veps^2),
\end{align}
where $\delta_{\rm sd} = \delta(z-\tilde{z}_{\rm cut}\theta^\beta)$,
yields\footnote{We include the argument $\overline \Theta_{\rm sd}(\veps)$ to indicate that the corresponding functions are evaluated with the shifted soft drop condition in \eq{ShiftSD}. We then indicate ${\cal O}(\veps)$ boundary corrections to various functions by including a subscript $\veps$ to distinguish them from the ${\cal O}(\veps^0)$ part.}
\begin{align}
\label{eq:ScBareShifted}
S_{c}^{\kappa[1], \rm bare} \Bigl [
\ell^+ \qcut^{\frac{1}{1+\beta}}, \frac{R_g}{2}\qcut^{\frac{1}{1+\beta}},\overline \Theta_{\rm sd}(\veps), \beta, \mu \Bigr ]
&= S_c^{\kappa[1], \rm bare}\Bigl[\ell^+ \qcut^{\frac{1}{1+\beta}}, \frac{R_g}{2}\qcut^{\frac{1}{1+\beta}}, \beta, \mu\Bigr]
\\
&+ \frac{ Q\veps}{\qcut} \, S_{c, \veps}^{\kappa[1], \rm bare} \Bigl [
\ell^+ \qcut^{\frac{1}{1+\beta}}, \frac{R_g}{2}\qcut^{\frac{1}{1+\beta}}, \beta, \mu \Bigr ]+{\cal O}(\veps^2)
\,. \nn
\end{align}
The ${\cal O}(\veps^0) $ term in \eq{sdShiftExpand} simply results in the usual soft drop condition, returning the bare one-loop version of the collinear-soft function in \eq{ScLargeRg}, while the ${\cal O}(\veps)$ correction is
\begin{align}\label{eq:ScShiftedBare0}
&\qcut^{\frac{1}{1+\beta}} S_{c, \veps}^{\kappa[1], \rm bare} \Bigl [
\ell^+ \qcut^{\frac{1}{1+\beta}}, \frac{R_g}{2}\qcut^{\frac{1}{1+\beta}}, \beta, \mu \Bigr ]
\\
&\qquad\equiv \frac{2}{2+\beta}
\frac{\alpha_s C_\kappa}{\pi}
\Big(\frac{\mu}{\qcut}\Big)^{2\eps}
\frac{(e^{\gamma_E})^\eps}{\Gamma(1-\eps)}
\int\frac{dp^+}{\qcut}
\Big(\frac{p^+}{\qcut}\Big)^{-\frac{2(1+\beta)}{2+\beta}(1+\eps)}
\nn
\\
&\qquad \qquad\times \Bigg[
\Theta \bigg ( \frac{R_g}{2} - \Big(\frac{p^+}{\qcut} \Big)^{\frac{1}{2+\beta}} \bigg)
\big[
\delta\big (\ell^+ - p^+ \big )
- \delta (\ell^+ )
\big]
- \delta(\ell^+)
\Theta \bigg ( \Big(\frac{p^+}{\qcut} \Big)^{\frac{1}{2+\beta}} - \frac{R_g}{2}
\bigg)
\Bigg]
\,.
\nn
\end{align}
We have used the $\delta_{\rm sd}$ to perform the integral over $p^-$, and have isolated the contribution from the soft-drop failing piece, proportional to $\delta(\ell^+)$, so as to regulate the integral for $p^+ \ra 0$. The term proportional to $\delta(\ell^+)$ in the third line is no longer scaleless as it is bounded from below. Solving the remaining integration, we find for $\beta = 0$ and $\beta > 0$ respectively:
\begin{align}\label{eq:ScLargeShiftb0}
\nn
& S_{c,\veps}^{\kappa[1], \rm bare}
\Bigl[\ell^+ \qcut^{\frac{1}{1+\beta}}, \frac{R_g}{2}\qcut^{\frac{1}{1+\beta}} , \beta = 0, \mu\Bigr]
\\
&\qquad =
\frac{\alpha_s C_\kappa}{\pi}
\Bigg (
- \frac{1}{\eps_{\rm UV}}\delta\big(\ell^+\qcut\big)
+
\frac{1}{\mu^2} {\cal L}_0 \Big(\frac{\ell^+\qcut}{\mu^2}\Big)
+
\Theta\bigg(\frac{\ell^{ +} \qcut }{\big(\frac{R_g}{2}\qcut \big)^{2 }} -1\bigg)
\frac{1}{\ell^+\qcut}
\Bigg )
\, ,
\nn
\\
&S_{c,\veps}^{\kappa[1], \rm bare} \Bigg[
\tilde \ell =
\frac{\ell^{ +} \qcut^{\frac{1}{1+\beta}}}{\big(\frac{R_g}{2}\qcut^{\frac{1}{1+\beta}}\big)^{2+\beta}}
, \frac{R_g}{2}\qcut^{\frac{1}{1+\beta}} ,
\beta > 0,\mu
\Bigg]
\nn
\\
&\qquad =
\frac{2}{2+\beta}	\frac{\alpha_s (\mu)C_\kappa}{\pi}
\frac{\big(\frac{R_g}{2}\qcut^{\frac{1}{1+\beta} }\big)^{-(2+\beta)}}{ \big(\frac{R_g}{2}\big)^{\beta}}
\Bigg(
{\cal L}_0^{-\frac{\beta}{2+\beta}}(\tilde \ell)
+
\frac{\Theta(\tilde \ell - 1) }{\tilde \ell^{\frac{2(1+\beta)}{2+\beta}}}
\Bigg)
\,.
\end{align}
We see that for $\beta = 0$ the $\eps > 0$ regulator results in a UV pole, while for $\beta > 0$ the result is a power correction and dimensional regularization is not needed.
Here, we have defined\footnote{We note that for $a>0$ the function ${\cal L}_0^a (x)$ is equivalent to $1/x^{1-a}$ which has an integrable singularity as $x \ra 0$.}
\begin{align}
{\cal L}_0^a (x)\equiv {\cal L}^a(x) + \frac{1}{a} \delta(x)
\, , \qquad
{\cal L}^a(x) = \Big[ \frac{\Theta(x)}{x^{1-a}}\Big]_+
\, , \qquad a \neq 0 \, .
\end{align}

We can carry out a similar calculation for the ${\cal O}(\veps)$ boundary corrections to the global-soft function. In full analogy with \eq{ScBareShifted}, we find that up to ${\cal O}(\veps^2)$ terms
\begin{align}
& S_G^{\kappa[1], \rm bare} \bigl( \qcut,R, \beta, \Theta_{\rm sd}(\veps) , \mu\bigr)
= S_G^{\kappa[1], \rm bare} \bigl( \qcut,R, \beta, \mu\bigr)
+
\frac{Q\veps}{\qcut}
S_{G, \veps}^{\kappa[1], \rm bare} \bigl( \qcut, R, \beta , \mu\bigr)
\,,
\end{align}
where the additional argument $ \Theta_{\rm sd}(\veps)$ denotes now the shift in the soft drop \emph{failing} condition. The result for the ${\cal O}(\veps)$ correction to the bare global-soft function with shifted soft drop reads
\begin{align}\label{eq:SGshift}
S_{G, \veps}^{\kappa[1], \rm bare} \bigl( \qcut, R, \beta , \mu\bigr)
&= \frac{\alpha_s C_\kappa}{\pi} \times
\left\{ \begin{array}{cc}
\bigg[
\frac{1}{\eps_{\rm UV}} + 2 \log \Big(\frac{\mu}{\qcut \tan\frac{R}{2}}\Big)
\bigg]
&\quad \big(\beta = 0 \big)
\\[10pt]
\bigg[
\frac{2}{\beta}\frac{1}{\sin^\beta\big(\frac{R}{2}\big)}
+ \ldots
\bigg]
&\quad \big(\beta > 0 \big)
\end{array}\right.
\, ,
\end{align}
where the additional terms not shown for $\beta > 0$ are ${\cal O}(1)$ and not relevant to the discussion here.
The details of this derivation and complete results for $\beta > 0$ are provided in \app{LargeRgShift}.
We see that the additional single logarithmic UV divergence for $\beta = 0$ case has the opposite sign to that in \eq{ScLargeShiftb0}, as required by RG consistency. Interestingly, we now find that for $\beta = 0$ the one-loop collinear-soft and global-soft functions develop a non-zero non-cusp anomalous dimension proportional to $\veps$. With conventions for prefactors in anomalous dimensions, summarized in \eqs{GammaBetaExpansion}{RGEall}, we find
\begin{align}\label{eq:noncuspEps}
\gamma_0^{S_G^\kappa} (\veps, \zcut ) = - \gamma_0^{S_c^\kappa} (\veps, \zcut ) \,\equiv\,\gamma_0(\veps, \zcut) \,=\, 8C_\kappa \frac{Q\veps}{\qcut}
\,
\qquad
(\beta = 0)
\, .
\end{align}
Hence, for $\beta = 0$, the NLL$'$ resummed cross section for shifted soft drop will involve additional running between the global-soft and collinear-soft virtualities due to this non-cusp anomalous dimension.
We now turn to addressing how the various ${\cal O}(\veps)$ corrections are combined with the other EFT matrix elements so as to arrive at the result for resummed cross section with shifted soft drop condition.

\tocless\subsubsection{Assembling the ingredients}

\noindent
Having found that the shift in the soft drop condition induces different ${\cal O}(\veps)$ modifications to the collinear-soft function on the l.h.s. of \eq{ScConvLargeRg},
we now generalize the decomposition in the r.h.s. of that equation to account for these additional boundary terms.
Since the correction for $\beta = 0$ consists of logarithms induced by the extra anomalous dimension in \eq{noncuspEps}, we find it appropriate to combine this term with $S_c^\kappa\big[\ell^+ \qcut^{\frac{1}{1+\beta}}, \beta, \mu\big]$ in the r.h.s. of \eq{ScConvLargeRg}, which is responsible for RG evolution. On the other hand, the non-logarithmic, $R_g$ dependent terms in \eq{ScLargeShiftb0} can be written as a function of the variable $\tilde \ell$ and are treated in the same way as the $\Delta S_c^{[C_\kappa]}$ piece in \eq{ScLargeRgFO3}.
Thus, we express the renormalized collinear-soft function with cumulant $R_g$ measurement and soft drop shift as follows:
\begin{align}\label{eq:ScLargeShiftConv}
&S_c^\kappa
\Bigl[\ell^+ \qcut^{\frac{1}{1+\beta}}, \frac{R_g}{2}\qcut^{\frac{1}{1+\beta}}, \overline \Theta_{\rm sd}(\veps), \beta , \mu\Bigr] =
\int d \ell^{\prime+}\,	 \qcut^{\frac{1}{1+\beta}}\,
S_c^\kappa \Bigl[(\ell^+ - \ell^{\prime+}) \qcut^{\frac{1}{1+\beta}}, \beta, \mu, \, \delta_{\beta,0}\gamma_0^{S_c^\kappa}(\veps, \zcut) \Bigr]
\nn \\
& \qquad \times \int d\ell''^+\Delta S_c^\kappa
\Bigl [(\ell^{\prime+}-\ell''^+)\qcut^{\frac{1}{1+\beta}}, \frac{R_g}{2}\qcut^{\frac{1}{1+\beta}}, \beta\Bigr]
\: \Delta S_{c,\veps}^{\kappa}
\Bigl [\ell''^+\qcut^{\frac{1}{1+\beta}}, \frac{R_g}{2}\qcut^{\frac{1}{1+\beta}}, \beta\Bigr]\,.
\end{align}
Here the collinear soft function for single differential jet mass measurement includes the additional logarithm that is generated for $\beta = 0$ in \eq{ScLargeShiftb0}, as signaled by the extra argument $\delta_{\beta,0} \gamma_0^{S_c^\kappa}(\veps \zcut)$, and is modified as follows:
\begin{align}\label{eq:ScBeta0}
&S_c^\kappa \Bigl[\ell^+ \qcut^{\frac{1}{1+\beta}}, \beta, \mu, \, \delta_{\beta,0}\gamma_0^{S_c^\kappa}(\veps, \zcut) \Bigr]
\equiv
S_c^\kappa \Bigl[\ell^+\qcut^{\frac{1}{1+\beta}}, \beta, \mu\Bigr]
+\delta_{\beta,0}\frac{Q\veps}{\qcut}
\frac{\alpha_s C_\kappa}{\pi}
\frac{1}{\mu^{2}}
{\cal L}_0 \Big(
\frac{\ell^+\qcut }{\mu^{2}}
\Big) \, ,
\end{align}
while the term $\Delta S_{c,\veps}^{\kappa}$ is defined analogously to $\Delta S_c^\kappa$ given in \eq{ScLargeRgFO}:
\begin{align}
\label{eq:ScLargeRgFOEps}
\Delta S_{c,\veps}^{\kappa}
\Bigl [\ell^+\qcut^{\frac{1}{1+\beta}}, \frac{R_g}{2}\qcut^{\frac{1}{1+\beta}}, \beta\Bigr]
&=
\delta\Bigl( \ell^+ Q_{\rm cut}^{\frac{1}{1+\beta}}\Bigr)
+
\int \frac{d \ell^{\prime +} \qcut^{\frac{1}{1+\beta}}}{\big(\frac{R_g}{2}\qcut^{\frac{1}{1+\beta}}\big)^{2+\beta}}
\:
\Delta S_c^{[\beta_0]}
\Bigl[(\ell^+ - \ell^{\prime+})\qcut^{\frac{1}{1+\beta}}, \beta, \mu\Bigr]
\nn
\\
&
\qquad \times
\frac{Q \veps}{\qcut}\frac{1}{\big(\frac{R_g}{2}\big)^\beta}
\Delta S_{c,\veps}^{[ C_\kappa]}
\Bigg[
\frac{\ell^{\prime +} \qcut^{\frac{1}{1+\beta}}}{\big(\frac{R_g}{2}\qcut^{\frac{1}{1+\beta}}\big)^{2+\beta}},
\beta,\mu
\Bigg]
\, .
\end{align}
The $\Delta S_c^{[\beta_0]}$ piece is the same as that defined in \eq{ScLargeRgFO2} and is again included to cancel the $\mu$ dependence due to the running coupling at NLL$'$.
Finally, from \eq{ScLargeShiftb0} we have
\begin{align}\label{eq:DeltaScLargeShift}
& \Delta S_{c,\veps}^{[ C_\kappa]}
\Bigg[
\tilde \ell =
\frac{\ell^{ +} \qcut^{\frac{1}{1+\beta}}}{\big(\frac{R_g}{2}\qcut^{\frac{1}{1+\beta}}\big)^{2+\beta}},\,
\beta ,\,\mu
\Bigg] =
\\
&
\qquad
\frac{2}{2+\beta}	\frac{\alpha_s (\mu)C_\kappa}{\pi}
\Bigg(
\Theta(\beta > 0)\:
{\cal L}_0^{-\frac{\beta}{2+\beta}}(\tilde \ell)
+
\frac{\Theta(\tilde \ell - 1) }{\tilde \ell^{\frac{2(1+\beta)}{2+\beta}}}
\Bigg)
\bigg(1
+ \frac{\alpha_s(\mu)}{\pi} a_{20,\veps}^{S_c}
\bigg)
+ {\cal O}(\alpha_s^2)
\, .\nn
\end{align}
Here the $\Theta (\beta > 0)$ signifies that the first term is only present for the $\beta > 0$ case, whereas the second term applies for all $\beta \ge 0$. As we shall see later, this second term contributes to the bulk of the boundary corrections, and hence to $C_2(m_J^2)$.

Note that in defining $\Delta S_{c,\veps}^{\kappa}$ in this way we factored out the $R_g^{-\beta}$ power present in \eq{ScLargeShiftb0} for $\beta >0 $. Comparing this scaling with the ${\cal O}(\veps)$ correction to the global-soft function in \eq{SGshift}, we see that the $R_g^{-\beta}$ power dominates the $\sin^{-\beta}(R/2)$ power in the $R_g \ll R$ approximation which is formally still valid in the large $R_g$ region. Thus we expect boundary contributions to the global-soft piece (for $\beta > 0$) to be numerically insignificant, and we will leave them out from our numerical predictions.
Finally, since we lack the knowledge of the non-logarithmic two-loop boundary corrections in $S_c^\kappa$ that cannot be predicted by the anomalous dimensions we have included an additional parameter $a_{20,\veps}^{S_c}$ in spirit of \eq{ScLargeRgFO3}, to facilitate estimating the uncertainty resulting from these terms. We will come back to this again in \sec{Num} when studying the perturbative uncertainties.

Taking the Laplace transform of \eq{ScLargeShiftConv}, performing an expansion up to {\cal O}($\veps$), and dropping terms that are ${\cal O} (\alpha_s^3)$ we explicitly get
\begin{align}\label{eq:ScLargeRgShift}
&\tilde S_c^\kappa\Bigl(s, \frac{R_g}{2}\qcut^{\frac{1}{1+\beta}}, \overline \Theta_{\rm sd}(\veps) , \beta,\mu \Bigr)
=
\tilde S_c^\kappa\Bigl(s, \frac{R_g}{2}\qcut^{\frac{1}{1+\beta}},\beta,\mu \Bigr)
\\
& \qquad
+\frac{Q\veps}{\qcut}\bigg\{
-\delta_{\beta,0}
\frac{\alpha_s C_\kappa}{\pi}
\log(s e^{\gamma_E} \mu^2)
\Delta \tilde S_c^{[ C_\kappa]} \Bigl(s\, \qcut^{\frac{2+\beta}{1+\beta}} \Big(\frac{R_g}{2}\Big)^{2+\beta} , \beta, \mu\Bigr)
\nn
\\
&\qquad +\frac{1}{\big(\frac{R_g}{2}\big)^\beta}\:
\tilde S_c^\kappa\bigl(s,\beta,\mu \bigr)
\Delta \tilde S_c^{[\beta_0]} \bigl (s,\beta,\mu \bigr)
\Delta \tilde S_{c,\veps}^{[ C_\kappa]} \Bigl(s\, \qcut^{\frac{2+\beta}{1+\beta}} \Big(\frac{R_g}{2}\Big)^{2+\beta} , \beta, \mu\Bigr)
\nn
\\
&\qquad +\frac{1}{\big(\frac{R_g}{2}\big)^\beta}\:
\Delta \tilde S_c^{[C_\kappa]} \Bigl(s\, \qcut^{\frac{2+\beta}{1+\beta}} \Big(\frac{R_g}{2}\Big)^{2+\beta} , \beta, \mu\Bigr)
\Delta \tilde S^{[C_\kappa]}_{c,\veps} \Bigl(s\, \qcut^{\frac{2+\beta}{1+\beta}} \Big(\frac{R_g}{2}\Big)^{2+\beta} , \beta, \mu\Bigr)
\bigg\}
\nn
\, ,
\end{align}
where we used \eqs{ScConvLargeRgLap}{ScConvLargeRgLap2} to expand the Laplace transform of the $\Delta S_c^\kappa$ term, while the Laplace transform of the $\Delta S_{c,\veps}^{[ C_\kappa]}$ term is given in \eq{DeltaScLargeLaplace}. The first line reproduces the ${\cal O}(\veps^0)$ result in \eq{ScConvLargeRg} and vanishes when taking the $\veps$ derivative in \eq{C1C2multiDiffDef}. The second line describes boundary corrections induced by the additional logarithm present for $\beta = 0$, while the remaining lines encode boundary corrections that are $R_g$ dependent. The product of the two $\Delta S_c^{[C_\kappa]}$ functions in the last line is important to ensure a smooth transition to the regime $R_g \ll \theta_g^\star$, which justifies including the boundary corrections $\Delta S_{c,\veps}^\kappa$ multiplicatively in \eq{ScLargeShiftConv}.

Likewise, we express the global-soft function with ${\cal O}(\veps)$ corrections as
\begin{align}\label{eq:SGShift}
S_G^\kappa \bigl(\qcut, R, & \overline \Theta_{\rm sd}(\veps), \beta, \mu\bigr)
=
S_G^\kappa\Big(\qcut\tan^{1+\beta}\frac{R}{2}, \beta, \mu_{gs},R\Big)
+ \delta_{\beta,0}\frac{Q\veps}{\qcut} \frac{2\alpha_s C_\kappa}{\pi}
\log \Big(\frac{\mu}{\qcut \tan\frac{R}{2}}\Big)
\nn
\\
&\qquad + \frac{Q \veps}{\qcut}
\frac{\Theta (\beta > 0)}{\sin^\beta \big(\frac{R}{2}\big)} \:
S_G^\kappa\Big(\qcut\tan^{1+\beta}\frac{R}{2}, \beta, \mu_{gs},R\Big)
\Delta S_{G,\veps}^{\kappa} \bigl( \qcut,R, \beta\bigr)
\, ,
\end{align}
with the result for $\mu$-independent $\Delta S_{G,\veps}^{\kappa}$ piece given in \eq{DeltaSGShift}.

\tocless\subsubsection{Cumulative boundary cross section}

\noindent
With all the ingredients for the shifted soft drop condition at hand, we are finally in the position to write resummed results for the cumulative cross section. \eq{FactLargeRgLaplace} is still formally valid, but the global soft and collinear-soft functions that appear there now include boundary corrections according to \eqs{ScLargeShiftConv}{SGShift}. As a consequence, the equivalent of \eq{xsecRGLargeRg} has two kind of terms:
\begin{align}
\label{eq:factLargeRgShift}
\frac{d \Sigma (R_g, \overline \Theta_{\rm sd}(\veps))}{d m_J^2}
&=\frac{d \Sigma\big (R_g, \delta_{\beta,0} \gamma_0(\veps, \zcut)\big)}{d m_J^2}
+ \frac{Q\veps}{\qcut} \frac{d\Delta \Sigma_\veps (R_g)}{d m_J^2 }
+{\cal O}(\veps^2)
\,.
\end{align}
They describe boundary corrections to the NLL RG evolution and the $R_g$ dependence respectively, and we now examine them in turn.
The first term reduces to the ${\cal O}(\veps^0)$ cross section for positive $\beta$ but, for $\beta = 0$, includes the additional one-loop fixed order logarithms and running due to the ${\cal O}(\veps)$ one-loop non-cusp anomalous dimension displayed in \eq{noncuspEps}.
Since these corrections do not depend on the jet mass, they can be absorbed in a generalization of the operator $d \sigma^q(\partial_\Omega)$ in \eq{mJOperator} yielding
\begin{align}
\label{eq:xsecRGLargeShiftb0}
\frac{d \Sigma\big (R_g,\delta_{\beta,0} \gamma_0(\veps, \zcut)\big)}{d m_J^2 } \bigg|_{R_g \lesssim \theta_g^\star}
& \equiv N^q_{\rm evol}\big (Q, \qcut,\beta, R, \mu ; \mu_N, \mu_{gs}\big)
\\
& \times
\frac{d \sigma^q\big [ \partial_\Omega , \delta_{\beta,0} \gamma_0(\veps, \zcut) \big]}{d m_J^2}\:
\Bigg(
\frac{e^{\gamma_E\Omega}}{\Gamma(- \Omega)}
+
{\cal Q}^q\Big [ \Omega, \frac{R_g}{\theta_g^\star}, \beta \Big]
\Bigg)
\bigg|_{\Omega = \tilde \omega(\mu_{cs}, \mu_J)}
\,,
\nn
\end{align}
while the evolution kernel in brackets is unchanged.
The operator is obtained from the following ${\cal O}(\veps)$ modifications to its analogue in \eq{mJOperator}:
\begin{enumerate}
\item The collinear-soft function in \eq{xsecRGLargeRg} includes the additional $\cal{O}(\veps)$ logarithm in \eq{ScLargeRgShift}.
\item The global-soft function includes the additional $\cal{O}(\veps)$ logarithm in \eq{SGShift}.
\item The collinear-soft function has additional running:
\begin{align}
K_{cs}\big(\mu, \mu_{cs}, \delta_{\beta,0}\gamma^{S_c^q}_0(\veps, \zcut)\big)
&=
K_{cs}(\mu,\mu_{cs}) - \delta_{\beta,0}\,
\eta\Big(
\gamma_0(\veps ,\zcut ) , \mu , \mu_{0}
\Big)
\, ,
\end{align}
where the LL single logarithmic evolution kernel $\eta(\Gamma, \mu, \mu_0)$ is given in \eq{w} from the one-loop non-cusp anomalous dimension.
\item The normalization factor $N^q_{\rm evol}$ has additional running due to the global soft function:
\begin{align}
K_{gs}\big(\mu, \mu_{gs}, \delta_{\beta,0}\gamma^{S_G^q}_0(\veps, \zcut)\big)
&=
K_{gs}(\mu,\mu_{gs}) + \delta_{\beta,0}\,
\eta\big(
\gamma_0(\veps ,\zcut ) , \mu , \mu_{0}
\big)
\, .
\end{align}
\end{enumerate}

The second term in \eq{factLargeRgShift} consists of the residual ${\cal O}(\veps)$ non-logarithmic, $R_g$-dependent corrections to the collinear-soft and global-soft functions in \eqs{ScLargeRgFOEps}{SGShift}, which apply for every $\beta \geq 0$.
Since these terms do not alter evolution, but have a nontrivial dependence on the jet mass and $R_g$, they require a different kernel than \eq{UJScOmega}:
\begin{align}\label{eq:UJScOmegaEps}
{\cal Q}^\kappa_{\veps}\bigg [ \Omega, \upsilon = \Big( \frac{R_g}{\theta_g^\star}\Big)^{2+\beta}, \beta \bigg]
&\equiv
e^{\gamma_E\Omega}
\Delta \tilde S_c^{[\beta_0]} \bigg [\partial_{\Omega} + \log\bigg( \frac{Q \mu_{cs}}{m_J^2} \Big(\frac{\mu_{cs}}{\qcut}\Big)^{\frac{1}{1+\beta}}\bigg) ,\beta, \alpha_s(\mu_{cs})\bigg ] \\
& \times
\int \frac{d\varrho }{2\pi i} \: e^{\varrho }
\varrho^{\Omega}
\Delta \tilde S_{c,\veps}^{[C_\kappa]}
\big(\varrho\vf, \beta, \mu_{cs}\big)
\Big(1 + \Delta \tilde S_{c}^{[C_\kappa]}
\big(\varrho \vf, \beta, \mu_{cs}\big)\Big)
\, ,
\nn
\end{align}
but use the same function of the derivative operator, $d \sigma^\kappa(\partial_\Omega)$ defined in \eq{mJOperator}, valid for $\veps = 0$:
\begin{align}
\label{eq:xsecRGLargeShiftDelta}
\frac{d\Delta \Sigma_\veps (R_g)}{d m_J^2}\bigg|_{R_g \lesssim \theta_g^\star}
&=
N^q_{\rm evol}\big (Q, \qcut,\beta, R, \mu ; \mu_N, \mu_{gs}\big)
\frac{d \sigma^q\big [\partial_\Omega \big]}{d m_J^2}
\Bigg(
\frac{1}{\big(\frac{R_g}{2}\big)^\beta}\:
{\cal Q}^q_{\veps}\Big [ \Omega, \frac{R_g}{\theta_g^\star}, \beta \Big]
\\
&\qquad + \frac{\Theta (\beta > 0)}{\sin^\beta \big(\frac{R}{2}\big)}
\:\Delta S_{G,\veps}^{q} \bigl (\qcut,R, \beta\bigr)
\frac{e^{\gamma_E\Omega}}{\Gamma(- \Omega)}
\Bigg)
\bigg|_{\Omega = \tilde \omega(\mu_{cs}, \mu_J)}
\nn
\, .
\end{align}
The explicit expression of the evolution kernel ${\cal Q}^\kappa_{\veps}$ between jet and soft scales for shifted soft drop is given in \eq{UJScOmegaEpsFull}.
Note that the term $\Delta S_{G,\veps}^{\kappa}$ for $\beta > 0$ at NLL$'$ accuracy contributes only to the boundary correction in the single differential jet mass cross section, and is thus independent of $R_g$. In the following, we will simply ignore this term as this is also power suppressed compared to the term in the first line in \eq{xsecRGLargeShiftDelta}.

We now briefly describe the generalization of the results above to $pp$ collisions. First off, the proportionality constant of ${\cal O}(\veps)$ terms is substituted as follows
\begin{align}\label{eq:ppC2Prefactor}
\frac{Q}{\qcut \big(\frac{R_g}{2}\big)^\beta} \ra \frac{1}{R_g^\beta \zcut'} \, .
\end{align}
The same applies for the ${\cal O}(\veps)$ one-loop non-cusp anomalous dimension in \eq{noncuspEps}. Thus generalization of the cross section in \eq{xsecRGLargeShiftb0} to the $pp$ case is simply given by \eq{ppmJOperator} after including the additional ${\cal O}(\veps)$ logarithms and single log kernels as outlined above. The result for the second term in \eq{factLargeRgShift} for $pp$ collisions can also be analogously derived from \eq{xsecRGLargeShiftDelta}:
\begin{align}
\label{eq:ppxsecRGLargeShiftDelta}
\frac{\veps}{\zcut'}\frac{d\Delta \Sigma_\veps (R_g)}{d m_J^2 d\Phi_J}\bigg|_{R_g \lesssim \theta_g^\star}
&=
\sum_{\kappa = q,g}
N^\kappa_{\rm evol}\big (\Phi_J, \qcut,\beta, R, \mu ; \mu_N, \mu_{gs}\big)
\\
&\qquad \times
\frac{d \sigma^\kappa\big [\partial_\Omega \big]}{d m_J^2 d\Phi_J}
\Bigg(
\frac{\veps}{\zcut' R_g^\beta}\:
{\cal Q}^\kappa_{\veps}\Big [ \Omega, \frac{R_g}{\theta_g^{(pp)\star}(m_J^2)}, \beta \Big]
\Bigg)
\bigg|_{\Omega = \tilde \omega(\mu_{cs}, \mu_J)}
\nn
\, .
\end{align}

At this point, we have the soft drop boundary cross section which is cumulant in the groomed jet radius. In order to evaluate $C_2$, following \eq{C1C2multiDiffDef}, we need to switch to the cross section differential in the groomed jet radius and further take the $\veps$ derivative. We perform this explicitly in \app{C2diff}, where we compile the results for the boundary cross section differential in $R_g$. The large $R_g$ regime discussed so far is the only relevant geometry in the definition of $C^q_2(m_J^2)$. For completeness, we discuss in \app{LargeRgShift} the boundary corrections in the intermediate and small $R_g$ regimes and show consistency with the results in the large $R_g$ regime as $R_g/\theta_g^\star(m_J^2) \ra 0$.

\subsection{Recovering the Leading Logarithmic coherent branching result} \label{sec:LLCB}
In this section we make use of the results derived above for the cumulative cross section and the boundary corrections to demonstrate that
the EFT that we have developed reproduce the computation of $C_1$ and $C_2$ at LL accuracy within the coherent branching formulation from~\cite{Hoang:2019ceu}.
This also makes explicit how the EFT encodes the strong angular ordering limit used there. For simplicity we limit ourselves to the case of $\ee$ collisions here.
The LL results for $C_1$ and $C_2$ were derived by considering a chain of emissions that are ordered in their respective contribution to the jet mass.\footnote{Formally, the emissions are to be angular ordered, but in the derivation the angular ordering is replaced by their ordering in the jet mass, i.e. in the variable $\rho_i = z_i \theta_i^2$ for the $i^{\rm th}$ emission. This approximation is essential to set the resolution of the jet mass measurement.} The emission at the largest angle is required to satisfy soft drop and set the jet mass. This leads to the following expressions for the two moments:
\begin{align} \label{eq:CBresult0}
C_1^q(m_J^2) &=
\bigg(\frac{C_0^q(m_J^2)}{m_J^2}\bigg)^{-1}
\int_0^1 dz
\int_{0}^R \frac{d\theta^2}{\theta^2} \: \frac{\theta}{2}\frac{C_F}{\pi}\alpha_s\Big( \frac{Qz\theta}{2}\Big)
p_{gq}(z) e^{-{\cal R}( z \theta^2)}
\\
&\qquad \times
\Theta \bigg(z - \frac{\qcut}{Q}\Big(\frac{\theta}{2}\Big)^\beta\bigg)
\delta \Big (m_J^2 -\frac{1}{4} z \theta^2 Q^2\Big )\,,\nn
\\
C_2^q(m_J^2) &= \frac{m_J^2}{Q^2} \bigg(\frac{C_0^q(m_J^2)}{m_J^2}\bigg)^{-1}
\int_0^1 dz
\int_{0}^R \frac{d\theta^2}{\theta^2} \: \frac{2}{\theta}\frac{C_F}{\pi}\alpha_s\Big( \frac{Qz\theta}{2}\Big)
p_{gq}(z) e^{-{\cal R}( z \theta^2)}
\nn \\
&\qquad \times
\delta \bigg(z - \frac{\qcut}{Q}\Big(\frac{\theta}{2}\Big)^\beta\bigg)
\delta\Big (m_J^2 -\frac{1}{4} z \theta^2 Q^2\Big )\,,\nn
\end{align}
where $C_0^q(m_J^2)/m_J^2$ is the single differential jet mass distribution and is given at this order by
\begin{align} \label{eq:CBresultC0}
\frac{1}{m_J^2}C_0^q(m_J^2) \equiv \frac{d\hat \sigma_{\rm LL}^q}{d m_J^2} &=
\frac{1}{m_J^2}\int_0^1 dz
\int_{0}^R \frac{d\theta^2}{\theta^2} \frac{C_F}{\pi}\alpha_s\Big( \frac{Qz\theta}{2}\Big)
p_{gq}(z) e^{-{\cal R}( z \theta^2)}
\\
&\qquad \times
\Theta \bigg(z - \frac{\qcut}{Q}\Big(\frac{\theta}{2}\Big)^\beta\bigg)
\delta \Big (m_J^2 -\frac{1}{4} z \theta^2 Q^2\Big )\,,\nn
\end{align}
where the result for $C_0(m_J^2)$ is valid up to LL.
In \eqs{CBresult0}{CBresultC0} we have rewritten the soft drop condition in \eq{SD} terms of $\qcut$. In the expression for $C_1^q$ and $C_0^q$, the condition is imposed on the last emission with energy fraction $z$ and angle $\theta$.
In case of $C_2^q$ we see, however, that the soft drop $\Theta$ function is replaced by $\delta$ function in order to capture the corrections at the soft drop boundary.
The radiator, $\mathcal{R}_q(\theta_c^2)$, results from all the emissions at larger angles that are either vetoed by the jet mass measurement or that fail soft drop (and the virtual emissions). The radiator resums double logarithms of $\theta_g^\star/\theta_c$, which however cancel out in the ratio in \eqs{CBresult}{CBresultC0}. The splitting kernel $p_{gq}$ is given by
\begin{align} \label{eq:splittingKernel}
p_{gq}(z) = \frac{1+(1-z)^2}{2z} \underset{z\to 0}{\sim} \frac{1}{z}\,.
\end{align}
Terms that are power corrections in the $z\to 0$ limit are formally beyond LL accuracy, and we will drop them in the following; in the EFT formulation, these are captured by the NLO corrections to the jet function. Simplifying \eq{CBresult0} and using the LL limit of $p_{gq}(z)$ in \eq{splittingKernel} we get
\begin{align}\label{eq:CBresult}
C_0^q(m_J^2) \big|_{\rm LL}&= e^{-\mathcal{R}_q(\theta_c^2)}
\int_{\theta_c}^R \frac{d\theta^2}{\theta^2} \frac{C_F}{\pi}\alpha_s\Big(\frac{m_J\theta_c}{\theta}\Big) 
\Theta(\theta_g^\star -\theta)\,, \\
C_1^q(m_J^2)\big|_{\rm LL}	&=\frac{e^{-\mathcal{R}(\theta_c^2)}}{C_0^q(m_J^2)}
\int_{\theta_c}^R \frac{d\theta^2}{\theta^2} \frac{\theta}{2}\frac{C_F}{\pi}\alpha_s\Big(\frac{m_J\theta_c}{\theta}\Big) 
\Theta(\theta_g^\star -\theta)\,, \nn
\\
C_2^q(m_J^2) \big|_{\rm LL} &= \frac{e^{-\mathcal{R}(\theta_c^2)}}{C_0^q(m_J^2)}
\int_{\theta_c}^R \frac{d\theta^2}{\theta^2} \: \frac{2}{\theta}
\frac{C_F}{\pi}\alpha_s\bigg(\qcut\Big(\frac{\theta}{2}\Big)^{1+\beta} \bigg)
\frac{Q}{\qcut} \Big(\frac{\theta}{2}\Big)^{-\beta}
\delta\left(1 - \Big(\frac{\theta}{\theta_{g}^\star}\Big)^{ 2+\beta}\right)
\, ,
\nn
\end{align}
where we have expressed the results in terms of the angular bounds $\theta_c$ and $\theta_g^\star$ defined in \eq{angularBoundsExplicit}.

To establish a connection with our results, we must compare with the large $R_g$ regime of \sec{LargeRgFact}, since due to the associated geometry for power corrections in the SDOE region of the single differential jet mass distribution, this is the primary regime that is relevant for defining the coefficients $C_1$ and $C_2$.\footnote{We note that the coherent branching description is based on ordering in a single variable, the jet mass contribution of each emission, $\rho_i = z_i\theta_i^2$.
This is in correspondence with the mode picture in the large $R_g$ regime where the measurement of jet mass provides a distinction between the various EFT modes (see \fig{modePicture}, left). This is, however, not the case with small and intermediate $R_g$ regimes:
in the case of small $R_g$ regime, the primary measurement is the groomed jet radius which distinguishes the modes based on their angular separation. In the intermediate $R_g$ regime an analogous treatment in coherent branching would presumably involve ordering in two different variables.} Thus, the EFT prediction for the LL double differential cross section in this regime is obtained from the cumulant in \eq{xsecRGLargeRg},
\begin{align} \label{eq:towardsCB}
\frac{d^2 \hat{\sigma}_{\rm LL}^q}{d m_J^2 d\theta_g }\bigg|_{R_g \lesssim \theta_g^\star}
&=
\frac{d}{d R_g}\bigg[\frac{d\sigma^q_{\rm LL}\big [\partial_\Omega \big]}{dm_J^2}
\bigg(
\frac{e^{\gamma_E\Omega}}{\Gamma(- \Omega)}
+
{\cal Q}^q\Big [ \Omega, \frac{R_g}{\theta_g^\star}, \beta \Big]
\bigg)\bigg]_{R_g=\theta_g}
\nn \\
&= e^{-\mathcal{R}_q} \frac{d}{d R_g}
{\cal Q}^q \Big [ \Omega, \frac{R_g}{\theta_g^\star}, \beta \Big] \bigg|_{R_g=\theta_g}\,.
\end{align}
In the second line we have used that the evolution in this regime is independent of the groomed jet radius and identified the LO differential operator with the exponential of the radiator $\mathcal{R}^q$. This follows by setting the Laplace space fixed order ingredients in \eq{mJOperator} to unity.

We now notice that the fixed-order corrections in \eqs{CBresult}{CBresultC0} are described in the coherent branching formalism by the single splitting function $p_{gq}$, in contrast with the convolution of the jet and collinear-soft functions in \eq{FactLargeThetaCS}. This implies that the coherent branching result ignores some subleading terms in the kernel $\mathcal{Q}^q$, which are induced by performing the RG evolution between $\mu_J$ and $\mu_{cs_m}$ in Laplace space rather than distribution space. These terms are instead included in our EFT approach; we can remove them by setting $\Omega\to 0$ in \eq{UJScOmegaFull}, obtaining
\begin{align} \label{eq:Qexpand}
{\cal Q}^q \Big [ \Omega\to 0 , \frac{R_g}{\theta_g^\star}, \beta \Big] = \Theta(\theta_g^\star-\theta_g)\,\frac{2\alpha_s(\mu_{cs})C_F}{\pi}\,\ln\Big(\frac{\theta_g^\star}{R_g}\Big)\,,
\end{align}
where we have used that under the integral
\begin{align} \label{eq:distribExpansion}
x^{-1-\Omega} = -\frac{1}{\Omega}\delta(x) + \mathcal{O}(\Omega^0)\,.
\end{align}
Next, we adopt the natural scale choice for $\mu_{cs}$ in \eq{muCSNatural}, noting that
\begin{align}
m_J^2\frac{\theta_c}{\theta_g^\star} = \mu_{cs}^{\rm can.} \, .
\end{align}
This corresponds to setting $\theta = \theta_g^\star$ in \eq{CBresult}, with the difference between the two expressions being beyond LL accuracy.
Substituting \eq{Qexpand} in \eq{towardsCB} gives the integrand of \eq{CBresult}. The LL coherent branching result for $C_1$ then trivially follows by including the additional factor $\theta/2$ in the angular integration.

To obtain the LL coherent branching result for $C_2$ from the EFT formalism we can follow the same steps. We first notice that the anomalous evolution induced by boundary corrections at $\beta=0$ affects the cross section beyond LL, and we can thus ignore this correction. The relevant cumulant cross section for large $R_g$ is then given by \eq{xsecRGLargeShiftDelta}. Differentiating in $R_g$, we get up to $\mathcal{O}(\veps^2)$ corrections
\begin{align}
\label{eq:DDiffRGLargeShiftDelta}
\frac{d \hat{\sigma}_{\rm LL}^q(\veps)}{d m_J^2 d\theta_g}\bigg|_{R_g \lesssim \theta_g^\star}
&= e^{-\mathcal{R}_q} \frac{d}{d R_g} \bigg[ \Big(\frac{R_g}{2}\Big)^{-\beta}
{\cal Q}_\veps^q \Big ( \Omega, \frac{R_g}{\theta_g^\star}, \beta \Big)\bigg]_{R_g=\theta_g}\, \\ \nn
&= e^{-\mathcal{R}_q} \frac{Q\veps}{\qcut}
\frac{\alpha_s(\mu_{cs}) C_F}{\pi}\Big(\frac{\theta_g}{2}\Big)^{-(1+\beta)}
\frac{e^{\gamma_E\Omega}}{\Gamma(- \Omega)}
{\cal L}_0^{-\Omega} \bigg(1 - \Big(\frac{\theta_g}{\theta_{g}^\star(m_J^2)}\Big)^{ 2+\beta}\bigg)
\, .
\nn
\end{align}
The first line is obtained from the same manipulations as in \eq{towardsCB}, while the explicit expression in the second line follows from differentiating the kernel $\mathcal{Q}^q_\veps$ given in \eq{UJScOmegaEps}, which reduces to \eq{UJScOmegaEpsFull} when discarding NLO corrections.
In order to compare with the coherent branching result, we again expand out this result in the limit $\Omega \rightarrow 0$. Using the expansion in \eq{distribExpansion},
\begin{align}
\frac{1}{\Gamma(-\Omega)}{\cal L}_0^{-\Omega} \bigg(1 - \Big(\frac{\theta_g}{\theta_{g}^\star}\Big)^{ 2+\beta}\bigg)
\,\rightarrow\,
\delta\left(1 - \Big(\frac{\theta_g}{\theta_{g}^\star}\Big)^{ 2+\beta}\right)
\end{align}
so that
\begin{align}
\frac{d}{d\veps}\bigg|_{\veps =0}\frac{d \hat{\sigma}_{\rm LL}^q(\veps)}{d m_J^2 d\theta_g}\bigg|_{R_g \lesssim \theta_g^\star}
\,\rightarrow\,
\frac{Q}{\qcut}\frac{\alpha_s(\mu_{cs}) C_F}{\pi}\Big(\frac{\theta_g}{2}\Big)^{-(1+\beta)} \delta\left(1 - \Big(\frac{\theta_g}{\theta_{g}^\star(m_J^2)}\Big)^{ 2+\beta}\right)\, .
\end{align}
Including the factor of $m_J^2/Q^2$, according to our definition of $C_2^\kappa$ from \eq{C1C2multiDiffDef}, this yields the LL coherent branching result of \eq{CBresult}. Finally, setting $\theta = \theta_g^\star$ using the $\delta$ function and noting that
\begin{align}
\qcut \Big(\frac{\theta_g^\star}{2}\Big)^{1+\beta} = \mu_{cs}^{\rm can.}\, ,
\end{align}
we find that the arguments of the running coupling in the two expressions also match.

\section{Matching the three regimes}
\label{sec:match}
Here we describe our strategy for consistently matching the results in various regimes so as to obtain a smooth interpolation across the entire phase space of $m_J^2$-$R_g$ measurement.
Which of the three EFTs come into play will crucially depend on the range over which the groomed jet radius varies for given values of $\zcut$, $\beta$ and the jet mass.
In the following we first describe the basic strategy for implementation of the matched cross section in \sec{matchImplement}
using profile functions for scale choices~\cite{Ligeti:2008ac,Abbate:2010xh} and weight functions which control the contribution from different EFTs~\cite{Lustermans:2019plv}.
In \sec{trans} we generalize the weight function construction to depend on two variables that track where we are in our two dimensional phase space so that the appropriate factorized formula can be used.
Next in \sec{prof} we describe the profile functions for each of the large, intermediate, and small $R_g$ regimes, and then in \sec{vary} how we vary them to obtain an estimate for the perturbative uncertainty.

\subsection{Implementation of matched cross section}
\label{sec:matchImplement}

Here we describe our implementation of the complete cumulative-$R_g$ and differential jet mass soft drop cross section. This will then set up the groundwork for evaluation of the moments $C_{1,2}^\kappa(m_J^2)$.
An important ingredient are the profile functions for our different EFT regions, which are the functions used for the renormalization $\mu_i$ scales in the factorization formulae, and which depend on both $m_J$ and $R_g$.
First we recall the set of scales that are relevant for the various kinematic regions:
\begin{align}	\label{eq:profAll}
&\text{Large $R_g$ region}:
& & R_g \lesssim \theta_g^\star (m_J^2, \qcut, \beta) < R
& &\mu_{\rm large} \equiv \{\mu_J, \mu_{gs}, \mu_{cs} \} \, ,
\nn \\
&\text{Intermediate $R_g$ region}:
& & \theta_c(m_J^2) \ll R_g \ll \theta_g^{\star}(m_J^2, \qcut, \beta)
& &\mu_{\rm int} \equiv \{\mu_J, \mu_{gs}, \mu_{cs_m}, \mu_{cs_g}\}
\,, \nn
\\
&\text{Small $R_g$ region}:
& & \theta_c(m_J^2) \lesssim R_g
& &\mu_{\rm small} \equiv \{\mu_c, \mu_{gs}, \mu_{cs_g} \}
\, ,
\nn
\\
&\text{Plain jet mass region}:
& & R_g \lesssim R
& &\mu_{\rm plain} \equiv \{\mu_J, \mu_s \}
\,.
\end{align}
In this analysis we will focus exclusively on the groomed region described by our EFT, although we will also extrapolate our results into the ungroomed region. In the groomed region, we construct the interpolation between different EFT regimes via the following formula:
\begin{align}\label{eq:SigMatched}
d \Sigma_{\rm matched} \equiv d\Sigma_{\rm int} |_{\mu_{\rm hyb}}
+ \big(d\Sigma_{\rm large} - d\Sigma_{\rm int} \big)|_{\mu_{\rm large}}
+ \big(d\Sigma_{\rm small} - d\Sigma_{\rm int} \big)|_{\mu_{\rm small}}
\, ,
\end{align}
where $d\Sigma_i(m_J^2)$ for $i = $\{large, int, small\} stand for the cumulative-$R_g$ factorized cross section formulae in the large, intermediate and small $R_g$ regimes respectively. We have also indicated the choice of profile scales used for jet, global-soft and collinear-soft scales
used in these factorization formulae utilizing the notation introduced in \eq{profAll}.
In \eq{SigMatched} the cross sections in the large and small $R_g$ regimes are evaluated with their respective profile scales, whereas the intermediate $R_g$ uses either the profile scales in \eq{profAll} (in the two subtraction terms), or a set of hybrid profiles denoted by $\mu_{\rm hyb}$. The hybrid $\mu_{\rm hyb}$ profiles are designed such that they interpolate between the $\mu_{\rm small}$, $\mu_{\rm int}$ and $\mu_{\rm large}$ sets as follows:
\begin{align}
\label{eq:muHyb}
\mu_{\rm hyb} \equiv \big(\mu_{\rm small}\big)^{w_{\rm small}} \,
\big(\mu_{\rm int}\big)^{w_{\rm int}}
\big(\mu_{\rm large}\big)^{w_{\rm large}}
\, , \qquad w_{\rm small} + w_{\rm int} + w_{\rm large} = 1 \, ,
\end{align}
where the exponents $0 \leq w_i \equiv w_i(m_J, R_g) \leq 1$ should turn on/off in the appropriate regions and add up to one~\cite{Lustermans:2019plv}. We give an explicit construction of these weight functions below for our case.

For now, we note that in the large $R_g$ region we will have $w_{\rm large} = 1$ (and hence $w_{\rm small} = w_{\rm int} = 0$) such that the $\mu_{\rm hyb}$ profiles are the same as the set $\mu_{\rm large}$. The term $d\Sigma_{\rm int} |_{\mu_{\rm large}}$ in \eq{SigMatched} corresponds to evaluating the intermediate $R_g$ cross section with large $R_g$ profile scale settings, i.e. by setting $\mu_{cs_g}, \mu_{cs_m}\ra \mu_{cs}$. This leads to exact cancellation of the terms $d\Sigma_{\rm int} |_{\mu_{\rm hyb}}$ and $d\Sigma_{\rm int} |_{\mu_{\rm large}}$ in the $R_g \lesssim \theta_g^\star$ region where $w_{\rm large} = 1$, and hence the matched cross section is given by
\begin{align}\label{eq:SigMatchedLarge}
d \Sigma_{\rm matched} = d\Sigma_{\rm large} |_{\mu_{\rm large}} + \big(d\Sigma_{\rm small} - d\Sigma_{\rm int} \big)|_{\mu_{\rm small}} \, , \qquad R_g \lesssim \theta_g^\star \,, \quad (w_{\rm large} = 1) \, .
\end{align}
The term $d\Sigma_{\rm int} |_{\mu_{\rm small}}$ here denotes the intermediate $R_g$ cross section evaluated with the $\mu_{\rm small}$ profiles, i.e. by replacing $\mu_{cs_m} , \mu_J\ra \mu_c$. Thus the difference $\big(d\Sigma_{\rm small} - d\Sigma_{\rm int} \big)|_{\mu_{\rm small}}$
supplements \eq{SigMatchedLarge} with the power corrections that are missing in both the intermediate and large $R_g$ factorizations in the $\theta_c \lesssim R_g $ region.

Similarly, in the small $R_g$ region we will have $w_{\rm small} = 1$ (and thus $w_{\rm large} = w_{\rm int} = 0$). As a result, the $\mu_{\rm hyb}$ profiles will coincide with $\mu_{\rm small}$ and the terms $d\Sigma_{\rm int} |_{\mu_{\rm hyb}}$ and $d\Sigma_{\rm int} |_{\mu_{\rm small}}$
in \eq{SigMatched} will cancel exactly. Hence, the $d \Sigma_{\rm matched}$ in this region will be given by
\begin{align}
d \Sigma_{\rm matched} = d\Sigma_{\rm small} |_{\mu_{\rm small}} + \big(d\Sigma_{\rm large} - d\Sigma_{\rm int} \big)|_{\mu_{\rm large}} \,,
\qquad \theta_c \lesssim R_g \,, \qquad (w_{\rm small} = 1) \, .
\end{align}
By using $\mu_{\rm large}$ profiles in the intermediate $R_g$ cross section we turn off the additional resummation relative to the large $R_g$ regime, so that the second term $\big(d\Sigma_{\rm large} - d\Sigma_{\rm int} \big)|_{\mu_{\rm large}}$ captures the power corrections that are missing in the intermediate and small $R_g$ factorizations in the $R_g \lesssim \theta_g^\star$ region.

Finally, if there is a contribution purely from the intermediate region where $w_{\rm int} = 1$, the $\mu_{\rm hyb}$ profiles will be the same as the bulk intermediate $R_g$ profiles in \eq{profAll}. Here the term $d\Sigma_{\rm int} |_{\mu_{\rm hyb}}$ in \eq{SigMatched} provides for the essential intermediate $R_g$ resummation, whereas the other two terms contribute the power corrections from the large and small $R_g$ regions.
We will find that the intermediate regime only has a dominant weight for larger values of $\beta$, and even then in a rather small range of phase space.
When the intermediate regime is not relevant, the formalism we develop simply implements a smooth transition
between the small and large $R_g$ EFTs with $w_{\rm int} = 0$ throughout the entire range of $R_g$ (for a fixed $m_J$).
The smooth variation of $w_i(m_J^2,R_g)$ will ensure consistent interpolation between all these regions.

\subsection{Weight functions}
\label{sec:trans}

Here we describe the implementation of weight functions $w_i(m_J^2, R_g)$ that appear in \eq{muHyb}. We first recall that the transition from the large $R_g$ EFT, $ R_g \lesssim \theta^\star_g(m_J^2,\qcut,\beta)$, to the intermediate EFT corresponds to the canonical \CSg and \CSm scales becoming well separated. On the other hand, transitioning from the small $R_g$ EFT, $\theta_c(m_J^2) \lesssim R_g$, to the intermediate EFT happens when $R_g \gg \theta_c(m_J^2)$, resulting from separation of the canonical \CSm and the collinear scales. We can therefore characterize the regions with two power counting parameters defined as
\begin{align}\label{eq:IntermFactExpn}
\textrm{${\cal C^\kappa}$ factorizes into $J_\kappa$ and $S_{c_m}^\kappa$}: &\qquad \lambda_{\rm min}(R_g) \equiv \frac{\mu_{cs_m}}{\mu_J} \ll 1 \, ,\\
\textrm{$S_c^\kappa$ factorizes into $S_{c_g}^\kappa$ and $S_{c_m}^\kappa$}:&\qquad \lambda_{\rm max}(R_g)\equiv \frac{\mu_{cs_g}}{\mu_{cs_m}} \ll 1\, . \nn
\end{align}
Here the $\mu$ scales stand for the virtuality of the modes determined by $\{m_J^2, R_g, Q, \qcut,\beta\}$ such that
\begin{align}
\label{eq:LambdaCgCS}
\lambda_{\rm min}(R_g) = \frac{m_J}{QR_g/2} = \frac{\theta_c(m_J^2)}{R_g}
\, ,\quad
\lambda_{\rm max}(R_g) = \frac{Q \qcut}{m_J^2}\Big(\frac{R_g}{2}\Big)^{2+\beta}
= \Big(\frac{R_g}{\theta_g^\star(m_J^2,\qcut, \beta)}\Big)^{2+\beta}
,
\end{align}
which is consistent with \eqs{CFact}{CSrefactorization}.\footnote{Although the power corrections between the small and intermediate $R_g$ regimes are of the form $\theta_c^2/R_g^2$, it is the power counting parameter $\lambda_{\rm min} = \theta_c/R_g$ that is the important one, since it controls the size of the logarithms in the transition between leading power contributions in different regimes.} The $pp$ case involves precisely the same expressions but with the $pp$ versions of the angles $\theta_c$ and $\theta_g^\star$ used in \eq{angularBoundspp}. Furthermore, since the power counting parameters are defined via ratios of angles, the factors of $\cosh\eta_J$ cancel, and thus all the formulae derived in this section will apply for the $pp$ case with the corresponding definition of $\qcut$ in \eq{qcutDef}.

Note that $\lambda_{\rm min}$ grows with $R_g$ while $\lambda_{\rm max}$ decreases with $R_g$, and in particular they become equal when $R_g=\theta_t$, where
\begin{align}
\theta_{t} (m_J^2, \qcut, \beta)\equiv \big[ \theta_c\, (\theta_g^\star)^{2+\beta} \big]^{\frac{1}{3+\beta}}
= 2 \bigg[ \Big ( \frac{m_J}{Q}\Big )^3 \frac{Q}{\qcut}\bigg]^{\frac{1}{3+\beta}}
\, .
\end{align}
Whether or not the additional resummation in the intermediate $R_g$ region is essential is given by the condition $\lambda_{\rm min}(\theta_t) = \lambda_{\rm max}(\theta_t) \ll 1$. In our numerical analysis below we replace the strong inequality by a factor of 3, such that for intermediate EFT resummation to be significant we require
\begin{align}\label{eq:IntValid}
\lambda_{\rm min}(\theta_t) = \lambda_{\rm max}(\theta_t)
= \bigg[\Big(\frac{m_J}{Q}\Big)^\beta \frac{\qcut}{Q}\bigg]^{\frac{1}{3+\beta}}
\leq \frac{1}{3}\,.
\end{align}
When this condition is not satisfied, we directly match the cross section in region $R_g \lesssim \theta_{g}^\star$ to $\theta_c\lesssim R_g$. We refer to this as the ``two-EFT scheme'' since it does not include the intermediate $R_g$ regime. In the opposite situation, when the condition in \eq{IntValid} is valid, then the additional resummation in the intermediate $R_g$ region is needed, thus defining a ``three-EFT'' scheme. For a fixed value of $\{Q, \qcut, \beta\}$, this happens at smaller jet masses. In the three-EFT scheme we can further identify the two angles where respectively the small-$R_g$ and large-$R_g$ power corrections are equal to $1/3$,
\begin{align}
\theta_{t, \rm min}(m_J^2) \equiv 3 \theta_c(m_J^2) \, , \qquad
\theta_{t, \rm max} (m_J^2, \qcut, \beta) \equiv 3^{\frac{-1}{2+\beta}}\theta_g^\star(m_J^2, \qcut, \beta)
\, .
\end{align}
This suggests that intermediate $R_g$ resummation is important for $\theta_{t, \rm min} \leq R_g\leq \theta_{t, \rm max}$.

Finally, we introduce a transition function which smoothly interpolates between $0$ and $1$,
\begin{align}\label{eq:Xtrans}
X(\theta, \theta_t) \equiv \frac{1}{2} \bigg[
1 + \tanh \Big( r_t \frac{\theta - \theta_t}{\theta_{\rm max} - \theta_{\rm min}}\Big)
\bigg]
\, ,
\end{align}
where $r_t$ controls the rate of the transition and is normalized with respect to the range of integration. In our numerical analysis in \sec{Num} we use $r_t =20$. This transition function is used to construct the weights $w_{\rm int}$ and $w_{\rm large}$ for the two- and three-EFT schemes:
\begin{align}\label{eq:weightFunc}
&\text{2-EFT scheme:}&
&w_{\rm large} = X(R_g, \theta_t) \, ,&
&w_{\rm int}=0 \, ,&
\\
&\text{3-EFT scheme:}&
&w_{\rm large} = X(R_g, \max \{ \theta_{t, \rm max}, \theta_t\}) \, ,&
&w_{\rm int} = \big(1 - X(R_g, \theta_{t, \rm max})\big)X(R_g, \theta_{t, \rm min}) \, ,&
\nn
\end{align}
and $w_{\rm small} = 1 - w_{\rm large} - w_{\rm int}$ in either of these cases.

In the 2-EFT scheme, the large $R_g$ EFT is employed for $R_g > \theta_t$ and small $R_g$ for $R_g < \theta_t$. Here the intermediate $R_g$ resummation is not present and the piece $d\Sigma_{\rm int} |_{\mu_{\rm hyb}}$ in \eq{SigMatched} serves mainly to provide an interpolation between these two regimes. In the 3-EFT scheme, the intermediate $R_g$ resummation is effective in the range $\theta_{t, \rm min} < R_g < \theta_{t, \rm max}$.
Note that $w_{\rm large}$ in this scheme reduces to the corresponding weight in the 2-EFT scheme when \eq{IntValid} is valid with equality sign. The transition function $X(R_g, \theta_{t, \rm min})$ in $w_{\rm int}$ turns off the intermediate $R_g$ resummation for $R_g \ll \theta_{t, \rm min}$. Furthermore, in the 2-EFT scheme, we have $\theta_{t, \rm min} > \theta_{t, \rm max}$ which automatically drives $w_{\rm int}$ to $0$.

\subsection{Profile functions}
\label{sec:prof}

With the exception of the hard and global-soft functions, the logarithms that appear in the various matrix elements in the soft drop cross sections depend on the jet mass and groomed jet radius, and thus vary along the spectrum. Hence, to minimize these logarithms, the renormalization scales must also vary accordingly in the resummation region as discussed in \sec{resummation}. Additionally, for larger jet masses where $R_g\lesssim R$ we enter the ungroomed region where the soft drop resummation must be turned off. For smaller jet masses, when we approach the nonperturbative region the soft scale must be frozen to an ${\cal O}(1)$ GeV value. This requires some care since the boundary of the nonperturbative region is a non-trivial function of both $m_J$ and $R_g$ measurements. Hence, in order to consistently describe the entire jet mass spectrum, we make use of profile functions~\cite{Abbate:2010xh,Ligeti:2008ac} which depend on the jet mass and groomed jet radius. In the following we will state results for both $\ee$ and $pp$ cases.

\tocless\subsubsection{General strategy and canonical relations}
In this section we first review the canonical relations between the scales that form the basis for implementation of the profile functions. To set up the notation first consider a relatively simple scenario of plain jet mass resummation for a jet with radius $R$ in $\ee$ collisions. The relevant canonical scales are the hard, jet, and the (ungroomed) soft scales:
\begin{align}
\label{eq:canPlain}
\mu_N^{\rm can.} = Q \tan\Big(\frac{R}{2}\Big) \, ,
\qquad
\mu_J^{\rm can.}= m_J \,,
\qquad
\mu_s^{\rm can.} = \frac{m_J^2}{Q\tan\big(\frac{R}{2}\big)}
\, ,
\end{align}
such that the endpoint of the spectrum is at $m_J = m^{ee}_{\rm max} = Q \tan(R/2)$, where the jet and the soft scales merge with the hard scale. Thus, the seesaw relation
\begin{align}\label{eq:SeeSawPlain}
\mu_J^2 = \mu_N\: \mu_s
\end{align}
between the three scales is at the heart of the soft-collinear factorization in what is often called \SCETa.

When we include soft drop grooming, the canonical hard and the jet scales remain the same, whereas only the soft sector is modified, splitting into the global-soft and collinear-soft scales:
\begin{align}\label{eq:canMax}
\mu_{gs}^{\rm can.} = \qcut \tan^{1+\beta}\Big(\frac{R}{2}\Big) \, ,
\qquad
\mu_{cs}^{\rm can.}= \Big(\frac{m_J^2}{Q}\Big)^{\frac{1+\beta}{2+\beta}}\qcut^{\frac{1}{2+\beta}}
\, .
\end{align}
We see that $\mu_{cs}$ is independent of the jet radius $R$. These scales are relevant for the single differential jet mass distribution as well as in the large $R_g$ region.
For the single differential jet mass case the canonical scales were first discussed in Ref.~\cite{Frye:2016aiz}, with the generalization to include $R$ dependence given in Refs.~\cite{Kang:2018jwa,Chien:2019osu}.

We note that the collinear-soft and global-soft scales are related by the following condition:
\begin{align}\label{eq:SeeSawMax}
(\mu_{cs})^{2+\beta} &=\big(\mu_s\big)^{1+\beta}\: \mu_{gs} \nn \\
&= \Big(\frac{\mu_J^2}{\mu_N}\Big)^{1+\beta} \mu_{gs}\, .
\end{align}
This relation can be understood from RG consistency, noting that the $\zcut$ and $\beta$ dependence must cancel within the soft sector, as jet and hard functions do not depend on these parameters.
The relation in the second line is obtained using \eq{SeeSawPlain} and we will use it below in implementing the profile for our jet scale.

In the intermediate $R_g$ region, the collinear-soft sector further splits into \CSg and \CSm modes, with the canonical values of the corresponding scales being
\begin{align}\label{eq:canInt}
\mu_{cs_m}^{\rm can.}= \frac{m_J^2}{Q (R_g/2)} \, ,
\qquad
\mu_{cs_g}^{\rm can.}= \qcut \Big(\frac{R_g}{2}\Big)^{1+\beta} \, .
\end{align}
They satisfy the following see-saw relation:
\begin{align}\label{eq:SeeSawInt}
\big(\mu_{cs}\big)^{2 + \beta} = \big(\mu_{cs_m}\big)^{1+\beta} \mu_{cs_g} \, ,
\end{align}
which holds for all values of $R_g$. This relation will be used below to define the $\mu_{cs_m}$ profile. It is analogous to the relation in \eq{SeeSawMax} as can be seen by setting $R_g = R$. We will elaborate on this more below.

Lastly, we consider the canonical scales in the small $R_g$ region. In addition to the $\mu_{gs}$ and $\mu_{cs_g}$ scales given above, the jet and the CS$_m$ scale in the intermediate region merge into a single collinear scale. As we saw in \sec{Small}, the logarithms in the collinear function ${\cal C}^\kappa$ are minimized for $\mu \sim Q R_g/2$, hence the canonical scale is given by
\begin{align}\label{eq:canMin}
\mu_c^{\rm can.} = \frac{Q R_g}{2}
\, ,
\end{align}
and the canonical relation between three scales in the small $R_g$ region reads
\begin{align}\label{eq:SeeSawMin}
\mu_c = \mu_N \: \bigg(\frac{\mu_{cs_g}}{\mu_{gs}}\bigg)^{\frac{1}{1+\beta}} \, .
\end{align}
In deriving this relation we have replaced $R_g/2\ra \tan(R_g/2)$, capturing some of the power corrections in $R_g/2$ that are important in this region. Again, we will use the see-saw relation in \eq{SeeSawMin} to fix the $\mu_c$ profile. Furthermore, given the consistency between the small and intermediate regime, the following canonical relation also holds:
\begin{align}\label{eq:SeeSawMin2}
\mu_J^2 = \mu_c\: \mu_{cs_m} \, .
\end{align}
This relation is relevant in the ungroomed region, as can be seen by substituting $R_g \ra R$ in \eq{SeeSawPlain}.

\subsubsection*{Consistency of the intermediate $R_g$ with the small and large $R_g$ regions}

It is worthwhile to explore the behavior of the scales in the intermediate $R_g$ region as $R_g$ approaches its upper and lower kinematic bounds. At the upper limit $R_g\ra \theta_g^\star (m_J^2,\qcut,\beta)$ (in the groomed region) the scales merge into the $\mu_{cs}$ scale,
\begin{align}\label{eq:muCSgLargeRg}
\mu_{cs_m}(m_J^2, R_g=\theta_g^\star) = \mu_{cs_g}(\qcut, R_g=\theta_g^\star)
&= \mu_{cs}(m_J^2, \qcut)
\, , \qquad\quad \text{for $m_J< m_J^0$}
\,,
\end{align}
as can be seen by substituting $R_g = \theta_g^\star(m_J^2,\qcut, \beta)$ in \eq{canInt}, with $\theta_g^\star$ defined in \eq{angularBoundsExplicit}. In the ungroomed region the upper limit is set by the jet radius, and the collinear-soft scales for $R_g\ra R < \theta_g^\star(m_J^2,\qcut,\beta)$ behave differently:
\begin{align}\label{eq:muCSgUngroomed}
\mu_{cs_m}(m_J^2, R_g = R) &= \mu_{s}(m_J^2)\,,
\qquad\hspace{1.0pt} \qquad\quad \text{for $m_J> m_J^0$}
\,, \\
\mu_{cs_g}(\qcut, R_g = R) &= \mu_{gs}(\qcut, R)
\, , \qquad\quad \text{for $m_J>m_J^0$} .
\end{align}
This suggests that, unlike for groomed to ungroomed transition for the single differential jet mass distribution, there is a smooth transition between the intermediate and the ungroomed region owing to the additional $R_g$ measurement.
Here the soft drop resummation is automatically turned off as $\mu_{cs_g}$ approaches $\mu_{gs}$, whereas $\mu_{cs_m} \ra \mu_s$ provides for the essential plain jet mass resummation in the ungroomed region. We stress, however, that the intermediate $R_g$ regime still lacks the power corrections of ${\cal O}(m_J^2/(Q\qcut))$ in the ungroomed region.

Next, we note that upon decreasing $R_g$ the $\mu_{cs_m}$ scale meets the jet scale as $R_g$ approaches the lower limit of the angle of the collinear mode:
\begin{align}\label{eq:muCSmSmallRg}
\mu_{cs_m}(m_J^2, R_g = \theta_c) = \mu_J(m_J^2) \, .
\end{align}
On the other hand, from \eq{canInt} we see that $\mu_{cs_g}$ keeps decreasing on reducing $R_g$. Being the smallest of all the scales, for a given jet mass it becomes nonperturbative first for $R_g \sim (R_g)_{\rm NP}$ defined in \eq{RgNP}. Thus we will first implement the $\mu_{cs_g}$ profile so as to consistently freeze it to an ${\cal O}(1 {\rm GeV})$ value in the nonperturbative region.

\subsubsection*{Rescaled variables}

To simplify the expressions we will work with the following unit-interval variables corresponding to jet mass and $R_g$. For the case of $\ee$ collisions, we have
\begin{align}\label{eq:psiDef}
\xi(m_J^2) \equiv \frac{m_J^2}{Q^2 \tan^2\frac{R}{2}} = \frac{m_J^2}{(\mu_N^{\rm can.})^2}
\, , \qquad
\psi(R_g) \equiv \frac{\tan\big(\frac{R_g}{2}\big)}{\tan\big(\frac{R}{2}\big)} \,
\qquad
(\text{$\ee$ case}) \, .
\end{align}
Here we use $\mu_N^{\rm can.} = Q\tan(R/2)$ to keep it distinct from $\mu_N$ which we will vary to study perturbative uncertainty.
Analogously, the corresponding variables in the $pp$ case are defined as
\begin{align}
\xi(m_J^2) \equiv \frac{m_J^2}{p_T^2 R^2}
\, ,\qquad
\psi(R_g) \equiv \frac{R_g}{R} \,
\qquad
(\text{$pp$ case})
\, .
\end{align}

From \eq{m0} we see that for $m_J^2 > m_0^2$ we enter the ungroomed resummation region. Here we turn off the soft-drop dependent resummation in the soft sector by merging the collinear-soft and global-soft scales to the ungroomed soft scale. In terms of $\xi$, the transition point is given by
\begin{align}\label{eq:txi0def}
\tilde \xi_0 \equiv \frac{m_0^2}{\mu_N^2} =
\frac{\qcut}{Q} \tan^\beta\Big(\frac{R}{2}\Big) \cos^{(2+\beta)} \Big(\frac{R}{2}\Big)
= \frac{\mu_{gs}}{\mu_N}\cos^{(2+\beta)} \Big(\frac{R}{2}\Big)
\, .
\end{align}
We note that as a result of the $\cos^{2+\beta}(R/2)$ the transition to ungroomed region happens slightly earlier than the value of $\xi = \xi_0$ for which the global-soft and collinear-soft scales merge:
\begin{align} \label{eq:xi0def}
\xi_0 \equiv \xi\Big|_{\mu_{gs} = \mu_{cs}} = \frac{\mu_{gs}}{\mu_N} =
\frac{\qcut}{Q}\tan^\beta\Big(\frac{R}{2}\Big) \,
\qquad
(\text{$\ee$ case}) \, .
\end{align}
In the following we will, however, ignore the difference between $\tilde \xi_0$ and $\xi_0$ and take $\xi_0$ in \eq{xi0def} to correspond to the groomed and ungroomed transition point. For the $pp$ case we have
\begin{align}
\xi_0 \equiv \zcut' R^\beta \, \qquad (\text{$pp$ case}) \, .
\end{align}
From \eq{angularBounds} the phase space in the variables $\xi$ and $ \psi$ is then given by
\begin{figure}[t!]
\centering
\includegraphics[width=0.6\textwidth]{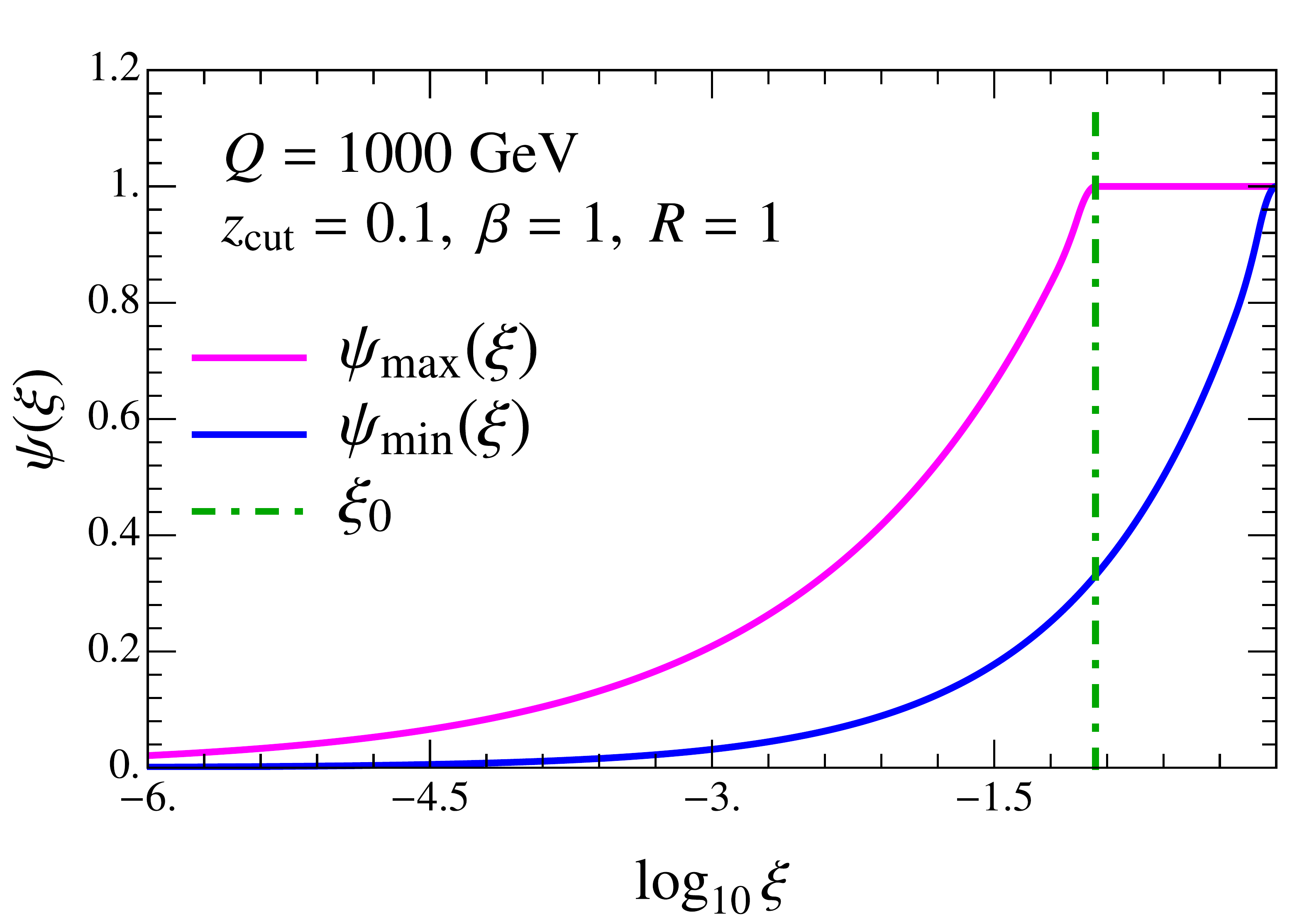}\\
\vspace{-0.2cm}
\caption{The range of values of allowed $\psi$ values as a function of $\xi$ are shown by the region between the solid (magenta and blue) lines. The dot-dashed (green) line indicates where the transition to the ungroomed regime takes place. The maximum value of $\psi$ corresponds to $R_g = \theta_g^\star(\xi)$, or $\psi=R$ in the ungroomed regime.}
\label{fig:psiMinMax}
\vspace{-0.2cm}
\end{figure}
\begin{align}
\label{eq:psiGroomedRange}
\psi_{\rm min}(\xi) \equiv \sqrt{\xi} \: \leq \: \psi \: \leq \: \psi_{\rm max}(\xi)
\equiv \min \big \{ \psi^{\star}(\xi),\: 1 \big \}\:
\, , \qquad
0\leq \xi \leq 1
\,,
\end{align}
and is shown in \fig{psiMinMax}.
Here $\psi^{\star}(\xi)$ corresponds to $\psi$ for $R_g = \theta_g^\star(m_J^2, \qcut, \beta)$, and is given by
\begin{align}\label{eq:psiStarDef}
\psi^{\star}(\xi) \equiv\Big( \frac{\xi}{\xi_0}\Big)^{\frac{1}{2+\beta}}\,.
\end{align}
For $\xi > \xi_0$, i.e. in the ungroomed region, $R_g$ is bounded above by the jet radius $R$, which limits $\psi \leq 1$. We note that \eqs{psiGroomedRange}{psiStarDef} also hold for the $pp$ case.

We now summarize the canonical scales in terms of $\xi$ and $\psi$ variables:
\begin{enumerate}
\item Canonical scales for plain jet mass:
\begin{align} \label{eq:muNSummary}
&\mu_N^{\rm can.} = Q \tan\frac{R}{2} \,&
&(\text{$\ee$ case})\,,& \\
&\mu_N^{\rm can.} = p_T R\, &
&(\text{$pp$ case}) \, ,&\nn
\end{align}
\begin{align}
\mu_s^{\rm can.}= \mu_N^{\rm can.} \xi \, ,
\qquad\qquad
\mu^{\rm can.}_J = \mu_N^{\rm can.}\sqrt{\xi} \, .
\end{align}
\item Canonical scales for large $R_g$ region (with $\mu_N$ and $\mu_J$ same as above):
\begin{align}\label{eq:muCSCan}
\mu_{gs}^{\rm can.} = \mu_N^{\rm can.} \xi_0 \, , \qquad
\mu_{cs}^{\rm can.}
= \mu_N^{\rm can.} \xi_0^\frac{1}{2+\beta}\xi^{\frac{1+\beta}{2+\beta}}
= \mu_{gs}^{\rm can.} \big[ \psi^{\star}(\xi)\big]^{1+\beta}
\, , \qquad
\xi < \xi_0 \, ,
\end{align}
where $\xi_0$ is defined as
\begin{align} \label{eq:xi0def}
&\xi_0 =\frac{\qcut}{Q}\tan^\beta\Big(\frac{R}{2}\Big)\, &
&(\text{$\ee$ case})\, , &\nn \\
&\xi_0 =\zcut' R^\beta\,&
&(\text{$pp$ case}) \, . &
\end{align}
\item Canonical scales for intermediate $R_g$ region (with $\mu_N$, $\mu_J$ and $\mu_{gs}$ same as above):
\begin{align}
\mu_{cs_m}^{\rm can.} = \mu_N^{\rm can.} \frac{\xi}{\psi} \, ,
\qquad \mu^{\rm can.}_{cs_g} = \mu_{gs}^{\rm can.} \psi^{1+\beta}
\, .
\end{align}
\item Canonical scales for the small $R_g$ region (with $\mu_N$ and $\mu_{gs}$ same as above):
\begin{align}
\mu_c^{\rm can.} = \mu_N^{\rm can.} \psi \, .
\end{align}
\end{enumerate}
We will find the expression of $\mu_{cs}$ in terms of $\psi^\star$ in \eq{muCSCan} useful to construct the $\mu_{cs}$ profile below.

Lastly, we note that for the case of $\ee$ collisions in the small angle limit, we have $\psi\tan(R/2) = \tan(R_g/2) \sim R_g/2$. Since using the variable $\psi$ allows us to smoothly connect all the profiles from small to large values of $R_g$, we also additionally replace all the appearances of $R_g/2$ in the factorization formulae above by $\psi\tan(R/2)$. The difference between the two is a power correction, and when $R_g$ is ${\cal O}(1)$ it automatically results in the correct expressions. This includes all the fixed-order results for factorization functions in \app{Oneloop}, the appearances of $R_g$'s in the logarithms (which are simply the canonical scales displayed above), the pre-factors in soft drop boundary cross section (see \eq{xsecRGLargeShiftDelta}), in the ratio $\vf = (R_g/\theta_g^\star)^{2+\beta}$ that shows up in the resummation kernels in the large $R_g$ region (see \eqs{UJScOmega}{UJScOmegaEps}), in the weight functions in \eq{weightFunc}, and as well as any non-logarithmic dependences on $R_g$ (such as in the collinear function in the small $R_g$ region in \eq{CqFiniteMJ}).

\tocless\subsubsection{Profile functions for double differential cross section}

\noindent
We now turn to implementing the profile functions based on these canonical scale choices and relations. Here we state results solely for $\ee$ collisions with their generalizations to $pp$ case being straightforward.
Our goal is twofold: accounting for freezing of any perturbative scale that approaches $\Lambda_{\rm QCD}$ and providing a way to estimate perturbative uncertainty though scale variation. We consider profile variations in the next section.
We base the freezing behavior of all the scales on the the $\mu_{cs_g}$ profile, defined as follows:
\begin{align}\label{eq:muCSg}
\mu_{cs_g}(\psi) \equiv \mu_{gs} \, f_{\rm run} \big ( \psi(R_g)^{1+\beta}\big) \, ,
\end{align}
where $ f_{\rm run}(\psi) $ is defined by
\begin{align}
\label{eq:frunRg}
f_{\rm run}(\psi)
&\equiv \left\{
\begin{array}{ll}
y_0 \Big(1+ \frac{\psi^2}{4y_0^2}\Big)	& ~~~~~~~~~~\psi \leq 2 y_0\\[4pt]
\psi & ~~~~~~~~~~ 2 y_0 < \psi \leq 1
\end{array}
\right.\;.
\end{align}
In the resummation region $\mu_{cs_g}(\psi) \sim \mu_{gs}\psi^{1+\beta}$, consistent with \eq{canInt}. The transition of the $\mu_{cs_g}$ scale into the NP region is determined by
\begin{align}\label{eq:y0}
y_0 \equiv \frac{n_0}{(\mu_{gs}/1\, {\rm GeV})}
>\frac{\Lambda_{\rm QCD}}{Q_{\text{cut}}}
\, ,
\end{align}
where $n_0$ is an ${\cal O}(1)$ number that controls the value for the perturbative scale we freeze to which is $> \Lambda_{\rm QCD}$. Recall that it is mandatory for the scales to be frozen in this manner, since the entire formalism is based on perturbative expansions of things like the anomaous dimensions. In the numerical analysis presented in \sec{Num} we use $n_0 = 0.75$ which from \eq{frunRg} corresponds to onset of the freezing of the coupling at $\mu = 1.5$ GeV.

Note that $\mu_{cs_g}$ has the lowest virtuality, so it is always the first scale to enter the NP region when the jet mass grows smaller.
To freeze in turn the scales with larger virtuality as we venture into the NP region while maintaining the canonical scaling relations, we now simply insert \eq{muCSg} in the canonical relations derived above. First, as seen from \eq{muCSCan}, the collinear-soft scale can be derived using $\mu_{cs_g}(\psi)$ by setting $R_g = \theta_g^\star (m_J^2, \qcut, \beta)$. Thus we have
\begin{align}\label{eq:muCS}
\mu_{cs} (\xi ) \equiv \mu_{gs} \, f_{\rm run} \Big ( \big[ \psi^{\star}(\xi)\big]^{1+\beta} \Big) =
\mu_{gs} \, f_{\rm run} \bigg ( \Big(\frac{\xi}{\xi_0}\Big)^{\frac{1+\beta}{2+\beta}}\bigg) \, ,
\qquad \xi < \xi_0\,.
\end{align}
We note that, compared to the plain jet-mass soft scale in \eq{canPlain}, the collinear-soft scale should become nonperturbative at yet smaller jet masses in the SDNP region defined by \eq{mJSDNP}. This is correctly captured in \eq{muCS} because from \eq{frunRg} we see that the $\mu_{cs}(\xi)$ profile freezes for $\xi \leq \xi_{\rm SDNP}$ where
\begin{align}\label{eq:mJNP}
\xi_{\rm SDNP} \equiv \xi_0 (2y_0)^{\frac{2+\beta}{1+\beta}}
>
\frac{(m_J^2)_{\rm SDNP}}{\mu_N^2} =
\frac{ \Lambda_{\rm QCD}}{\mu_N}\Big(\frac{\Lambda_{\rm QCD}}{\mu_{gs}}\Big)^{\frac{1}{1+\beta}} \, ,
\end{align}
which is consistent with \eq{mJSDNP}.

Having defined the $\mu_{cs}(\xi)$ and $\mu_{cs_g}(\psi)$ profiles we can now use \eq{SeeSawInt} to implement the $\mu_{cs_m}$ profile as follows:
\begin{align}\label{eq:muCSm}
\mu_{cs_m} (\xi, \psi) &\equiv
\mu_{cs}(\xi) \Bigg[\frac{\mu_{cs}(\xi)}{\mu_{cs_g}(\psi)}\Bigg]^{\frac{1}{1+\beta}}
\, , \qquad \xi < \xi_0
\,,
\end{align}
while using \eqs{SeeSawPlain}{SeeSawMax} we implement the profile for the jet scale as follows:
\begin{align}\label{eq:muJ}
\mu_J (\xi)
\equiv \Bigg[
\mu_N\: \mu_{cs}(\xi) \: \bigg(\frac{\mu_{cs}(\xi)}{\mu_{gs}}\bigg)^{\frac{1}{1+\beta}}
\Bigg]^\frac{1}{2}
\,,
\qquad \xi < \xi_0
\, .
\end{align}
Finally, using \eq{SeeSawMin} the $\mu_c(\psi)$ profile is given by
\begin{align}\label{eq:muC}
\mu_c (\psi) \equiv \mu_N \Bigg(\frac{\mu_{cs_g}(\psi)}{\mu_{gs}}\Bigg)^{\frac{1}{1+\beta}} \, .
\end{align}
\begin{figure}[!t]
\includegraphics[width=0.52\textwidth]{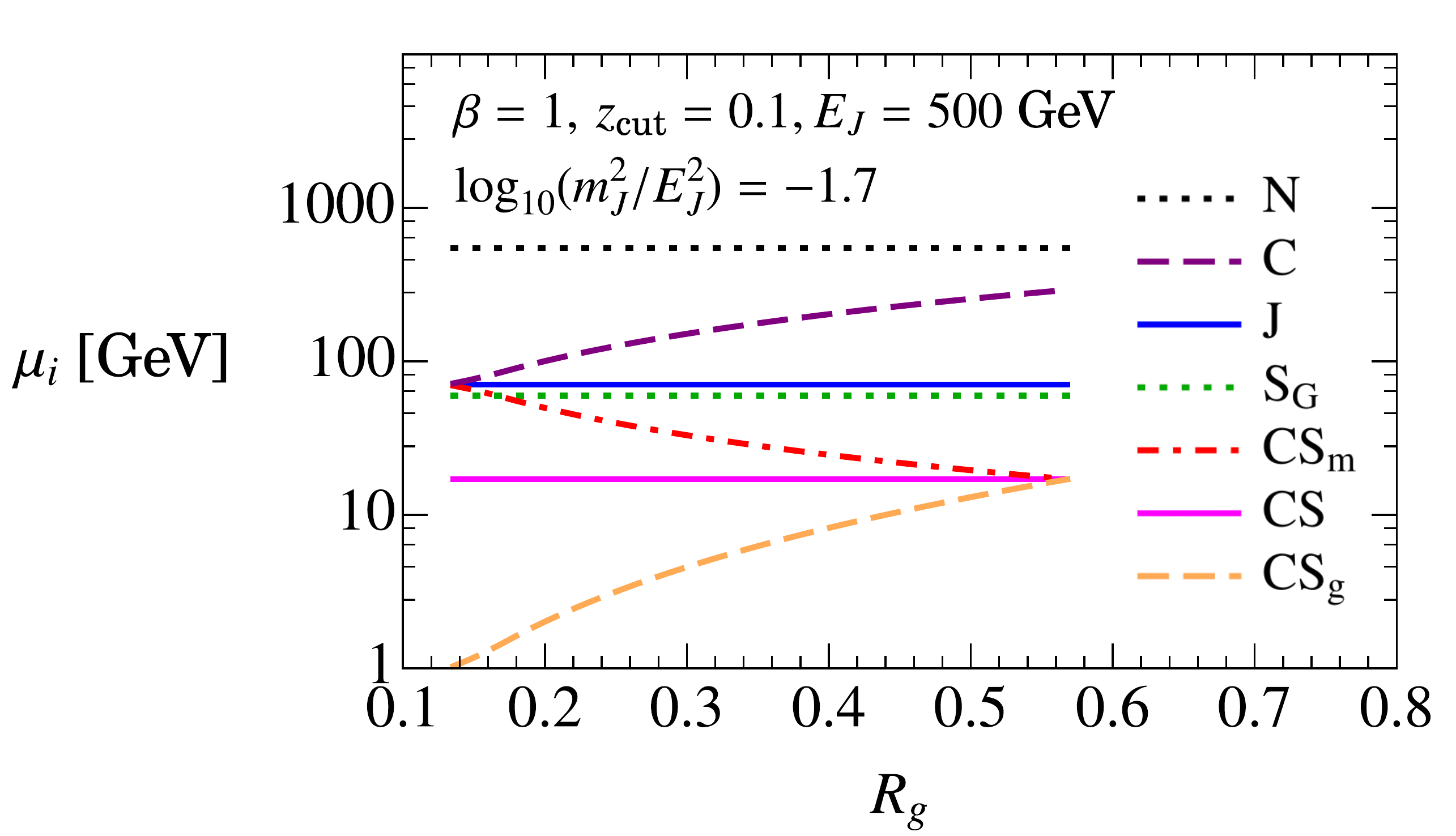}
\hspace{0.01\textwidth}
\includegraphics[width=0.47\textwidth]{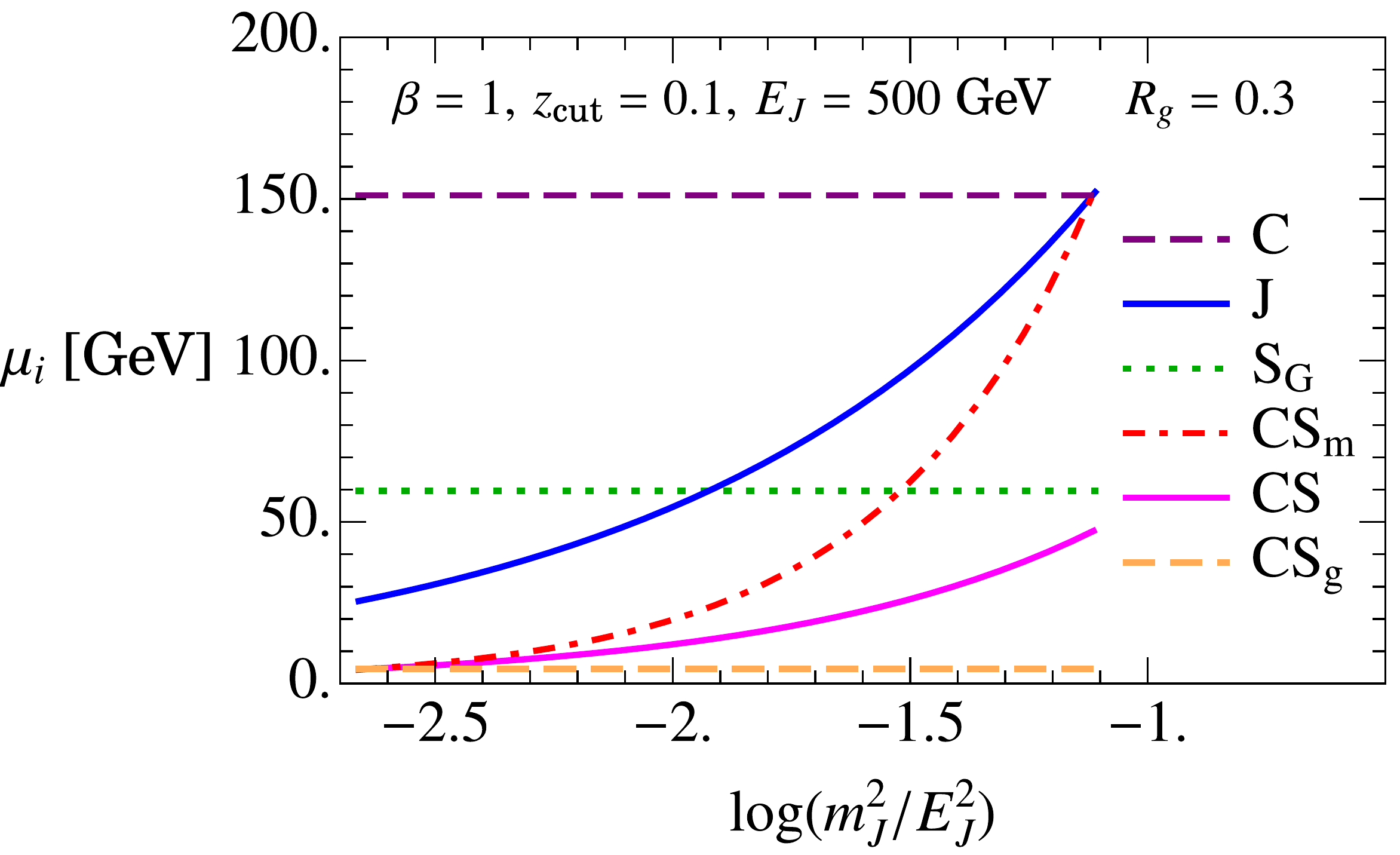}
\caption{Profile scales $\mu_i$ as a function of the groomed radius at fixed jet mass (top left) and as a function of the mass at fixed radius (top right), for $\beta=1$. For more clarity the former plot uses a logarithmic scale, while we omit the hard scale $\mu_N$ in the latter. At small groomed radii (large jet masses) $\mu_{cs_m}$ and $\mu_{J}$ merge into $\mu_C$, while at large radii (small masses) $\mu_{cs_m}$ and $\mu_{cs_g}$ become a single soft-collinear scale $\mu_{cs}$, consistent with the mode picture.}
\label{fig:profilePlot}
\end{figure}

We have so far defined the profile functions only in the groomed region for $\xi < \xi_0$.
A complete analysis of the distribution in the ungroomed region is beyond the scope of this work.
We will, however, extend our results into the ungroomed region making use of the following extensions of the collinear-soft profile scales:
\begin{align}\label{eq:muNoSD}
\text{Large $R_g$}:& &\mu_{cs} = \mu_{gs} = \mu_s(\xi) = \mu_N \xi \, & &(\xi > \xi_0) \, ,& \\
\text{Intermediate/Small $R_g$}:& &\mu_{cs_g} = \mu_{gs}\, ,
\quad \mu_{cs_m}= \mu_s(\xi) = \mu_N \xi & &(\xi > \xi_0) \,. &
\end{align}
The large $R_g$ transition occurs when we cross the dot-dashed green line in \fig{psiMinMax} for values of $\psi$ that place us near the upper solid magenta line, whereas the intermediate/small $R_g$ transitions occur when we cross the dot-dashed green line when we are far from all solid lines, or close to the lower blue line, respectively.
Note that in the large $R_g$ regime the resummation related to jet-grooming must be immediately turned off as $\xi > \xi_0$.
This is the same as the groomed to ungroomed profile transition for the single differential jet mass case, which was discussed in Ref.~\cite{Chien:2019osu}.
In case of intermediate $R_g$ regime, however, this happens more smoothly owing to its more factorized nature. Finally, the jet and the collinear scales remain the same in the ungroomed region.

In \fig{profilePlot} we show the profile scales resulting from the above formulae picking $\beta = 1$. In the left panel we show them as a function of the groomed jet radius for a fixed jet mass. The scales satisfy the joining condition for the end points of the $R_g$ spectrum: $\mu_{cs}$, $\mu_{cs_g}$, and $\mu_{cs_m}$ merge at the end point $R_g = \theta_g^\star(m_J^2, \qcut, \beta)$, whereas $\mu_J$ and $\mu_{cs_m}$ merge into $\mu_c$ at the lower end point $R_g = \theta_c(m_J^2)$. In the right panel of \fig{profilePlot} we fix the $R_g$ to a representative value of $R_g = 0.3$ and plot the scales as a function of the jet mass. Our choice of $R_g = 0.3 < R$ implies that the entire allowed mass range lies within the groomed region. Once again we see the expected behavior when the scales join. The profile functions for other values of $\beta$ look qualitatively similar.

\begin{figure}[!t]
{\includegraphics[width=0.5\textwidth]{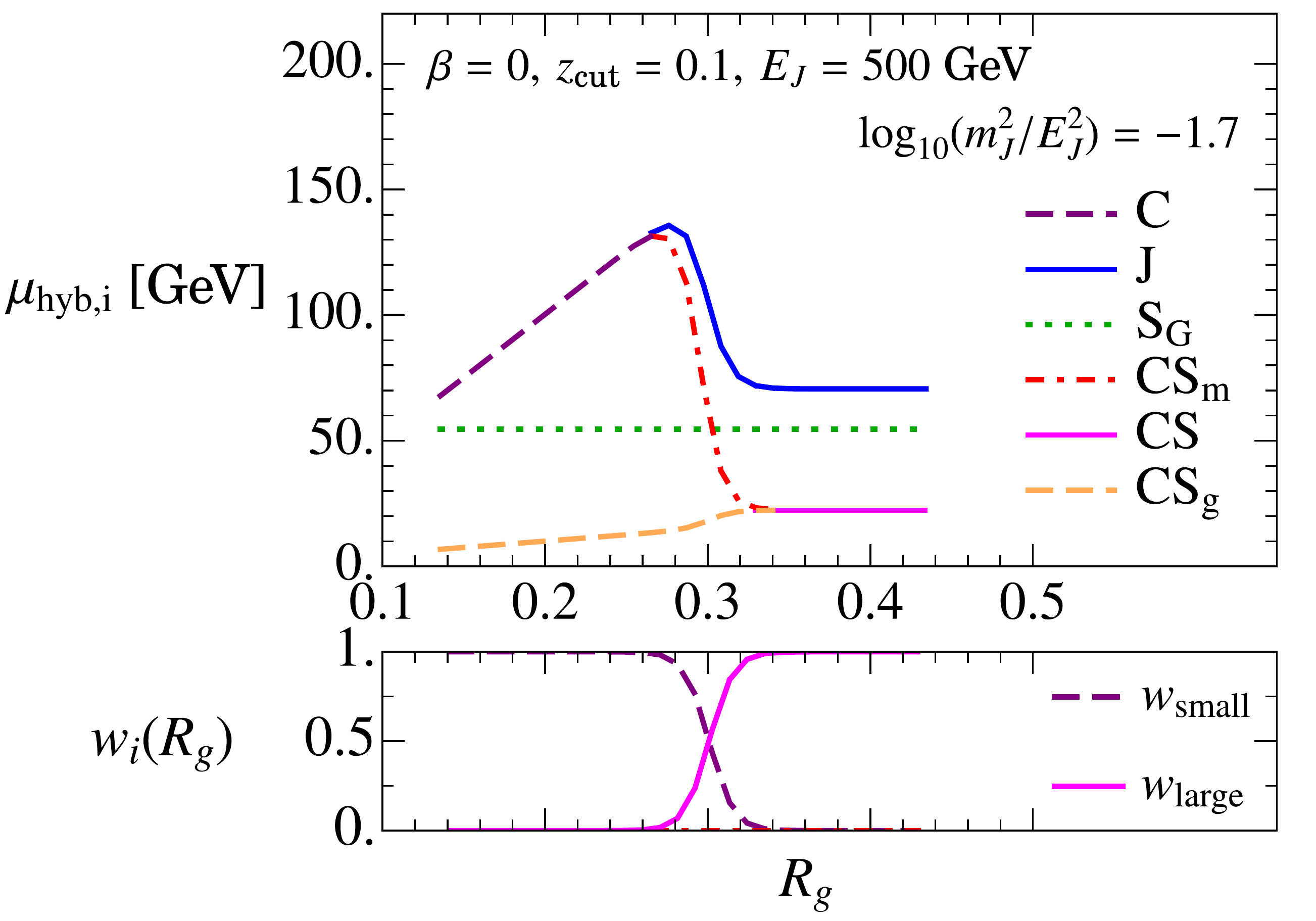}}
{\includegraphics[width=0.5\textwidth]{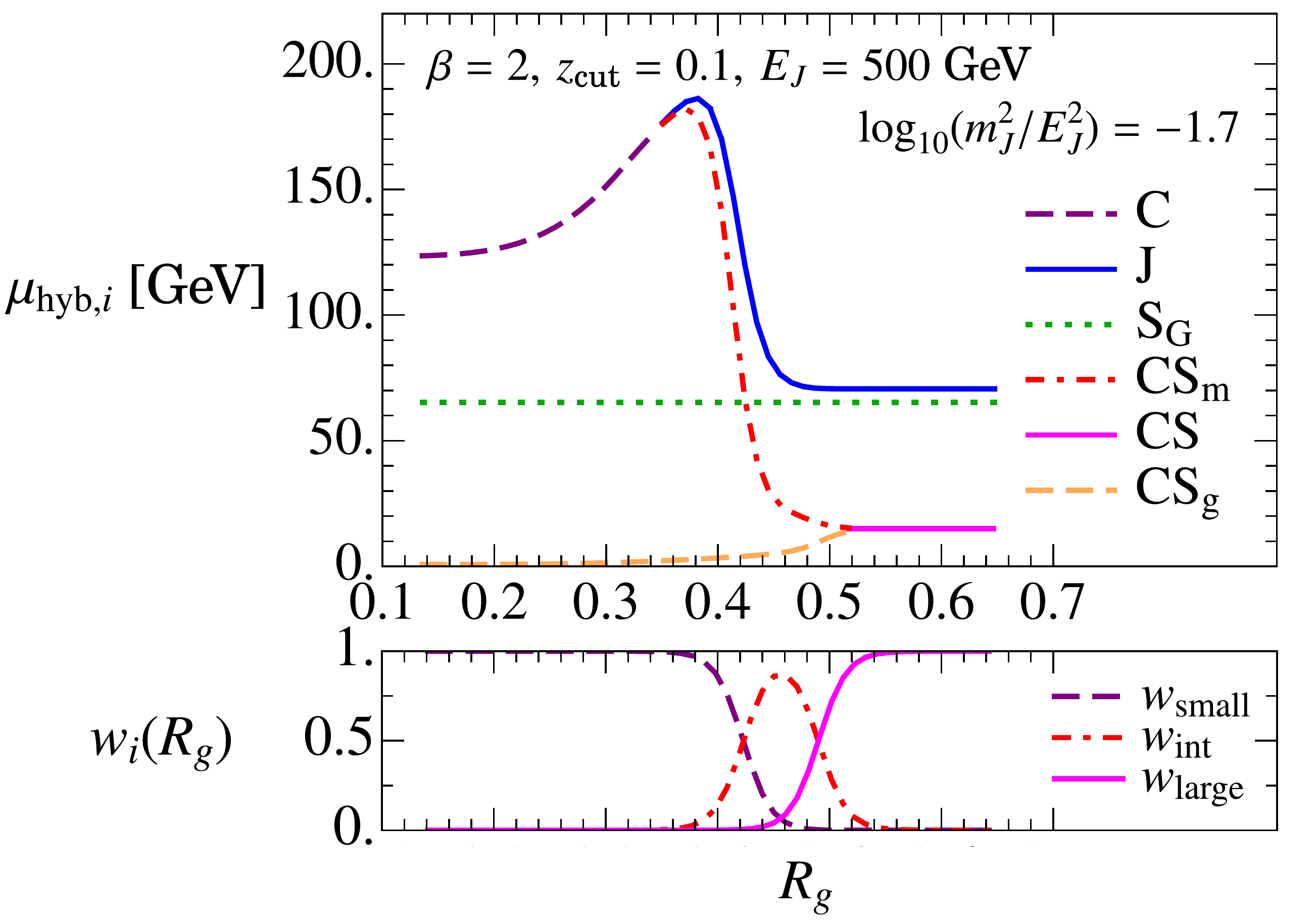}}
\caption{ Hybrid profile scales $\mu_{{\rm hyb},i}$ for $\beta=0$ and $\beta = 2$, shown as a function of the groomed radius $R_g$ at a fixed jet mass. The bottom panels show the transition functions that give weight to the various regimes. In the left panel we have a situation involving the 2-EFT situation, whereas in the right panel we have a 3-EFT case.}
\label{fig:profilePlotHyb}
\end{figure}
As we discussed above in \sec{matchImplement}, the implementation of the matched cumulative $R_g$ cross section involves constructing hybrid profiles $\mu_{\rm hyb}$, that select the appropriate set of scales from each regime following \eq{muHyb}. These involve the weight functions $w_i(m_J^2, R_g)$ that we constructed in \sec{trans}. In \fig{profilePlotHyb} we show the hybrid profiles for $\beta = 0$ and $\beta =2$ for a given jet mass value in the upper left and right panels respectively. In both these plots we show the scales only in the region where they are valid (except for the global-soft scale that is common to all).
The $\mu_c$ scale splits off into $\mu_J$ and $\mu_{cs_m}$ as we transition out of the small $R_g$ regime, and the $\mu_{cs_m}$ and $\mu_{cs_g}$ scales merge into the single collinear soft scale $\mu_{cs}$ as we enter the large $R_g$ regime. In the bottom panels we show the weight functions that correspond to that specific choice of jet mass and and soft drop parameters. In the case of $\beta = 0$ we transition directly between the small and large $R_g$ regimes, thus providing an example of the 2-EFT scheme. In this case the scale $\mu_{cs_m}$ only serves to provide an interpolation into the large $R_g$ region. On the other hand, we see that for $\beta = 2$ there is a significant region where $w_{\rm int} > 0$ and the intermediate $R_g$ resummation is active.

\vspace{0.2cm}
\subsection{Perturbative uncertainty}
\label{sec:vary}

We now briefly discuss our prescription to assess the perturbative uncertainty of the EFT results by means of scale variations and varying the unknown ${\cal O}(\alpha_s^2)$ non-logarithmic terms necessary for NLL$'$ resummation in the large and small $R_g$ regimes through the parameters $a_{20}^{\cal C}$, $a_{20}^{S_c}$, and $a_{20,\veps}^{S_c}$ that appear in Eqs.~(\ref{eq:CqFiniteMJ}), (\ref{eq:ScLargeRgFO3}) and (\ref{eq:DeltaScLargeShift}) respectively.
Note that we will not consider variations in the ungroomed region, leaving a more detailed analysis of this transition to future work, and hence we often cutoff our plots with uncertainty bands prior to hitting this transition.

\tocless\subsubsection{Scale variations}

In this analysis we will consider the following three types of scale variations:
\begin{enumerate}
\item Vary the hard and global-soft scales by a multiplicative factor, and let this variation propagate to other scales through the canonical relations:
\begin{align} \label{eq:eHdef}
\mu_{N}^{\rm vary}(e_{gs}) &\equiv e_{gs} \mu_{N}^{\rm can.} \, ,
\qquad
\mu_{gs}^{\rm vary}(e_{gs}) \equiv e_{gs} \mu_{gs}^{\rm can.} \, .
\end{align}
Here we will vary $e_{gs}$ from the default value of $e_{gs} = 1$ to $e_{gs} = 1/2$ and $2$. Note that we vary the $\mu_N$ and $\mu_{gs}$ scales simultaneously. This is to ensure that the variation does not affect the groomed to ungroomed transition point $\xi_0$ defined in \eq{xi0def}. This variation will affect all the scales in Eqs.~(\ref{eq:muCSg}), (\ref{eq:muCS}), (\ref{eq:muCSm}), (\ref{eq:muJ}), and (\ref{eq:muC}).
\item A variation of the jet mass and $R_g$ dependent scales from their canonical values in Eqs.~(\ref{eq:canPlain}), (\ref{eq:canMax}), (\ref{eq:canInt}), and (\ref{eq:canMin}) while maintaining the various see-saw relations in Eqs.~(\ref{eq:SeeSawPlain}), (\ref{eq:SeeSawMax}), (\ref{eq:SeeSawInt}), and (\ref{eq:SeeSawMin}). This is done such that the variations are frozen to the default value at the end point. More specifically, we first define the variation of the $\mu_{cs_g}$ scale as
\begin{align} \label{eq:alphadef}
\mu_{cs_g}^{\rm vary}(\psi,\alpha) &\equiv
\mu_{gs} \Big[ f_{\rm vary}\big ( \psi^{1+\beta}\big) \Big]^{\alpha}
\, f_{\rm run} \big ( \psi^{1+\beta}\big) \,
\end{align}
and use it to derive variations of other scales. Here we make use of the following trumpet function $f_{\rm vary}(\psi)$ to turn off the variation at the end point $\psi = 1$:
\begin{equation}\label{eq:fVary}
f_{\rm vary}(\psi)
= \left\{ \begin{array}{l r}
2(1 - \psi^2)\,, & \qquad \psi < 0.5 \\
1 + 2(1- \psi)^2\,, & \qquad 0.5 \leq \psi \leq 1
\end{array}
\right. .
\end{equation}
The choice $\alpha = 0$ returns the default profile, and varying $\alpha = \pm 1$ allows for variation in the resummation region up to a factor of $2$.
In accordance with \eq{muCS}, the variation of $\mu_{cs}$ scale is given by
\begin{align} \label{eq:alphadef2}
\mu_{cs}^{\rm vary}(\xi, \alpha) &\equiv \mu_{gs} \Bigg[ f_{\rm vary}\bigg ( \Big(\frac{\xi}{\xi_0}\Big)^{\frac{1+\beta}{2+\beta}}\bigg) \Bigg]^{\alpha}
\, f_{\rm run} \bigg ( \Big(\frac{\xi}{\xi_0}\Big)^{\frac{1+\beta}{2+\beta}}\bigg) \, , \qquad \xi < \xi_0
\, .
\nn
\end{align}
Having defined $\alpha$-variations of $\mu_{cs_g}$ and $\mu_{cs}$, the corresponding variations of $\mu_J$, $\mu_{cs_m}$ and $\mu_c$ scales are derived using 	Eqs.~(\ref{eq:muJ}), (\ref{eq:muCSm}) and (\ref{eq:muC}) by replacing $\mu_{cs_g, cs} \ra \mu_{cs_g, cs}^{\rm vary}$.
\item A variation that relaxes the canonical see-saw relations that we used to derive the $\mu_{cs_m}$, $\mu_J$, and $\mu_{c}$ scales. The two sets of canonical relations, \eqs{SeeSawMax}{SeeSawInt}, and \eqs{SeeSawPlain}{SeeSawMin2}, are a consequence of soft drop and cumulative-$R_g$ constraints respectively. The two sets are equivalent when $R_g \ra R$, where both constraints are lifted.
For convenience, we define an auxiliary plain-jet mass soft scale $\mu_s(\xi, \alpha)$ using \eq{SeeSawMax}:
\begin{align}
\widetilde \mu_s(\xi) \equiv \mu_{cs}(\xi) \Bigg[\frac{\mu_{cs}(\xi)}{\mu_{gs} }\Bigg]^{\frac{1}{1+\beta}}
\, , \qquad \xi < \xi_0 \, .
\end{align}
We now parameterize the deviations in the $\mu_{cs_m}$ scale upon modifying the soft-drop see-saw relations in \eqs{SeeSawMax}{SeeSawInt} with a small exponent $\rho$:
\begin{align} \label{eq:rhodef}
\Big[\widetilde \mu_{cs}(\xi, \rho)\Big]^{2+\beta}	= \Big[\widetilde \mu_s (\xi) \Big]^{1+\beta + \rho } \big[\mu_{gs}\big]^{1 - \rho}
&=
\Big[\mu^{\rm vary}_{cs_m} (\xi, \psi , \rho)\Big]^{1+\beta + \rho} \Big[\mu_{cs_g}\Big]^{1- \rho}
\, .
\end{align}
Solving for $\mu_{cs_m}$ yields the variation
\begin{align}\label{eq:muCSmVariation}
\mu^{\rm vary}_{cs_m} (\xi, \psi , \rho) &\equiv
\mu_{cs}(\xi) \Bigg[\frac{\mu_{cs}(\xi)}{\mu_{gs} }\Bigg]^{\frac{1}{1+\beta}} \Bigg[\frac{\mu_{cs_g}(\psi)}{\mu_{gs}}\Bigg]^{\frac{-(1 - \rho)}{1+ \rho + \beta}}
\, .
\end{align}
Next, we apply the same strategy to the $\mu_J$ and $\mu_{cs_m}$ scales by entering deviations in \eqs{SeeSawPlain}{SeeSawMin2} through a small exponent $\gamma$:
\begin{align} \label{eq:gammadef}
\big[\mu^{\rm vary}_J (\xi, \gamma)\big]^2 &= \big[ \mu_N\big]^{1 + \gamma} \Big[\widetilde\mu_{s}(\xi )\Big]^{1-\gamma}
\\
&= \Big[ \mu^{\rm vary}_c (\psi, \gamma ) \Big]^{1+ \gamma} \Big[\mu_{cs_m}(\xi,\psi)\Big]^{1-\gamma}
\, .
\nn
\end{align}
Note that we have set $\rho = 0$ in \eq{muCSmVariation} as these two variations are independent. Solving for the scales $\mu_J$ and $\mu_c$ yields the variations
\begin{align}\label{eq:muJLargeVary}
\mu_{J}^{\rm vary}(\xi, \gamma)
&
\equiv
\big[\mu_N\big]^{\frac{1}{2} + \gamma}
\Bigg[
\mu_{cs}(\xi) \: \bigg(\frac{\mu_{cs}(\xi)}{\mu_{gs}}\bigg)^{\frac{1}{1+\beta}}
\Bigg]^{\frac{1}{2} - \gamma} \, ,
\\
\mu^{\rm vary}_c (\psi, \gamma ) &\equiv
\mu_N \: \Bigg[ \frac{\mu_{cs_g}(\psi)}{\mu_{gs}}\Bigg]^{\frac{1-\gamma}{1+\gamma}\frac{1}{1+\beta}}
\nn \, .
\end{align}
\end{enumerate}

\tocless\subsubsection{Parameterizing variation of the ${\cal O}(\alpha_s^2)$ non-logarithmic terms}

\noindent
We now consider the variations of the two-loop parameters $a_{20}^{\cal C}$, $a_{20}^{S_c}$, and $a_{20,\veps}^{S_c}$ introduced in Eqs.~(\ref{eq:CqFiniteMJ}), (\ref{eq:ScLargeRgFO3}) and (\ref{eq:DeltaScLargeShift}) respectively, in order to assess uncertainty of our lack of knowledge of these terms. We first note that in combining the cross section through the recipe in \eq{SigMatched} we are relying on the fact that the intermediate $R_g$ cross section reproduces the EFT result on either end of the $R_g$ spectrum up to power corrections. This is guaranteed to hold from \eqs{CSrefactorization}{CFact} when all the functions are expanded to a given order. Given our complete knowledge of ${\cal O}(\alpha_s)$ pieces, we can use the construction in \eq{SigMatched} to NLL accuracy. However, when we include ${\cal O}(\alpha_s^2)$ pieces in the small and large $R_g$ regimes to account for NLL$'$ resummation, the missing non-logarithmic terms parameterized by $a_{20}^i$ spoil this delicate cancellation when they are extrapolated into the intermediate regime. We solve this issue by explicitly multiplying the ${\cal O}(\alpha_s^2)$ fixed-order pieces in the large and small $R_g$ regimes by the weight factors $w_i(m_J^2,R_g)$ so that they are set to zero when extrapolated into other regimes:
\begin{align}\label{eq:a20Vary}
d\Sigma_{\rm small} &\equiv d\Sigma^{[1]}_{\rm small} (a_{10}^{\cal C}) + w_{\rm small}(m_J^2, R_g)\, d\Sigma^{[2]}_{\rm small}(a_{20}^{\cal C}) \, \qquad (m_J > 0)\, , \\
d\Sigma_{\rm large} &\equiv d\Sigma^{[0]}_{\rm large} + d\Sigma^{[1]}_{\rm large}(a_{10}^{S_c}) + w_{\rm large}(m_J^2, R_g)\, d\Sigma^{[2]}_{\rm large} (a_{20}^{S_c})\, , \nn
\end{align}
where the terms $d\Sigma^{[i]}$ are ${\cal O}(\alpha_s^i)$. The $ d\Sigma^{[0]}_{\rm large}$ term in the large $R_g$ regime is simply the single differential jet mass cross section, and does not contribute to the differential $R_g$ measurement. By demanding $m_J>0$, the ${\cal O}(\alpha_s^0)$ $\delta$-function piece in the small $R_g$ regime does not contribute. The ${\cal O}(\alpha_s)$ terms depend on the LO expressions in \eqs{a10C}{a10Sc} in the large and small $R_g$ regimes respectively. The ${\cal O}(\alpha_s^2)$ terms are the additional pieces required to achieve NLL$'$ accuracy which include the cross terms shown in \tab{RGE}, as well as the parameters $a_{20}^{\cal C}$ and $a_{20}^{S_c}$ that are included for uncertainty estimation. Thus, we achieve the reasonable outcome that the variations induced by these parameters will be limited to the region where the corresponding weight function $w_i(m_J^2,R_g)$ is non-zero.

Having constructed the matched cumulative $R_g$ cross section, we will use it directly to evaluate $C_1^q(m_J^2)$. The uncertainties resulting from scale variations and 2-loop parameter variations will then be directly translated to uncertainties in predictions for $C_1^q(m_J^2)$. For computing $C_2^q(m_J^2)$ we will, however, only make use of the large $R_g$ boundary cross section at the differential level. Since we have no contribution from the small and intermediate $R_g$ regime, following \eq{a20Vary} we adopt the prescription
\begin{align} \label{eq:C2Tx}
\frac{d}{d \veps}
\frac{d \sigma(\veps)}{d m_J^2 d\theta_g}\bigg|_{\rm large}
\equiv
w_{\rm large} (m_J^2, \theta_g)\frac{d}{d \veps}\,\Bigg(
\frac{d \sigma^{[1]}(\veps)}{d m_J^2 d\theta_g}\bigg|_{\theta_g \sim \theta_g^\star}
+
\frac{d \sigma^{[2]}(\veps)}{d m_J^2 d\theta_g}\bigg|_{\theta_g \sim \theta_g^\star}
\Bigg)
\, .
\end{align}
From this equation we see explicitly that the cutoff on the large $R_g$ cross section appearing in the determination of $C_2^q$ is directly determined by $w_{\rm large}$.

\vspace{0.3cm}
\section{Numerical results and comparison with Monte Carlo simulations}
\label{sec:Num}

Having set up the formalism, provided the method to match across different regimes and described our prescription for quantifying the perturbative uncertainty, we are now in position to present the numerical results for the matched cumulative-$R_g$ cross section and the moments $C_1^q(m_J^2)$ and $C_2^q(m_J^2)$. We build our \cpp code based on the core of the \SCETlib package~\cite{scetlib}, extending it for soft drop observables, and cross check the numerical results using \Mathematica.
We first present the results for the matched double differential cross section in \sec{matched}, and then use it to calculate $C_1^q(m_J^2)$ in \sec{C1results} and $C_2^q(m_J^2)$ in \sec{C2results} in the manner described in earlier sections.
For definiteness, we fix $\zcut = 0.1$, $E_J = 500$ GeV and $R=1$ throughout the whole analysis, but we sample different values of the soft drop angular weight $\beta$. Along with the numerical results for the moments we will also present a comparison with parton shower Monte Carlo predictions.

\begin{figure}[!t]
\begin{center}
{\includegraphics[width=0.485\textwidth]{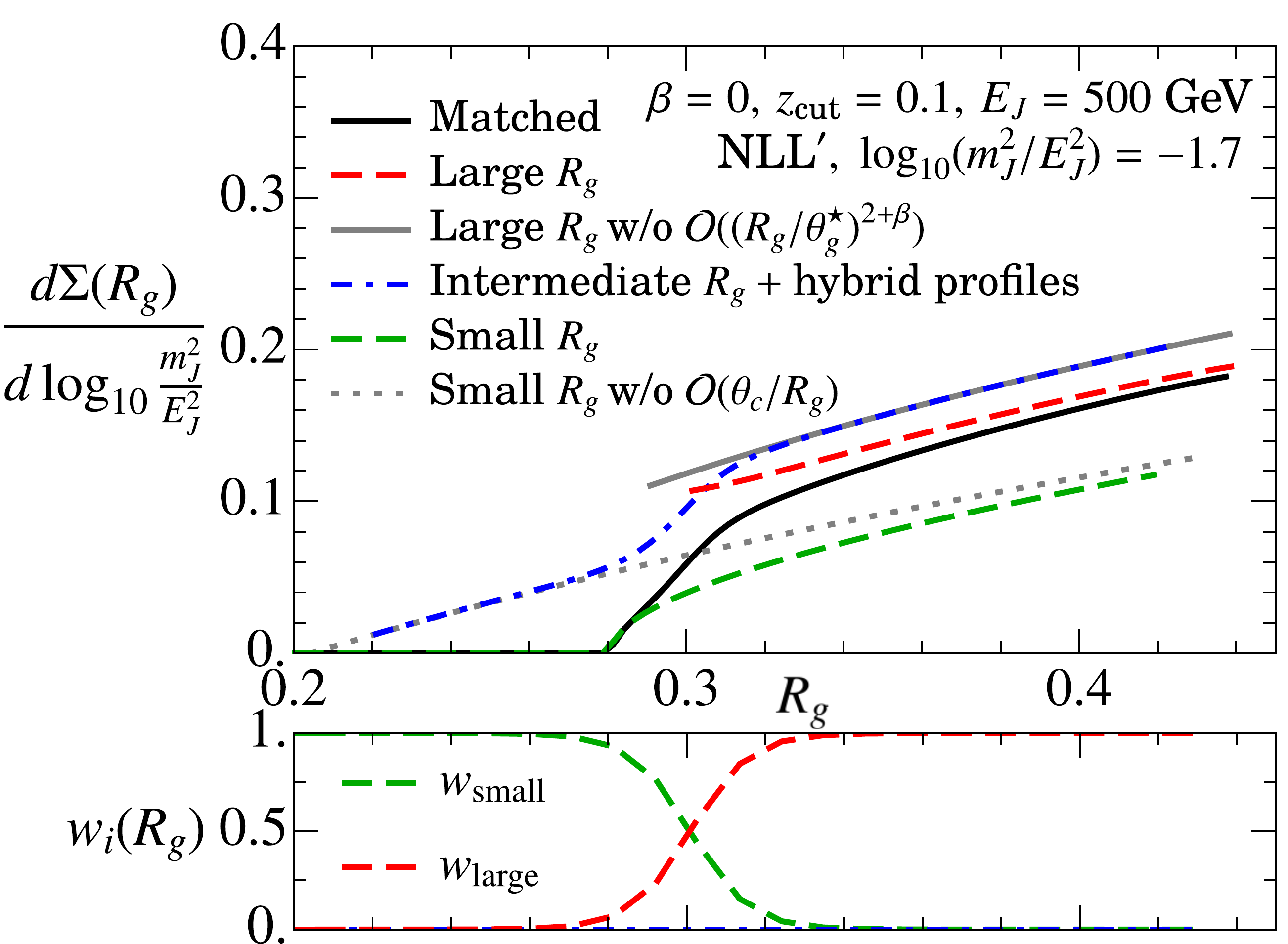}}
\hspace{0.01\textwidth}
{\includegraphics[width=0.485\textwidth]{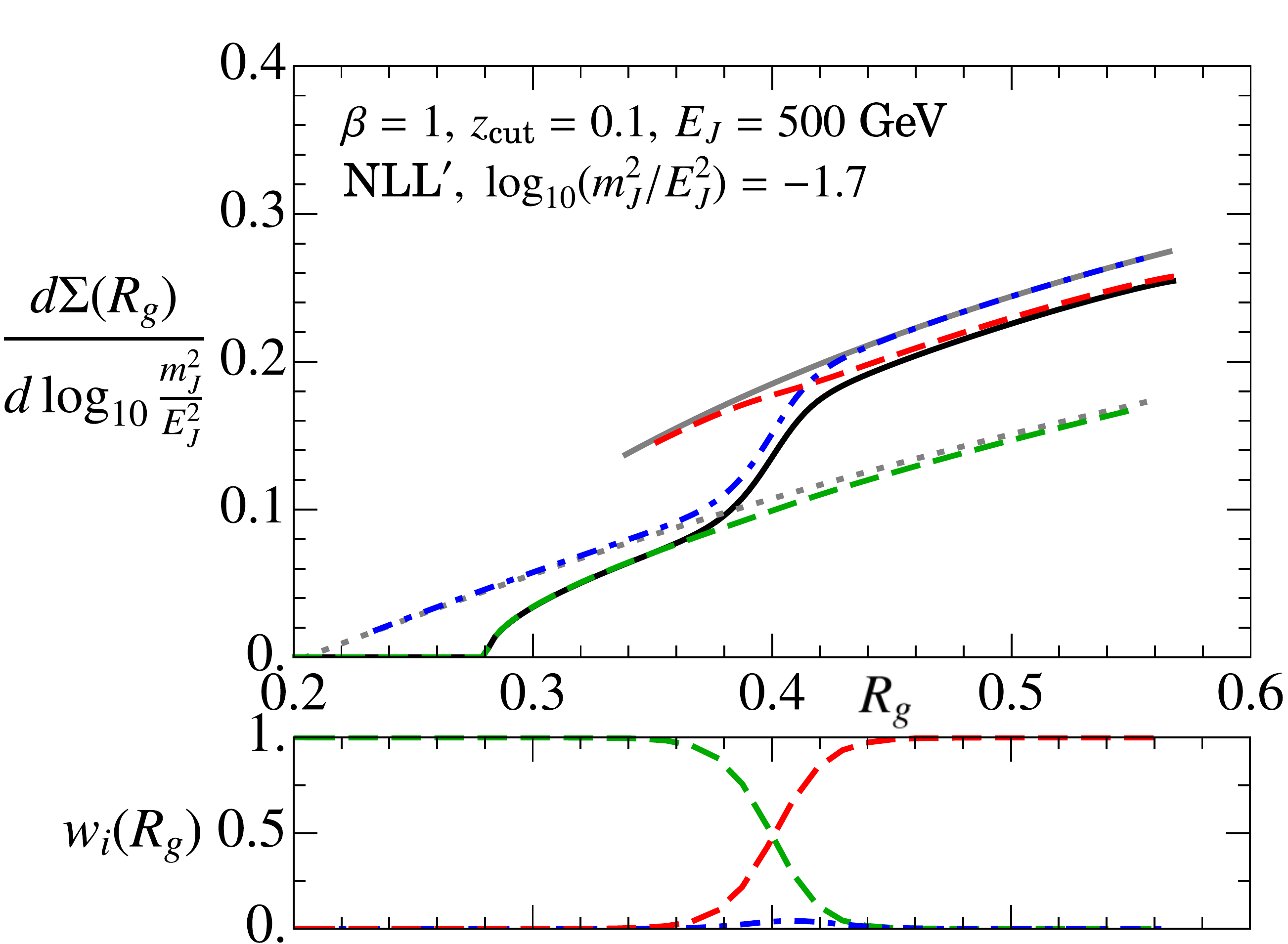}}
{\includegraphics[width=0.485\textwidth]{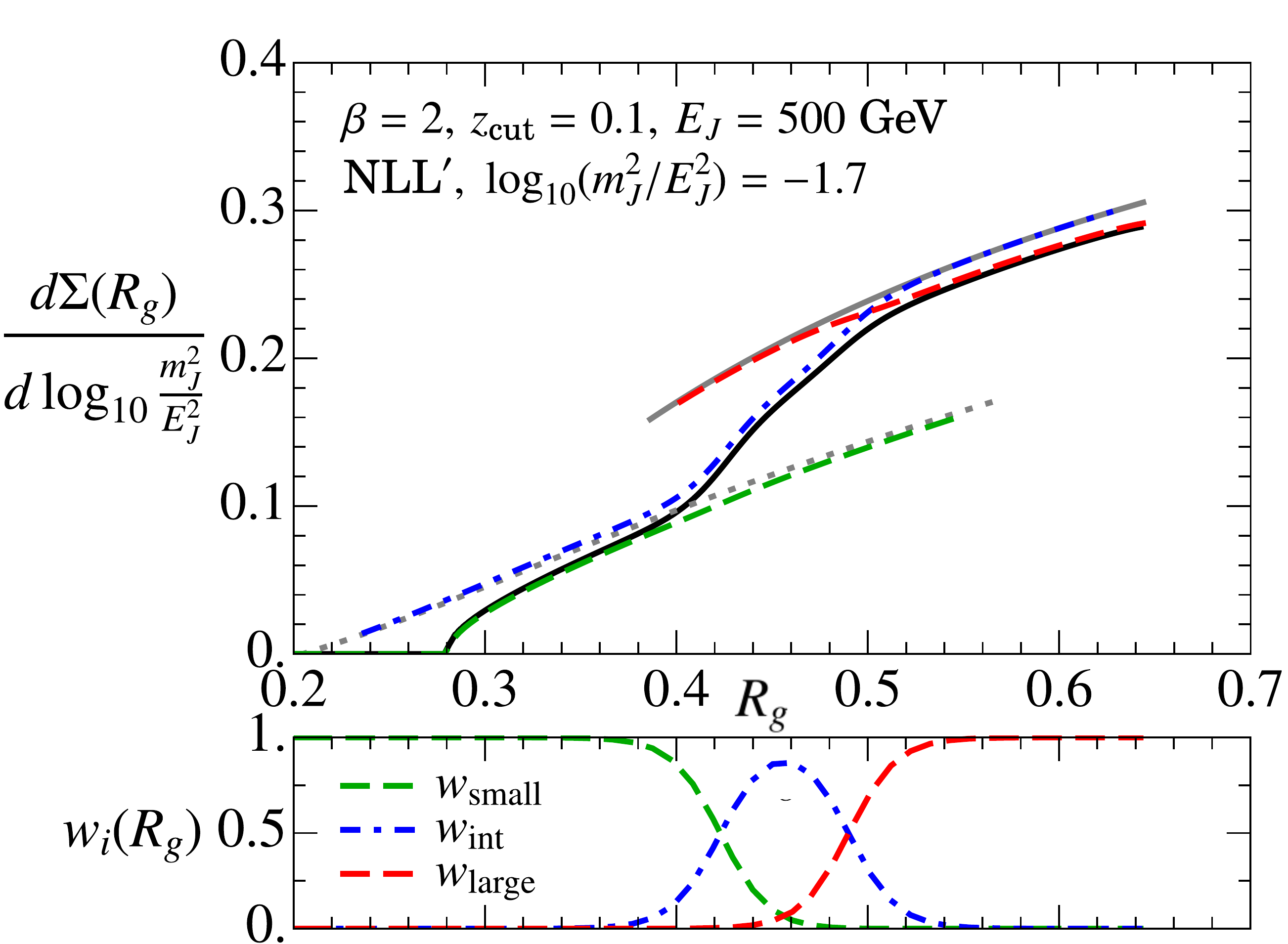}}
\caption{ The matching prescription smoothly interpolating between the three EFT regimes is shown for the distribution cumulative in $R_g$, at NLL$^\prime$ accuracy, for different values of $\beta$ and a fixed value of $m_J$. The large $R_g$ (red dashed) and small $R_g$ (green dashed) predictions include fixed-order corrections that are relevant in the respective plotted range; the profile-improved intermediate prediction (blue, dot-dashed) does not yet include these corrections while resumming large logarithms in the central region. We obtain the matched result (solid black) by weighting the three curves with the exponents $w_i(R_g)$ shown in the bottom panels, and subtracting the overlap (grey) as described in the text.}
\label{fig:matchingPlot}
\end{center}
\end{figure}

\subsection{Results for the matched cross section} \label{sec:matched}

Following the strategy outlined in \sec{matchImplement}, we now utilize the profile and transition functions described in \sec{trans} and \sec{prof} to smoothly match the cross section across all our EFT regimes. Here we consider explicitly the cross section cumulative in $R_g$ which determines $C^q_1$ as well as the normalization factor $C_0^q$ it depends on, whereas $C^q_2$ will be calculated at the differential level using the prescription in \eq{C2Tx}.
Since our focus is on making predictions for $C_{1,2}^q$, where NGL effects in the global soft sector do not contribute, we will not include the function $\Xi_G^q$ in our numerical results for the double differential cross sections presented in this section (despite the fact that the resummation of these NGLs may be important to determine the overall normalization). However, we do include the NGLs described by $\Xi_S^q$, and $\Xi_C^q$, which affect both the shape and normalization of the distribution.

In \fig{matchingPlot} we show the various components that enter the calculation of the matched cross section for $\beta = 0,1,$ and $2$ plotted as a function of $R_g$ for a representative jet mass value giving $\log_{10} (m_J^2/E_J^2) = -1.7$,
which with our choice for $E_J$ is $m_J=70.6\,{\rm GeV}$.
In order to explain the legend and the meaning of each curve, consider the first panel with $\beta = 0$.
The solid black line is the matched cross section that is evaluated after combining the cross sections in the three regimes according to \eq{SigMatched}. The results for cross section in the large $R_g$ EFT, $d\Sigma_{\rm large}|_{\mu_{\rm large}}$, and in the small $R_g$ EFT, $d\Sigma_{\rm small}|_{\mu_{\rm small}}$, evaluated with their corresponding profiles are shown by the red dashed and green dashed curves respectively. The overlap of the intermediate $R_g$ with either of these regimes are shown in gray-solid for $d\Sigma_{\rm int}|_{\mu_{\rm large}}$ (labeled as ``Large $R_g$ w/o ${\cal O}\big((R_g/\theta_g^\star)^{2+\beta}\big)$''), and gray-dotted for $d\Sigma_{\rm int} |_{\mu_{\rm small}}$ (labeled as ``Small $R_g$ w/o ${\cal O}\big(\theta_c/R_g)$''). We see that the $d\Sigma_{\rm int} |_{\mu_{\rm large}}$ term matches with the $d\Sigma_{\rm large}|_{\mu_{\rm large}}$ term as $R_g \ra 0$, such that the power correction ${\cal O}\big((R_g/\theta_g^\star)^{2+\beta}\big)$ becomes small. Likewise, the $d\Sigma_{\rm int} |_{\mu_{\rm small}}$ cross section merges with the small $R_g$ result as $R_g \ra \theta_g^\star$, up to the ${\cal O}\big(\theta_c^2/R_g^2)$ power correction.
Finally, the blue dot-dashed curve that spans the entire range of $R_g$ values is the interpolation $\Sigma_{\rm int}|_{\mu_{\rm hyb}}$ that makes use of the hybrid profiles described in \eq{muHyb}. The hybrid profiles are constructed to reproduce the large-$R_g$ and small-$R_g$ profiles in their respective regions. Thus, we see that $\Sigma_{\rm int}|_{\mu_{\rm hyb}}$ matches exactly with the $d\Sigma_{\rm int} |_{\mu_{\rm large}}$ (gray-solid) in the large $R_g$ regime and $d\Sigma_{\rm int} |_{\mu_{\rm small}}$ (gray-dotted) in the small-$R_g$ regime.
The weight functions $w_i(m_J^2, R_g)$ given in \eq{weightFunc} demarcate the boundary of each region and govern the behavior of the hybrid profiles,
and are shown below each cross section panel.

\begin{figure}[!t]
{\includegraphics[width=0.48\textwidth]{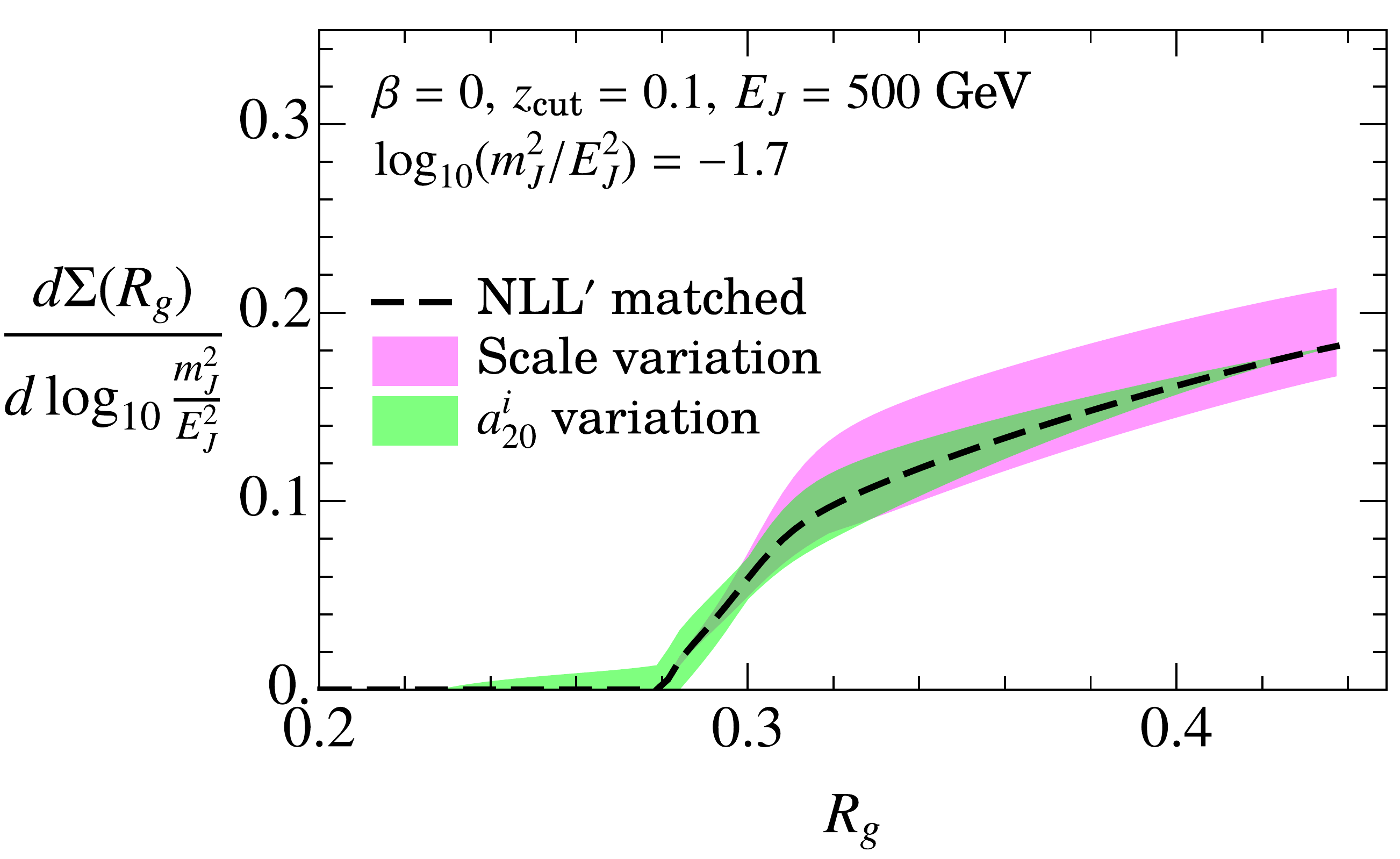}}
{\includegraphics[width=0.48\textwidth]{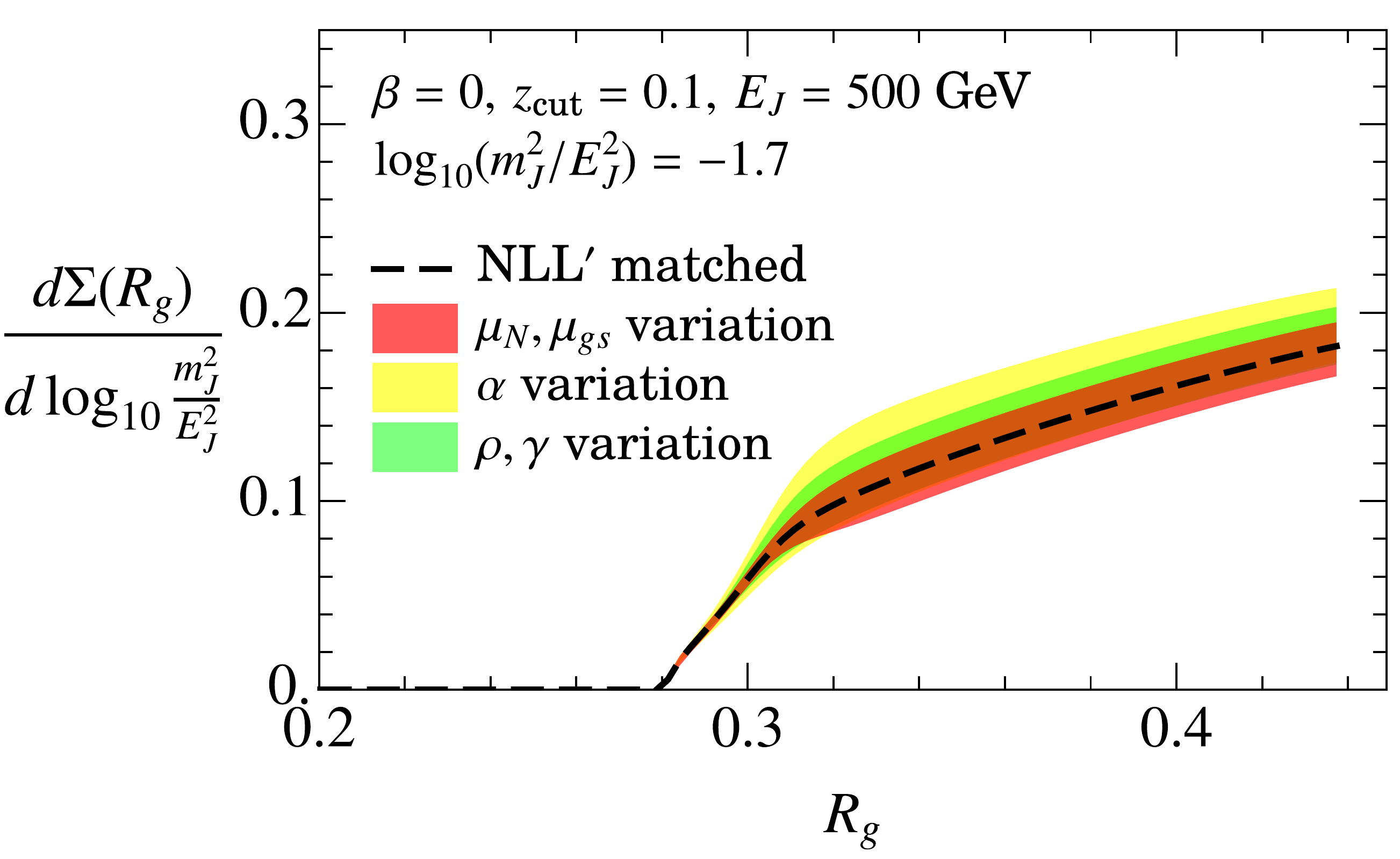}}
{\includegraphics[width=0.48\textwidth]{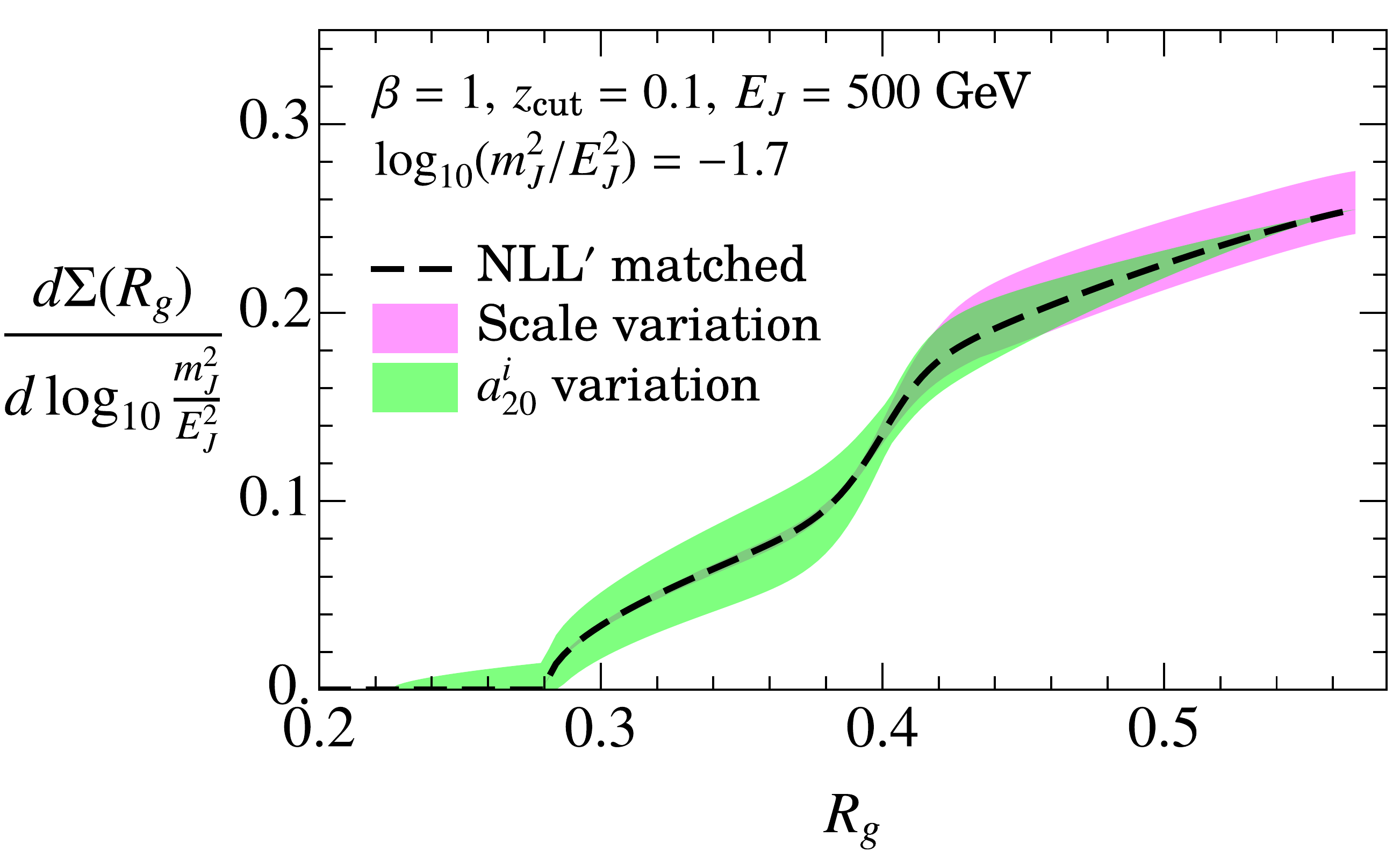}}
\hspace{0.02\textwidth}
{\includegraphics[width=0.48\textwidth]{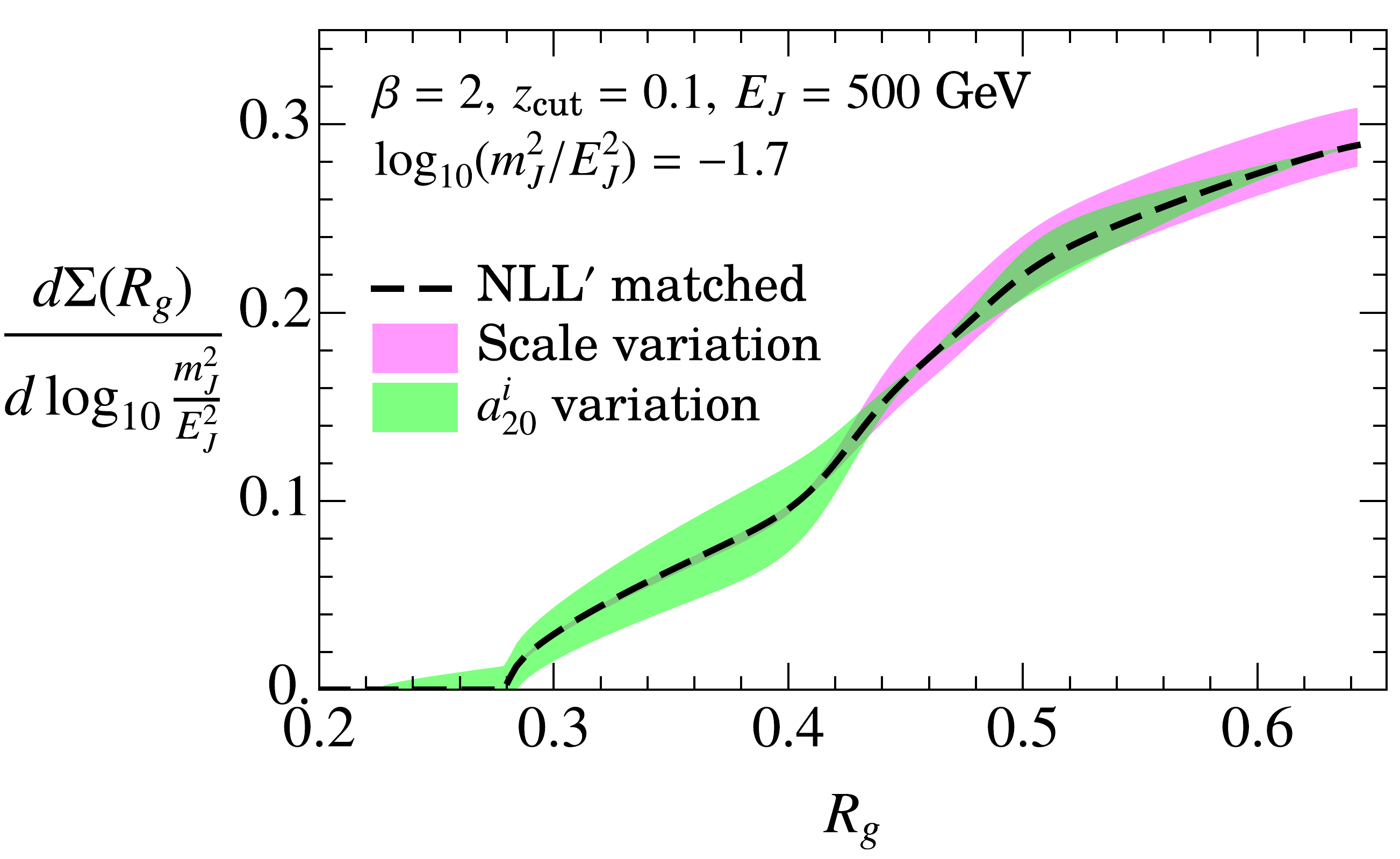}}
\caption{Theoretical uncertainty on the cross section cumulant in $R_g$, for $\log_{10}(m_J^2/E_J^2)=-1.7$ and different values of $\beta$. The central curve (black dashed) is obtained with the matching prescription described in the text, while the colored bands probe respectively scale variation (magenta), and variation of the fixed order two-loop constants $a_{20}^{S_c}$ and $a_{20}^{\cal C}$, collectively denoted as $a^i_{20}$ variation (green). In the top right plot, the envelope of scale variations for $\beta=0$ is broken down into various components: overall variation by a common factor (red), variation by a trumpet factor preserving canonical relations (yellow), breaking of canonical relations (green).}
\label{fig:cumulVaryPlot}
\end{figure}

As mentioned in \sec{trans}, for the cases $\beta = 0, 1$ the intermediate $R_g$ resummation is not present (for the jet mass value shown in \fig{matchingPlot}) as there is simply no room for this EFT to be valid according to \eq{IntValid}, whereas for $\beta = 2$ there is a significant region where the intermediate $R_g$ EFT is valid. For each of the three $\beta$ values that we show, we see that our choice of
the locations where the weight functions turn themselves on and off agrees with the locations where the power corrections become visible. This can be seen from the difference of the large and small $R_g$ cross sections with the corresponding intermediate $R_g$ overlaps.

Next we show in \fig{cumulVaryPlot} our estimate of perturbative uncertainty on the NLL$^\prime$ cumulant cross section, again for $\beta = 0$, $1$, and $2$. The uncertainty bands in the top left plot are obtained by varying the profile scales as described in \sec{vary} (magenta) and by varying the quantities $a_{20}^{{\cal C}}$ and $a_{20}^{S_c}$ that parametrize our ignorance of the two-loop constants terms in respectively the small $R_g$ and large $R_g$ EFTs (green). Specifically, we vary both $a_{20}^{ S_c}$ and $a_{20}^{{\cal C}}$ in the range $[-2\pi,2\pi]$, and take the envelope of the two independent variations. We notice that scale variation is the dominant source of uncertainty at large $R_g$, while at small $R_g$ the perturbative uncertainty is mainly set by the $a_{20}^{{\cal C}}$ term.
As discussed below \eq{a10C}, the lower endpoint of the distribution receives perturbative corrections starting at NLO. It is clear from \eq{CqFiniteMJ} that this effect is captured by variations of $a_{20}^{{\cal C}}$, resulting in a non-vanishing uncertainty band below the endpoint of the central value.
This effect has no equivalent in the scale variations, where by construction we set $a_{20}^{{\cal C}} = 0$ and use the same LO endpoint as for the central value.

The top right plot in \fig{cumulVaryPlot} shows for $\beta=0$ the breakdown of the scale uncertainty into the three independent variations listed in \sec{vary}: overall scale uncertainty obtained by simultaneously varying the hard and soft scales by a factor 2 (red band); variations of collinear-soft scales by a trumpet factor, setting $\alpha = \pm 1$ in \eqs{alphadef}{alphadef2} (yellow band); and uncertainty due to relaxing the canonical relations between scales, achieved by varying $\rho,\gamma \in [-0.1,0.1]$ in \eqs{rhodef}{gammadef} (green band). We see that the three sources of uncertainty have similar size, with the trumpet factor playing the most relevant role. A similar qualitative behavior occurs for $\beta = 1, 2$, so for simplicity we do not plot the breakdown for these cases, and only show the envelope of scale variations and the fixed-order uncertainty due to the parameters $\{a^i_{20}\}$ in the lower panels of \fig{cumulVaryPlot}. In these plots we do not show the relative impact of the NGLs we have included in our cross section, but we note that here their contribution is much smaller than the theoretical uncertainty across the whole range in $R_g$.

Before moving on to the numerical results for $C_1^q$ and $C_2^q$, we anticipate that the theoretical uncertainty for these two observables will look rather different than the one just shown for the double differential cumulative cross section. In particular, the part of uncertainty that only depends on the jet mass but not on $R_g$ will cancel in the ratio of the angular moments in \eq{C1C2multiDiffDef} and the cumulant distribution, since effectively this means they are normalized point by point in the $m_J^2$ spectrum. We anticipate that this will lead to smaller uncertainties for both the $C_1^q$ and $C_2^q$ coefficients.

\subsection{Results for $C_1^q$} \label{sec:C1results}

We now show predictions for $C_1^q$, based on the NLL$'$ expressions for $M_1^q$ obtained by evaluating \eq{C1C2def} on the resummed cross section described in \sec{resum} and shown in \sec{matched}. We then compare these EFT results against Monte Carlo simulations and numerical results from coherent branching. We explicitly present predictions for quark jets, expecting gluon jets to follow the same qualitative behavior. We set again $E_J = 500$ GeV, $R=1$, and $\zcut =0.1$, and explore the three cases $\beta=0,1,2$ across the mass range $-3.5 < \log_{10}(m_J^2/E_J^2) < -0.5$. At low jet masses, the double differential distribution on which we base our predictions becomes nonperturbative. However, these NP effect should not be included in $C^q_1$, which is by definition a purely perturbative coefficient. What portion of the jet mass range is significantly affected by NP corrections crucially depends on the value of $\beta$.
Our method of calculating $C_1^q$ automatically deals with this as discussed in \sec{prof}.

\begin{figure}[t!]
\centering
\includegraphics[width=0.48\textwidth]{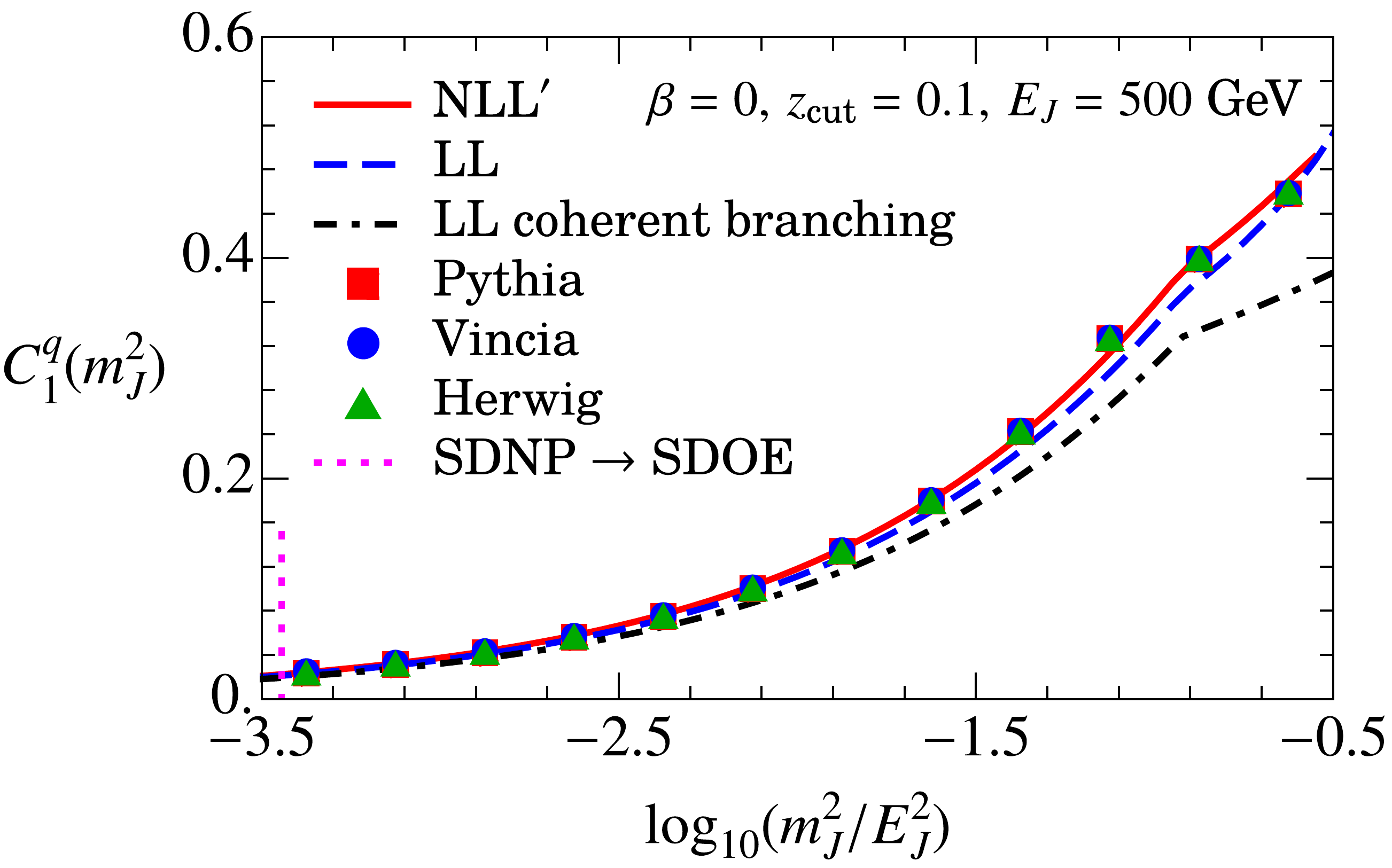}
\hspace{0.01\textwidth}
\includegraphics[width=0.48\textwidth]{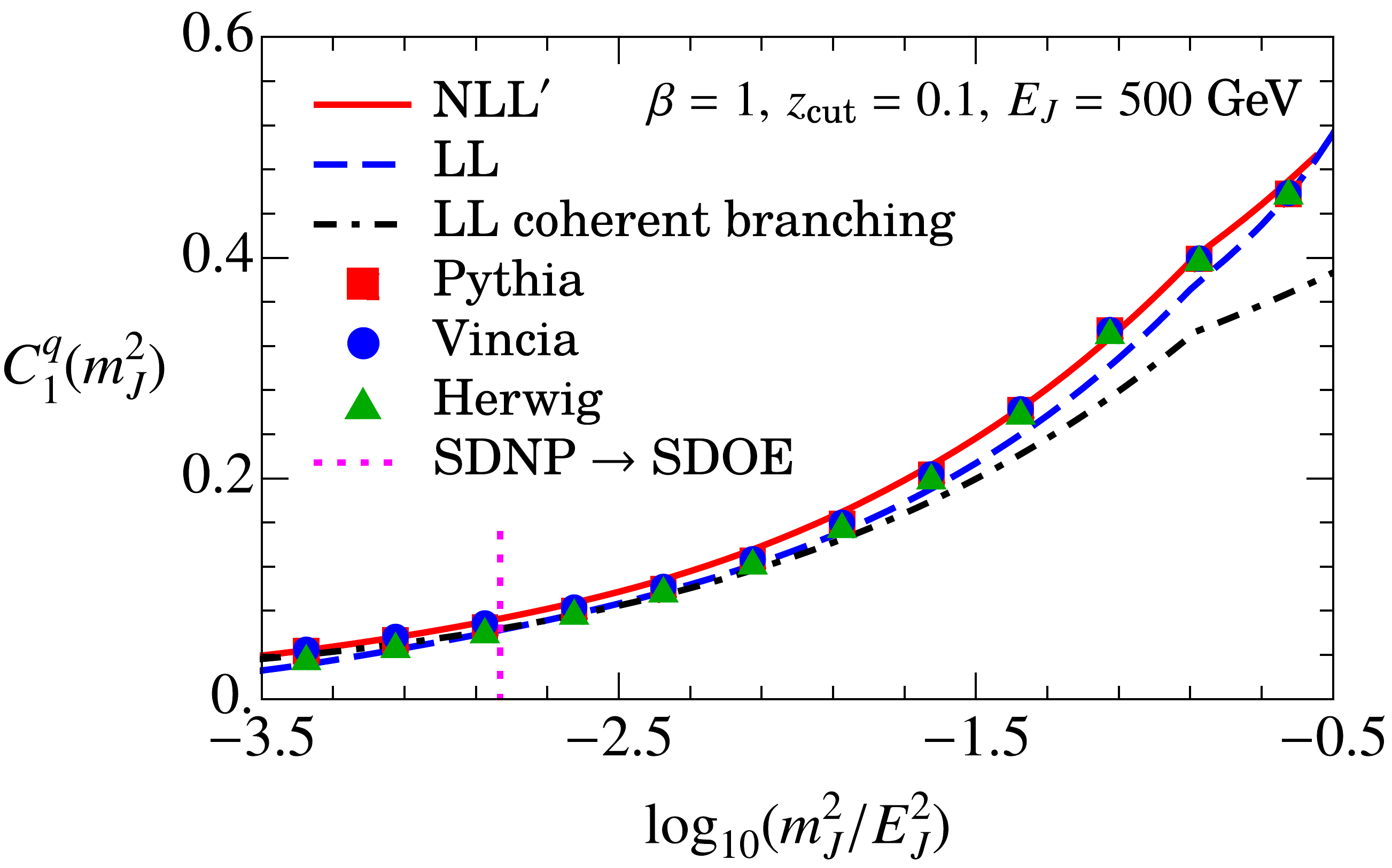}
\includegraphics[width=0.48\textwidth]{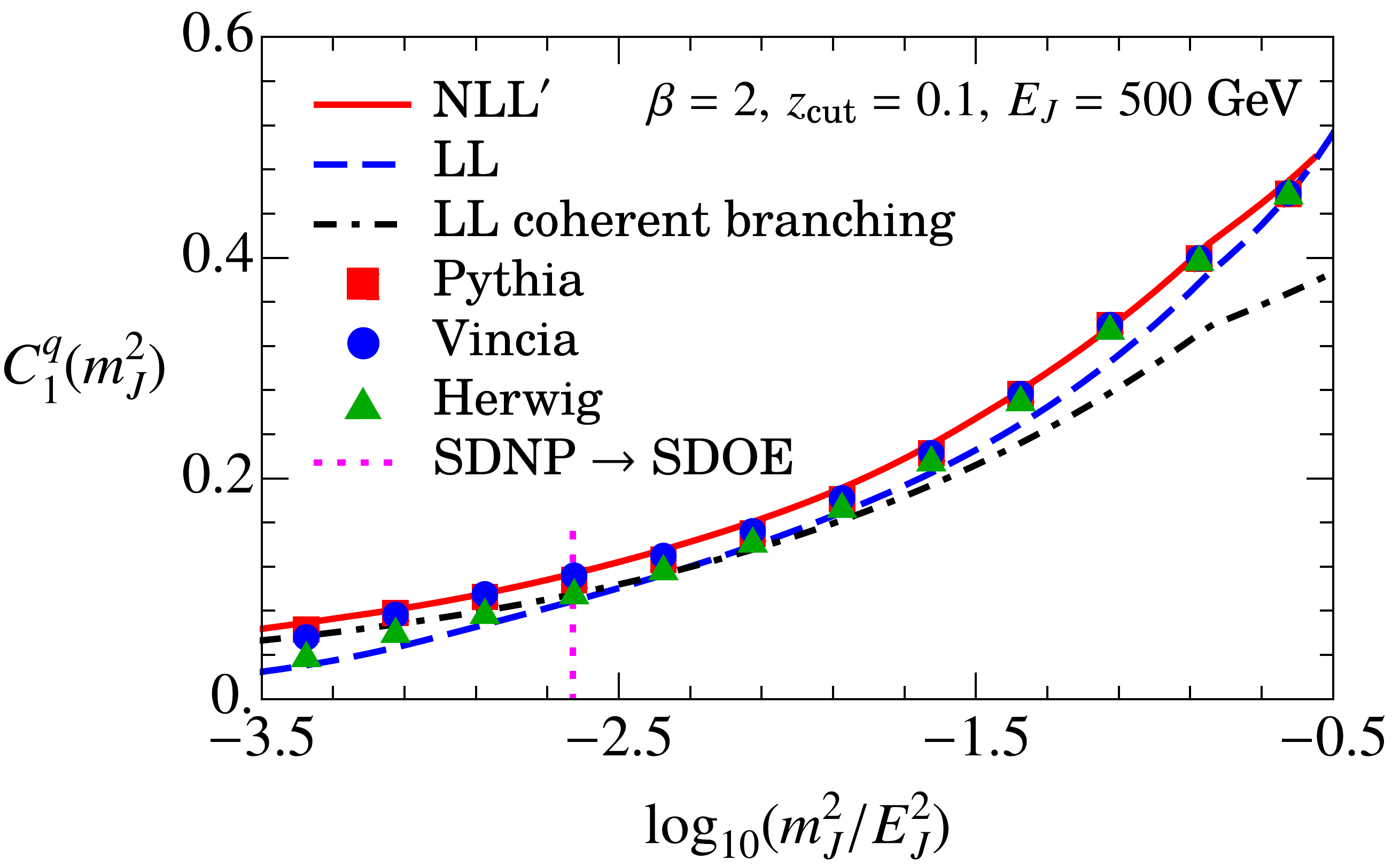}
\caption{Comparison between the LL (blue dashed) and NLL$'$ (red) effective theory predictions for the perturbative coefficient $C_1$, the LL predictions from coherent branching formalism (black dot-dashed) and Monte Carlo simulations (colored markers), for different values of $\beta$. The vertical dotted magenta line signals the transition from the SDNP region at small $m_J$ to the perturbative SDOE region at larger $m_J$.}
\label{fig:C1vsMC}
\end{figure}

In \fig{C1vsMC} we present central values for our EFT results for $C_1^q$ at both LL and NLL$'$ orders. These results are compared to results from coherent branching and from Monte Carlo simulations. For the latter, we consider predictions from \Pythia~8.2~\cite{Sjostrand:2014zea}, \Vincia~2.2~\cite{Fischer:2016vfv} and \Herwig~7.1~\cite{Bellm:2015jjp}, where the jet reconstruction and soft drop are performed using \Fastjet~3.3~\cite{Cacciari:2011ma}. We signal the region where NP effects are relevant by showing the transition from SDNP to SDOE region
by a vertical dotted magenta line,
as defined in \eq{SDOE}. Note that \Pythia, \Herwig and \Vincia return very similar results for $C_1^q$~\cite{Hoang:2019ceu}. At intermediate and large masses the NLL$'$ predictions differ significantly from the ones from coherent branching, in better agreement with the Monte Carlo, while the LL-EFT prediction lies in-between these two results. In the SDNP region the NLL$'$ and LL coherent branching curves agree very well with each other, while the LL-EFT result shows some deviation. The fact that our LL-EFT predictions differ from the LL coherent branching derived in \cite{Hoang:2019ceu} is not too surprising: as we discussed in \sec{LLCB}, the EFT description is more refined in that it involves matching the three regimes via profile scales which includes a resummation of additional large logarithms, but on the other hand the LL-EFT result lacks some of the power corrections that are present in the splitting kernel $p_{gq}(z)$. These power corrections have a small effect for $C_1^q$, thus the difference seen in \fig{C1vsMC} is driven by the additional resummation included in the LL-EFT results.

\begin{figure}[t!]
\centering
\includegraphics[width=0.48\textwidth]{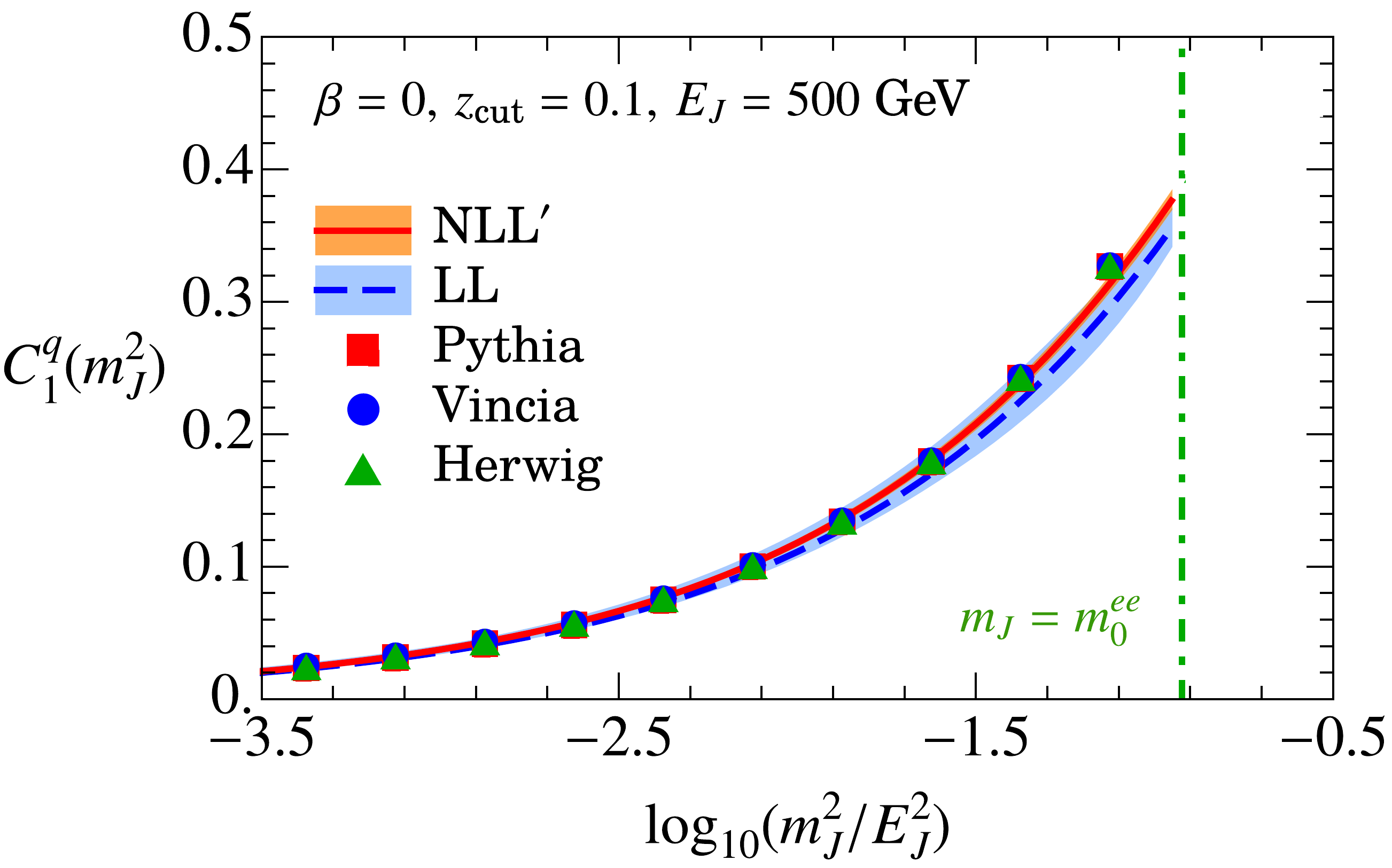}
\hspace{0.02\textwidth}
\includegraphics[width=0.48\textwidth]{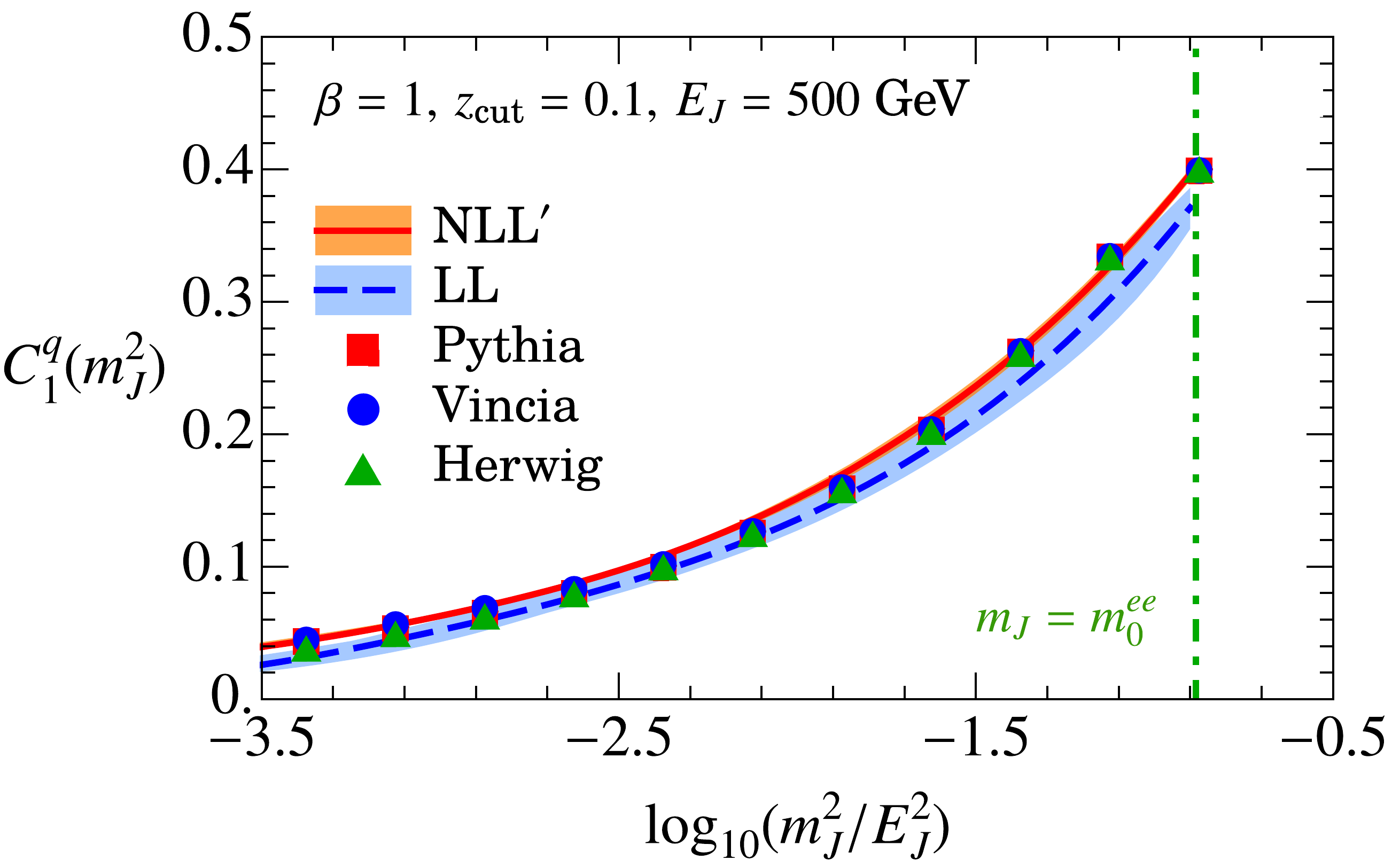}
\includegraphics[width=0.48\textwidth]{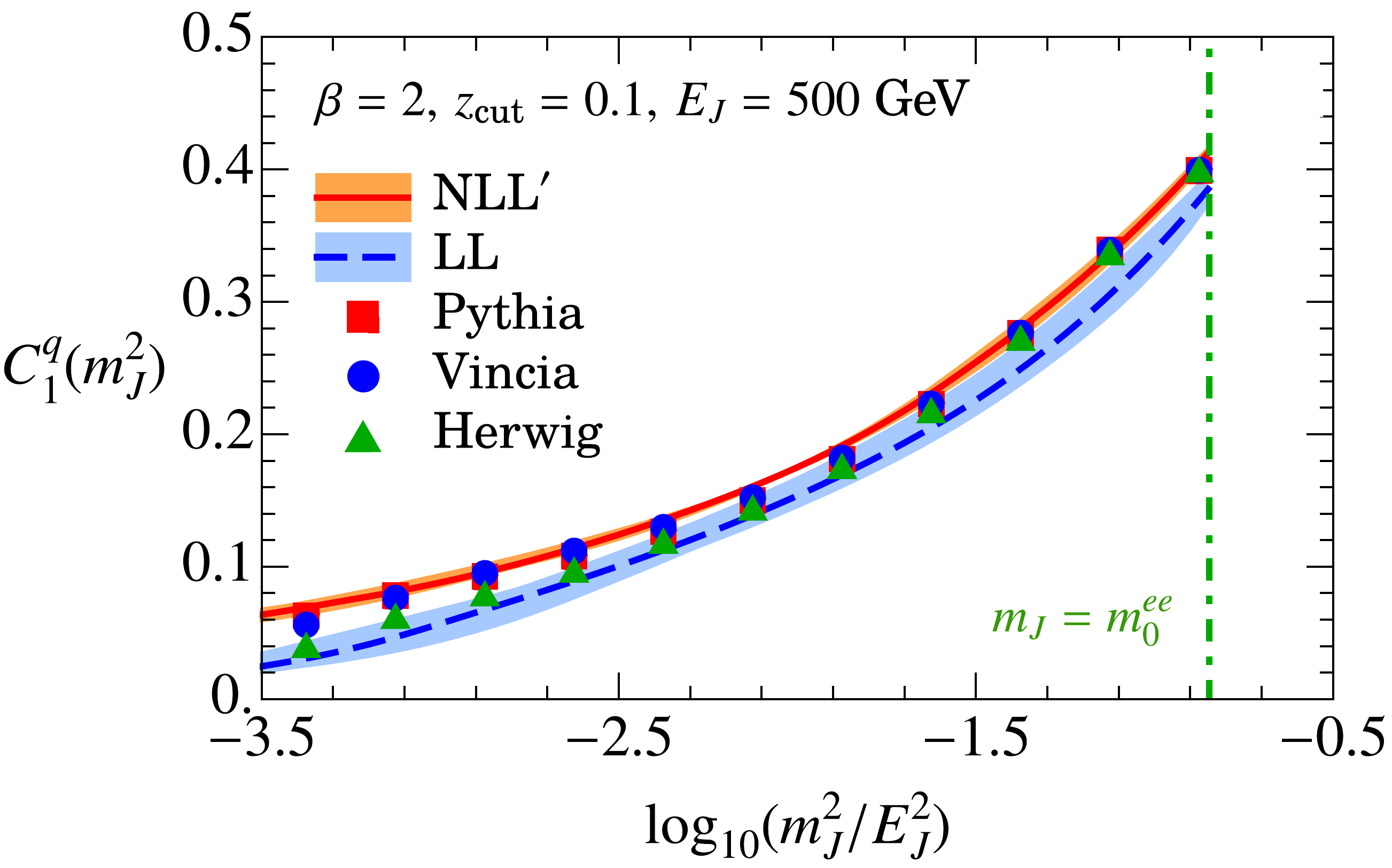}
\caption{Uncertainty estimate for the NLL$'$ and LL effective theory predictions for the perturbative coefficient $C_1^q$. We do not estimate LL uncertainty in the ungroomed region, where the EFT description loses validity. NLL$'$ uncertainty bands are tiny and examined in more detail in \fig{C1vsMCvary}.}
\label{fig:C1vsMCbands}
\end{figure}

In \fig{C1vsMCbands} we reconsider the EFT predictions, now including also our estimate of their perturbative uncertainties (repeating the Monte Carlo simulation results for reference).
While the LL uncertainty band (blue) is obtained purely from scale variation, the NLL$'$ uncertainty band (orange) is obtained from the envelope of scale variation and variation of the two-loop fixed-order constants $\{a^i_{20}\}$.
Since our estimates for the perturbative uncertainty does not cover the ungroomed region, we exclude values $m_J>m_0^{ee}$ in the plotted range.
We observe good convergence between the two orders, with only the $\beta=2$ results being slightly outside the LL uncertainty band.
The most striking feature is the very small size of the NLL$'$ uncertainty band, whose width becomes noticeable only for $\beta = 2$ (low panel).
Given that for the NLL$'$ cumulant cross section the variations are rather large (see \fig{cumulVaryPlot}), the tiny uncertainty bands for $C_1^q$ are due to cancellations between numerator and denominator of \eq{C1intermediateExplicit}. Indeed, we know that the leading double logarithms cancel in the ratio to all orders in perturbation theory.

\begin{figure}[t!]
\centering
\includegraphics[width=0.48\textwidth]{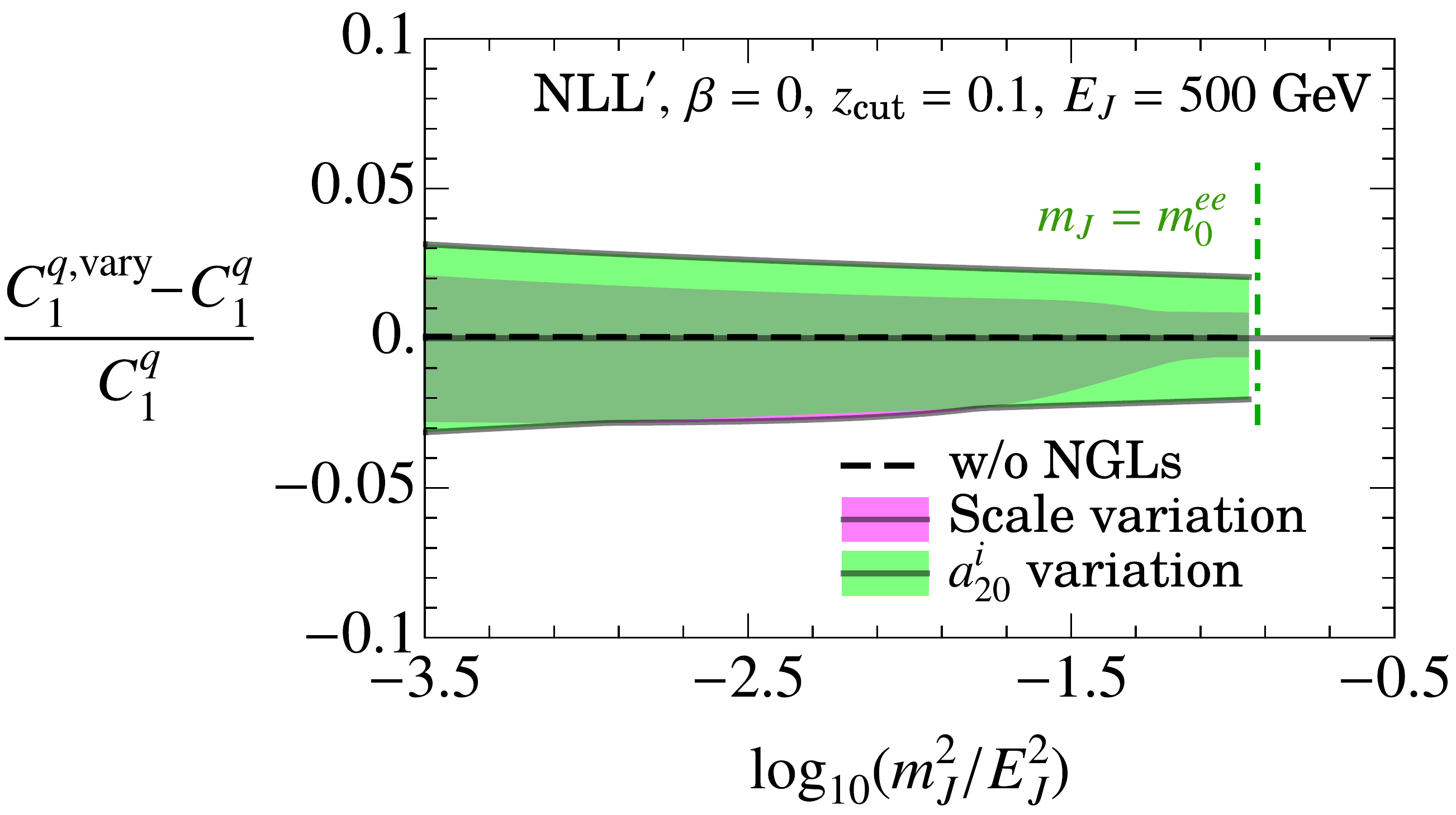}
\hspace{0.02\textwidth}
\includegraphics[width=0.48\textwidth]{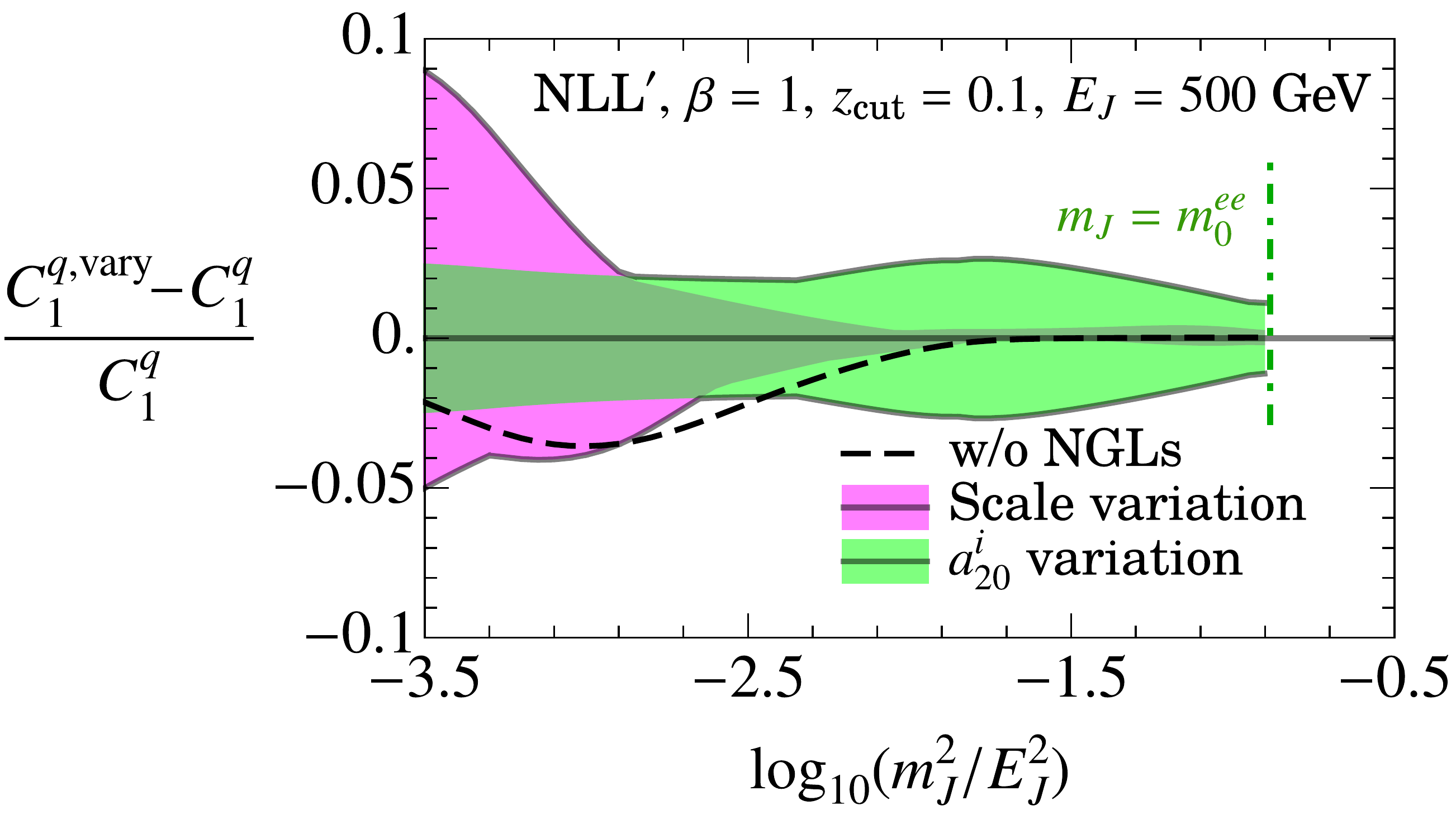}
\includegraphics[width=0.48\textwidth]{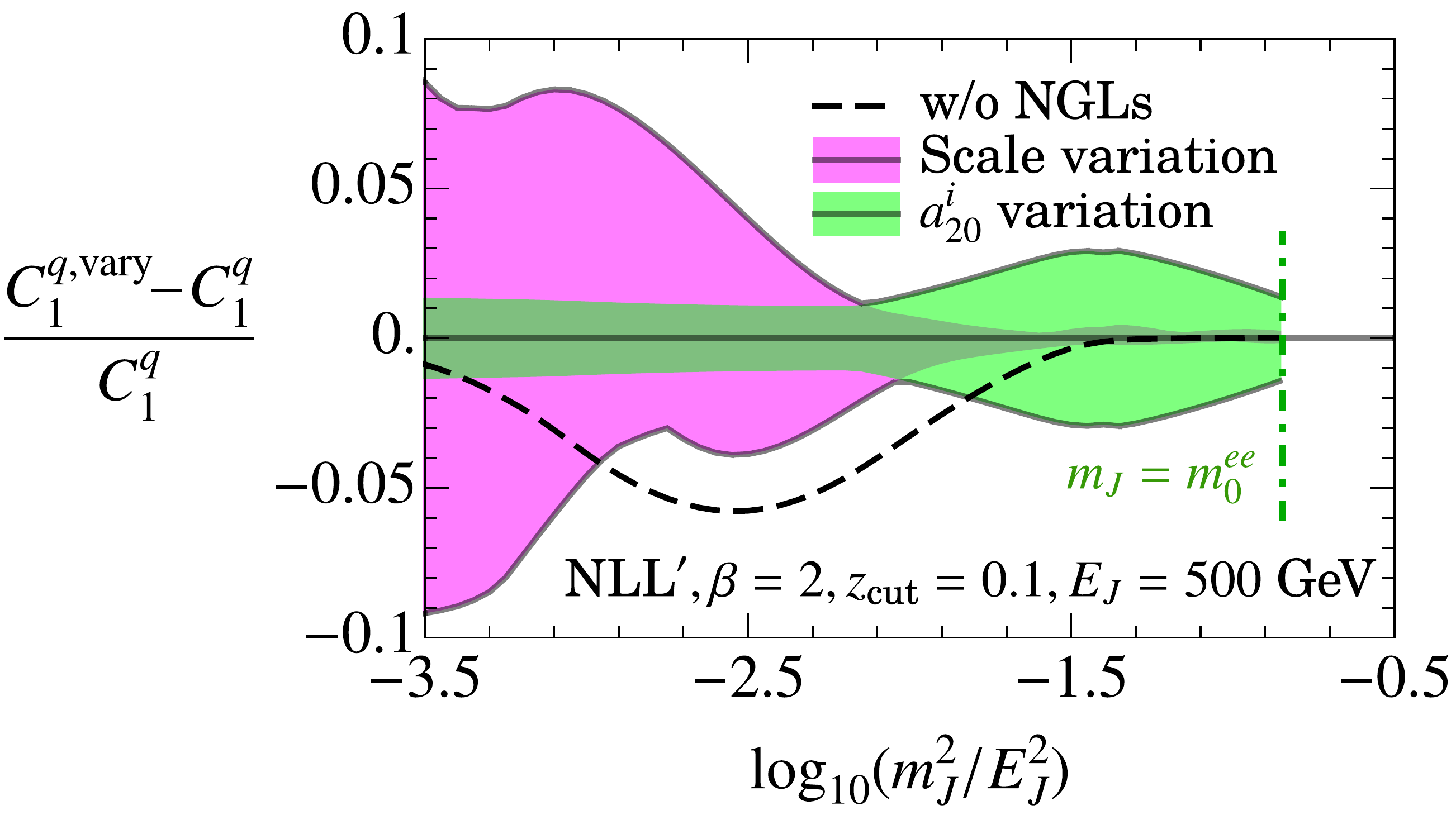}
\hspace{0.02\textwidth}
\includegraphics[width=0.48\textwidth]{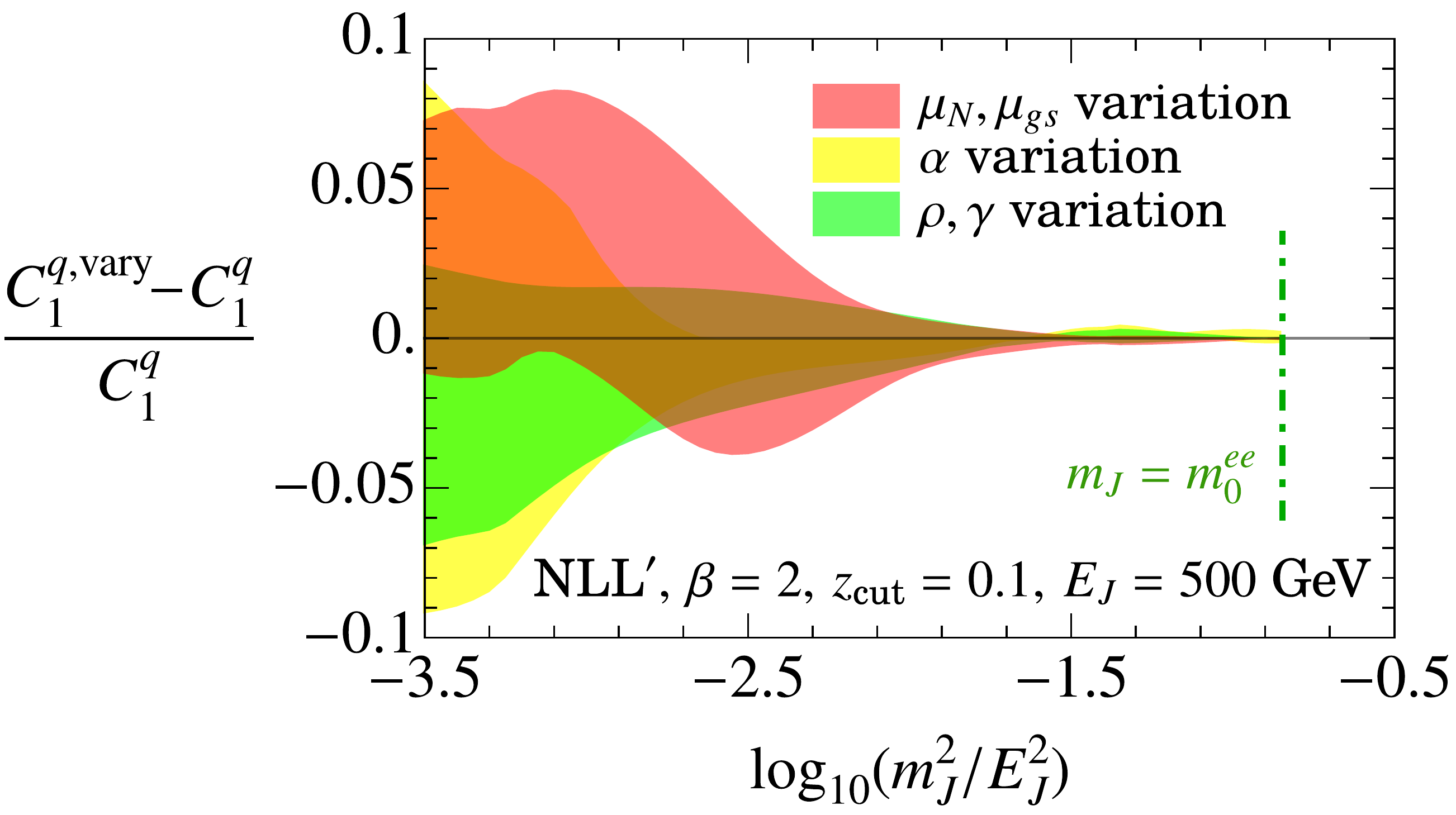}
\caption{Theoretical uncertainly on the NLL$'$ predictions for $C_1$, estimated through scale variation (magenta) and by varying the two-loop fixed-order constant (green). When the jet mass is perturbative, both methods return very small uncertainty bands, thus we normalize variations to the central value for clarity. The black dashed curve shows the shift to the central value when NGLs are ignored. The bottom right panel shows the breakdown of the envelope of scale variation for $\beta=2$, labeled as in \fig{cumulVaryPlot}.}
\label{fig:C1vsMCvary}
\end{figure}

To further study the small NLL$'$ theoretical uncertainty, in \fig{C1vsMCvary} we show the relative difference of the variations to the central curve. As we did for the cumulant in \fig{cumulVaryPlot}, we separate scale variation (magenta) and the variation of the two-loop constants (green). Again, we do not show the perturbative uncertainty beyond the groomed region. We first notice that in the SDOE region the uncertainty, of a few percent, is dominated by the unknown two-loop term. Scale variation gradually takes over in the SDNP region, which becomes larger as $\beta$ increases. For $\beta=1,2$ the enlarged uncertainty at high jet mass comes from varying the two-loop constant $a_{20}^{\cal C}$ in the small $R_g$ region, whereas variations of the large $R_g$ term $a_{20}^{S_c}$ are roughly constant across the mass range, and effectively negligible. The situation is different for $\beta = 0$, where varying $a_{20}^{S_c}$ gives the dominant contribution. That the variations from the small $R_g$ region are less relevant for this $\beta$ was already clear from \fig{cumulVaryPlot}. In the bottom right plot of \fig{C1vsMCvary} we further break down the scale variations for the $\beta=2$ case into the three different sources, where the same descriptions made for \fig{cumulVaryPlot} apply. A similar picture holds for the other values of $\beta$, except that the overall scale variation (light red) becomes less and less relevant as $\beta$ decreases.

In \fig{C1vsMCvary} we also investigate the relative contribution of NGLs to $C_1^q$, by showing the shift to the central curve obtained when omitting the leading
NGLs and abelian clustering logarithms (black dashed). For $\beta=0$ they are entirely negligible, while for larger $\beta$ their contribution is small but comparable to the theoretical uncertainty. This further justifies including the leading NGLs in our central value predictions for $C_1^q$ with the method described above.
Finally, we tested the dependence of our results on the choice of weight functions discussed in \sec{trans}. Specifically, we modified the inequality in \eq{IntValid}
from $\leq \frac13$ to $\leq \frac{1}{2}$ and $\leq \frac{1}{4}$
(thus varying the range where the intermediate $R_g$ EFT is included) and changed the transition rate $r_t$ between EFTs in \eq{Xtrans} by a factor 2. We find that in both cases the variations lie within our uncertainty bands.

\subsection{Results for $C_2^q$} \label{sec:C2results}

\begin{figure}[t!]
\centering
\includegraphics[width=0.48\textwidth]{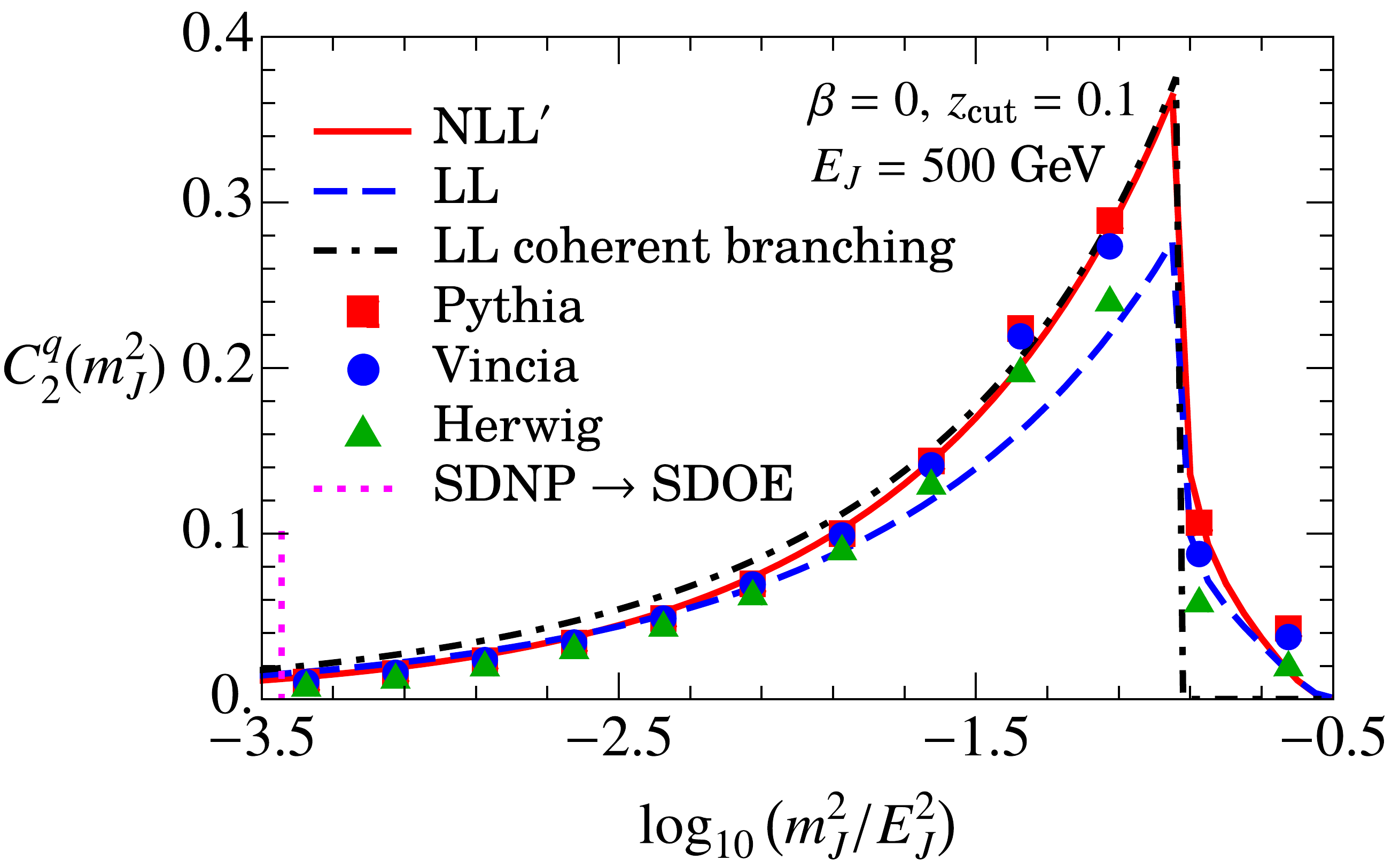}
\includegraphics[width=0.48\textwidth]{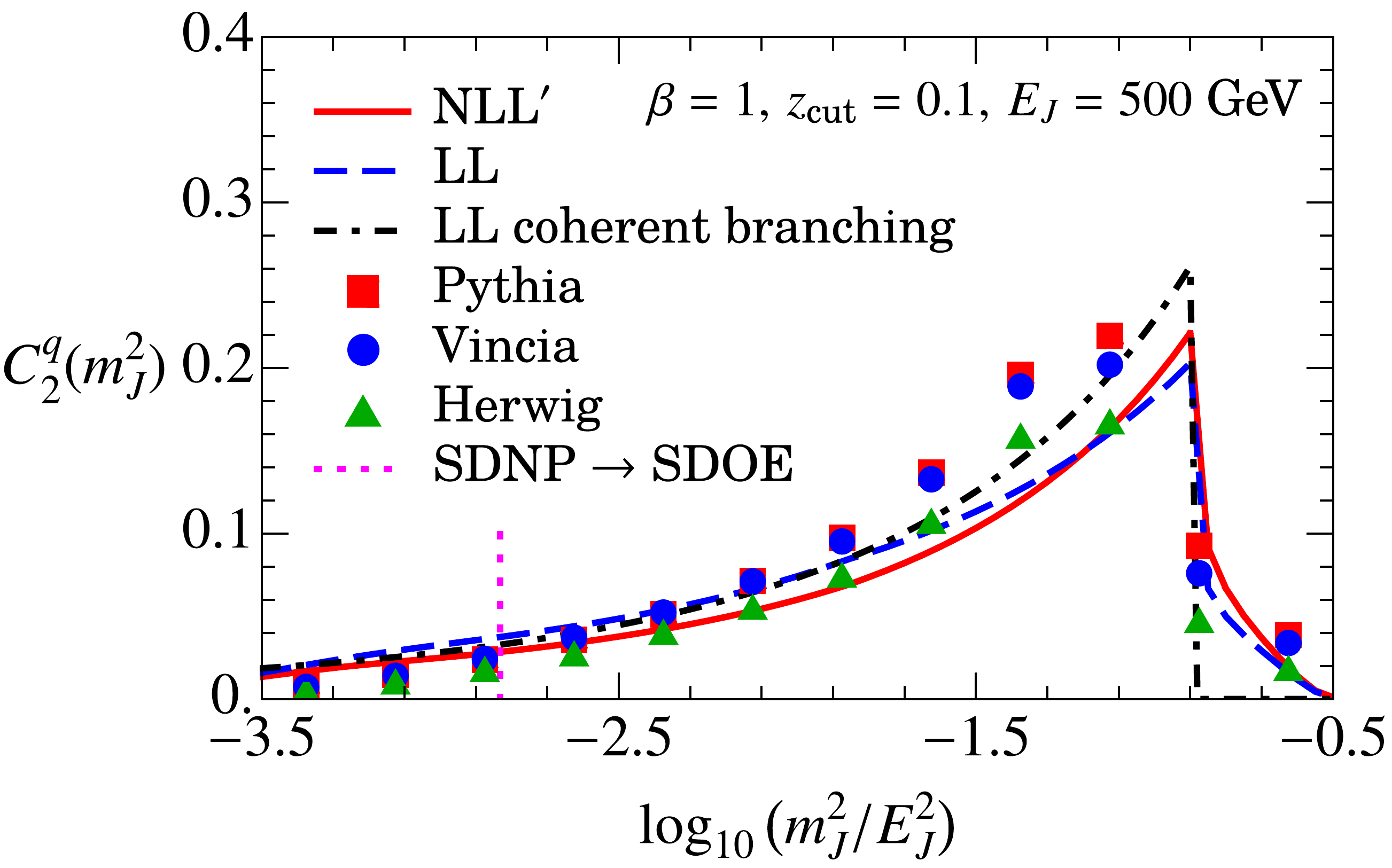}
\hspace{0.02\textwidth}
\includegraphics[width=0.48\textwidth]{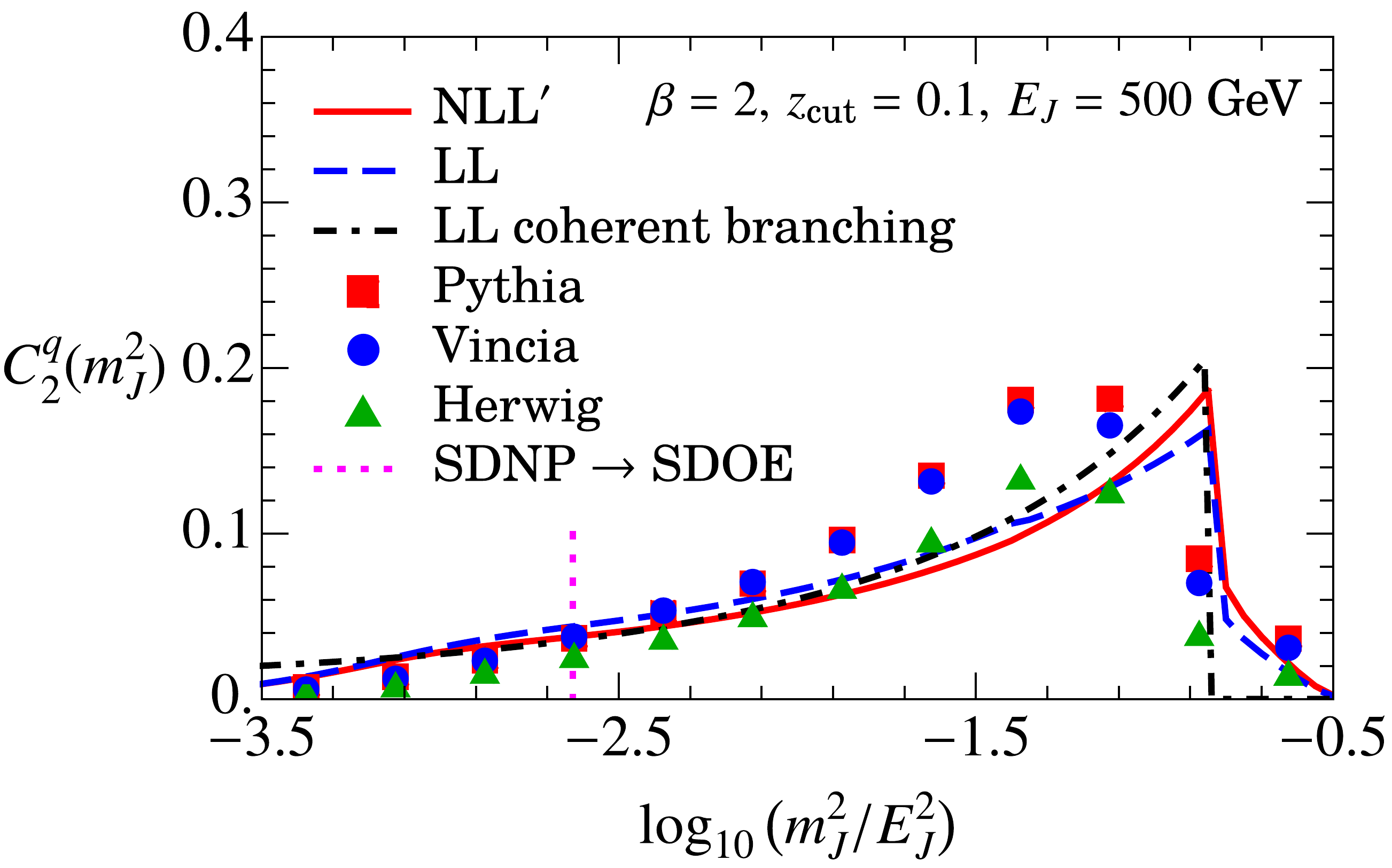}
\includegraphics[width=0.48\textwidth]{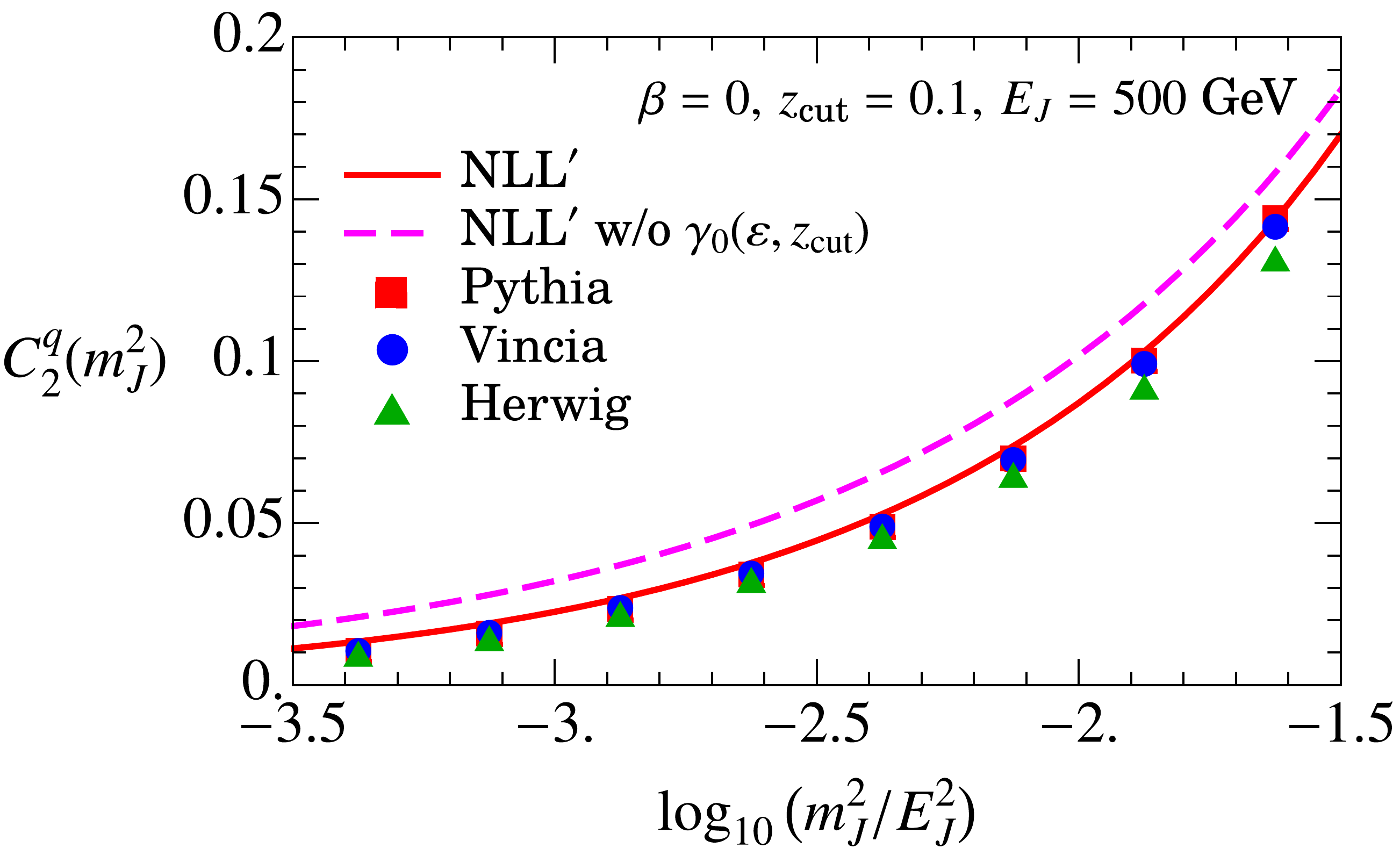}
\caption{Comparison between the NLL$'$/ LL effective theory predictions for the perturbative coefficient $C_2$ (red/blue dashed), the analogous predictions from coherent branching formalism (black dotdashed) and Monte Carlo simulations (colored markers). The last panel shows the impact of the additional anomalous dimension $\gamma_0(\veps,\zcut)$ present in the boundary distribution in the case $\beta=0$, where we zoom on the range of validity of our EFT.}
\label{fig:C2}
\end{figure}
Based on the definition in \eq{C1C2multiDiffDef}, we obtain numerical results for $C_2^q(m_J^2)$ by using only the large $R_g$ EFT, setting up the calculation at the differential level, weighting the large $R_g$ cross section according to \eq{C2Tx} and following the explicit construction given in \app{C2diff}. Note that since the large $R_g$ regime is free of NGLs in contributions modifying the spectrum, so are our predictions for $C^q_2$.
We consider the same kinematical values as for $C_1^q$, and in \fig{C2} plot the comparison between our LL and NLL$'$ EFT results, the LL coherent branching prediction, and MC data. For this observable the spread between different MC results is much larger, with \Herwig predictions being systematically lower than \Pythia and \Vincia. We see that our calculations are in good agreement with the MC in the region of validity of the EFT,
with a preference for the \Herwig results for larger $\beta$. Unlike the treatment in \cite{Hoang:2019ceu}, here we do not make the small jet radius approximation in the coherent branching LL predictions, i.e. we distinguish $\tan(R_g/2) \neq R_g/2$.
This ensures that the coherent branching and EFT calculations both have their transitions to the ungroomed region at the same value of the jet mass.
Although the impact of these corrections for $C_1^q$ were minimal, here they induce considerable corrections to $C_2^q$ at large mass, shifting the peak of the distribution towards larger values and improving the agreement with MC data. This happens because the boundary cross section, on which the calculation of $C_2^q$ is based, is highly peaked at larger $R_g$, where the $R_g/2\ll 1$ approximation is less justified.
We stress, however, that a more accurate description of this region would involve a
an extension of our EFT to account for power corrections ${\cal O}(m_J^2/(Q\qcut))$ that become relevant at the groomed-ungroomed transition point, $m_0^{(ee)}$ defined in \eq{m0}, which we do not attempt here.
We further notice in \fig{C2} that the LL coherent branching result is systematically larger than the LL-EFT result. Compared to NLL$'$, the LL coherent branching result is similar for $\beta=0$ but systematically larger for $\beta=1,2$.
As noted in \sec{LLCB}, our LL-EFT predictions include extra resummation in Laplace space, which is absent in the LL coherent branching results, while the latter include some additional power corrections through the splitting function $p_{gq}$. We find that at the large jet masses ($\log_{10}(m_J^2/E_J^2)\gtrsim -1.5$) the disagreement is almost entirely due to the power corrections, while at smaller jet masses the additional evolution in the EFT becomes more important.
Finally, in the bottom right plot we show the effect of including the extra anomalous dimension $\gamma_0(\veps,\zcut)$ from \eq{noncuspEps}, which is present only in the case $\beta=0$. The NLL$'$ result is shown before (dashed magenta) and after (red) including the fixed-order logarithms and additional running induced by $\gamma_0$. For more clarity, we zoom in on the range $[-3.5,-1.5]$ in the logarithmic jet mass variable, leaving out the ungroomed region. We observe that the $\gamma_0$ contribution is important across the whole plotted range,
and the improved agreement with MC predictions shows that the parton showers considered here likely do incorporate this NLL effect.

\begin{figure}[t!]
\centering
\includegraphics[width=0.48\textwidth]{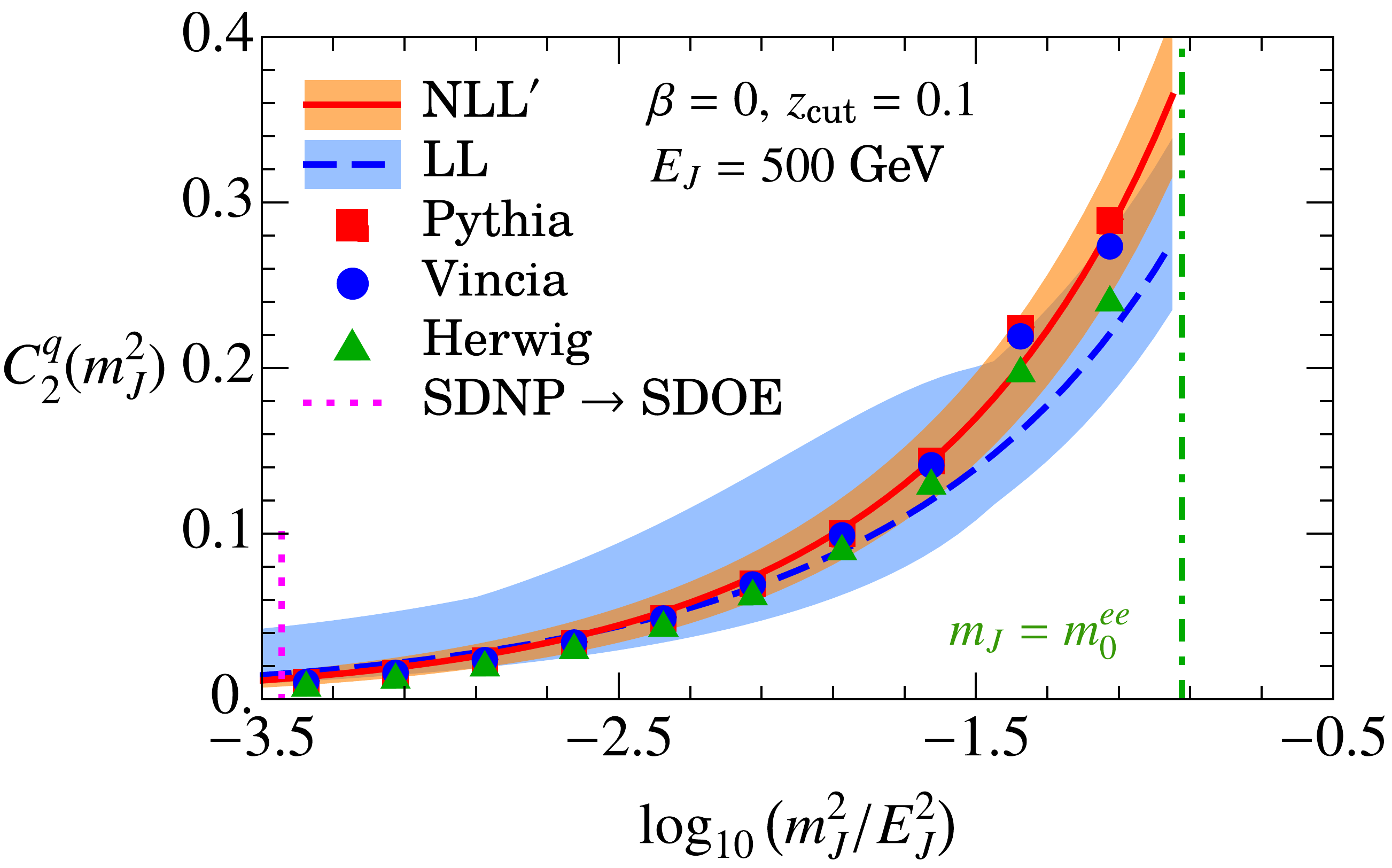}
\hspace{0.02\textwidth}
\includegraphics[width=0.48\textwidth]{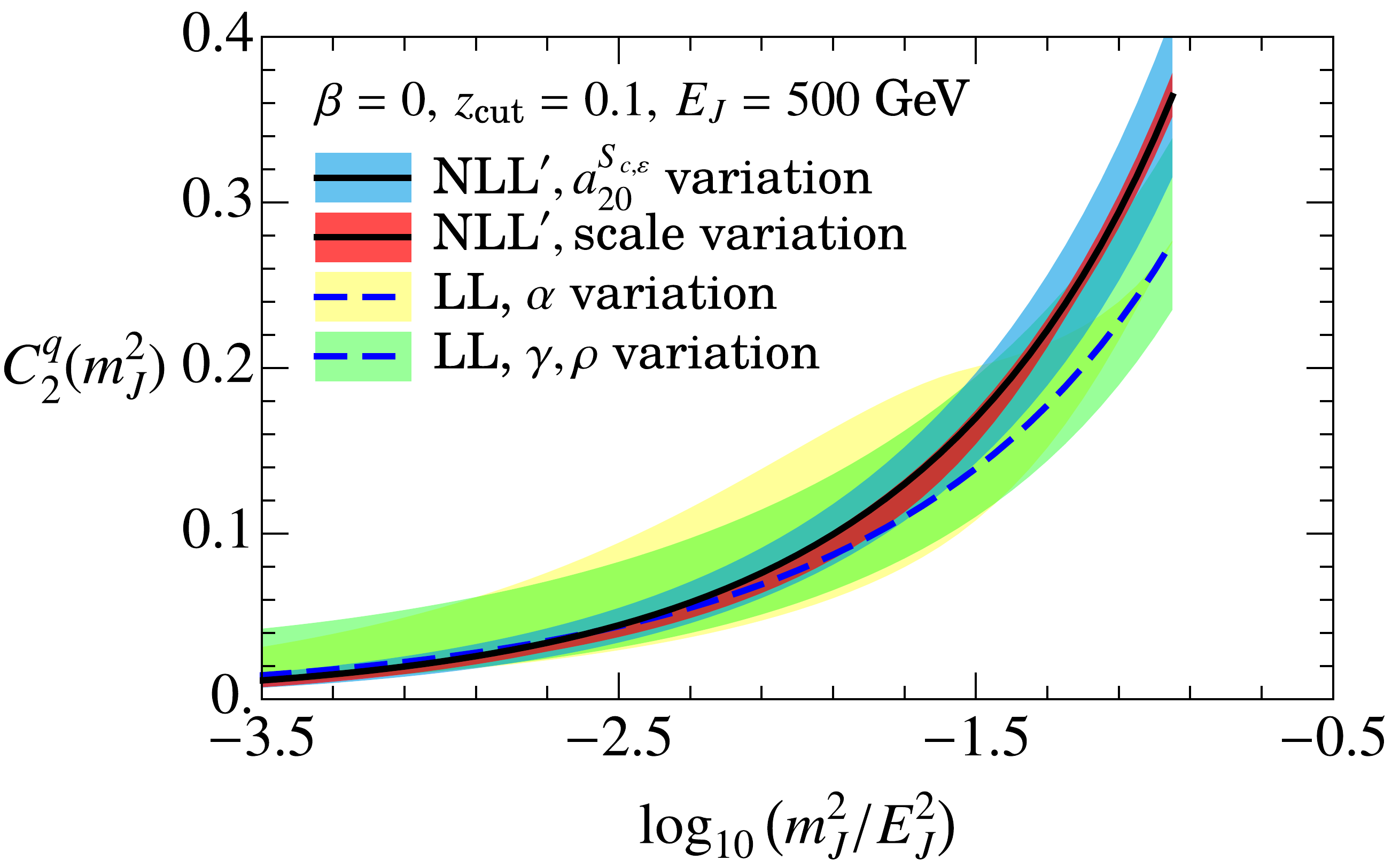}
\includegraphics[width=0.48\textwidth]{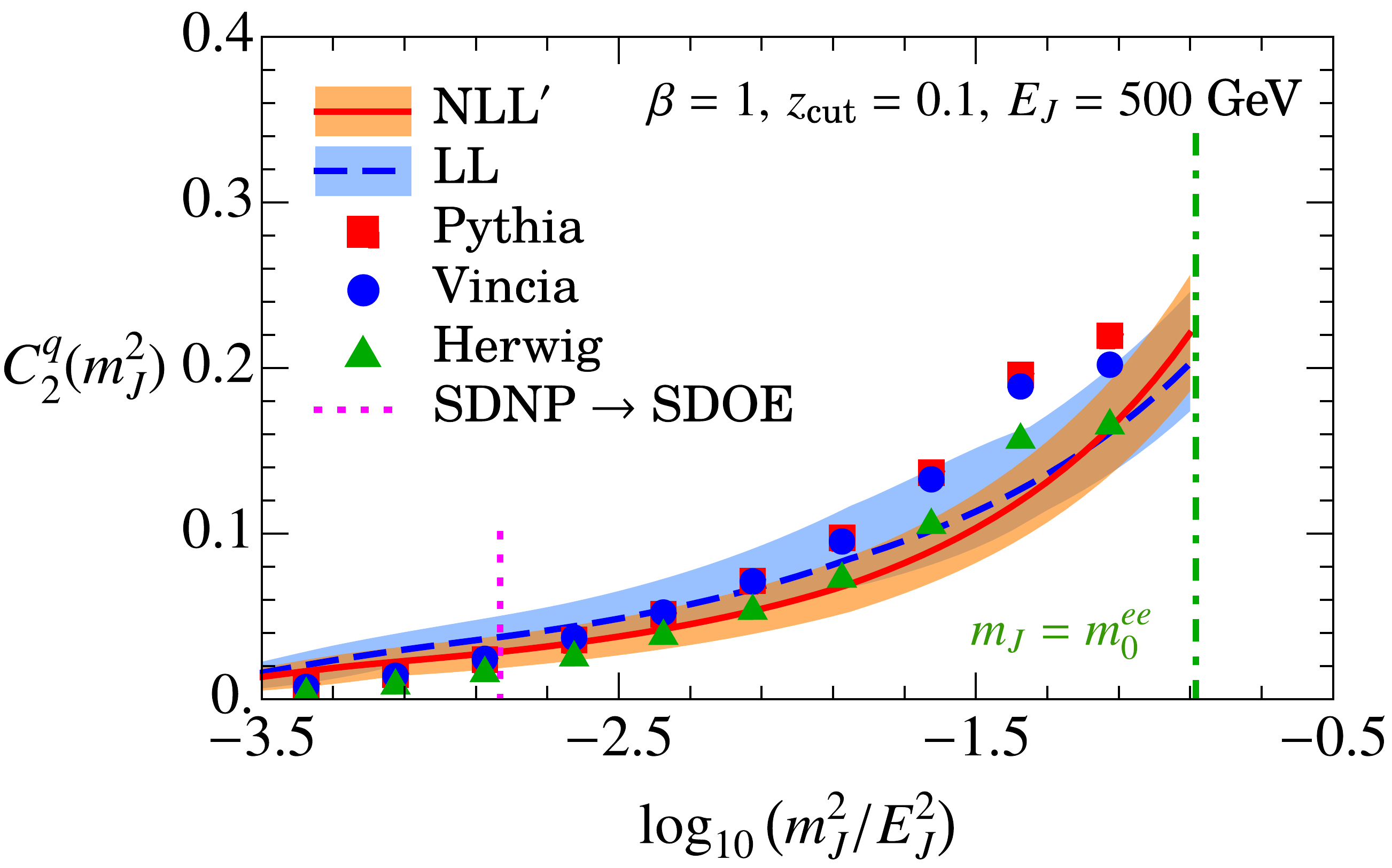}
\hspace{0.02\textwidth}
\includegraphics[width=0.48\textwidth]{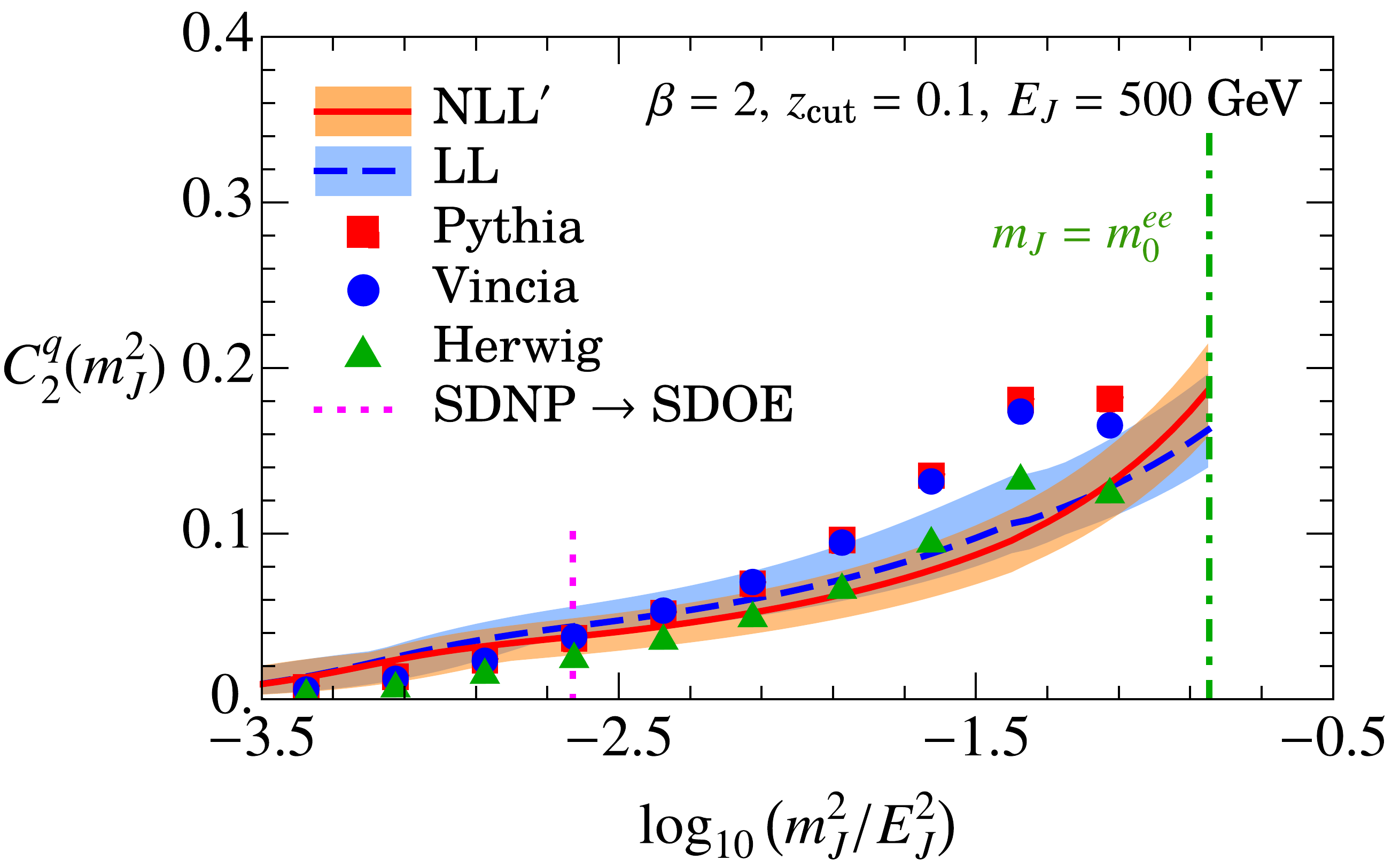}
\caption{Uncertainty estimate for the NLL$'$ (orange band) and LL (blue band) effective theory predictions for the perturbative coefficient $C_2^q$, at different values of $\beta$. The right top panel shows the breakdown of the LL envelope into the various sources of variation, showing that the large uncertainty is induced by the trumpet variation (yellow) rather than by breaking the canonical relations (green). The same panel shows that at NLL$'$ varying the two-loop constant (blue) dominates over the scale uncertainty (red).}
\label{fig:C2bands}
\end{figure}
We also show the LL and NLL$'$ EFT results with theoretical uncertainties in figure \fig{C2bands}, again restricting to the region $m_J<m_0^{ee}$ where grooming is active. The NLL$'$ uncertainties are estimated again by combining scale variation with variations of the two-loop non-logarithmic term via $a_{20}^{S_c,\veps}$ introduced in \eq{DeltaScLargeShift}. We see that the uncertainty bands shrink from LL to NLL$'$, but unlike $C_1^q$, their size is still significant and visible at NLL$'$. For this reason we do not show plots of relative variations here. For $\beta=0$ the NLL$'$ curves have excellent agreement with the MC data, while $\beta=1$ and $\beta=2$ show some tension at larger jet masses, especially with \Pythia and \Vincia. This could be due
to power corrections that we have not added to the EFT results, as inferred from the comparison with the LL coherent branching results for $\beta=1,2$ in \fig{C2}.
Note that since the $C_2^q$ calculation is carried out with the large $R_g$ cross section, that the NGL and clustering effects present for the corresponding double differential cross section only modify the normalization, and hence cancel out in the ratio of moments that determine $C_2^q$.

While the NLL$'$ uncertainty in \fig{C2bands} has roughly the same size for all $\beta$, the LL band becomes larger as $\beta$ decreases. To investigate this further, we show the breakdown of variations for $\beta=0$ in the top right panel of \fig{C2bands}. The LL uncertainty, entirely determined by scale variation, is once more decomposed into the variation of the collinear-soft scale by a trumpet factor (yellow) and the breaking of the canonical relations (green). We find that the overall variation of the hard and soft scales is irrelevant for $C_2^q$, and hence have not displayed it here. In the same plot, the NLL band is decomposed into scale variation (red) and variation of the two-loop constant $a_{20}^{S_c,\veps}$ (blue),
where we vary $a_{20}^{S_c,\veps} \in [-\pi,\pi]$.
We see that the variation of the two-loop constant dominates over scale variation across the whole range in the jet mass. This is because the denominator in \eq{C1C2multiDiffDef} does not involve $a_{20}^{S_c,\veps}$ (which parametrizes uncertainty in the boundary cross section),
and hence unlike other variations does not have a cancellation in the ratio of moments used to compute $C_2^q$.
We also checked that variations of the weight functions through the parameters in \eqs{IntValid}{Xtrans} have a small impact compared to the theoretical uncertainty. (Note that even though $C_2^q$ uses only the large $R_g$ regime, it still involves the weight function as in \eq{C2Tx}.)

\section{Conclusion}
\label{sec:conclusion}

In this paper we have set up an effective field theory description of processes involving jets which are groomed using soft drop, where the jet mass and groomed jet radius are simultaneously measured.
We presented factorization formulae formulated using SCET in order to resum large logarithms in the cross section which is differential in these two variables. We also combined the three EFT regimes that span the phase space of the double differential measurement to obtain a coherent description of the double differential cross section across the whole range of kinematics where grooming is active.
We performed a detailed numerical analysis of results for $\ee$ collisions in dijet limit.

As a first application of this formalism, we computed two angular moments of the double differential cross section: the average value of the groomed jet radius, $M_1^\kappa(m_J^2)$, and the average value of the inverse groomed jet radius for emissions at the soft drop boundary, $\Mb(m_J^2)$. Apart from being interesting observables in their own right, they are closely related to the Wilson coefficients $C_1^\kappa(m_J^2)$ and $C_2^\kappa(m_J^2)$ which describe the leading nonperturbative corrections to the soft drop jet mass cross section~\cite{Hoang:2019ceu}.
By making the correspondence $C_1^\kappa\simeq M_1^\kappa$ and $C_2^\kappa\simeq \Mb$, we use our NLL$'$ calculation of the moments to determine these Wilson coefficients with higher precision than previously available.
This leads in turn to a more accurate description of the leading nonperturbative corrections to the groomed jet mass.

We determined the regions of validity of three different EFT regimes based on their value of the groomed jet radius $R_g$, combined them using profile scales, and estimated the theoretical uncertainties on the double differential cross section and its angular moments via scale variation. The NLL$'$ improved computation involved, in certain regimes, fixed order ingredients at ${\cal O}(\alpha_s^2)$ accuracy. We parameterized unknown 2-loop terms that are not predicted by the renormalization group equations, and included them in our estimate of perturbative uncertainty.
We also compared our predictions with parton shower Monte Carlo simulations and found that the two agree within uncertainty in the region of validity of the formalism.
Although we did not resum the nonglobal logarithms in our factorization formulae, we showed that they cancel in $C_2^q$. We computed the leading effects on $C_1^q$ finding them to be not larger than other missing ${\cal O}(\alpha_s^2)$ effects included in our uncertainty estimate.

While we have focused on the application of simultaneous measurement of the groomed jet radius and the jet mass, we expect the EFTs explored in this work to have a much wider applicability, and our work to pave the way to analogous computations for other groomed jet observables. Single differential jet based observables, such as the jet mass and jet angularities, have exciting prospects for measurements of standard model parameters such as the strong coupling constant $\alpha_s$ and the top mass $m_t$, with grooming playing a key role in reducing contamination in jets. Having a double differential cross section prediction at hand offers complementary ways of extracting these parameters.

The double differential cross section includes a resummation of large logarithms of the groomed jet radius, and this resummation remains relevant even when $m_J$ is taken to be in the ungroomed region.
The study of power corrections for the multi-differential cross section in this region are also rather interesting. In the future work, we envision extending our double differential predictions to the ungroomed region, to achieve a more complete description of the double differential cross section.

One of the greatest advantages of jet grooming is the decrease to the impact of nonperturbative corrections compared to those in ungroomed jets. Grooming also significantly reduces contamination from multi-parton interactions, and is thus an invaluable tool for carrying out phenomenology with jets, and for analyzing heavy ion collisions.

In hadronic collisions, measuring observables on groomed jets also allows for much cleaner access to the initial state hadron structure, which is one of the principal goals of nuclear physics. By bringing under control the small nonperturbative effects from final state radiation, we are one step closer to the accurate extraction of hadron structure functions such as the transverse momentum dependent parton distributions.
Additional natural areas for future explorations of multi-differential groomed jets include for transverse momentum observables and for the phenomenology of jets in heavy ion collisions.

\begin{acknowledgments}
This work was supported in part by FWF Austrian Science Fund under the Project No. P28535-N27 and by the Office of Nuclear Physics of the U.S. Department of Energy under the Grant No. DE-SCD011090.
A.P. is a member of the Lancaster-Manchester-Sheffield Consortium for Fundamental Physics, which is supported by the UK Science and Technology Facilities Council (STFC) under grant number ST/T001038/1.
I.S. was also supported by the Simons Foundation through the Investigator grant 327942. L.Z. was supported in part by the ERC grant ERC-STG-2015-677323 and in part by the Deutsche Forschungsgemeinschaft (DFG, German Research Foundation) -- Research Unit FOR 2926, grant number 409651613.
A.P. also thanks FWF Austrian Science Fund under the Doctoral Program Particles and Interactions No.W1252-N27 for partial support. L.Z. would like to thank the Center for Theoretical Physics of MIT for hospitality during the early stage of this project. Figures are made using the \Mathematica package \PLHot~\cite{plhot}.
\end{acknowledgments}

\appendix
\section{Formulae}
\label{app:formulae}

\subsection{Laplace transforms and RG Evolution}
\label{app:transforms}

Here we present the details of the resummation of large logarithms via RG evolution. The earliest papers exploring evolution equations, with the type of momentum space convolutions encountered here, were \Refs{Korchemsky:1993uz,Balzereit:1998yf}, and further advances to the approach were made in \Ref{Neubert:2005nt} which presented formulas valid to all orders in perturbation theory. These rely on the equations being multiplicative in Laplace or Fourier space. Solutions to these equations can also be obtained functionally to all orders, both directly in momentum space by formulas combining plus functions~\cite{Fleming:2007xt,Ligeti:2008ac}, or with derivatives to compactly transform logarithms from Laplace to momentum space~\cite{Becher:2006nr,Becher:2006mr}.
\Ref{Almeida_2014} provides a review and we follow the notation used there.

We consider a generic $\overline {\rm MS}$ renormalized SCET matrix element ${\cal F}(q,\mu)$, with $j_{\cal F}$ the mass dimensions of $q$.
When ${\cal F}(q, \mu)$ enters factorization theorems in a Laplace convolution over the momentum space variable $q$, its renormalization group equation (RGE) reads
\begin{align}
\label{eq:momRGE}
\mu\frac{d}{d\mu} {\cal F}(q, \mu) = \int d q'\, \gamma_{\cal F}(q - q', \mu)\, {\cal F}(q, \mu) \, ,
\end{align}
where $\gamma_{\cal F}$ is the anomalous dimension in momentum space. In SCET such functions are ubiquitous and result from the fact that UV renormalization of soft and collinear matrix elements accounts for double poles in the dimensional regulator, that are in turn reminiscent of the double IR poles of the corresponding QCD matrix elements. The presence of double poles in counterterms, and the convolution nature of the factorization theorem, lead to nontrivial running such as the one shown in \eq{momRGE}. To solve of \eq{momRGE}, we consider the Laplace transform of ${\cal F}$ given by
\begin{align}
\widetilde {\cal F}(x,\mu) \equiv \int_{0}^{\infty} dq \, e^{- q x} {\cal F}(q, \mu) \, ,
\end{align}
such that $x$ has mass dimensions $-j_{\cal F}$. The formula for inverse Laplace transform reads
\begin{align}
{\cal F}(q,\mu) = \int_{c-i\infty}^{c+i\infty} \frac{dx}{2 \pi i}\, e^{qx} \widetilde{\cal F} (x,\mu) \, ,
\end{align}
where the real number $c$ determines a vertical contour in the complex plane that leaves any pole of $\widetilde{\cal F}$ to its left.
In position space, the RGE is simply multiplicative:
\begin{align} \label{eq:posRGE}
&\mu \frac{d}{d\mu} \widetilde{\cal F} (x,\mu) =
\widetilde \gamma_{\cal F}(x, \mu)\, \widetilde{{\cal F}}(x, \mu) \equiv \Big[\Gamma_{\cal F}[\alpha_s]\, \ln (e^{\gamma_E} x \mu^{j_{\cal F}}) + \gamma_{\cal F}	[\alpha_s] \Big]\, \widetilde{{\cal F}}(x, \mu)\,,
\end{align}
where the anomalous dimension\footnote{Had there been higher powers of the logarithm in \eq{posRGE} the factorization into soft and collinear matrix elements would not have been possible. This has been shown to be true explicitly up to four loops~\cite{Becher:2019avh}.} $\widetilde\gamma_{\cal F}(x, \mu) $ consists of a cusp piece $\Gamma_{\cal F}[\alpha_s]$, responsible for resumming double logarithms, and a non-cusp part $\gamma_{\cal F}[\alpha_s]$. The cusp part, $\Gamma_{\cal F}[\alpha_s]$, is proportional to the cusp-anomalous dimension such that $\Gamma_{\cal F}[\alpha_s] = \kappa_{\cal F} C_\kappa \Gamma^{\rm cusp}[\alpha_s]$, where $C_\kappa = C_F$ or $C_A$.
Equation \eqref{eq:posRGE} has a formal solution
\begin{align}
&{\cal F}(q, \mu) = \int\! d q^\prime\, U_{\cal F}(q - q^\prime, \mu, \mu_0) \, {\cal F}(q^\prime, \mu_0) = \int_{c-i\infty}^{c+i\infty}\! \frac{dx}{2 \pi i} \,e^{q x}\,\widetilde U_{\cal F}(x, \mu, \mu_0) \, \widetilde{\cal F}(x, \mu_0) \, ,
\end{align}
in terms of the evolution operator
\begin{align}\label{eq:UCalF}
\widetilde{U}_{\cal F}(x, \mu, \mu_0) = e^{K_{\cal F}(\mu, \mu_0)} \, (e^{\gamma_E} x \mu_0^{j_{\cal{F}}} )^{\omega_{\cal F}(\mu, \mu_0)} \, .
\end{align}
The evolution kernels $K_{\cal F}$ and $\omega_{\cal F}$ are obtained from integrals of the anomalous dimensions,
\begin{align}
\label{eq:evolKernels}
K_{\cal F}(\mu, \mu_0)
&= K^{\Gamma}_{\cal F}(\mu, \mu_0) + K^{\gamma}_{\cal F}(\mu, \mu_0) \\
K^{\Gamma}_{\cal F} (\mu, \mu_0)
&=j_{\cal{F}} \int_{\alpha_s(\mu_0)}^{\alpha_s(\mu)} \frac{d\alpha}{\beta [\alpha]} \Gamma_{\cal F}[\alpha] \int_{\alpha_s(\mu_0)}^{\alpha} \frac{d\alpha^\prime}{\beta [\alpha^\prime]} \nn \\
K^{\gamma}_{\cal F} (\mu, \mu_0)
&= \int_{\alpha_s(\mu_0)}^{\alpha_s(\mu)} \frac{d\alpha}{\beta [\alpha]} \gamma_{\cal F}[\alpha] \, , \nn\\
\omega_{\cal F}(\mu, \mu_0) &= \int_{\alpha_s(\mu_0)}^{\alpha_s(\mu)} \frac{d\alpha}{\beta [\alpha]} \Gamma_{\cal F}[\alpha] \,, \nn
\end{align}
involving the QCD beta function $\beta[\alpha_s]$, whose perturbative expansion we give in \eq{GammaBetaExpansion}.
We also state the results for functions $F(\mu, \mu_F)$ that enter the factorized cross section as multiplicative factors (and not through convolutions) with $\mu_F$ being the energy scale specific to $F$. Here the RGE is directly given by
\begin{align}
\label{eq:normRGE}
&\mu \frac{\df}{\df\mu} F(\mu,\mu_F) =
\gamma_F(\mu,\mu_F) F(\mu,\mu_F) =
\Big[\Gamma_F[\alpha_s] \log\Big(\frac{\mu}{\mu_F}\Big) + \gamma_F(\alpha_s) \Big] F(\mu, \mu_0) \, ,
\end{align}
with the solution for evolution from scale $\mu_0$ to $\mu$ being
\begin{align}\label{eq:UF}
F(\mu,\mu_F) &= U_F(\mu,\mu_0; \mu_F) F(\mu_0, \mu_F) \equiv e^{K_F(\mu, \mu_0) } \, \Big(\frac{\mu_0}{\mu_F}\Big)^{\omega_F(\mu,\mu_F)} F(\mu_0, \mu_F) \, ,
\end{align}
where the evolution kernels $K_F$ and $\omega_F$ are again given by \eq{evolKernels}. The kernels in \eq{evolKernels} depend on the specific function $\cal{F}$ (or $F$) only through the anomalous dimensions: we can thus rewrite them in terms of two universal kernels
\begin{align}\label{eq:KOmega}
K^{\Gamma}_{\cal F}(\mu,\mu_0) =j_{\cal{F}} K(\Gamma_{\cal F}[\alpha_s], \mu, \mu_0) \, , \quad K^{\gamma}_{\cal F}(\mu,\mu_0) = \eta(\gamma_{\cal F}[\alpha_s], \mu, \mu_0) \, , \quad \omega_{\cal F}(\mu,\mu_0) = \eta(\Gamma_{\cal F}[\alpha_s], \mu, \mu_0) \, ,
\end{align}
whose explicit NLL expressions are
\begin{align}
\label{eq:w}
\eta(\Gamma,\mu,\mu_0) &=
-\frac{\Gamma_0}{2\beta_0}
\bigg[\ln r+\frac{\alpha_s(\mu_0)}{4\pi}\bigg(\frac{\Gamma_1}{\Gamma_0}-\frac{\beta_1}{\beta_0}\bigg)(r-1) \bigg]\,,\\
\label{eq:K}
K(\Gamma,\mu,\mu_0) &=
\frac{\Gamma_0}{4\beta_0^2}\Bigg[\frac{4\pi}{\alpha_s(\mu_0)} \Big(\ln r+\frac{1}{r}-1\Big)
+ \bigg(\frac{\Gamma_1}{\Gamma_0}-\frac{\beta_1}{\beta_0}\bigg)(r-1-\ln r)
-\frac{\beta_1}{2\beta_0}\ln^2 r \Bigg]\,,
\end{align}
where $r=\alpha_s(\mu)/\alpha_s(\mu_0)$. Here the perturbative expansion of a generic anomalous dimension $\Gamma[\alpha_s]$ and $\beta[\alpha_s]$ is written as
\begin{align} \label{eq:GammaBetaExpansion}
& \Gamma[\alpha_s]=\sum_{n=0}^{\infty}\Gamma_n\Big(\frac{\alpha_s}{4\pi}\Big)^{n+1}
\,,\qquad
\beta[\alpha_s]=-2\,\alpha_s\,\sum_{n=0}^{\infty}\beta_n\Big(\frac{\alpha_s}{4\pi}\Big)^{n+1}
\,
\end{align}
and the terms relevant for NLL resummation read
\begin{align}
\Gamma_0^{\rm cusp} &= 4 \, , \qquad
\Gamma^{\rm cusp}_1
=4 C_A \bigg(\frac{67}{9} - \frac{\pi^2}{3}\bigg) - \frac{80 n_fT_R}{9}
\, ,
\end{align}
\begin{align}
\beta_0 &= \frac{11}{3}C_A - \frac{4}{3}T_R n_f
\, , \qquad
\beta_1 = \frac{34}{3} C_A^2 - 4 T_R n_f \bigg (C_F + \frac{5}{3}C_A \bigg)
\, .
\end{align}

\subsection{One-loop anomalous dimensions}
\label{app:anomDim}

We state here one-loop results for the anomalous dimensions of the various functions appearing in \eq{xsecRG} with the cusp and noncusp pieces defined as in \eqs{posRGE}{normRGE}. The RGEs of these functions are given by
\begin{align}
\label{eq:RGEall}
\mu\frac{d}{d \mu}\log N^\kappa(Q, R, \mu)
&= \Gamma_{N^\kappa}[\alpha_s] \log\Big(\frac{\mu}{Q\tan\frac{R}{2}}\Big) + \gamma_{N^\kappa}[\alpha_s] \, ,
\\
\mu\frac{d}{d\mu} \log {\cal C}^\kappa\bigg (\frac{m_J^2}{Q^2R_g^2}, QR_g,\mu\bigg )
&= \Gamma_{{\cal C}^\kappa}[\alpha_s] \log\Big(\frac{\mu}{Q\frac{R_g}{2}}\Big) + \gamma_{{\cal C}^\kappa}[\alpha_s] \, ,
\nn
\\
\mu\frac{d}{d \mu}\log \tilde J_\kappa(x, \mu)
&= \Gamma_{J_\kappa}[\alpha_s] \log\big(e^{\gamma_E} x\mu^2\big) + \gamma_{J_\kappa}[\alpha_s] \, ,
\nn\\
\mu\frac{d}{d \mu}\log \tilde S^\kappa_{c_m}(y, \mu)
&= \Gamma_{S^\kappa_{c_m}}[\alpha_s] \log\big(e^{\gamma_E}y \mu\big) + \gamma_{S^\kappa_{c_m}}[\alpha_s] \, ,\nn
\nn \\
\mu\frac{d}{d \mu}\log \tilde S^\kappa_{c}(z, \beta, \mu)
&= \Gamma_{S^\kappa_{c}}[\alpha_s] \log\big(e^{\gamma_E} z \mu^{\frac{2+\beta}{1+\beta}}\big) + \gamma_{S^\kappa_{c}}(\beta)[\alpha_s] \, ,	\nn \\
\mu\frac{d}{d \mu}\log S_G^\kappa\Big(\qcut\tan^{1+\beta}\frac{R}{2},R, \beta, \mu\Big)
&=\frac{1}{1 + \beta} \Gamma_{S^\kappa_G}[\alpha_s] \log\Big(\frac{\mu}{\qcut\tan^{1+\beta}\frac{R}{2}}\Big) + \gamma_{S^\kappa_G}(\beta)[\alpha_s] \, ,
\nn \\
\mu\frac{d}{d \mu}\log S^\kappa_{c_g}\Big (\frac{R_g}{2}\qcut^{\frac{1}{1 + \beta}} , \beta, \mu\Big)
&= \frac{1}{1 + \beta}\Gamma_{S^\kappa_{c_g}}[\alpha_s] \log\bigg [
\bigg(\frac{\mu}{\qcut\big(\frac{R_g}{2}\big)^{1+\beta}}\bigg) \bigg]
+ \gamma_{S^\kappa_{c_g}}(\beta)[\alpha_s] \, .\nn
\end{align}
and the inverse Laplace transforms for $J_\kappa$, $S_{c_m}^\kappa$ and $S_{c}^\kappa$ are defined as follows,
\begin{align}\label{eq:LapDefAll}
J_\kappa(m_J^2, \mu) &= \int_{c-i\infty}^{c+i\infty} \frac{dx}{2\pi i} \: e^{ x\,m_J^2} \tilde J_\kappa(x,\mu) \, ,
\\
S^\kappa_{c_m}\Big(\frac{\ell^+}{R_g/2}, \mu\Big) &= \int_{c-i\infty}^{c+i\infty} \frac{dy}{2\pi i} \: e^{y\,\frac{\ell^+}{R_g/2}} \:
\tilde S_{c_m}^\kappa(y, \mu) \, ,
\nn \\
S_c^\kappa\big(\ell^+\qcut^{\frac{1}{1+\beta}}, \mu\big) &= \int_{c-i\infty}^{c+i\infty} \frac{dz}{2\pi i}\: e^{z\,\ell^+\qcut^{\frac{1}{1+\beta}}} \:
\tilde S_c^\kappa(z, \mu) \, ,\nn
\end{align}
such that the mass dimensions of $x$, $y$, and $z$ are respectively
\begin{align}
-j_{J_\kappa} = -2\,,\qquad
-j_{S^\kappa_{c_m}} = -1\,,\qquad
-j_{S_c^\kappa} = -(2+\beta)/(1+\beta)\,.
\end{align}
The cusp pieces in \eq{RGEall} are multiples of the universal cusp anomalous dimension $\Gamma^{\rm cusp}[\alpha_s]$ and are given by
\begin{align}
\Gamma_{N^\kappa} [\alpha_s]&
= -\Gamma_{S_G^\kappa}[\alpha_s]
= -\Gamma_{J_\kappa}[\alpha_s]
= -\Gamma_{{\cal C}^\kappa}[\alpha_s]
=\Gamma_{S^\kappa_{c_m}}[\alpha_s]
= \Gamma_{S_{c}^\kappa}[\alpha_s]
= \Gamma_{S_{c_g}^\kappa}[\alpha_s]
=-2 C_\kappa \,\Gamma^{\rm cusp}[\alpha_s] \,,
\end{align}
and the one-loop noncusp pieces are
\begin{align}
\label{eq:noncuspAll}
\gamma^{N^q}_0&= -\gamma^{J_q}_0=-\gamma^{{\cal C}^q}_0= -6 C_F \,,
\quad
\gamma^{N^g}_0= -\gamma^{J_g}_0 = -\gamma^{{\cal C}^g}_0 = -2\beta_0 \,,
\quad
\gamma^{S^\kappa_{c_m}}_0 = \gamma^{S_{c}^\kappa}_0 = \gamma^{S_G^\kappa}_0 = \gamma^{S_{c_g}^\kappa}_0 = 0\, .
\end{align}

We can now check the RG consistency relation for the factorized cross section for doubly differential $m_J^2$ and $R_g$ measurements.
Written in position space, the dijet cross section for jet masses is given in the intermediate regime by \eq{LaplaceDbleDiff}. By differentiating with respect to the renormalization scale, it is immediate to show that the sum of the cusp terms must vanish,
\begin{align}
&\Gamma_{N^\kappa} [\alpha_s] \log\Big(\frac{\mu}{Q\tan\frac{R}{2}}\Big)
+ \frac{1}{2}\frac{1}{1+\beta}\Gamma_{S^\kappa_G}[\alpha_s] \log\Big(\frac{\mu}{\qcut\tan^{1+\beta}\frac{R}{2}}\Big)
+ \Gamma_{J_\kappa}[\alpha_s] \log\big(e^{\gamma_E} x\mu^2\big) \\
&\qquad + \Gamma_{S^\kappa_{c_g}}[\alpha_s]
\log\bigg [
\Big(\frac{\mu}{\qcut}\Big)^{\frac{1}{1 + \beta}} \frac{2}{R_g}
\bigg]
+ \Gamma_{S^\kappa_{c_m}}[\alpha_s] \log\Big(e^{\gamma_E}x \frac{QR_{g_L}}{2}\mu\Big) = 0 \, , \nn
\end{align}
and that the noncusp pieces must also add to zero from RG consistency,
\begin{align}
\gamma_{N^\kappa}[\alpha_s] + \gamma_{J_\kappa}[\alpha_s] + \gamma_{S^\kappa_G}[\alpha_s] + \gamma_{S^\kappa_{c_m}}[\alpha_s] + \gamma_{S^\kappa_{c_g}}[\alpha_s] = 0 \, ,
\end{align}
as can be easily verified at one loop from \eq{noncuspAll}. This puts nontrivial constraint on the $\beta$ dependence of the global-soft and collinear-soft non cusp anomalous dimensions. As a consequence of the refactorization in \eq{CSrefactorization} in the regime of large groomed jet radius, the relation
\begin{align}
\Gamma_{S^\kappa_{c_g}}[\alpha_s]
\log\bigg (
\Big(\frac{\mu}{\qcut}\Big)^{\frac{1}{1 + \beta}} \frac{2}{R_g}
\bigg)
+ \Gamma_{S^\kappa_{c_m}}[\alpha_s] \log\Big(e^{\gamma_E}x \frac{QR_{g}}{2}\mu\Big) =
\Gamma_{S^\kappa_{c}}[\alpha_s] \log\big(e^{\gamma_E} x Q \qcut^{\frac{-1}{1+\beta}} \mu^{\frac{2+\beta}{1+\beta}}\big)
\end{align}
is also enforced by RG consistency.

\section{One-loop results}
\label{app:Oneloop}

In this appendix we collect the one-loop results for the ingredients appearing in the double differential cross section. We start from the functions that enter the factorized expression for the cross section cumulant in $R_g$ and describe their evolution, then we move on to describing the ingredients of the boundary cross section that enters the definition of $C_2^q$.

\subsection{Results for the cumulant distribution}
\label{app:LargeRg}
\tocless\subsubsection{Global-soft function}
The global soft function accounts for the soft radiation which is groomed away, namely lies within the original jet but fails the soft drop test. The one loop correction for $e^+e^-$ collisions requires computing
\begin{align}\label{eq:SGBare}
& S_G^{\kappa[1], \rm bare} \bigl( \qcut,R, \beta, \mu\bigr) =
\frac{\alpha_s C_\kappa}{\pi}
\frac{(\mu^2 e^{\gamma_E})^{\eps}}{\Gamma(1-\eps)}
\int \frac{dp^+ dp^-}{(p^+p^-)^{1+\eps}} \,\Theta_{\rm sd} \,\overline \Theta_R\,,
\end{align}
where $C_q=C_F$ and $C_g=C_A$, the condition for failing soft drop is
\begin{align}\label{eq:ThetaLarge}
\Theta_{\rm sd}
&\equiv
\Theta
\bigg(
\zcut \Big(2\sin \frac{\theta_p}{2}\Big)^\beta
-
\frac{p^++p^-}{Q}
\bigg)\,,
\end{align}
with $\sin^2(\theta_p/2) = p^+/(p^+ + p^-)$, and $\overline \Theta_R$ constrains radiation within a jet of radius $R$,
\begin{align}\label{eq:ThetaR}
\overline \Theta_R \equiv \Theta \bigg(\tan \frac{R}{2} - \sqrt{\frac{p^+}{p^-}} \bigg)\,.
\end{align}
Solving the integration in \eq{SGBare} we get
\begin{align}\label{eq:SG1loop}
S_G^\kappa\Big(\qcut\tan^{1+\beta}\frac{R}{2},R, \beta, \mu\Big) =
1 + \frac{\alpha_s(\mu)C_\kappa}{\pi}
&\bigg[
\frac{1}{(1+\beta)} \log^2\Big(\frac{\mu}{\qcut\tan^{1+\beta}\frac{R}{2}}\Big)
-\frac{\pi^2}{24}
\Big(
\frac{1}{1+\beta}
\Big)
\\
&
-\frac{(2+\beta)}{4}
\Big(
2{\rm Li}_2\Big[\sin^2\Big(\frac{R}{2}\Big) \Big]
+ \log^2\Big [\cos^2\frac{R}{2}\Big]
\Big)
\bigg]
\nn
\, .
\end{align}
For $pp$ collisions with small jet radius, $R \ll 1$, one retrieves the familiar expression
\begin{align}\label{eq:SG1loopExpand}
S_G^\kappa\bigl( p_T \zcut' R^{1+\beta}, \beta, \mu\bigr) =
1 + \frac{\alpha_s(\mu)C_\kappa}{\pi} \frac{1}{1+\beta}
&\bigg[ \log^2\Big(\frac{\mu}{p_T \zcut' R^{1+\beta}}\Big)
-\frac{\pi^2}{24}\bigg]
\, .
\end{align}

\tocless\subsubsection{Jet function}

The one-loop result for the jet function is well known~\cite{Bauer:2003pi,Bosch:2004th} and given by
\begin{align}\label{eq:JMomDef}
J_\kappa (m_J^2, \mu) \equiv \delta(m_J^2) + \frac{\alpha_s(\mu)}{\pi} \Bigg[
\frac{C_\kappa}{\mu^2}{\cal L}_1\Big(\frac{m_J^2}{\mu^2}\Big) - \frac{\gamma_0^{J_\kappa} }{8} \frac{1}{\mu^2}{\cal L}_0\Big(\frac{m_J^2}{\mu^2}\Big)
-\delta(m_J^2)\Big(C_\kappa \frac{\pi^2}{4}- c_1^\kappa \Big)
\Bigg] \, .
\end{align}
The Laplace transform is defined as (see \eq{LapDefAll})
\begin{align}\label{eq:JLapTrans}
\tilde J_\kappa(x,\mu) &= \int_0^\infty d m_J^2 \: e^{ - x\,m_J^2} 	J_\kappa(m_J^2, \mu) \, ,
\end{align}
such that \eq{JMomDef} gives
\begin{align}\label{eq:JLapDef}
\tilde J_\kappa \bigl (x, \mu \bigr )
&\equiv
1 + \frac{\alpha_s(\mu)}{\pi} \bigg[
\frac{C_\kappa}{2} \log^2\big(e^{\gamma_E}x \mu^2 \big) + \frac{\gamma_0^{J_\kappa} }{8} \log \big(e^{\gamma_E} x \mu^2 \big)
- C_\kappa \frac{\pi^2}{6}
+ c_1^\kappa
\bigg]
\, ,
\end{align}
where the one-loop non-cusp anomalous dimension for the quark and gluon jet functions are
\begin{align}
\gamma_0^{J_q} &= 6 C_F \, ,\qquad
\gamma_0^{J_g} = 2 \beta_0 \, ,
\end{align}
the one-loop constant terms being
\begin{align}
c_1^q = \frac{7}{4} \, , \qquad
c_1^g = \frac{\Gamma^{\rm cusp}_1}{8}
= \frac{C_A}{2} \bigg(\frac{67}{9} - \frac{\pi^2}{3}\bigg) - \frac{10 n_fT_R}{9}
\, .
\end{align}
Equivalently, following to the alternative notation in \eq{altNotation}, we write \eq{JLapDef} as
\begin{align}\label{eq:altNotationJet}
\tilde J_\kappa \big[\ell_J, \alpha_s(\mu_J)\big]
&=
1 + \frac{\alpha_s(\mu_J)}{\pi} \bigg[
\frac{C_\kappa}{2} \ell_J^2 + \frac{\gamma_0^{J_\kappa} }{8} \ell_J- C_\kappa \frac{\pi^2}{6} + c_1^\kappa
\bigg]
\, .
\end{align}

\tocless\subsubsection{Collinear-soft functions at intermediate $R_g$}

\noindent
The \CSg function describes emissions that live at the groomed jet boundary and pass soft drop. At one loop, it requires computing
\begin{align}\label{eq:SCgBare}
& S_{c_g}^{\kappa[1], \rm bare} \Big(\frac{R_g}{2}\qcut^{\frac{1}{1+\beta}},\beta, \mu\Big)=
\frac{\alpha_s C_\kappa}{\pi}
\frac{(\mu^2 e^{\gamma_E})^{\eps}}{\Gamma(1-\eps)}
\int \frac{dp^+ dp^-}{(p^+p^-)^{1+\eps}} \,\overline \, \Theta\bigg(\frac{R_g}{2}-\sqrt{\frac{p_+}{p_-}}\bigg) \overline \Theta_{\rm sd}\,,
\end{align}
with
\begin{align}\label{eq:ThetaLargeBar}
\overline \Theta_{\rm sd} \,=\,
1 - \Theta_{\rm sd} \, = \,
\Theta
\bigg(
\frac{p^-}{Q}-\zcut\Big(\frac{p^+}{p^-}\Big)^{\beta/2}
\bigg)\,,
\end{align}
where the last equality uses that for collinear-soft radiation $p^+ \ll p^-$. This leads to the $e^+e^-$ equivalent of the one-loop result of~\cite{Kang:2019prh},
\begin{align}
S_{c_g}^{\kappa}\Big(\frac{R_g}{2}\qcut^{\frac{1}{1+\beta}},\beta, \mu\Big) &=
1 - \frac{\alpha_s C_\kappa}{ \pi}\frac{1}{1+\beta} \bigg[ \log^2 \bigg(\frac{\mu}{\qcut (R_g/2)^{(1+\beta)}}\bigg) - \frac{\pi^2}{24}\bigg] \, ,
\end{align}

The \CSm function describes emissions at the groomed jet boundary that set the jet mass (therefore, they automatically pass soft drop in this regime). From the bare one-loop expression
\begin{align}\label{eq:SCmBare}
& S_{c_m}^{\kappa[1], \rm bare} \Big (\frac{R_g}{2}\qcut^{\frac{1}{1+\beta}},\beta, \mu\Big) =
\frac{\alpha_s C_\kappa}{\pi}
\frac{(\mu^2 e^{\gamma_E})^{\eps}}{\Gamma(1-\eps)}
\int \frac{dp^+ dp^-}{(p^+p^-)^{1+\eps}} \,\overline \, \Theta\bigg(\frac{R_g}{2}-\sqrt{\frac{p^+}{p^-}}\bigg) \delta\Big(\frac{\ell^+}{p^+}p^- - 1\Big)\,,
\end{align}
we get
\begin{align}
S_{c_m}^\kappa \Big(\frac{2\ell^+}{R_g}, \mu\Big) &=
\delta\Big(\frac{2\ell^+}{R_g}\Big) + \frac{\alpha_s C_\kappa}{\pi} \bigg[\frac{-2}{\mu} {\cal L}_1\bigg(\frac{1}{\mu}\frac{2\ell^+}{R_g}\bigg) + \frac{\pi^2}{24} \delta \Big(\frac{2\ell^+}{R_g}\Big)\bigg] \,,
\end{align}
which is the equivalent to the result for the soft function for ungroomed jets of radius $R \ra R_g$ computed in \Ref{Ellis:2010rwa} (up to a different choice in normalization). Following \eq{LapDefAll} the Laplace transform of $S_{c_m}$ is defined as
\begin{align}\label{eq:ScmLapTrans}
&\tilde S_{c_m}^\kappa \bigl( u, \mu\bigr)
=
\int \frac{d \ell^+}{R_g/2}
e^{-u \frac{\ell^+}{R_g/2}}
S_{c_m}^\kappa \Bigl[ \frac{\ell^+}{R_g/2}, \mu\Bigr]
\, &
&(\text{$\ee$ case})\,,&
\\
&\tilde S_{c_m}^\kappa \bigl( u, \mu\bigr)
=
\int \frac{d r^+}{R_g}
e^{-u \frac{r^+}{R_g}}
S_{c_m}^\kappa \Bigl[ \frac{r^+}{R_g}, \mu\Bigr]
\, &
&(\text{$pp$ case})\,,&
\nn
\end{align}
such that at one loop (for either scenario)
\begin{align}\label{eq:ScmLap}
\tilde S_{c_m}^\kappa\bigl(u,\mu \bigr)
&=
1 + \frac{\alpha_s C_\kappa}{\pi}
\bigg[
-
\log^2 \big( u e^{\gamma_E}\mu \big)
-\frac{\pi^2}{8}
\bigg]
\, .
\end{align}
In the notation of \eq{altNotation}, useful in our final expressions for the resummed cross sections, we get
\begin{align} \label{eq:SCmAlterNotation}
\tilde S_{c_m}^\kappa\bigl[L_{cs_m},\alpha_s(\mu_{cs_m} )\bigr]
& =
1 + \frac{\alpha_s(\mu_{cs_m}) C_\kappa}{\pi}
\bigg[
-
L_{cs_m}^2
-\frac{\pi^2}{8}
\bigg]\,.
\end{align}

\tocless\subsubsection{Collinear-soft functions at large $R_g$}

\noindent
In the large $R_g$ case, the collinear-soft function knows about both the mass measurement and the soft-drop condition. At one loop, we need
\begin{align}
S_c^{\kappa[1],{\rm bare}}
\Big[
\ell^+ \qcut^{\frac{1}{1+\beta}}, \frac{R_g}{2}\qcut^{\frac{1}{1+\beta}}, \beta, \mu
\Big] =
\frac{\alpha_s C_\kappa}{\pi}
\frac{(\mu^2 e^{\gamma_E})^{\eps}}{\Gamma(1-\eps)}
\!\int\!\! \frac{dp^+ dp^-}{(p^+p^-)^{1+\eps}} \,\overline \, \Theta\bigg(\frac{R^2_g}{4}-\frac{p^+}{p^-}\bigg) \delta\Big(\frac{\ell^+p^-}{p^+} - 1\Big)
\overline \Theta_{\rm sd} \,,
\end{align}
yielding the result presented in \eq{ScLargeRg}. Given our multiplicative treatment of this function, see \eq{ScConvLargeRg}, we need to separate the terms that contribute to RG evolution from the ones that depend on $R_g$. The $R_g$ independent function coincides with the collinear-soft function for single differential jet mass distribution. Its one-loop result in the $\overline{\rm MS}$ scheme is
\begin{align}
\label{eq:sc1loop}
S_c^\kappa\Bigl[\ell^+\, Q_{\rm cut}^{\frac{1}{1+\beta}}, \beta,\mu \Bigr]
&= \delta\Bigl( \ell^+ Q_{\rm cut}^{\frac{1}{1+\beta}}\Bigr)
\\
&+ \frac{C_\kappa\alpha_s(\mu)}{\pi} \Biggl\{ \!
\frac{-2(1+\beta)}{(2\!+\!\beta)\mu^{\frac{2+\beta}{1+\beta}}}
\,{\cal L}_1\!\biggl[ \frac{\ell^+ Q_{\rm cut}^{\frac{1}{1+\beta}}}{\mu^{\frac{2+\beta}{1+\beta}}} \biggr]
+ \frac{\pi^2}{24} \frac{2\!+\!\beta}{1\!+\!\beta} \,
\delta\Bigl( \ell^+ Q_{\rm cut}^{\frac{1}{1+\beta}}\Bigr)
\!\Biggr\}
,\nn
\end{align}
where ${\cal L}_1(x)= \big[\Theta(x) \frac{\ln x}{x} \big]_+$ is the standard plus function which integrates to zero on $x\in [0,1]$.
Following the convention for Laplace transform in \eq{ScTransformDef} we have at one loop
\begin{align}\label{eq:LapSc}
\tilde S_c^\kappa\bigl(s, \beta,\mu \bigr)
&=
1 + \frac{\alpha_s C_\kappa}{\pi}
\bigg[
- \Big(
\frac{1+\beta}{2+\beta}
\Big)
\log^2 \big( s e^{\gamma_E}\mu^{\frac{2+\beta}{1+\beta}} \big)
-\frac{\pi^2}{24}
\frac{\beta(3\beta+ 4)}{(1+\beta)(2+\beta)}
\bigg]
\, ,
\end{align}
Likewise, Laplace transforms of other pieces appearing in \eq{ScConvLargeRgLap} for NLL$'$ resummation in the large $R_g$ region are also defined with respect to the momentum variable $\ell^+$ following \eq{ScTransformDef}, such that at one-loop we find
\begin{align}\label{eq:ScCalLaplace}
\Delta \tilde S_c^{[\beta_0]} \bigl(s, \beta,\mu \bigr)
&= 1 + \frac{\alpha_s(\mu)\beta_0}{2\pi} \Big(\frac{1+\beta}{2+\beta}\Big)
\log \big( s e^{\gamma_E}\mu^{\frac{2+\beta}{1+\beta}} \big)
\, ,
\\
\Delta \tilde S_c^{[C_\kappa]} \Bigl( z = s\, \qcut^{\frac{2+\beta}{1+\beta}} \Big(\frac{R_g}{2}\Big)^{2+\beta} , \beta, \mu\Bigr)
&=
-\frac{\alpha_s C_\kappa}{\pi} \frac{2}{2+\beta}
\frac{\partial}{\partial b} z^{-b}\, \Gamma[b, z]\Bigg|_{b = 0}
\, ,
\nn
\end{align}
where the incomplete Gamma function is defined by
\begin{align}\label{eq:IncompleteGammaDef}
\Gamma[b,z] \equiv \int_z^\infty dt\: t^{b-1} e^{-t} = z^b \int_1^\infty dt \: t^{b-1} e^{-zt} \, ,
\qquad
\frac{\partial}{\partial b} z^{-b}\, \Gamma[b, z]\Bigg|_{b = 0}
= \int_1^\infty dt \: \frac{\log(t)}{t} e^{-zt} \, .
\end{align}
In our alternative Laplace space notation, we have for \eqs{LapSc}{ScCalLaplace} respectively
\begin{align} \label{eq:ScLap2}
\tilde S_c^\kappa \big[L_{cs} , \beta , \alpha(\mu_{cs})\big] & =
1 + \frac{\alpha_s (\mu_{cs})C_\kappa}{\pi}
\bigg[
- \Big(
\frac{1+\beta}{2+\beta}
\Big)
L_{cs}^2
-\frac{\pi^2}{24}
\frac{\beta(3\beta+ 4)}{(1+\beta)(2+\beta)}
\bigg]
\, , \\
\label{eq:ScBeta0Lap}
\Delta \tilde S_c^{[\beta_0]} \big[L_{cs}, \beta, \alpha_s(\mu_{cs})\big] & = 1 + \frac{\alpha_s(\mu)\beta_0}{2\pi} \Big(\frac{1+\beta}{2+\beta}\Big) L_{cs} \, .
\end{align}

\subsection{Derivation of the large $R_g$ resummed cross section} \label{app:deriveQ}

\noindent
In order to derive \eq{xsecRGLargeRg} and obtain an explicit expression for the kernel ${\cal Q^\kappa}$ defined in \eq{UJScOmega} we first write down the cross section in Laplace space in \eq{FactLargeRgLaplace} with resummation made explicit using \eqs{UCalF}{UF}:
\begin{align}\label{eq:FactLargeRgLaplace2}
&\int_0^\infty \! dm_J^2\, \,e^{-x\,m_J^2} \frac{d \Sigma(R_g)}{d m_J^2}\bigg|_{R_g \lesssim \theta_g^\star}\!
\!\!=
N^q_{\rm evol}\big (Q, \qcut,\beta, R, \mu ; \mu_N, \mu_{gs}\big)\:
e^{K_{cs} (\mu, \mu_{cs} ) + K_{J} (\mu, \mu_{J} )}
\nn
\\
&\quad \times
\Big(\frac{\mu_J^2}{m_{J}^2}\Big)^{\tilde \omega_J(\mu, \mu_J)}
\bigg( \frac{Q \mu_{cs}}{m_J^2} \Big(\frac{\mu_{cs}}{\qcut}\Big)^{\frac{1}{1+\beta}}\bigg)^{\omega_{cs} (\mu, \mu_{cs})}
\big(m_J^2 xe^{\gamma_E}\big)^{\tilde \omega(\mu_{cs}, \mu_J)}
\tilde{J}_q (x, \mu_{J}) \,
\tilde S_c^q\Bigl( x Q \qcut^{-\frac{1}{1+\beta}}, \beta , \mu_{cs}\Bigr) \,
\nn \\
&\quad \times \bigg(1+
\Delta \tilde S_c^{[\beta_0]} \bigl( x Q \qcut^{-\frac{1}{1+\beta}},\beta,\mu_{cs} \bigr)
\Delta \tilde S_c^{[C_F]} \Bigl( x Q \qcut \Big(\frac{R_g}{2}\Big)^{2+\beta} , \beta, \mu_{cs}\Bigr)\bigg)\,.
\end{align}
Here we have decomposed the collinear-soft function as in \eqs{ScConvLargeRgLap}{ScConvLargeRgLap2}. We recognize in the first two lines of \eq{FactLargeRgLaplace2} the resummed, single differential jet mass cross section, described in distribution space by the differential operator in \eq{mJOperator}. We now write $x m_J^2 = \varrho $ such that the functions $\tilde{J}_q$ and $\tilde S_c^q$ depend on their arguments only through the logarithms
\begin{align} \label{eq:LapLogs}
L_J = \log \Big(\frac{\mu_J^2}{m_J^2}\Big) + \log(\varrho) + \gamma_E \, ,
\qquad
L_{S_c} = \log \bigg( \frac{Q \mu_{cs}}{m_J^2} \Big(\frac{\mu_{cs}}{\qcut}\Big)^{\frac{1}{1+\beta}}\bigg) + \log(\varrho) + \gamma_E
\, ,
\end{align}
and note that the argument of the $R_g$-dependent piece $\Delta \tilde S_c^{[C_F]}$ in \eq{FactLargeRgLaplace2} reduces to the combination

\begin{align}
x Q \qcut \Big(\frac{R_g}{2}\Big)^{2+\beta} = \varrho \Big(\frac{R_g}{\theta_g^\star}\Big)^{2+\beta}
\, .
\end{align}
Abbreviating $\tilde \omega (\mu_{cs}, \mu_J) = \Omega $ and $ (R_g/\theta_g^\star)^{2+\beta} = \upsilon $, the inverse Laplace transform of \eq{FactLargeRgLaplace2} in terms of dummy variable $r$ conjugate to $\varrho$ for $r = 1$, yields
\begin{align}\label{eq:xsecRGLargeRg2}
\frac{d \Sigma(R_g)}{d m_J^2 }\bigg|_{R_g \lesssim \theta_g^\star}
& =
N^q_{\rm evol}\big (Q, \qcut,\beta, R, \mu ; \mu_N, \mu_{gs}\big)
\\
&\times
\frac{d \sigma^q\big [\partial_\Omega \big]}{d m_J^2 } e^{\gamma_E\Omega } \int \frac{d\varrho}{2\pi i} \: e^{\varrho} \varrho^\Omega
\bigg(1+
\Delta \tilde S_c^{[\beta_0]} ( L_{S_c},\beta, \alpha_s( \mu_{cs}) )
\Delta \tilde S_c^{[C_F]} ( \rho \upsilon , \beta, \mu_{cs})\bigg)\,.
\nn
\end{align}
Noting from \eq{ScCalLaplace} that $\Delta \tilde S_c^{[\beta_0]}$ depends on its argument only through the logarithm defined in \eq{LapLogs}, written in the notation of \eq{ScBeta0Lap}, allows us to rewrite it as a function of $\partial_\Omega$ and recognize in the second term of \eq{xsecRGLargeRg2} the definition of ${\cal Q}^q$ from \eq{UJScOmega}. This completes the derivation of \eq{xsecRGLargeRg}. The same steps carry over to the $pp$ case, leading to the expression in \eq{xsecRGLargeRgPP}.

To find an explicit expression for ${\cal Q^\kappa}$, valid for both $\kappa = q,g$, we take the inverse Laplace transform of $\Delta \tilde S_c^{[C_\kappa]}$ from \eq{ScCalLaplace}, which gives the integral
\begin{align}
\int \frac{d\varrho}{2\pi i} \: e^{\varrho}
\varrho^\Omega \big(\varrho \upsilon \big)^{-b} \Gamma \big[b, \varrho \upsilon \big]
\nn
&= \,\int_1^\infty dt \: t^{b-1}\: \int \frac{d\varrho}{2\pi i} \: e^{\varrho(1-\upsilon t)} \varrho^\Omega
\nn
\\
&= 	\,\frac{\Theta(1-\upsilon )}{\Gamma[-\Omega]} \upsilon^{-b} B[1-\upsilon ; -\Omega, b] \, .
\end{align}
Here we used the incomplete Beta function
\begin{align}\label{eq:BetaDef}
B[1 - \upsilon, -\Omega, b] \equiv \int_0^{1-\upsilon} dx \: x^{-\Omega - 1} (1- x)^{b - 1}
\, .
\end{align}
Including the constant factors, we get the result
\begin{align}\label{eq:UJScOmegaFull}
&{\cal Q}^\kappa\Big [ \Omega,
\upsilon = \Big( \frac{R_g}{\theta_g^\star}\Big)^{2+\beta}, \beta \Big]
=
\Delta \tilde S_c^{[\beta_0]} \bigg [\partial_{\Omega} + \log\bigg( \frac{Q \mu_{cs}}{m_J^2} \Big(\frac{\mu_{cs}}{\qcut}\Big)^{\frac{1}{1+\beta}}\bigg) ,\beta, \alpha_s(\mu_{cs})\bigg ]
\\
&\qquad \times
\frac{e^{\gamma_E\Omega}}{\Gamma(- \Omega)}
\frac{2}{2+\beta}
\frac{\alpha_s(\mu_{cs}) C_\kappa}{\pi}
\bigg(
-\Theta(1-\upsilon )\frac{\partial}{\partial b}
\upsilon^{-b}
B\big[1 - \upsilon ; \, -\Omega, b\big] \bigg|_{b = 0} \bigg)
\, ,	\nn
\end{align}
where we can equivalently write
\begin{align}
\partial_b \: \upsilon^{-b} B[1-\upsilon ; -\Omega , b ] \Big|_{b = 0}=
\int_0^{1- \upsilon} \frac{dx}{1-x} \: x^{-\Omega - 1} \log\frac{1-x}{\upsilon} \, .
\end{align}

\subsection{Results for the boundary cross section}
\label{app:LargeRgShift}

We now provide results and derivations for boundary corrections to soft matrix elements. We first consider the case of global-soft function. We then present results for the boundary corrections to the $S_{c_g}$ function, that would be needed to compute the boundary cross section in the small and intermediate $R_g$ regions. Finally, we give an explicit result for the kernel ${\cal Q}^\kappa_\veps$ defined in \eq{UJScOmegaEps}.

\tocless\subsubsection{Global-soft function with shifted soft drop}
\label{app:GlobShift}
Here we derive the one-loop expression of the global-soft function with shifted soft drop condition, following the analogous derivation preformed for the collinear-soft function in \sec{LargeRgShift}. In analogy with \eq{ScBareShifted} we can write
\begin{align}\label{eq:SGBareShifted}
& S_G^{\kappa[1], \rm bare} \bigl( \qcut,R, \beta, \Theta_{\rm sd}(\veps) , \mu\bigr) =
\frac{\alpha_s C_\kappa}{\pi}
\frac{(\mu^2 e^{\gamma_E})^{\eps}}{\Gamma(1-\eps)}
\int \frac{dp^+ dp^-}{(p^+p^-)^{1+\eps}} \Theta_{\rm sd}(\veps) \overline \Theta_R \nn
\\
&\qquad= S_G^{\kappa[1], \rm bare} \bigl( \qcut,R, \beta, \mu\bigr)
+
\frac{Q\veps}{\qcut}
S_{G,\veps}^{\kappa[1], \rm bare} \bigl( \qcut , R, \beta , \mu\bigr)
+{\cal O}(\veps^2)
\, ,
\end{align}
where the first term is the global soft function for the cross section cumulant in $R_g$ that we computed in \eq{SGBare}, the constraint $\overline\Theta_R$ was given in \eq{ThetaR}, while $\Theta_{\rm sd}(\veps)$ refers to the soft drop failing condition with $\veps$ shift (without small angle approximation):
\begin{align}\label{eq:ThetaShiftLarge}
\Theta_{\rm sd}(\veps)
&\equiv
\Theta
\bigg(
\tzcut \Big(\sin \frac{\theta_p}{2}\Big)^\beta
-
\frac{p^++p^-}{Q}
-\veps
\bigg)
\\
&= \Theta_{\rm sd} - \veps \delta\bigg(
\frac{p^++p^-}{Q}
-\tzcut \Big(\sin \frac{\theta_p}{2}\Big)^\beta
\bigg) + {\cal O}(\veps^2)
\nn
\, .
\end{align}
The ${\cal O}(\veps)$ correction in \eq{SGBareShifted} is then given by substituting the soft drop condition with $- \delta_{\rm sd}$ term in \eq{ThetaShiftLarge}, which fixes $p^-$ to the following value:
\begin{align}\label{eq:pMinusEps}
p^- = \qcut^{\frac{2}{2+\beta}} (p^+)^{\frac{\beta}{2+\beta}} \bigg[1 -\Big( \frac{p^+}{\qcut}\Big)^{\frac{2}{2+\beta}}\bigg]
\, ,
\qquad
p^+ < \qcut \, .
\end{align}
This leads to
\begin{align}\label{eq:SGShift0}
S_{G, \veps}^{\kappa[1], \rm bare} &\bigl(\qcut, R, \beta , \mu\bigr) =
-	\frac{2}{2+\beta}
\frac{\alpha_s C_\kappa}{\pi}
\Big(\frac{\mu}{\qcut}\Big)^{2\eps}
\frac{(e^{\gamma_E})^\eps}{\Gamma(1-\eps)}
\\
&\times
\int\frac{dp^+}{\qcut}
\Big(\frac{p^+}{\qcut}\Big)^{-\frac{2(1+\beta)}{2+\beta}(1+\eps)}
\bigg[1 -\Big( \frac{p^+}{\qcut}\Big)^{\frac{2}{2+\beta}}\bigg]^{-(1+\eps)}
\Theta \bigg ( \sin \frac{R}{2} - \Big(\frac{p^+}{\qcut} \Big)^{\frac{1}{2+\beta}} \bigg)
\nn
\, .
\end{align}
We first consider the case of $\beta = 0$ for which the result is given by
\begin{align}\label{eq:SGShiftb0bare}
S_{G, \veps}^{\kappa[1], \rm bare} &\bigl(\qcut, R, \beta = 0, \mu\bigr) + \text{virtual term}=
\frac{\alpha_s C_\kappa}{\pi}
\bigg[
\frac{1}{\eps_{\rm UV}} + 2 \log \Big(\frac{\mu}{\qcut \tan\frac{R}{2}}\Big)
\bigg]\,.
\end{align}
Note that we only considered the real emission term in the measurement function in \eq{SGBareShifted}. We then demand that the virtual term, which we have not explicitly included, appropriately turns the IR pole in \eq{SGShiftb0bare} into a UV pole.

Next we consider the case of $\beta > 0$. Here, we note that the integral is power law IR divergent and will be set to zero in dimensional regularization. This results from the fact that we are trying to factorize the soft function at subleading power but without considering the full set of operators that appear at this order.
Hence, in order to resolve this ambiguity it is helpful to consider the combined measurement resulting from the shifted soft drop condition in \eq{ThetaShiftLarge}, the $R_g$ constraint in \eq{ThetaRg} (without small-angle approximation), the jet radius constraint in \eq{ThetaR}, along with $p^+$ measurement applied to the matrix element defined by a single soft mode (i.e. not factorizing it into a global-soft and a collinear-soft mode). In this case, the measurement function is given by
\begin{align}\label{eq:deltaPlainRg}
\delta_{\rm full} \big(p^+, \ell^+, R_g\big)
&\equiv \overline \Theta_R \bigg[\overline \Theta_{\rm sd}(\veps)
\overline \Theta_{R_g}
\delta(\ell^+ - p^+) + \Theta_{\rm sd}(\veps)\delta(\ell^+)
\bigg] + \Theta_R \delta(\ell^+) - \delta(\ell^+)
\nn\\
&= \overline \Theta_{R} \overline\Theta_{\rm sd}(\veps)
\big[\overline \Theta_{R_g}\delta(\ell^+ - p^+) - \delta(\ell^+)\big]
\, ,
\end{align}
where the three terms in the first line correspond to emissions that pass the jet radius constraint, that fail and the virtual term. The ones that are clustered in the jet radius are then tested for (shifted) soft drop condition and the $R_g$ constraint as in \eq{ScBareShifted0}.
Writing $\overline \Theta_{\rm sd}(\veps) = \overline \Theta_{\rm sd} +\veps \delta_{\rm sd}$ and using the $\delta$ function to fix $p^-$ to \eq{pMinusEps}, we find
\begin{align}
&\int_0^\infty d p^- \:\overline \Theta_{R}
\big[\overline \Theta_{R_g}\delta(\ell^+ - p^+) - \delta(\ell^+)\big]
\delta \bigg(
p^- - \qcut^{\frac{2}{2+\beta}} (p^+)^{\frac{\beta}{2+\beta}} \bigg[1 -\Big( \frac{p^+}{\qcut}\Big)^{\frac{2}{2+\beta}}\bigg]
\bigg)
\\
&=
\Theta \bigg ( \sin \frac{R_g}{2} - \Big(\frac{p^+}{\qcut} \Big)^{\frac{1}{2+\beta}} \bigg)
\big [\delta(\ell^+ - p^+) - \delta(\ell^+)\big]
+\Theta \bigg ( \sin \frac{R_g}{2} < \Big(\frac{p^+}{\qcut} \Big)^{\frac{1}{2+\beta}} < \sin \frac{R}{2} \bigg)
\delta (\ell^+)
\, .
\nn
\end{align}
The first term in the second line when expanded in small-angle approximation matches with the corresponding term in the measurement function that we used above for the $\ell^+$-differential collinear-soft function $S_c$ in \eq{ScBareShifted0}. The second term proportional to $\delta(\ell^+)$ is of interest to us and when integrated over $p^+$ leads precisely to the bare ${\cal O}(\veps)$ correction in \eq{SGShift0} but now with $p^+$ integral bounded below due to $R_g$ measurement, which then renders it IR finite. This $R_g$-dependent term in the factorized cross section is captured in the ${\cal O}(\veps)$ modification to the collinear-soft functions: the $\ell^+$-differential $S_c$, as seen in \eq{ScShiftedBare0}, or equivalently in the cumulative-$R_g$ $S_{c_g}^\kappa$ function if we are considering intermediate $R_g$ regime as we show below.
Thus, we can consider the integration in \eq{SGShift0} including the lower bound on $p^+$ due to $R_g$ measurement, and identify the pieces depending on $R$ and $R_g$ as parts of the global-soft and the collinear-soft functions respectively.
We then arrive at the following result for $\beta>0$:
\begin{align}\label{eq:SGshiftbgt0}
S_{G, \veps}^{\kappa[1], \rm bare} &\bigl(\qcut, R, \beta , \mu\bigr)= \\
\Big(\frac{\beta}{2} \not \in \mathbb{N} \Big)\qquad
&=
-\frac{\alpha_s C_\kappa}{\pi}
B\Big[\sin^2\Big(\frac{R}{2}\Big), - \frac{\beta}{2}, 0\Big]
\nn
\\
\Big(\frac{\beta}{2} \in \mathbb{N} \Big)\qquad
&=
-\frac{\alpha_s C_\kappa}{\pi}
\Bigg[\log \Big[\tan^2 \Big(\frac{R}{2}\Big)\Big]+\sum_{k=1}^{\beta/2}\frac{1}{k}\Big[\sin^2\Big(\frac{R}{2}\Big)\Big]^{-k}
\Bigg]
\nn
\, .
\end{align}
The result for $S_{c_g,\veps}^{\kappa[1]\rm bare}$ is simply negative of the result for $S_{G,\veps}^{\kappa[1], \rm bare}$ with $R$ replaced by $R_g$ and dropping the subleading terms in $R_g \ll R$. This leads to the result presented below in \eq{ScEpsIntBare}.

The terms in \eq{SGshiftbgt0} are included in the $\Delta S_{G,\veps}^{\kappa}$ piece in \eq{SGShift} for NLL$'$ resummation as follows:
\begin{align}\label{eq:DeltaSGShift}
&\Delta S_{G,\veps}^{\kappa} \bigl(\qcut, R, \beta \bigr)=
\Theta(\beta > 0)
\Bigg[
1 + \frac{\alpha_s (\mu_{gs})\beta_0}{2\pi} \log \Big(\frac{\mu}{\qcut \tan^{1+\beta}\frac{R}{2}}\Big)
\Bigg]\\
&\qquad \times
\frac{\alpha_s(\mu_{gs}) C_\kappa}{\pi}
\Bigg[
\frac{2}{\beta}- \sin^{\beta}\frac{R}{2} \bigg(
\Theta (0 < \beta < 2)\: B \Big[\sin^{2}\frac{R}{2}, 1- \frac{\beta}{2}, 0 \Big]
+ \delta_{\beta, 2} \log \tan^2 \frac{R}{2}
\bigg)
\Bigg]
\, ,\nn
\end{align}
where as in the case of $R_g$-dependent collinear-soft function in \eq{ScLargeRgFO} we have included a $\beta_0$ term to cancel the $\mu$ dependence due to the running coupling.

\tocless\subsubsection{Boundary corrections to the collinear-soft kernel}

\noindent
Here we explicitly solve the Laplace transforms in \eq{UJScOmegaEps}, which define the evolution kernel ${\cal Q}^\kappa_{\veps}$ governing boundary corrections to the cross section in the large $R_g$ region.
The $\Delta S_{c,\veps}^{[C_\kappa]}$ piece defined in \eq{DeltaScLargeShift} transforms into
\begin{align}\label{eq:DeltaScLargeLaplace}
&\Delta \tilde S_{c,\veps}^{[C_\kappa]} \Bigl( z = s\, \qcut^{\frac{2+\beta}{1+\beta}} \Big(\frac{R_g}{2}\Big)^{2+\beta} , \beta, \mu\Bigr)
\\
&\qquad =
\frac{2}{2+\beta}\frac{\alpha_s (\mu)C_\kappa}{\pi}
z^{\frac{\beta}{2+\beta}}
\bigg[\Theta(\beta > 0)\:
\Gamma\Big(\!- \frac{\beta}{2+\beta}\Big) -
\Gamma\bigg(\!-\frac{\beta}{2+\beta} ,z \bigg)
\bigg]
\, ,
\nn
\end{align}
where the incomplete gamma function was defined in \eq{IncompleteGammaDef}.
Using this result we find
\begin{align}\label{eq:UJScOmegaEpsFull}
&\int \frac{d\varrho }{2\pi i} \: e^{\varrho }
\varrho^{\Omega}
\Delta \tilde S_{c,\veps}^{[C_\kappa]}
(\varrho\vf, \beta, \mu_{cs})
=
\frac{2}{2+\beta} \frac{\alpha_s(\mu_{cs})C_\kappa}{\pi}
\frac{1}{\Gamma(- \Omega)}
\nn
\\
&\qquad \times
\upsilon^{\frac{\beta}{2+\beta}}
\Bigg(\Theta(\beta > 0)\:B\Big[ -\Omega , -\frac{\beta}{2+\beta}\Big] -
\Theta(1-\upsilon )
B\Big[1 - \upsilon; \, -\Omega , -\frac{\beta}{2+\beta}\Big]
\Bigg)
\, .
\end{align}
We now state the result for the second term in \eq{UJScOmegaEps}, which involves again \eq{ScCalLaplace}
\begin{align}\label{eq:DeltaUJScOmegaEps}
&\int \frac{d\varrho }{2\pi i} \: e^{\varrho }
\varrho^{\Omega}
\Delta \tilde S_{c,\veps}^{[C_\kappa]}
(\varrho\vf, \beta, \mu_{cs} )
\Delta \tilde S_{c}^{[C_\kappa]}
(\varrho \vf, \beta, \mu_{cs})
=
-\Bigg(\frac{2}{2+\beta}\frac{\alpha_s (\mu)C_\kappa}{\pi} \Bigg)^2
\frac{1}{\Gamma(- \Omega)}
\\
&\qquad\times
\upsilon^{-b}
\Bigg(
\Theta(\beta > 0)
B[-\Omega, b]
\Theta(1-\upsilon)
\int_0^{1- \upsilon} \frac{dx}{1-x} \: x^{b - \Omega - 1} \log\frac{1-x}{\upsilon}
\nn \\
&\qquad\qquad- \Theta\Big(\frac{1}{2} - \upsilon \Big)
\int_\upsilon^{1-\upsilon} \frac{dx}{1-x} \: x^{b - \Omega - 1} \log\frac{1-x}{\upsilon}
\: B\Big[1-\frac{\upsilon}{x} , -\Omega, b\Big]
\Bigg)
\Bigg|_{b = \frac{-\beta}{2+\beta}}
\nn
\, .
\end{align}

\tocless\subsubsection{Boundary Corrections for intermediate and small $R_g$}
\label{app:IntShift}
\noindent
Here we discuss the soft drop boundary cross section in the intermediate and small $R_g$ regimes.
In both of the regimes where $R_g \ll \theta_g^\star$ it is the $S_{c_g}^\kappa$ function that receives ${\cal O}(\veps)$ corrections upon shift to the soft drop condition. This allows to treat the two cases where on equal footing and results in considerable simplifications, since none of the boundary corrections now depend on the jet mass.
In this work we only provide the details of the boundary corrections in the soft sector in these regimes and leave further studies of the corresponding resummation to future work.

As commented below \eq{SGshiftbgt0}, the results for boundary corrections to $S_{c_g}^\kappa$ are closely related to the corresponding corrections to the global-soft function. Following the reasoning described there, we find
\begin{align}\label{eq:ScEpsIntBare}
S_{c_g}^{\kappa [1], \rm bare} & \Bigl[ \frac{R_g}{2}\qcut^{\frac{1}{1+\beta}} ,\overline \Theta_{\rm sd}(\veps) , \beta, \mu \Bigr]
- S_{c_g}^{\kappa [1], \rm bare} \Bigl [\frac{R_g}{2}\qcut^{\frac{1}{1+\beta}} , \beta, \mu\Bigr ]
\\
&(\beta = 0) \qquad=
-\frac{Q\veps}{\qcut } \frac{ \alpha_s(\mu) C_\kappa}{\pi}
\bigg[
\frac{1}{\eps} + 2 \log \Big(\frac{\mu}{\qcut (\frac{R_g}{2})} \Big)
\bigg]\, ,
\nn
\\
&(\beta > 0) \qquad =
-\frac{Q\veps}{\qcut}\Big(\frac{R_g}{2}\Big)^{-\beta} \frac{2}{\beta}\frac{\alpha_s(\mu) C_\kappa}{\pi}
\nn
\, .
\end{align}
Again, as required by consistency with RGE and the large $R_g$ case, for $\beta = 0$ we find a UV pole, which cancels against the global-soft function in \eq{SGshift}. Hence, we have an additional running between the scales $\mu_{gs}$ and $\mu_{cs_g}$ resulting from the non-cusp anomalous dimension
\begin{align}\label{eq:noncuspEpsScg}
\gamma_0^{S_{c_g}^\kappa} (\veps, \zcut )= -\gamma_0 (\veps, \zcut ) = -8C_\kappa \frac{Q\veps}{\qcut}
\, .
\qquad
(\beta = 0)
\end{align}
In analogy with the treatment of the boundary corrections to the global-soft function in \eq{SGShift}, we write the ${\cal O}(\veps)$ corrections to the CS$_g$ function as
\begin{align}
\label{eq:ScGShift}
S_{c_g}^\kappa \Bigl[\frac{R_g}{2}\qcut^{\frac{1}{1+\beta}} , \overline \Theta_{\rm sd}(\veps), \beta, \mu \Bigr]
&= S_{c_g}^{\kappa} \Bigl [\frac{R_g}{2}\qcut^{\frac{1}{1+\beta}} , \beta, \mu \Bigr] -2\delta_{\beta,0}\frac{Q\veps}{\qcut } \frac{ \alpha_s(\mu) C_\kappa}{\pi}
L_{cs_g}
\\
&+\frac{Q \veps}{\qcut}
\frac{\Theta (\beta > 0)}{ \big(\frac{R_g}{2}\big)^\beta}
S_{c_g}^{\kappa} \Bigl [\frac{R_g}{2}\qcut^{\frac{1}{1+\beta}} , \beta, \mu \Bigr]
\Delta S_{c_g,\veps}^{\kappa}\Bigl [\frac{R_g}{2}\qcut^{\frac{1}{1+\beta}} , \beta \Bigr]
\nn
\, ,
\end{align}
where
\begin{align}
\label{eq:ScGDelta}
\Delta S_{c_g,\veps}^{\kappa}\Bigl [\frac{R_g}{2}\qcut^{\frac{1}{1+\beta}} , \beta \Bigr]
&= \Theta(\beta > 0)\frac{-2}{\beta}
\frac{\alpha_s(\mu)C_\kappa}{\pi}
\bigg (
1 + \frac{\alpha_s(\mu)\beta_0}{2\pi} L_{cs_g}
\bigg)
\end{align}
and the logarithm being
\begin{align}
L_{cs_g} \equiv \log \Bigg(\frac{\mu}{\qcut \big(\frac{R_g}{2}\big)^{1+\beta}}\Bigg)
\, .
\end{align}
Like in \eq{ScLargeRgFO} for the large-$R_g$ collinear-soft function, \eq{ScGShift} includes an additional $\beta_0$ term to compensate for the $\mu$ dependence of the running coupling, such that the non-logarithmic correction $\Delta S_{c_g,\veps}^{\kappa}\Bigl [\frac{R_g}{2}\qcut^{\frac{1}{1+\beta}} , \beta \Bigr]$ is explicitly $\mu$ independent at NLL$'$. This result consistently matches with the boundary correction in the large $R_g$ region stated above in \eq{UJScOmegaEpsFull}. To see this we can take $\upsilon \ra 0$ limit in \eq{UJScOmegaEpsFull} finding
\begin{align}
\lim\limits_{\upsilon \ra 0} \int \frac{d\varrho }{2\pi i} \: e^{\varrho }
\varrho^{\Omega}
\Delta \tilde S_{c,\veps}^{[C_\kappa]}
(\varrho\vf, \beta, \mu_{cs})
&= \frac{-2}{\beta} \frac{\alpha_s(\mu_{cs})C_\kappa}{\pi}
\frac{1}{\Gamma(- \Omega)} \, ,
\end{align}
which results in the right coefficient so as to match onto the boundary correction in \eq{ScGDelta}.

\vspace{0.1cm}
\subsection{Calculation of $C_2^q$ from the differential boundary cross section} \label{app:C2diff}

Here we set up the calculation of $C_2^q$ based on the cross section differential in groomed jet radius, i.e. by taking the $R_g$ derivative of the large $R_g$ cumulative cross section with boundary corrections described in \sec{LargeRgShift}. Although the extensions to $pp$ collisions is straightforward, here we limit ourselves to the $\ee$ case. We first consider the case of $\beta >0$ and then include the additional terms that appear for $\beta = 0$ at NLL. Based on the relation with the moment $\Mbq$ defined in \eq{C1C2multiDiffDef} adapted for the $\ee$ case considered here and following our prescription for computing the large $R_g$ differential boundary cross section in \eq{C2Tx}, $C_2^q(m_J^2)$ is given by
\begin{align} \label{eq:C2howTo}
C^q_2(m_J^2 ) \frac{d\sigma}{dm_J^2}
&= \frac{m_J^2}{Q^2}
\int_{\theta_{\rm min}}^{\theta_{\rm max}}
\!\!\!\!d \theta_g\, \frac{2}{\theta_g}
w_{\rm large}(m_J^2, \theta_g)
\frac{d}{d \veps}
\: \frac{d \sigma(\overline \Theta_{\rm sd}(\veps))}{d m_J^2 d\theta_g}\bigg|_{{\rm large }\, R_g,\veps\to 0}
\, ,
\end{align}
where the double differential boundary cross section is computed in the large $R_g$ theory and is matched to zero at small angle using the weight function $w_{\rm large}$ as described in \sec{match}.
We switch to the auxiliary variable $\vf = (\theta_g/\theta_g^\star)^{2+\beta}$ introduced in \eq{UJScOmegaEps},
\begin{align} \label{eq:C2howToUps}
C^q_2(m_J^2 ) \frac{d\sigma}{dm_J^2}
&= \frac{m_J^2}{Q^2}
\int_{\vf_{\rm min}}^{\vf_{\rm max}}
\!\!\!\!d \vf\, \frac{2}{\theta_g(\vf)}
w_{\rm large}(m_J^2, \theta_g(\vf))
\frac{d}{d \veps}
\: \frac{d \sigma(\overline \Theta_{\rm sd}(\veps))}{d m_J^2 d\vf}\bigg|_{{\rm large }\, R_g,\veps\to 0}
\, ,
\end{align}
and following the decomposition of the cumulant in \eq{factLargeRgShift} we write
\begin{align}
\label{eq:DDiffLargeRgShift}
\frac{d \sigma(\overline \Theta_{\rm sd}(\veps))}{d m_J^2 d\vf} \equiv
\frac{d}{d \vf} \frac{d \Sigma(R_g(\vf), \overline \Theta_{\rm sd}(\veps))}{dm_J^2}
&= \frac{d}{d\vf}\frac{d \Sigma\big (R_g(v),\delta_{\beta,0} \gamma_0(\veps, \zcut)\big)}{d m_J^2}
+ \frac{Q\veps}{\qcut} \frac{d\Delta \sigma_\veps}{d m_J^2 d\vf}
\, ,
\end{align}
where each of the two terms in the r.h.s. is defined as the $\vf$-derivative of the equivalent contribution in \eq{factLargeRgShift}. The first term is only present for $\beta=0$, while the second contributes for each $\beta$.

We now compute the two terms in turn, starting from the second. This requires differentiating \eq{xsecRGLargeShiftDelta}, where we note that the term involving $S^q_G$ is independent of $\theta_g$ (or $\vf$),
\begin{align}\label{eq:CumulantBoundary}
\frac{d\Delta \sigma_\veps}{d m_J^2 d\vf}
&=
\Big(\frac{2}{\theta_g^\star}\Big)^{\beta} 	N^q_{\rm evol}
\frac{d \sigma^q \big[\partial_\Omega \big]}{d m_J^2 }
\frac{d}{d \vf}\Big(v^{-\frac{\beta}{2+\beta}}{\cal Q}_\veps^q[\Omega, \vf, \beta]\Big)
\\
&=
v^{\frac{-\beta}{2+\beta}}\Big(\frac{2}{\theta_g^\star}\Big)^{\beta} 	N^q_{\rm evol}
\frac{d \sigma^q \big[\partial_\Omega \big]}{d m_J^2 }
\tilde S_c^{[\beta_0]} \big[\partial_{\Omega}\big]
\bigg[ \frac{e^{\gamma_E\Omega}}{\Gamma(- \Omega)}
\sum_{i = 1, 2}
\Big(\frac{2}{2+\beta} \frac{\alpha_s(\mu_{cs})C_F}{\pi}\Big)^{i}
{\cal R}^{[i]}_{\veps}\big ( \Omega, \upsilon , \beta \big)
\bigg]\,.
\nn
\end{align}
In the second step we extracted the operator $\tilde S_c^{[\beta_0]} \big[\partial_{\Omega}\big]$ from the definition on ${\cal Q}_\veps^q$ in \eq{UJScOmegaEps} (omitting for brevity the remaining arguments) and we organized the remaining $\vf$ dependent terms as a perturbative expansion in $\alpha_s$.
For simplicity we suppress the arguments of the normalization factor $N^q_{\rm evol}$ defined in \eq{HSGevol}.
By taking derivatives of the inverse Laplace transforms of respectively \eqs{UJScOmegaEpsFull}{DeltaUJScOmegaEps}, we determine the coefficients ${\cal R}^{[1]}_\veps$ and ${\cal R}^{[2]}_\veps$:
\begin{align}\label{eq:Qprime1}
{\cal R}^{[1]}_{\veps}\big ( \Omega, \upsilon , \beta \big)
&=
{\cal L}_0^{-\Omega} (1 -\vf)\,,
\\
{\cal R}^{[2]}_{\veps}\big ( \Omega, \vf , \beta \big)
&=
B[-\Omega, b] \Theta(1-\vf) \vf^{-b} B[1-\vf, b - \Omega, 0]
- \Theta(\beta>0)\Theta \Big(\frac{1}{2} - \vf\Big)
\nn\\
\label{eq:Qprime2}
&\hspace{-30pt}\times \int_\vf^{1-\vf}\!\!\!
\frac{dx}{1-x} \: \frac{1}{x^{1+\Omega}}
\Bigg(
\log\frac{1-x}{\vf}
\Big(1 - \frac{\vf}{x}\Big)^{-(1+\Omega)}+
\Big(\frac{\vf}{x}\Big)^{-b}
B\Big[1-\frac{\vf}{x} , -\Omega, b\Big]
\Bigg)
\Bigg|_{b = \frac{-\beta}{2+\beta}}\,.
\end{align}

Let us now move on to the first term in \eq{DDiffLargeRgShift}, which is the $\vf$-derivative of \eq{xsecRGLargeShiftb0},
\begin{align} \label{eq:secondPieceDiffC2}
& \frac{d}{d\vf}\frac{d \Sigma\big (R_g(v),\delta_{\beta,0} \gamma_0(\veps, \zcut)\big)}{d m_J^2} =
N^q_{\rm evol}
\frac{d \sigma^q\big [ \partial_\Omega , \delta_{\beta,0} \gamma_0(\veps, \zcut) \big]}{d m_J^2 }\:
\frac{d}{d\vf}{\cal Q}^q\big [ \Omega, \vf, \beta \big]
\\ &\qquad \nn =
N^q_{\rm evol}
\frac{d \sigma^q\big [ \partial_\Omega , \delta_{\beta,0} \gamma_0(\veps, \zcut) \big]}{d m_J^2 }\:
\tilde S_c^{[\beta_0]} \big[\partial_{\Omega}\big]
\frac{e^{\gamma_E\Omega}}{\Gamma(- \Omega)}
\bigg(\frac{2}{2+\beta}
\frac{\alpha_s(\mu_{cs}) C_F}{\pi}\bigg)
\frac{1}{\vf}B\big[1-\vf , -\Omega, 0\big]
\, .
\end{align}
Here we used that $d \sigma\big [ \partial_\Omega , \delta_{\beta,0} \gamma_0(\veps, \zcut) \big]$ does not depend on the groomed jet radius, extracted $S_c^{[\beta_0]} \big[\partial_{\Omega}\big]$ from the definition of ${\cal Q}^q$ in \eq{UJScOmegaFull}, and took the $\vf$ derivative of the inverse Laplace transform in \eq{ScCalLaplace}.
Expanding the operator $d \sigma\big [\partial_\Omega,\delta_{\beta,0} \gamma_0(\veps, \zcut)\big]$ to ${\cal O}(\veps)$, we get
\begin{align}\label{eq:ExpnLargeRgOperator}
&\frac{d \sigma^q\big [\partial_\Omega,\delta_{\beta,0} \gamma_0(\veps, \zcut)\big]}{d m_J^2 }
=
\frac{d \sigma^q\big [\partial_\Omega \big]}{d m_J^2 }
\bigg[1+
\delta_{\beta,0}\frac{Q\veps}{\qcut}
\eta\big(	\gamma_0(\veps ,\zcut ) , \mu_{cs} , \mu_{gs}	\big)
\bigg] \\
&\qquad
+ \delta_{\beta,0}\frac{Q\veps}{\qcut}
\frac{d \sigma^q\big [\partial_\Omega \ra 0\big]}{d m_J^2 } \bigg [
\frac{2\alpha_s(\mu_{gs}) C_F}{\pi}
\log \Big(\frac{\mu_{gs}}{\qcut \tan\frac{R}{2}}\Big)
-	\frac{\alpha_s(\mu_{cs}) C_F}{\pi} \partial_\Omega
\bigg]
\,,
\nn
\end{align}
where the kernel $\eta\big(\gamma_0(\veps ,\zcut )\big)$ in \eq{w} arises from the ${\cal O}(\veps)$ expansion of the Sudakov exponent, and the additional terms in the second line result from the ${\cal O}(\veps)$ logarithms in the global-soft and collinear soft function in \eqs{SGShift}{ScBeta0} respectively. Note that after taking inverse Laplace transform the log in the collinear-soft function is replaced by the derivative $\partial_\Omega$. Additionally, the two terms are evaluated at different $\mu$ arguments. Furthermore, $d \sigma\big [\partial_\Omega \ra 0\big]$ denotes setting to zero ${\cal O}(\alpha_s)$ (or higher) corrections in the term $d \sigma[\partial_\Omega]$ so as to account purely for the evolution factor in \eq{mJOperator}.
We can now finally combine \eqs{CumulantBoundary}{secondPieceDiffC2} into the double differential boundary cross section in \eq{DDiffLargeRgShift}. Substituting in \eq{C2howToUps} yields our final formula
\begin{align} \label{eq:C2final}
\nn C^q_2(m_J^2 )
\frac{d\sigma}{dm_J^2}
&= \frac{m_J^2}{Q \qcut}
N^q_{\rm evol}
\int_{\vf_{\rm min}}^{\vf_{\rm max}}
\frac{d \vf}{\vf}\, \frac{2}{\theta_g(\vf)}
w_{\rm large}(m_J^2, \theta_g(\vf))
\nn\\
\times&\Bigg(\delta_{\beta,0} \frac{d \sigma\big [\partial_\Omega \ra 0\big]}{d m_J^2 }
\bigg [
\frac{2\alpha_s(\mu_{gs}) C_F}{\pi}
\log \Big(\frac{\mu_{gs}}{\qcut \tan\frac{R}{2}}\Big)
- \frac{\alpha_s(\mu_{cs}) C_F}{\pi} \partial_\Omega \bigg]
B\big[1-\vf , -\Omega, 0\big]
\nn\\&
+\frac{2}{2+\beta} \frac{\alpha_s(\mu_{cs})C_F}{\pi}
\frac{d \sigma\big [\partial_\Omega \big]}{d m_J^2 }\tilde S_c^{[\beta_0]}
\big[\partial_{\Omega}\big]
\bigg\{
\delta_{\beta,0}\,\eta\big(\gamma_0(\veps ,\zcut ) , \mu_{cs} , \mu_{gs} \big)
B\big[1-\vf , -\Omega, 0\big]
\nn\\ &
+ v^{\frac{1+\beta}{2+\beta}}\Big(\frac{2}{\theta_g^\star}\Big)^{\beta}
\Big[ {\cal R}_\veps^{[1]}(\Omega,\vf,\beta) +
\Big(\frac{2}{2+\beta} \frac{\alpha_s(\mu_{cs})C_F}{\pi}\Big)
{\cal R}_\veps^{[2]}(\Omega,\vf,\beta) \Big] \bigg\}\Bigg) \frac{e^{\gamma_E\Omega}}{\Gamma(- \Omega)}\,.
\end{align}

As discussed in the introduction, the geometry relevant to the non-perturbative corrections multiplying $C_2^q$ requires computing this Wilson coefficient using only the large $R_g$ theory, which is enforced by the weight function $w_{\rm large}(m_J^2,\theta(\vf))$ in \eq{C2final}. This is consistent with the idea that $C_2^q$ should describe only perturbative physics. We note however that the integrand has an overall behavior $\sim 1/\vf$, which enhances NP effects at low jet mass. To make sure that there are no residual NP effects in $C_2^q$, we adopt a prescription analogous to the one used for freezing the scale $\mu_{c_g}$ in \eq{muCSg},
\begin{align}
\frac{1}{\vf} \simeq
\Big(\frac{\psi^\star}{\psi}\Big)^{2+\beta}
\;\to\;
\frac{(\psi^\star)^{2+\beta}}{\big [f_{\rm run} ( \psi^{1+\beta}) \big]^{\frac{2+\beta}{1+\beta}}}
\,,
\end{align}
where $\psi, \psi^\star$ are the auxiliary variables defined in \eqs{psiDef}{psiStarDef}, and the first equality is exact in the $R_g \ll 1$ limit. The profile function $f_{\rm run}$ defined in \eq{frunRg} smoothly freezes its argument when approaching non-perturbative values.

As a final remark, we notice that in the region $m_J^2 < m_0^2$ where grooming is active, $\vf_{\rm max} = 1$ and \eq{Qprime1} requires solving the class of integrals
\begin{align}
I_n(\vf_{\rm min},\Omega) =
\partial_\Omega^n \big[I_0(\vf_{\rm min},\Omega)\big] \equiv
\int_{\vf_{\rm min}}^1 \df \vf\frac{\big(-\ln(1-\vf)\big)^n}{(1-\vf)^{\Omega + 1}} f_n(\vf)\,,
\end{align}
where the derivatives arise from the operators $d \sigma\big [\partial_\Omega \big]$ and $\tilde S_c^{[\beta_0]} [\partial_\Omega \big]$ in \eq{C2final}.
To avoid numerical instabilities, we deal with these integrals using the subtraction
\begin{align}
I_n(\vf_{\rm min},\Omega) = \int_{\vf_{\rm min}}^1 \df \vf\frac{\big(-\ln(1-\vf)\big)^i}{(1-\vf)^{\Omega + 1}} \big[f_i(\vf) - f_i (1)\big]
+ f_i(1) \bigg ( \int_{\vf_{\rm min}}^1 \df \vf\frac{\big(-\ln(1-\vf)\big)^i}{(1-\vf)^{\Omega + 1}} \bigg)\,.
\end{align}
Similarly, we handle the singular integrand in the first term in \eq{Qprime2} using
\begin{align}
{\cal I}(v,\Omega) &\equiv \int_\vf^{1-\vf}
\frac{dx}{1-x} \: \frac{1}{x^{1+\Omega}}
\log\frac{1-x}{\vf}
\Big(1 - \frac{\vf}{x}\Big)^{-(1+\Omega)}
\nn \\
& = \int_\vf^{1-\vf}
\frac{dx}{1-x} \: \frac{1}{(x-v)^{1+\Omega}}
\log\frac{1-x}{\vf}
\nn \\
& = \int_\vf^{1-\vf}
\frac{f(x) - f(\vf)}{(x-\vf)^{1+\Omega}} - f(\vf)\frac{(1-2\vf)^{-\Omega}}{(-\Omega)}\,,
\end{align}
where
\begin{align}
f(x ) \equiv \frac{1}{1-x} \log\Big(\frac{1-x}{\vf}\Big) \, .
\end{align}

\section{Calculation of NGLs and Clustering effects}

\subsection{Nonglobal logarithms}
\label{app:NGLcluster}

The functions $\Xi_S^q, \Xi_C^q$ introduced in \eqs{NGLInt}{FactSmallThetaCS} can be expanded as a perturbative series in $\alpha_s$. Here we compute the first nontrivial correction, which arise at ${\cal O} (\alpha_s^2)$, starting from the function $\Xi_S^q$ that describes the NGLs in the intermediate $R_g$ region. To avoid plus distributions we switch to the function cumulative in jet mass. Following \cite{Kang:2019prh}, when the clustering effects are ignored we can write
\begin{align}
&
\Xi_S^q(t_S)
= 1- C_FC_A \left(\frac{\alpha_s}{2\pi}\right)^2\int \frac{dx_1}{x_1}\int\frac{dx_2}{x_2}\int_{1\in J}dc_1\frac{d\phi_1}{2\pi}\int_{2 \not\in J}dc_2\frac{d\phi_2}{2\pi}\nn\\
&\qquad\times \Theta(x_1-x_2)\,\Theta(x_2-\tzcut R_g^{\beta})\,
\Theta\left(\frac{\theta_c^2}{4}-\frac{x_1}{2}(1-c_1)\right)\frac{\cos \phi_2}{(1-c_1c_2-s_1s_2\cos \phi_2)s_1s_2}
\end{align}
where $c_i= \cos \theta_i$, $s_i = \sin \theta_i$ and the dependence on the (cumulative) jet mass variable is implicit in $(\theta_c/2)^2=(\theta_g^\star/2)^{2+\beta}\qcut/Q$.
Here the integration variables $x_i, \theta_i$ and $\phi_i$ represent respectively the energy fractions, colatitudes, and azimuths of the strongly ordered emissions. The geometry relevant to the NGLs requires the largest emission to lie inside the groomed jet and the second emission to lie out of it.

The energy integrals can be done easily and yield
\begin{align}
\Xi_S^q(t_S)
&=
1- C_FC_A \left(\frac{\alpha_s}{2\pi}\right)^2\int_{1\in J}dc_1\frac{d\phi_1}{2\pi}\int_{2 \slashed{\in} J}dc_2\frac{d\phi_2}{2\pi}\Theta\left( \frac{\theta_c^2}{2\tzcut R_g^{\beta}}-(1-c_1)\right)\nn\\
&\quad\times \frac{\cos \phi_2}{(1-c_1c_2-s_1s_2\cos \phi_2)s_1s_2}\frac{1}{2}\ln^2\bigg( \frac{\theta_c^2}{2(1-c_1)\tzcut R_g^{\beta}}\bigg)\,.
\end{align}
Since the gluon 1 is restricted to be inside the jet, $\theta_1/2 \leq R_g/2 \ll 1$, we can approximate the factor of $1-c_1$ inside the logarithm as $\theta_1^2/2$. At the same time, taking into account clustering effects as in \cite{Kang:2019prh}, the dominant term in the $R_g/2 \ll1$ limit gives the following result at two loops:
\begin{align} \label{eq:NGLFull2}
\Xi_S^{q}(t_S)	\,= \, 1- \frac{4}{9}C_FC_A \frac{\pi^2}{3} t_S^2 \, .
\end{align}

The leading NGLs in the other two regions can be obtained from the intermediate $R_g$ expression in \eq{NGLFull} by taking appropriate limits. For the large groomed jet radius regime $R_g = \theta_g^\star$ and the NGL contribution goes to 0 as was expected. For the small groomed jet radius regime we can formally set $\theta_c = R_g$ and the dominant contribution in this EFT is
\begin{align}\label{eq:XiCumulative}
\Xi_C^q(t_C) &= 1- \frac{4}{9}C_FC_A \frac{\pi^2}{3} t_C^2 \,,
\end{align}
which is larger than the contribution for the case of intermediate groomed jet radius. Written with $t_C$ see that \eq{XiCumulative} has the same form as \eq{NGLFull2}.

\subsection{Abelian clustering logarithms}
\label{app:Abel}

In this section we account for the leading logarithms arising from clustering effects of the C/A algorithm while implementing soft drop for independent emissions.
To this aim, we first consider a naive implementation of the soft drop algorithm where clustering logarithms are absent and see how clustering effects modify this result. The naive implementation simply tests for the SD condition for each particle independently, i.e. the allowed phase space for each emission is
\begin{align} \label{eq:naiveSD}
\Theta(z_i-\tzcut\theta_i^{\beta})\Theta(R_g-\theta_i)+\Theta(\tzcut\theta_i^{\beta}-z_i) \, .
\end{align}
However, we know that the C/A algorithm recursively clusters partons with minimum angular separation; since the SD algorithm works backwards through the history of the clustered jet, the grooming condition is then effectively applied on subjets instead of individual partons. Therefore, (in the case of 2 collinear-soft emissions that gives the leading configuration) we can have a scenario where one of the partons would fail soft drop and the other would pass if tested independently, but the two get clustered with each other first and with the collinear parton later. The SD condition is now tested on the entire collinear-soft subjet, which will pass the test. In this case the softest parton has been pulled inside the groomed jet, violating the naive measure in \eq{naiveSD}. Instead, in case the most energetic collinear-soft emission is clustered with the collinear parton first, then the naive implementation of the measurement is correct.

Let us denote the portion of phase space where the naive measure is violated by
\begin{align}
\Theta_{C/A}=\Theta(d_1-R_g^2)\Theta(R_g^2-d_2)\Theta(d_2-d_{12})\,,
\end{align}
where $d_{12}$ is the relative distance of the collinear soft emissions and $d_i$ their distance from the jet axis, so that parton 1 is at a distance greater then $R_g$ from the jet axis while parton 2 is within $R_g$ from it. For 1 and 2 to be clustered first, we then need $d_2>d_{12}$. Expressed in the small angle approximation, this becomes
\begin{align}
\Theta_{C/A}=\Theta(\theta^2_1-R_g^2)\Theta(R_g^2-\theta_2^2)\Theta(-\theta_1^2+2\theta_1\theta_2\cos \phi_2)\,,
\end{align}
where we have assumed without loss of generality that $\phi_{12}= \phi_2$.
The allowed phase space requires that parton 1 must fail SD while parton 2 can either pass or fail (with energy fractions $x_2\gg x_1$). However, for parton 1 to be pulled inside the jet, we need that parton 2 passes soft drop. If we ignore for the time being the mass measurement, and consider only a cumulative $R_g$ measurement, the naive result without clustering effects reads
\begin{align}\label{eq:deltaM}
M_{\rm indep.}(R_g) &\equiv \int d\Pi_2\, \Theta(\tzcut\theta_1^{\beta}-x_1) \Theta(x_2-\tzcut\theta_2^{\beta})\Theta_{C/A}\,
,
\end{align}
where we have abbreviated the two-parton phase space in the strongly ordered limit as
\begin{align}\label{eq:Pi2Strong}
\int d\Pi_2 = \left(\alpha_s \frac{C_\kappa}{\pi}\right)^2 \int \frac{dx_1}{x_1}\int \frac{dx_2}{x_2}\int \frac{d\theta_1}{\theta_1}\int \frac{d\theta_2}{\theta_2}\int \frac{d\phi_1}{2\pi} \int\frac{d\phi_2}{2\pi}\,.
\end{align}
Let us now account for clustering effects. The idea is that since parton 1 is pulled inside the jet, it never gets individually tested for SD and can in fact also be allowed to pass SD in this region of phase space. Therefore the measurement in \eq{deltaM} gets modified to give us
\begin{align}
M_{\rm cluster}(R_g) &\equiv \int d\Pi_2\, \Theta(x_2-\tzcut\theta_2^{\beta})\Theta_{C/A} \, .
\end{align}
We now express the result as a correction on top of the naive result by subtracting it out the from the full result to isolate the clustering effects:
\begin{align}
\Delta M_{\rm alg} (R_g)&\equiv M_{\rm cluster}- M_{\rm indep.} = \frac{1}{2!}\int d \Pi_2\,
\Theta(x_1-\tzcut\theta_2^{\beta})\Theta(x_2-\tzcut\theta_2^{\beta})\Theta_{C/A}
\, ,
\end{align}
which agrees with the integrand quoted in \cite{Kang:2019prh}. The factor of $1/2!$ takes into account the scenario where we interchange the labels 1 and 2 in the strongly ordered phase space measure in \eq{Pi2Strong}.

We can now extend this result to our case where we also have a jet mass measurement on the groomed jet. The naive value in \eq{deltaM} with a cumulative jet mass measurement becomes
\begin{align}
M_{\rm indep.}(m_J^2, R_g) &\equiv \int d \Pi_2\, \Theta(\tzcut\theta_1^{\beta}-x_1) \Theta(x_2-\tzcut\theta_2^{\beta})
\Theta\Big( (\theta_c(m_J^2))^2-x_2\theta_2^2\Big)
\Theta_{C/A}\, ,
\end{align}
which upon including clustering corrections gives
\begin{align}
M_{\rm cluster}(m_J^2, R_g) &\equiv \int d\Pi_2\, \Theta(\theta_c^2-x_1\theta_1^2)\Theta(x_2-\tzcut\theta_2^{\beta})\Theta(\theta_c^2-x_2\theta_2^2)\Theta_{C/A}\,,
\end{align}
where we note that the parton 1 will contribute to the jet mass regardless of whether it passes or fails since it is part of our groomed jet. Therefore the correction from the algorithm becomes
\begin{align}\label{eq:DeltaMalg}
\Delta M_{\rm alg}(m_J^2, R_g) \equiv \int d\Pi_2\,
\Big[\Theta(\theta_c^2-x_1\theta_1^2)-\Theta(\tzcut\theta_1^{\beta}-x_1)\Big] \Theta(x_2-\tzcut\theta_2^{\beta})\Theta(\theta_c^2-x_2\theta_2^2)\Theta_{C/A}\,.
\end{align}
To extract out the leading logarithmic behavior, we notice that none of the terms in \eq{DeltaMalg} are sensitive to the difference between $\theta_1$ and $\theta_2$, and hence can approximate $\theta_1 \sim \theta_2 \sim R_g$ in the integrand, such that
\begin{align}
\Theta(\theta_c^2-x_1\theta_1^2)-\Theta(\tzcut\theta_1^{\beta}-x_1) &\approx \Theta(\theta_c^2-x_1R_g^2)-\Theta(\tzcut R_g^{\beta}-x_1)\,,\\
\Theta(x_2-\tzcut\theta_2^{\beta})\,\Theta(\theta_c^2-x_2\theta_2^2) &\approx \Theta(x_2-\tzcut R_g^{\beta})\,\Theta(\theta_c^2-x_2 R_g^2)\,,
\nn
\end{align}
which allow us to factorize the angular and energy integrals,
\begin{align}
\Delta M_{\rm alg}(m_J^2, R_g)&= \frac{1}{2!}\left( \frac{\alpha_s C_\kappa}{\pi}\right)^2 \int \frac{d\theta_1}{\theta_1}\int \frac{d\theta_2}{\theta_2} \int \frac{d\phi_1}{2\pi} \frac{d\phi_2}{2\pi}
\Theta_{C/A} \\
&\times\int \frac{dx_1}{x_1}\Big[\Theta(\theta_c^2-x_1R_g^2)-\Theta(\tzcut R_g^{\beta}-x_1)\Big]\int\frac{dx_2}{x_2}\ \Theta(x_2-\tzcut R_g^{\beta})\Theta(\theta_c^2-x_2R_g^2)
\nn \\
&= \frac{1}{2!}\left( \frac{\alpha_s C_\kappa}{\pi}\right)^2 \ln^2 \bigg(\frac{\theta_c^2}{\tzcut R_g^{\beta+2}}\bigg)
\int \frac{d\theta_1}{\theta_1}\int \frac{d\theta_2}{\theta_2} \int \frac{d\phi_1}{2\pi} \int\frac{d\phi_2}{2\pi}\Theta_{C/A} \, .
\nn
\end{align}
The angular integral is now identical to the one in \cite{Kang:2019prh}, which can be solved to yield
\begin{align}
\Delta M_{\rm alg}(m_J^2, R_g)&= \frac{1}{2!}\left( \frac{\alpha_s C_\kappa}{\pi}\right)^2 \frac{\pi^2}{54}
\ln^2 \left(\frac{\big(\theta_c(m_J^2)\big)^2}{\tzcut R_g^{\beta+2}}\right)\,.
\end{align}
We see that the leading abelian clustering logarithms have the same functional form as the NGLs in \eq{NGLFull}, but are further suppressed by the factor $\frac{1}{36}$.
Furthermore, this result shows that just like the NGLs, the leading clustering logarithms also vanish in the large $R_g$ regime, consistent with the existence of a single collinear-soft mode rather than two modes that know about $R_g$ separated in virtuality. On the other hand, they are maximized in the small $R_g$ limit, $\theta_c = R_g$, where the effect of the mass measurement drops out and we find back the result of \cite{Kang:2019prh}.

We also note that in this calculation we have not considered the scenario when both partons naively fail the SD condition but pass if clustered together. This is evaluated in \cite{Frye:2016aiz} and would lead to a sub-leading single logarithmic contribution $\alpha_s^2 \log X$ compared to the correction explicitly considered here.

\bibliography{../sd}
\end{document}